\documentclass[11pt, twoside]{Thesis} % The default font size and one-sided printing (no margin offsets)

\graphicspath{{Pictures/}} % Specifies the directory where pictures are stored

\usepackage[square, numbers, comma, sort&compress]{natbib} % Use the natbib reference package - read up on this to edit the reference style; if you want text (e.g. Smith et al., 2012) for the in-text references (instead of numbers), remove 'numbers' 
\usepackage{slashed}
\usepackage{float}
\usepackage{caption}
\usepackage{subcaption}
\usepackage{appendix}
\usepackage{hyperref}
\usepackage{multirow, bigdelim}		    % For delim-command
\usepackage{mathtools}                            % For starred environment
\usepackage{bibentry}                               % Bibentries outside bibliography
\nobibliography*                                         % Bibentries are no bibliography
\makeatletter                                              % Bibentries are not linked to
\renewcommand\bibentry[1]{\nocite{#1}{\frenchspacing
     \@nameuse{BR@r@#1\@extra@b@citeb}}}
\makeatother
\usepackage{enumitem}                            % customize list environments
\makeatletter                                              % custom labels
\newcommand{\plabel}[2]{%
   \protected@write \@auxout {}{\string \newlabel {#1}{{#2}{\thepage}{#2}{#1}{}} }%
   \hypertarget{#1}{#2}
}
\makeatother
\hypersetup{urlcolor=blue, colorlinks=true} % Colors hyperlinks in blue - change to black if annoying
\title{\ttitle} % Defines the thesis title - don't touch this

\def\beq{\begin{equation}}
\def\eeq{\end{equation}}
\def\bea#1\eea{\begin{align}#1\end{align}}

\def\td{{\widetilde \delta}}
\def\qt{{\bf q}_{\perp}}

\def\beeq{\begin{eqnarray}}
\def\eeeq{\end{eqnarray}}
\def\nn{\nonumber}
\def\res{{\rm Res}_{ \{{\rm Im}\; \ell_0 < 0 \}}}
\def\b{{\rm Bub}}
\def\ep{\epsilon}
\def\sg{{\rm sign}}
\def\Eq#1{Eq.~(\ref{#1})}

\newcommand{\la}{\langle}
\newcommand{\ra}{\rangle}
\def\ket#1{|{#1}\ra}
\def\bra#1{\la{#1}|}

\def\bom#1{{\mbox{\boldmath $#1$}}}
\def\sp{{\bom {Sp}}}

\newenvironment{ropmatrix}
  {\array{@{}c@{}}}
  {\endarray}

\begin{document}

\frontmatter % Use roman page numbering style (i, ii, iii, iv...) for the pre-content pages

\setstretch{1.3} % Line spacing of 1.3

% Define the page headers using the FancyHdr package and set up for one-sided printing
%\fancyhead{} % Clears all page headers and footers
%\rhead{\thepage} % Sets the right side header to show the page number
%\lhead{} % Clears the left side page header

%\pagestyle{fancy} % Finally, use the "fancy" page style to implement the FancyHdr headers
%\fancyhead[LE,RO]{\thepage}
%\fancyhead[RE,LO]{\leftmark}

\newcommand{\HRule}{\rule{\linewidth}{0.5mm}} % New command to make the lines in the title page

% PDF meta-data
\hypersetup{pdftitle={\ttitle}}
\hypersetup{pdfsubject=\subjectname}
\hypersetup{pdfauthor=\authornames}
\hypersetup{pdfkeywords=\keywordnames}

%----------------------------------------------------------------------------------------
%	TITLE PAGE
%----------------------------------------------------------------------------------------

\begin{titlepage}
\begin{center}

\textsc{\LARGE \univname}\\[1.5cm] % University name
\textsc{\Large Doctoral Thesis}\\[0.5cm] % Thesis type

\HRule \\[0.4cm] % Horizontal line
{\huge \bfseries \ttitle}\\[0.4cm] % Thesis title
\HRule \\[1.5cm] % Horizontal line
 
\begin{minipage}{0.4\textwidth}
\begin{flushleft} \large
\emph{Author:}\\
{\authornames\\
\null} % Author name - remove the \href bracket to remove the link
\end{flushleft}
\end{minipage}
\begin{minipage}{0.4\textwidth}
\begin{flushright} \large
\emph{Supervisor:} \\
\href{http://ific.uv.es/~rodrigo/enter.html}{\supname} % Supervisor name - remove the \href bracket to remove the link  
\end{flushright}
\end{minipage}\\[2cm]

\begin{figure}[H]
\centering
\includegraphics[scale=0.5]{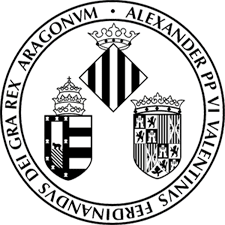}
\end{figure}
\vspace{1em}
\large \textit{3026 Programa Oficial de Doctorado en F\'isica}\\[0.3cm] % University requirement text
\vspace{1em}
\deptname\\\groupname\\[2.5cm] % Research group name and department name
 
%{\large \today}\\[4cm] % Date
{\large June 2015}\\[4cm] % Date

\vfill
\end{center}

\end{titlepage}

\clearpage

%----------------------------------------------------------------------------------------
%	FIRST PAGE
%----------------------------------------------------------------------------------------

\thispagestyle{plain}
\null
\begin{center}
\begin{minipage}{0.8\textwidth}
\begin{flushleft} \large
I certify that I have read this dissertation and that, in\\
my opinion it is fully adequate in scope and quality as a\\
dissertation for the degree Doctor of Philosophy.
\end{flushleft}
\end{minipage}
\end{center}
\vspace{1em}
\begin{center}
\begin{minipage}{0.8\textwidth}
\begin{flushleft} \large
\rule[1em]{25em}{0.5pt}
\begin{center}
\vspace{-1em}
Germ\'an Rodrigo\\ 
(Supervisor)
\end{center}
\end{flushleft}
\end{minipage}
\end{center}

\vfill
%\vspace{10em}

\begin{center}
%\begin{minipage}{1\textwidth}
% \Large{Ph.D program of the Unicersity of Valencia\\}
 %\end{minipage}
 \end{center}
 \begin{center}
\begin{minipage}{0.49\textwidth}
\begin{flushleft} \large
\emph{Author:}\\
\emph{Supervisor:}\\
\emph{Co-Supervisors:}\\
\null
\end{flushleft}
\end{minipage}
\begin{minipage}{0.49\textwidth}
\begin{flushleft} \large
Sebastian Buchta \\
Germ\'an Rodrigo\\
Stefano Catani\\
Grigorios Chachamis
\end{flushleft}
\end{minipage}\\[3cm]
%\vfill
\end{center}

%\vspace{1em}

\clearpage

%----------------------------------------------------------------------------------------
%	DECLARATION PAGE
%	Your institution may give you a different text to place here
%----------------------------------------------------------------------------------------
\addtocontents{toc}{\vspace{2em}}

\Declaration{

\addtocontents{toc}{\vspace{1em}} % Add a gap in the Contents, for aesthetics

I, \authornames, declare that this thesis titled, '\ttitle' and the work presented in it are my own. I confirm that:

\begin{itemize} 
\item[\tiny{$\blacksquare$}] This work is original and was done wholly or mainly while in candidature for a research degree at this University.
\item[\tiny{$\blacksquare$}] Where any part of this thesis has previously been submitted for a degree or any other qualification at this University or any other institution, this has been clearly stated.
\item[\tiny{$\blacksquare$}] Where I have consulted the published work of others, this is always clearly attributed.
\item[\tiny{$\blacksquare$}] Where I have quoted from the work of others, the source is always given. With the exception of such quotations, this thesis is entirely my own work.
\item[\tiny{$\blacksquare$}] I have acknowledged all main sources of help.
\item[\tiny{$\blacksquare$}] Where the thesis is based on work done by myself jointly with others, I have made clear exactly what was done by others and what I have contributed myself.\\
\end{itemize}
 
Signed:\\
\rule[1em]{25em}{0.5pt} % This prints a line for the signature
 
Date:\\
\rule[1em]{25em}{0.5pt} % This prints a line to write the date
}

\clearpage % Start a new page

%----------------------------------------------------------------------------------------
%	PUBLICATIONS PAGE
%----------------------------------------------------------------------------------------

\thispagestyle{simple}
\null
\smallskip

This thesis is based on the author's work conducted at the IFIC (Instituto de 
F\'isica Corpuscular, Universitat de Val\`{e}ncia, Consejo Superior de Investigaciones Cient\'ificas). 
Parts of it have already been published in articles and proceedings earlier.

\bigskip

{\bf Articles}

%\bigskip

\begin{itemize}
\item[\cite{Bierenbaum:2012th}] \bibentry{Bierenbaum:2012th}
\item[\cite{Buchta:2014dfa}] \bibentry{Buchta:2014dfa}
\item[\cite{paper}] \bibentry{paper}
\end{itemize}

\bigskip

{\bf Proceedings}

\begin{itemize}
\item[\cite{Bierenbaum:2013nja}] \bibentry{Bierenbaum:2013nja}
\item[\cite{Buchta:2014fva}] \bibentry{Buchta:2014fva}
\item[\cite{submitted}] \bibentry{submitted}
\end{itemize}

\clearpage

%----------------------------------------------------------------------------------------
%	QUOTATION PAGE
%----------------------------------------------------------------------------------------

%\pagestyle{empty} % No headers or footers for the following pages

%\null\vfill % Add some space to move the quote down the page a bit

%\textit{``Daß ich erkenne, was die Welt\\
%Im Innersten zusammenhält"}\\
%- Faust

%\begin{flushright}t
%Johann Wolfgang von Goethe
%\end{flushright}

%\vfill\vfill\vfill\vfill\vfill\vfill\null % Add some space at the bottom to position the quote just right

%t\clearpage % Start a new page

%----------------------------------------------------------------------------------------
%	ABSTRACT PAGE
%----------------------------------------------------------------------------------------

\thispagestyle{plain}

\addtotoc{Abstract} % Add the "Abstract" page entry to the Contents

\abstract{\addtocontents{toc}{\vspace{1em}} % Add a gap in the Contents, for aesthetics

The Loop-Tree Duality (LTD) is a novel perturbative method in QFT that establishes a relation between loop--level and tree--level 
scattering amplitudes. This is achieved by directly applying the Residue Theorem to the loop-energy-integration. The 
result is a sum over all possible single cuts of the Feynman diagram in consideration integrated over a modified phase-space. 
These single-cut integrals, called Dual contributions, 
are in fact tree-level objects and thus give rise to the opportunity of bringing loop-- and tree--contributions together, 
treating them simultaneously in a common Monte Carlo event generator.
Initially introduced for one--loop scalar integrals, the applicability of the LTD has been expanded ever since. In this thesis, 
we show how to deal with Feynman graphs beyond simple poles
by taking advantage of Integration By Parts (IBP) relations.
Furthermore, we investigate the cancellation of singularities among Dual contributions as well as 
between real and virtual corrections. 
For the first time, a numerical implementation of the LTD was done in the form of a computer program 
that calculates one--loop scattering diagrams. We present details on the contour deformation employed 
alongside the results for scalar integrals up to the pentagon- and tensor integrals up to the hexagon-level.
}

\resumen{

La Dualidad Loop-\'Arbol (LTD) representa un nuevo m\'etodo perturbativo en Teoria Cu\'antica de Campos que establece una relaci\'on entre 
amplitudes de dispersi\'on virtuales y de \'arbol. Se logra hacer esto por aplicaci\'on directa del Teorema de los Residuos a la 
integrati\'on de la componente de energ\'ia. El resultado es la suma de todos los cortes simples posibles 
del diagrama de Feynman considerado integrada sobre un espacio f\'asico modificado. Estas integrales de corte simple, 
denominadas Contribuciones Duales, de hecho son objetos de tipo \'arbol y 
por lo tanto dan lugar a la oportunidad de combinar las contribuciones virtuales y de \'arbol con el motivo de tratarlas simult\'aneamente 
en un generador de eventos de Monte Carlo. 
A pesar de ser introducido inicialmente para 
integrales escalares de un loop, la practicabilidad de la LTD fue extendida tremendamente. En esta tesis demonstramos 
como aplicar la LTD a diagramas con polos de multiplicidad elevada 
utilizando relaciones de Integraci\'on Por Partes (IBP). 
Adem\'as, examinamos la cancelaci\'on de singularidades entre Contribuciones 
Duales tanto como entre correcciones reales y virtuales. 
Por primera vez una implementaci\'on num\'erica de la LTD fue realizada en forma de un programa 
de ordenador que calcula diagramas de dispersi\'on. Presentamos detalles sobre la deformaci\'on de 
contorno empleada y los resultados de integrales escalares hasta el nivel de pentágono y de integrales 
tensoriales hasta el nivel de hexágono.
}

\clearpage % Start a new page

%----------------------------------------------------------------------------------------
%	ACKNOWLEDGEMENTS
%----------------------------------------------------------------------------------------

\setstretch{1.3} % Reset the line-spacing to 1.3 for body text (if it has changed)

\acknowledgements{\addtocontents{toc}{\vspace{1em}} % Add a gap in the Contents, for aesthetics
\vspace{1cm}
First and foremost, I would like to express my deep gratitude to Dr. Germ\'an Rodrigo, my research supervisor, 
for his patient guidance and astute advice throughout this project. In particular, I am grateful for his continuous support, encouragement and his way of thinking outside the box.

Especially I would like to thank my co-supervisor Dr. Grigorios Chachamis for investing a lot of time into discussions and detailed explanations as well as answering countless questions inside and outside of physics.

I further thank my colleagues Dr. Petros Draggiotis and Dr. Ioannis Malamos for frequently helping me on with the difficulties I faced at work.

I would also like to thank Daniel Götz, who explained me how to use the Cuba library in C++.

Special thanks go to Dr. Sophia Borowka and Dr. Gudrun Heinrich for all the help they provided me while I was using SecDec to produce reference values to compare to.

I do appreciate the financial support that I received and which allowed me to freely carry out my studies. In particular, this has been 
\begin{itemize}
\item the position as an Early Stage Researcher within the LHCPhenoNet during the first year, and
\item the JAE Predoc 2011 fellowship for the rest of the Ph.D.
\end{itemize}
Also in this context, I highly value the unique possibilities that being a part of the LHCPhenoNet offered me in terms of conferences, summer schools and general exposure to the scientific community.

Finally, I am grateful for the unconditional support of my family during the entirety of my Ph.D.
}
\clearpage % Start a new page

%----------------------------------------------------------------------------------------
%	LIST OF CONTENTS/FIGURES/TABLES PAGES
%----------------------------------------------------------------------------------------

\pagestyle{list} % The page style headers have been "empty" all this time, now use the "fancy" headers as defined before to bring them back
%\fancyhead[LE,RO]{\thepage}
%\lhead{\emph{Contents}} % Set the left side page header to "Contents"
\tableofcontents % Write out the Table of Contents

\addtocontents{toc}{\vspace{1em}} % Add a gap in the Contents, for aesthetics 

%\lhead{\emph{List of Figures}} % Set the left side page header to "List of Figures"
\listoffigures % Write out the List of Figures

%\addtocontents{toc}{\vspace{1em}} % Add a gap in the Contents, for aesthetics 

%\lhead{\emph{List of Tables}} % Set the left side page header to "List of Tables"
\listoftables % Write out the List of Tables

\pagestyle{intro}

\mainmatter % Begin numeric (1,2,3...) page numbering

\pagestyle{fancy} % Return the page headers back to the "fancy" style

% Include the chapters of the thesis as separate files from the Chapters folder
% Uncomment the lines as you write the chapters

%\addtocontents{toc}{\vspace{2em}} % Add a gap in the Contents, for aesthetics 

% Chapter Template

\chapter{Introduction} % Main chapter title

\label{Chapter1} % Change X to a consecutive number; for referencing this chapter elsewhere, use \ref{ChapterX}

%\lhead{Chapter 1. \emph{Introduction}} % Change X to a consecutive number; this is for the header on each page - perhaps a shortened title
%\fancyhead[LE,RO]{\thepage}

%LHC wiki
%Paragraph: has reached a level of consensus; where up to which level valid; whether higgs-like particle exists; + new physics
%lhc has offered a new particle which fits expected higgs description, waiting for second run which is hoped to reveal new interactions, to validaete or falisfy bsm-models by experimental data, set of models the commmunity was proposing extra dimensions, ...,
%still the main theory that everybody agrees on is the SM (great theoretical achievement, leaving gravity out, no theory for it): interactions are based on symmetries
%SM has gaps: dark matter

The Large Hadron Collider (LHC), the largest and most complex machine ever built by mankind, gives particle physicists a powerful tool at hand to verify existing models and probe the fundamental laws at very high energies. The aim of the LHC is: First, to investigate whether the Standard Model (SM) is still valid at the collider's energies. Second, to shed light on the electroweak symmetry breaking mechanism and to find or exclude a particle that fits the SM description of the Higgs boson \cite{Englert:1964et,Higgs:1964ia,Guralnik:1964eu}. Third, to search for Beyond Standard Model (BSM) physics like supersymmetry or extra-dimensions or particles that could be Dark Matter candidates. In the first run of the LHC, a Higgs-like particle has been found with a mass of 125 GeV. Apart from that, several other discoveries have been made including the first creation of a quark-gluon plasma or the rare $\text{B}_S$-decay. In the second run, in which the center-of-mass energies will be increased even further, signals of BSM physics are hoped to be detected and the properties of the Higgs further explored.\\
Despite the existence of observations which the SM cannot accommodate at the moment, e.g. the existence Dark Matter \cite{Peter:2012rz} or neutrino oscillations \cite{Fisher:1999fb}, it is still the only established theory of particle physics to date that describes experimental data, a great theoretical achievement of its own. Leaving out gravity, for which there is no quantum theory available yet, it successfully describes (almost) all relevant particle physical observables.\\
The SM is a relativistic quantum field theory with an $SU(3)\times SU(2)\times U(1)$ gauge symmetry. It describes three of the four fundamental forces of nature at microscopic distances, namely electromagnetism, strong and weak force.\footnote{For a full review of the Standard Model, see for example \cite{Peskin:1995ev,Beringer:1900zz}.} \\
The $SU(3)$ symmetry accounts for the strong force and the corresponding quantum field theory is called QCD (Quantum Chromodynamics, \cite{Fritzsch:1973pi}). It describes the interactions between quarks and gluons. Quarks are, besides leptons, the fundamental matter constituents and come in six different flavors. Gluons are massless spin 1 particles (bosons) carrying an $SU(3)$ colour charge and act as the mediators of the strong force. Because the gauge group is non-Abelian \cite{Yang:1954ek}, QCD features two remarkable properties: \textit{Confinement} \cite{Wilson:1974sk} and \textit{Asymptotic Freedom} \cite{Gross:1973id,Politzer:1973fx}.\\
Confinement accounts for the fact that physical objects are always colour-neutral at low energies, in particular no individual free quarks or gluons are observed. Nonetheless, an analytical proof of this property is still missing.\\
Asymptotic Freedom is a property of the theory. It means that due to the running of the strong coupling $\alpha_s$, the strong force becomes small at high energies (equivalentely, small distances), a fact that permits  the employment of perturbative techniques.\\
The weak force describes processes such as the decay of nuclei and the interaction of neutrinos with matter. In the modern context, it is better understood within the framework of the electroweak sector. By electroweak sector, we mean the unification of weak and electromagnetic forces \cite{Salam:1964ry,Weinberg:1967tq,Salam:1968rm,Glashow:1961tr} by Weinberg, Salam and Glashow, a major success which allowed to understand both forces in the common framework of the Electroweak Symmetry Breaking mechanism. Very similar to the strong force, its gauge group is the non-Abelian $SU(2)$. Contrary to QCD, the $W^{\pm}$- and $Z^0$-gauge bosons of the weak interaction are not only massive, but with masses of 80 GeV and 91 GeV respectively, they are quite heavy. Hence it is a short distance interaction and appears to be `weak'.\\
Finally, $U(1)$ is the gauge group of Quantum Electrodynamics (QED) \cite{Schwinger:1948yk} which is the quantum field theory describing the electromagnetic force. Its gauge boson, the photon, is massless and due to the Abelian nature of $U(1)$, photons do not interact with each other and the force is of unlimited range.\\ 
What we today know as the Standard Model of particle physics is QCD together with the electroweak sector. It has undergone countless checks and investigations over many different aspects and it has been exceptionally successful in making correct and accurate predictions for a wide range of physical observables.\\
%mathematical formulation is simple, believed to be low-energy approximation to the correct theory
%the sm-langrangian encodes all the symmetry principles that we believe to be required
%despite it being very simple one can not analytically solve it
However, the SM is believed to be a mere low-energy approximation to a still to be constructed unified field theory which would describe all four forces at all energy regimes. The SM-Lagrangian encodes all previosly known symmetry principles, for example conservation laws. Although its mathematical formulation is very simple, we cannot analytically solve the equations of motion. Only in certain parts of phase-space we can perform reliable theoretical calculations which are all based on perturbation theory. The underlying idea is that when the coupling of the interaction term is small, one can do a series expansion in which every term can be represented in a pictorial form. This is done by the so-called Feynman diagrams \cite{Feynman:1948fi,Feynman:1949zx}. Due to its clarity and predictive power, the diagrammatical approach is the most popular for theoretical calculations.
The first term of the expansion usually gives an estimate of the order of magnitude whereas in order to obtain the  first proper estimate, one has formally to go to next-to-leading order (NLO) precision. More and more processes  processes demand next-to-next-to-leading order (NNLO) precision to match the precision of the experimental data.\\
The discovery of the Higgs-like boson in 2012 has been a huge success \cite{Aad:2012tfa,Chatrchyan:2012ufa}. At the moment of writing this thesis, the LHC is warming up for its second phase with a center-of-mass energy of up to 13 TeV. In this second phase, as previously mentioned, it will be of great importance to measure as many properties of the discovered particle as possible and to continue the hunt for physics beyond the Standard Model. So far, the absence of signals hinting to BSM physics is a little disappointing as it does not define a clear direction for the theorists to follow.\\ 
The high quality of LHC data raises the need for high-precision theoretical predictions. The processes at the LHC are rather challenging to calculate because they typically involve many particles and because QCD plays the dominant role at the LHC. Furthermore, higher orders of the perturbation expansion have to be calculated in order to match the experimental precision. 
This has led to considerable progress in the analytical and numerical techniques for the calculation of Standard Model cross-sections. Apart from the usual diagrammatic approach, there are other methods, some of the most popular ones being Unitarity Methods \cite{Bern:1994cg,Bern:2007dw,Anastasiou:2006gt}, the OPP-Method \cite{Ossola:2006us,Ossola:2007bb}, Mellin-Barnes Representations \cite{Smirnov:1999gc,Tausk:1999vh} and Sector Decomposition \cite{Binoth:2000ps,Binoth:2003ak,Carter:2010hi,Borowka:2015mxa}. Thanks 
to these techniques, 2 $\to$ 4 processes at NLO are the standard nowadays, and even higher multiplicities are becoming more accessible \cite{Berger:2009zg,Melnikov:2009wh,Bevilacqua:2009zn,Bredenstein:2010rs}. They have achieved an incredible feat: The computation of Feynman graphs up to NNLO-level and in some cases even beyond. Still, many important issues remain. When calculating cross-sections one needs to consider tree- and loop-contributions separately. Thus, a lot of effort has to be put into cancelling infrared singularities between real and virtual corrections \cite{Catani:1996vz,Catani:1996jh,Frixione:1995ms,GehrmannDeRidder:2005cm,Catani:2007vq}. Additional difficulties arise from threshold singularities that lead to numerical instabilities. \\
Recently, a new method called the Loop--Tree Duality (LTD) \cite{Catani:2008xa} has been developed, which is designed to attack these problems. The basic concept at one-loop is to directly apply the Cauchy Residue Theorem to the Feynman integrals. The outcome is a sum of tree level objects in which each represents all possible single cuts of the considered diagram. This form is called the `dual integral', which closely resembles the real corrections. The idea is to then combine the dual integral with the tree-level contributions in order to treat them \textit{simultaneously} in a common Monte Carlo event generator. While initially the technique was limited to one-loop graphs, it has been greatly expanded since then. In \cite{Bierenbaum:2010cy} it has been shown how to extend it to diagrams with an arbitrary number of loops and in \cite{Bierenbaum:2012th}  how to deal with graphs which involve propagators that are raised to higher powers (higher order poles).

%----------------------------------------------------------------------------------------
%	Outline
%----------------------------------------------------------------------------------------

\section{Outline}

%Update chapter numbers
The remainder of this thesis is organised as follows: In Chapter \ref{Chapter2}, we establish the fundamentals of this work. In Chapter \ref{Chapter3} we introduce the Loop--Tree Duality method alongside some illustrative examples. In Chapter \ref{Chapter4}, we formalize the notation and extend the Loop--Tree Duality to double and higher order loop graphs. in Chapter \ref{Chapter5} we show how to deal with poles of higher multiplicities. In Chapter \ref{Chapter6} we report on the cancellation of singularities among dual contributions as well as between real and virtual corrections for massless internal lines. In Chapter \ref{Chapter7}, we present details on the numerical implementation of the Loop--Tree Duality for scalar one-loop integrals. This is the first time that the LTD has been applied to explicitly calculate Feynman diagrams and constitutes the main result of this thesis. In Chapter \ref{Chapter8}, we demonstrate that the computer program used in Chapter \ref{Chapter7} is also able to deal with tensor integrals. We conclude the thesis with Chapter \ref{Chapter9} in which we give a summary and our future plans.
 
% Chapter Template

\chapter{Standard Model Phenomenology} % Main chapter title

\label{Chapter2} % Change X to a consecutive number; for referencing this chapter elsewhere, use \ref{ChapterX}

The Standard Model is a theory of particle physics that has been refined continuously over the past, 
so that it is nowadays able to describe almost all particle reaction 
processes that we observe in the laboratory (particle colliders) as well as in space. Its tremendous 
success is threefold: 
\begin{itemize}
\item The SM has the ability to explain a wide variety of experimental results.  
\item The SM repeatedly predicted the existence of particles before their experimental 
discovery. This has been the case for the $W^{\pm}$ and $Z^0$ bosons \cite{Arnison:1983rp,Arnison:1983mk}, the gluon \cite{Brandelik:1979bd,Barber:1979yr,Berger:1979cj,Bartel:1979ut} and the charm \cite{Augustin:1974xw,Aubert:1974js} and 
top quarks \cite{Abe:1995hr,Abachi:1994td} and very recently the Higgs boson \cite{Chatrchyan:2012ufa,Aad:2012tfa}. For each of these particles, experiments later confirmed the predicted properties with good precision. 
\item The SM has passed a huge number of precision tests with flying colours, the most 
famous one being the anomalous magnetic dipole moment of the electron \cite{Hanneke:2008tm}: The agreement between theory 
and measurement is up to the 13th digit, which is a unique achievement not only in particle physics, but 
in all science. Other examples of precisely predicted quantities are the Neutron Compton Wavelength \cite{Kinoshita:1990kh} or 
the mass of the $Z^0$ boson \cite{ALEPH:2005ab}.
\end{itemize}
All of these accomplishments have strengthend the confidence in the SM as the proper theory for 
the description of the behaviour of elementary particles excluding gravity.\\
We dedicate the next section to an instructive review of the basic principles of the SM.

\newpage

\section{The Standard Model of Particle Physics}

The Standard Model of particle physics is a quantum field theory which nowadays allows us to correctly describe almost all physical processes with high precision on a fundamental level. It is based on the gauge group
\bea
SU(3)\times SU(2) \times U(1)
\eea
where $SU(3)$ is the gauge group of the strong interaction, $SU(2)$ the gauge group of the weak isospin, and $U(1)$ the gauge group of the hypercharge which is not identical with the gauge group of QED. To avoid confusion, the notations $U(1)_Y$ and $U(1)_{EM}$ with subscripts hinting to either the hypercharge ($Y$) or QED ($EM$) will be used.\\
Despite the interacting constituents being fields, the SM's perturbative description is carried out almost entirely in the particle picture. We imagine the force between two matter particles to be mediated via exchange particles. All interactions have in common that the matter fields are spin $\frac{1}{2}$ fermions and the exchange particles spin $1$ bosons, with the exception 
of the Higgs boson which has spin $0$. 

%-----------------------------------
%	QCD
%-----------------------------------

\subsection{QCD}
\label{sec:2qcd}

The QCD, Quantum ChromoDynamics, is a non-Abelian gauge theory with the action
\bea
S=\int d^4x\left\{\bar{\psi}(i\slashed{D}-m)\psi-\frac{1}{4}\text{tr}F_{\mu\nu}F^{\mu\nu}\right\}~,
\eea
and the symmetry group $SU(3)$. The fermion fields $\psi$ transform under its fundamental representation. They are, apart from the gluons, the dynamical degrees of freedom of the theory. There are six flavors of quarks, each with a different mass. $D$ is the so-called covariant derivative, defined as
\bea
D_{\mu}=\partial_{\mu}-ig_sA_{\mu}~,
\eea
with $A_{\mu}=A_{\mu}^aT^a$ being the gluon field, which lives in the adjoint representation of the gauge group. The group $SU(N)$ has $N^2-1$ generators, hence $SU(3)$ has eight of them. They span the Lie algebra $\mathfrak{su}(3)$ and are normalised by $\text{tr}(T^aT^b)=\delta^{ab}$ and $[T^a,T^b]=if^{abc}T^c$. $F_{\mu\nu}$ is the field strength tensor. It is an algebra-valued two-form defined by
\bea
F_{\mu\nu} =\frac{1}{ig_s} [D_{\mu},D_{\nu}]=\partial_{\mu}A_{\nu}-\partial_{\nu}A_{\mu}+ig_s[A_{\mu},A_{\nu}].
\label{eq:field-strength-tensor}
\eea
The first two terms are a four-dimensional rotation known from QED; the third term is typical for non-Abelian symmetry groups and the origin of the gluon self-interaction.\\
Considering QCD in the absence of quarks, the only thing left is the gluon self-interaction. This branch is called pure Yang-Mills theory. As we mentined in the Introduction, QCD has two special properties, Confinement and Asymptotic Freedom. They mainly have to do with the running coupling $\alpha_s(Q)$ of QCD. By running coupling we mean that the coupling is dependent on the energy at which the process happens \cite{Peskin:1995ev}:
\bea
\alpha_s(Q)=\frac{2\pi}{(11-2n_f/3)\ln(Q/\Lambda_{QCD})}~.
\eea
Confinement takes place at low energies at which the coupling is strong. Physical states are always color singlets: As a consequence of the non-Abelian gauge symmetry, the energy cost grows proportionally to the distance, if one tries to separate the particles of a color singlet. At some point it is energetically favorable for a new quark-antiquark pair to appear. Hence, neither free quarks nor free gluons are observed in nature.\\
Asymptotic freedom describes the phenomenon that the coupling of the theory becomes weaker as the energy increases. At very high energies, quarks and gluons are almost able to move as if they were free particles. In this energy regime, perturbation theory is applicable. However, one has to quantise the action first, a good method to do so is the one by Faddeev and Popov \cite{Faddeev:1967fc}.  An unwanted side effect of the  method is the introduction of ghost fields into the theory. They are called ghosts because of their incorrect spin-statistic relation and therefore are  unphysical.  To get rid of them, one can choose the axial gauge, in which ghosts and gluons decouple and the gauge of the action is fixed. From the gauge fixed action one can extract the propagators and vertices which allow to calculate the contributions to the perturbation series via the Feynman diagram approach. These diagarams are pictorial representations of the terms of the perturbation series. To calculate an amplitude of arbitrary order in $\alpha_s$, one draws all the Feynman diagrams that contribute to the process, contracts them with the outer polarisation vectors and spinors and finally sums them up.
\begin{figure}[h]
\centering
\begin{tabular}{p{6cm}p{5cm}}
\qquad\quad\,\,\,\includegraphics[scale=1]{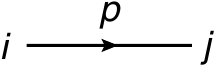} & \qquad {\Large \text{$\frac{i}{\slashed{p}-m}\delta_{ij}$}} \\
\qquad\includegraphics[scale=1]{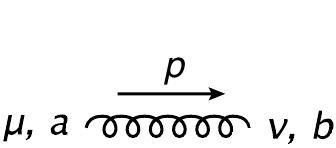} & \qquad {\Large \text{$-\frac{i}{p^2}g^{\mu\nu}\delta^{ab}$}}
\end{tabular}
\caption{QCD propagators in Feynman gauge, without ghosts.}
\end{figure}
\begin{figure}[h]
\centering
\begin{tabular}{p{7,5cm}p{8,5cm}}
\quad\qquad\qquad\quad\,\includegraphics[scale=1]{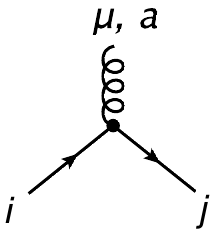} & \raisebox{5ex}[-5ex]{{\large \text{$ig\gamma^{\mu}T^a$}}} \\
\quad\qquad\qquad\includegraphics[scale=1]{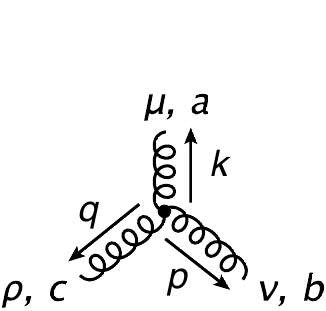} & \raisebox{9ex}[-9ex]{{\large \text{$gf^{abc}[g^{\mu\nu}(p-k)^{\rho}$}}}\newline\raisebox{9ex}[-9ex]{{\large \text{$\qquad\,\,\,\,+g^{\nu\rho}(q-p)^{\mu}$}}}\newline\raisebox{9ex}[-9ex]{{\large \text{$\qquad\,\,\,\,+g^{\rho\mu}(k-q)^{\nu}]$}}} \\
\quad\qquad\qquad\includegraphics[scale=1]{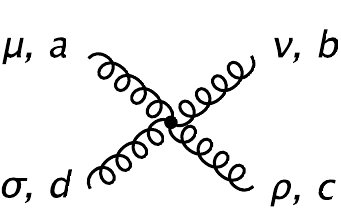} & \raisebox{9ex}[-9ex]{{\large \text{$-ig^2[f^{abe}f^{cde}(g^{\mu\rho}g^{\nu\sigma}-g^{\mu\sigma}g^{\nu\rho})$}}}\newline\raisebox{9ex}[-9ex]{{\large \text{$\,\,\,\quad+f^{ade}f^{bce}(g^{\mu\nu}g^{\rho\sigma}-g^{\mu\rho}g^{\nu\sigma})$}}}\newline\raisebox{9ex}[-9ex]{{\large \text{$\quad\,\,\,+f^{ace}f^{bde}(g^{\mu\nu}g^{\rho\sigma}-g^{\mu\sigma}g^{\nu\rho})]$}}} 
\end{tabular}
\caption{QCD vertices in Feynman gauge, without ghosts. All momenta outgoing.}
\end{figure}
A downside of this approach is that it consumes a lot of computational power. For example, the number of Feynman diagrams that contribute to the $n$-gluon amplitude grows faster than $n$!\cite{Weinzierl:2007vk}. Techniques like color-ordered amplitudes \cite{Bern:1990ux} and (tree-level) recurrence relations \cite{Berends:1987me,Cachazo:2004kj,Britto:2004ap} have been developed to deal with that problem.

%-----------------------------------
%	Electroweak Interaction
%-----------------------------------

\subsection{Electroweak Interaction}

Until 1967, QED and weak interaction were two separate theories. In that year, Salam, Weinberg and Glashow succeeded in understanding both interactions as special cases of one unified theory called Electroweak theory \cite{Salam:1964ry,Weinberg:1967tq,Salam:1968rm,Glashow:1961tr}. The reason why both interactions appear to be so different is the spontaneously broken symmetry of the Electroweak theory.\\
The corresponding action resembles the one of the QCD:
\bea
S=\int d^4x\left\{\bar{\psi}(i\slashed{D}-m)\psi-\frac{1}{4}\text{tr}W_{\mu\nu}W^{\mu\nu}-\frac{1}{4}\text{tr}B_{\mu\nu}B^{\mu\nu}\right\}+(D_{\mu}\phi)^{\dagger}D^{\mu}\phi-m^2\phi^{\dagger}\phi+\frac{\lambda}{4}(\phi^{\dagger}\phi)^2~.\nonumber
\eea
The matter fields $\psi$ of the electroweak interaction are the six leptons, but quarks do carry a weak charge as well. They can be organised into three generations. Electron, muon and tau participate in both the weak and the electromagnetic interactions. The corresponding neutrinos interact exclusively weakly. Within the SM, they are supposed to be massless, though recent experiments show that they actually have a small nonzero mass. This  allows to solve the solar neutrino problem using the concept of neutrino oscillations \cite{Fisher:1999fb}. 
The operator $D$ is defined as:
\bea
D_{\mu}=\partial_{\mu}-igW_{\mu}^a\sigma^a-\frac{ig'}{2}B_{\mu}Y~.
\label{eq:delweak}
\eea
The underlying symmetry group of the electroweak interaction is the $SU(2)\times U(1)_Y$. In \Eq{eq:delweak}, the $\sigma^a$ are the generators of the $SU(2)$ and the hypercharge $Y$ is the generator of the $U(1)_Y$. Because of the direct product of the two groups, they carry different coupling constants $g$ and $g'$.\\
The fields $W$ and $B$ are the gauge fields of the theory. $W$ is a field similar gluons of QCD with respect to their non-Abelian character. The corresponding field strength tensor is obtained by replacing $A$ with $W$ in equation (\ref{eq:field-strength-tensor}). The field $B$ transforms under the Abelian symmetry group $U(1)_Y$. Thus its field strength tensor does not have a commutator.\\
According to the way $B$ and $W$ appear in the Lagrangian, they are massless. However, experiments show that the $W$ and $Z$ bosons indeed have a mass. To get rid of this flaw of the theory, one introduces an additional field $\phi$, the so-called Higgs-field, that couples to the gauge bosons by the term $(D_{\mu}\phi)D^{\mu}\phi$. The a priori unphysical and massless fields $B$ and $W$ now obtain their masses via the Higgs-mechanism. By transforming the unphysical fields $B$ and $W$ in physical ones, the $SU(2)\times U(1)_Y$-symmetry is broken down to a $U(1)_{EM}$-symmetry. Simultaneously the Lagrangian picks up the correct mass terms for the $W$ and $Z$ bosons as well as the photon. The other terms of the Langrangian represent the potential of the Higgs field. This is illustrated in greater detail in the next subsection.\\
Despite the similarities between QCD and electroweak interaction, the latter one does not display the phenomenon ``Confinement''. The reason for this is that the exchange bosons of the electroweak interaction are very heavy, namely 80 GeV for the $W^{\pm}$ and 91 GeV for the $Z^0$, which does not allow for bound states. ``Asymptotic Freedom'', however, is observed for the weak interaction just as in QCD.

%-----------------------------------
%	Higgs boson and Electroweak Symmetry Breaking
%-----------------------------------

\subsection{Higgs Boson and Electroweak Symmetry Breaking}

In the Standard Model, the gauge group $SU(2)\times U(1)_Y$ is spontaneously broken. The left-over symmetry is $U(1)_{EM}$. One assumes the existence of an additional complex scalar field which transforms under SU(2) and has hypercharge $Y=1$. In the space of weak isospin, it can be represented by a two-component vector with complex entries \cite{lecture}.
\bea
\phi(x)=\begin{pmatrix} \phi^+(x) \\ \frac{1}{\sqrt{2}}(v+H(x)+i\chi(x))\end{pmatrix}~,
\eea
$\phi^+(x)$ is a complex field, i.e. it is composed of two components. The three components $\phi^+(x)$ and $\chi(x)$ are absorbed into the longitudinal modes of $W^{\pm}_{\mu}$ and $Z_{\mu}$. $H(x)$ is the actual physical Higgs-field. The Lagrangian of the Higgs-sector then reads
\bea
\mathcal{L}_{\text{Higgs}}=(D_{\mu}\phi)^{\dagger}(D^{\mu}\phi)\underbrace{-m^2\phi^{\dagger}\phi+\frac{\lambda}{4}(\phi^{\dagger}\phi)^2}_{V(\phi)}+\mathcal{L}_{\text{Yukawa}}~,
\eea
For the theory to be bound from below, $\lambda$ has to be greater than zero, $\lambda > 0$, and furthermore $m^2 < 0$ for the symmetry to be spontaneously broken. The covariant derivative is given by \Eq{eq:delweak}. 
%\bea
%D_{\mu}=\partial_{\mu}-\frac{ig}{2}W_{\mu}^a\sigma^a-\frac{ig'}{2}B_{\mu}Y
%\eea
The Higgs-dublet has hypercharge $Y=1$; from the minimization of the potential $V(\phi)$ follows
\bea
|\phi_{\text{min}}|^2=\frac{2m^2}{\lambda}=\frac{v^2}{2}\Rightarrow v=2\sqrt{\frac{m^2}{\lambda}}~.
\eea
In the next step, the physical contributions to $(D_{\mu}\phi)^{\dagger}D^{\mu}\phi$ will be calculated.
\bea
(D_{\mu}\phi)^{\dagger}(D^{\mu}\phi)|_{\text{phys}}&=\frac{1}{2}\partial_{\mu}H\partial^{\mu}H+\frac{1}{2}\left(\frac{gv}{2}\right)^2(W_{\mu}^1W^{1\mu}+W_{\mu}^2W^{2\mu})+\nonumber\\
&+\frac{1}{8}(B_{\mu},W_{\mu}^3)\begin{pmatrix} g'^2 & -gg' \\ -gg' & g^2\end{pmatrix}\begin{pmatrix} B^{\mu} \\ W^{3\mu}\end{pmatrix}
\label{eq:langrangian-ew-phys}
\eea
$H$ is already a physical field, but $B$ and $W$ are not. They are transformed to physical fields by virtue of the transformations
\bea
W_{\mu}^{\pm}=\frac{1}{\sqrt{2}}(W_{\mu}^1\mp W_{\mu}^2)
\eea
and
\bea
\begin{pmatrix} A_{\mu} \\ Z_{\mu} \end{pmatrix} = \begin{pmatrix} \cos \theta_W & \sin\theta_W \\ -\sin\theta_W & \cos\theta_W \end{pmatrix}\begin{pmatrix} B_{\mu} \\ W_{\mu}^3 \end{pmatrix}~.
\eea
Here, $\sin\theta_W$ and $\cos\theta_W$ are given by
\bea
\cos\theta_W=\frac{g}{\sqrt{g^2+g'^2}}\qquad\text{and}\qquad\sin\theta_W=\frac{g'}{\sqrt{g^2+d'^2}}~.
\eea
This way, one can rewrite the part of the Lagrangian of equation (\ref{eq:langrangian-ew-phys}) as
\bea
(D_{\mu}\phi)^{\dagger}(D^{\mu}\phi)|_{\text{phys}}&=\frac{1}{2}\partial_{\mu}H\partial^{\mu}H+\frac{1}{2}\left(\frac{gv}{2}\right)^2(W_{\mu}^{+*}W^{+\mu}+W_{\mu}^{-*}W^{-\mu})\nonumber\\
&+\frac{1}{2}\left(\frac{v}{2}\sqrt{g^2+g'^2}\right)^2(A_{\mu},Z_{\mu})\begin{pmatrix} 0 & 0 \\ 0 & 1\end{pmatrix}\begin{pmatrix} A^{\mu} \\ Z^{\mu}\end{pmatrix}~.
\eea
Now, the masses of the $W$- and $Z$-bosons can be read off easily:
\bea
m_W=\frac{v}{2}g\qquad\text{and}\qquad m_Z=\frac{v}{2}\sqrt{g^2+g'^2}~.
\eea
The couplings $g$ and $g'$ are related to the elementary charge $e$ by
\bea
e=\frac{gg'}{\sqrt{g^2+g'^2}} \quad\text{or}\quad g=\frac{e}{\sin\theta_W}\quad\text{and}\quad g'=\frac{e}{\cos\theta_W}\quad \text{respectively.}
\eea
Within this framework, it is possible to have massive gauge bosons which otherwise are forbidden by unbroken gauge symmetries.

%----------------------------------------------------------------------------------------
%	Particle phenomenology
%----------------------------------------------------------------------------------------

\section{Particle Phenomenology}

Phenomenology is the branch of particle physics that 
deals with the calculation of physical observables that can then be compared 
to experimental measurements. 
In order to test a theory, solutions have to be compared to experimental 
data. 
For the SM, there are two established ways to calculate them: Lattice calculations and perturbation 
theory. Since lattice techniques lie beyond the scope of this thesis, we focus on the latter. 
As mentioned in Section \ref{sec:2qcd}, perturbation theory (in the coupling) is employed 
at high energies when the coupling of the theory becomes small.
Usually, Feynman diagrams are used to calculate the individual contributions to the 
perturbation series. We distinguish betweeen two main types of diagrams: Tree 
and loop diagrams. The the leading order term is a tree-level contribution. Beyond 
leading order, a Feynman diagram gets dressed with more lines. These can connect 
two existing lines, thus leading to a loop or they connect to only one existing line 
resulting in an additional external leg. We use the terms 
next-to-leading order (NLO, one-loop) or next-to-next-to-leading order 
(NNLO, two-loops) to indicate the accuracy up to which a certain calculation has 
been performed. Today, the quality of data collected at colliders like the LHC has 
reached a level that makes the inclusion of terms beyond the leading 
order of perturbation theory necessary in order to provide theoretical 
estimates of similar precision. Apart from the cancellation of singularities between 
tree and loop corrections and phase-space integrations with many external legs, 
loop diagrams are the bottleneck when calculating scattering amplitudes. 
Hence, theorists are looking for ways to 
efficiently calculate NLO and NNLO corrections. The last decade in 
particular has seen a huge progress in the development of calculational 
techniques to tackle computations at NLO and also some progress at NNLO 
accuracy. In the LHC era, this is important, because 
of the availability of large amounts of data to which theoretical predictions can 
be compared to. Still, the task of obtaining 
theoretical estimates that can be compared with LHC-data is not easy. 
Tree and loop contributions have to be evaluated independently before 
bringing them together. The Loop--Tree Duality method, 
which will be introduced in Chapter \ref{Chapter3} and which is at the heart 
of this thesis, reexpresses loop-level objects in terms of tree-level 
objects. Thus it allows to directly combine the two, in order to evaluate them 
simultaneously. This evaluation is carried out using numerical integration. 
In the latter chapters of the thesis, we present results of the of the first 
implementation of the Loop--Tree Duality method as a computer progam. 
This program calculates one-loop integrals using a numerical integrator 
(Cuhre or VEGAS). Since the numerical integrator is employed as a 
black box, we recapitulate the basics of numerical integration in Section 
\ref{sec:numint}.

%From a certain 
%complexity level onwards, analytical methods become too expensive to still be viable. This 
%has lead to a shift to numerical methods. The Loop--Tree Duality method, 
%which will be introduced in Chapter \ref{Chapter3} and which is at the heart 
%of this thesis, is also a numerical method. It is expected to be very fast and 
%efficient, but it also offers a new feature. Due to way it works, it opens up the 
%possibility to combine real and virtual corrections in a natural way. In doing 
%so, it avoids the difficulties of cancelling infrared singularities among real and virtual 
%corrections. Since this thesis deals with the calculation of one--loop diagrams, 
%i.e. integrating them numerically, let us have an instructive look 
%at numerical integration.
%standard model
%comparing experimental data to theoertical predictions

%----------------------------------------------------------------------------------------
%	Numerical Integration
%----------------------------------------------------------------------------------------

\section{Numerical Integration Techniques\label{sec:numint}}

%-----------------------------------
%	Motivation for employing numerical techniques
%-----------------------------------

\subsection{Motivation for employing numerical techniques}

In Chapters \ref{Chapter7} and \ref{Chapter8}, we are going to calculate 
a certain type of three-dimensional integral.\\
Usually one is interested in knowing the value of an integral up to a 
certain accuracy and wants to obtain the result within an acceptable amount of time. 
Hence, a numerical integrator in the form of a computer program 
is the proper tool for this task.\\ %(see Chapter \ref{Chapter3})
Our long-term goal is to set up a fully automated program which is able to 
calculate (N)NLO cross-sections for arbitrary kinematics. 
In this thesis, we do the first step by addressing the virtual part.

%Furthermore, with the long-term goal of implementing such a 
%Monte-Carlo Event Generator in mind it is beneficial to go that route.

%-----------------------------------
%	Classical numerical integration
%-----------------------------------

\subsection{Classical numerical integration}

For the rest of this chapter, we closely follow the paper \cite{Weinzierl:2000wd} by Weinzierl, which is a nice introduction to the subject and is recommended to readers with further interest in the matter.

Quadrature rules have been known for a long time. The numerical integrator ``Cuhre'', part of the Cuba-library \cite{Hahn:2004fe}, uses quadrature rules to estimate the integral, therefore their idea shall be illustrated here. One distinguishes between formulae that evaluate the integrand at equally spaced abscissas (Newton-Cotes type) and formulae which evaluate the integrand at carefully selected, but non-equally spaced abscissas (Gaussian quadrature rules). The simplest example of a Newton-Cotes type rule is the so-called trapezoidal rule:
\beq
\int\limits_{x_0}^{x_0+\Delta x} dxf(x)=\frac{\Delta x}{2}[f(x_0)+f(x_0+\Delta x)]-\frac{(\Delta x)^3}{12}f''(\xi)~,
\eeq
where $x_0\leq \xi\leq x_0+\Delta x$. To approximate an integral over a finite interval $[x_0,x_n]$ with the help of this formula, one divides the interval into $n$ sub-intervals of length $\Delta x$ and applies the trapezoidal rule to each sub-interval. With the notation $x_j = x_0+j\cdot \Delta x$, one arrives at the compound formula
\beq
\int\limits_{x_0}^{x_n}dxf(x)=\frac{x_n-x_0}{n}\sum\limits_{j=0}^{n}w_jf(x_j)-\frac{1}{12}\frac{(x_n-x_0)^3}{n^2}\tilde{f}''
\label{eq:traprule}
\eeq
with $w_0=w_n=1/2$ and $w_j=1$ for $1\leq j \leq n-1$. Further 
\beq
\tilde{f}''=\frac{1}{n}\sum\limits_{j=1}^{n}f''(\xi_j),
\eeq
where $\xi_j$ is somewhere in the interval $[x_{j-1},x_j]$. Since the position of the $\xi_j$ cannot be known without knowing the integral exactly, the last term in  \Eq{eq:traprule} is usually neglected and introduces an error in the numerical evaluation. This error is proportional to $1/n^2$ and one has to evaluate the function $f(x)$ roughly $n$-times.\\
An improvement is given by Simpson's rule, which evaluates the function at three points:
\beq
\int\limits_{x_0}^{x_2}dxf(x)=\frac{\Delta x}{3}[f(x_0)+4f(x_1)+f(x_2)]-\frac{(\Delta x)^5}{90}f^{(4)}(\xi).
\eeq
This yields the compound formula
\beq
\int\limits_{x_0}^{x_2}dxf(x)=\frac{x_n-x_0}{n}\sum\limits_{j=0}^{n}w_jf(x_j)-\frac{1}{180}\frac{(x_n-x_0)^5}{n^4}\tilde{f}^{(4)}
\eeq
where $n$ is an even number, $w_0 = w_n=1/3$, and for $1 \leq j \leq n$ we have $w_j = 4/3$ if $j$ is odd and $w_j'$ if $j=2/3$ is even. The error estimate scales now as $1/n^4$.\\
\newline
As mentioned earlier, there are also rules involving non-equally spaced abscissas. A well known representative is the main formula of Gaussian quadrature.\\
If $w(x)$ is a weight function on $[a,b]$, then there exist weights $w_j$ and abscissas $x_j$ for $1 \leq j \leq n$ such that
\beq
\int\limits_{a}^{b}dx\,w(x)\,f(x) = \sum\limits_{j=1}^{n}w_jf(x_j)+\frac{f^{(2n)}(\xi)}{(2n)!}\int\limits_{a}^bdx\,w(x)\,[\Pi(x)]^2
\eeq
with
\bea
&\Pi(x)=(x-x_1)(x-x_2)\dots (x-x_n),\nonumber\\
&a\leq x_1 < x_2 < \dots < x_n \leq b, \quad a < \xi < b
\eea
The abscissas are given by the zeros of the orthogonal polynomial of degree $n$  associated to the weight function $w(x)$. In order to find them numerically, it is useful to know that they all lie in the interval $[a,b]$. The weights are given by the (weighted) integral over the Lagrange polynomials:
\bea
w_j=\int\limits_a^bdx\,w(x)l_j^n(x)
\eea
where the fundamental Lagrange polynomials are given by
\bea
l_i^n(x) = \frac{(x-x_0)\dots(x-x_{i-1})(x-x_{i+1})\dots(x-x_n)}{(x_i-x_0)\dots(x_i-x_{i-1})(x_i-x_{i+1})\dots(x_i-x_n)}
\eea

%-----------------------------------
%	Monte Carlo techniques
%-----------------------------------

\subsection{Monte Carlo techniques}

Monte Carlo integration is one of the two methods to perform the integrations in Chapters \ref{Chapter7} and \ref{Chapter8}, although we also present only the results of CI. This is common for the evaluation of multi-dimensional integrals. Generally speaking one is looking for an algorithm that meets a certain set of properties. It should
\begin{itemize}
\item give a numerical estimate of the integral together with an estimate of the error,
\item yield the result in a reasonable amount of time, e.g. at low computational cost,
\item be able to handle multidimensional integrals.
\end{itemize}
Monte Carlo integration delivers on all points of the list. In particular, as will be shown later, its error scales like $1/\sqrt{N}$, independent of the number of dimensions. This makes Monte Carlo integration the preferred method for integrals in high dimensions.

Another integrator provided by the Cuba-library \cite{Hahn:2004fe} is ``VEGAS''. It is a Monte Carlo integrator paired with variance reducing techniques. These two concepts shall be shown in the following subsection.\\
Quadrature rules are inefficient for multidimensional integrals. Therefore one resorts to Monte Carlo integration. On the plus side, its error scales like $1/\sqrt{N}$ independent of the number of dimensions. But, on the other hand, a convergence by a rate of $1/\sqrt{N}$ is pretty slow. To improve efficiency, variance reducing techniques like importance sampling are employed.\\
Consider the integral of a function $f(u_1,\dots , u_d)$, depending on $d$ variables $u_1,\dots , u_d$ over the unit hypercube $[0,1]^d$. Furthermore, $f$ is assumed to be square-integrable. From now on, the short-hand notation $x=(u_1,\dots,u_d)$ and for the function evaluated at that point $f(x)=f(u_1,\dots , u_d)$ will be used. The Monte Carlo estimate for the integral
\bea
I = \int dx f(x) = \int d^du f(u_1,\dots,u_d)
\eea
is given by
\bea
E=\frac{1}{N}\sum\limits_{n=1}^Nf(x_n)~.
\eea
The law of large numbers ensures that the Monte Carlo estimate converges to the true value of the integral:
\bea
\lim\limits_{N\rightarrow\infty}\frac{1}{N}\sum\limits_{n=1}^Nf(x_n) = I~.
\eea
The corresponding error $S$ reads
\bea
S^2=\frac{1}{N}\sum\limits_{n=1}^N(f(x_n))^2-E^2~.
\eea
%The main advantage of Monte Carlo integration is the fact, that the error estimate is independent of the dimension $d$ of the integral. However a price has to be paid in the form of a relatively slow convergence to the true value at the rate of $1/\sqrt{N}$. Two common techniques that help to alleviate this issue are stratified sampling and importance sampling.\\

\subsubsection{Stratified sampling}

This technique consists of dividing the full integration space into subspaces, performing a Monte Carlo integration in each subspace, and adding up partial results in the end. Mathematically, this is based on the fundamental property of the Riemann integral
\bea
\int\limits_0^1dx f(x) = \int\limits_0^a dxf(x)+\int\limits_{a}^1dxf(x), \quad 0 < a < 1.
\eea
More generally, one splits the integration region $M=[0,1]^d$ into $k$ regions $m_j$ where $j=1,\dots ,k$. In each region one performs a Monte Carlo integration with $n_j$ points. For the integral $I$, one obtains the estimate
\bea
E=\sum\limits_{j=1}^k\frac{\operatorname{vol}(M_j)}{N_j}\sum\limits_{n=1}^{N_j}f(x_{jn})
\eea
and instead of $S^2$ one has now the expression
\bea
\sum\limits_{j=1}^k\frac{\operatorname{vol}(M_j)}{N_j}\cdot \left. N\cdot S^2\right|_{M_j}
\eea
If the subspaces and the number of points in each subspace are chosen carefully, this can lead to a dramatic reduction in the error compared with crude Monte Carlo, but it should be noted, that it can also lead to a larger error if the choice is not appropriate. In general, the total variance is minimized when the number of points in each sub-volume is proportional to $\left. N\cdot S^2\right|_{M_j}$.

\subsubsection{Importance sampling}

Mathematically, importance sampling corresponds to a change of integration variables:
\bea
\int\limits dx\, f(x) = \int \frac{f(x)}{p(x)}p(x)dx = \int \frac{f(x)}{p(x)}dP(x)
\eea
with
\bea
p(x) = \frac{\partial^d}{\partial x_1\dots \partial x_d}P(x)
\eea
If one restricts $p(x)$ to be a positive-valued function $p(x) \geq 0$ and to be normalized to unity
\beq
\int dx \, p(x) = 1
\eeq
one may interpret $p(x)$ as a probability density function. If one has a random number generator corresponding to the distribution $P(x)$ at his disposal, one may estimate the integral from a sample $x_1,\dots ,x_N$ of random numbers distributed according to $P(x)$:
\beq
E=\frac{1}{N}\sum\limits_{n=1}^N\frac{f(x)}{p(x)}.
\eeq
The statistical error $S$ of the Monte Carlo integration is given by
\beq
S^2\left(\frac{f}{p}\right) =\frac{1}{N}\sum\limits_{n=1}{N}\left(\frac{f(x_n)}{p(x_n)}\right)^2-E^2.
\eeq
It becomes evident that the relevant quantity is now $f(x)/p(x)$ and it will be advantageous to choose $p(x)$ as close in shape to $f(x)$ as possible. In practice, one chooses $p(x)$ such that it approximates $|f(x)|$ reasonably well in shape and such that one can generate random numbers distributed according to $P(x)$.\\
One disadvantage of importance sampling is the fact, that it is dangerous to choose functions $p(x)$, which become zero, or which approach zero quickly. If $p$ goes to zero where $f$ is not zero, $S^2(f/p)$ may be infinite and the usual technique of estimating the variance from the sample points may not detect this fact if the region where $p=0$ is small. 

\subsubsection{The VEGAS-algorithm}

The techniques described before require some advanced knowledge of the behaviour of the function to be integrated. In many cases this information is not available and one prefers adaptive techniques, e.g. an algorithm which learns about the functions as it proceeds. In the following, the VEGAS-algorithm will be presented. It combines the basic ideas of importance sampling and stratified sampling into an iterative algorithm, which automatically concentrates evaluations of the integrand in those regions where the integrand is largest in magnitude. VEGAS starts by subdividing the integration space into a rectangular grid and performs an integration in each subspace. These results are then used to adjust the grid for the next iteration according to where the integral receives dominant contributions. In this way VEGAS uses importance sampling and tries to approximate the optimal probability density function
\beq
p_{\text{optimal}}(x) = \frac{|f(x)|}{\int dx|f(x)|}
\eeq
by a step function. Due to storage requirements one has to use a separable probability density function in $d$ dimensions:
\beq
p(u_1,\dots , u_d) = p_1(u_1)\cdot p_2(u_2) \cdot \dots \cdot p_d(u_d).
\eeq
Eventually after a few iterations the optimal grid is found. In order to avoid rapid destabilizing changes in the grid, the adjustment of the grid includes usually a damping term. After this initial exploratory phase, the grid may be frozen and in a second evaluation phase the integral may be evaluated with high precision according to the optimized grid. The separation in an exploratory phase and an evaluation phase allows one to use less integrand evaluations in the first phase and to ignore the numerical estimates from this phase (which will in general have a larger variance). Each iteration yields an estimate $E_j$ together with an estimate for the variance $S_j^2$:
\beq
E_j=\frac{1}{N_j}\sum\limits_{n=1}^{N_j}\frac{f(x_n)}{p(x_n)},\qquad S_j^2=\frac{1}{N_j}\sum\limits_{n=1}^{N_j}\left(\frac{f(x_n)}{p(x_n)}\right)^2-E^2_j.
\eeq
Here $N_j$ denotes the number of integrand evaluations on iteration $j$. The results of each iteration on the evaluation phase are combined into a cumulative estimate, weighted by the number of calls $N_j$ and their variances:
\beq
E=\left(\sum\limits_{j=1}^m\frac{N_j}{S_j^2}\right)^{-1}\left(\sum\limits_{j=1}^m\frac{N_jE_j}{S_j^2}\right).
\eeq
If the error estimates $S_j^2$ become unreliable (for example if the function is not square integrable), it is more appropriate to weight the partial results by the number $N_j$ of integrand evaluations alone. In addition VEGAS returns the $\chi^2$ per degree of freedom:
\beq
\chi^2/\operatorname{dof} = \frac{1}{m-1}\sum\limits_{j=1}^m\frac{(E_j-E)^2}{S_j^2}.
\eeq
This allows a check whether the various estimates are consistent. One expects a $\chi^2/\operatorname{dof}$ not much greater than one. 
% Chapter Template

\chapter{Loop--Tree Duality at One--Loop} % Main chapter title

\label{Chapter3} % Change X to a consecutive number; for referencing this chapter elsewhere, use \ref{ChapterX}

%\lhead{Chapter 3. \emph{Loop--Tree Duality at One--Loop}} % Change X to a consecutive number; this is for the header on each page - perhaps a shortened title

%----------------------------------------------------------------------------------------
%	Introduction
%----------------------------------------------------------------------------------------

%\section{Introduction}
%\label{sec:intro}

This chapter serves to introduce the Loop-Tree Duality method (LTD) 
\cite{Gleisberg:2007zz,Rodrigo:2008fp} which we will be using 
subsequently throughout this thesis. It is a technique to numerically 
calculate multi-leg one-loop cross-sections in perturbative field theories. 
In fact, it is applicable in any quantum field theory in Minkowsky space with an 
arbitrary number of space-time dimensions. 
At its core, it establishes a relation between loop-level and tree-level amplitudes 
similar to the Feynman Tree Theorem (FTT) \cite{Feynman:1963ax, Feynman:1972mt}. 
Both methods allow to write basisc 
loop Feynman diagrams in terms of tree-level phase-space integrals which 
are obtained by cutting the original loop-integral. This is achieved by directly 
applying the Residue Theorem\footnote{
	Within the context of loop integrals, the use of the residue 
	theorem has been considered many times in textbooks and in the literature.
}.  to the loop integrand. However, there are also some inportant differences 
between them: While the LTD produces only 
single cuts, the FTT also involves higher order cuts (double, triple, and so forth).

%The Feynman Tree Theorem (FTT) \cite{Feynman:1963ax, Feynman:1972mt} applies to any 
%(local and unitary) quantum field theories in Minkowsky space with an 
%arbitrary number $d$ of space-time dimensions. It relates perturbative 
%scattering amplitudes and Green's functions at the loop level with 
%analogous quantities at the tree level. This relation follows from a basic 
%and more elementary relation between loop integrals and phase-space integrals. 
%Using this basic relation loop Feynman diagrams can be rewritten in terms 
%of phase-space integrals of tree-level Feynman diagrams. The corresponding
%tree-level Feynman diagrams are then obtained by considering {\em multiple} 
%cuts (single cuts, double cuts, triple cuts and so forth) of the original 
%loop Feynman diagram.

%Recently, a new method \cite{Gleisberg:2007zz,Rodrigo:2008fp} to 
%numerically compute multi-leg one-loop cross sections in perturbative field 
%theories has been proposed. The starting point of this method, called 
%Loop--Tree Duality method (LTD), is a {\em duality} relation 
%between one-loop integrals and phase-space integrals. Although the analogy with
%the FTT is quite close, there are important differences. The key difference 
%is that the Loop--Tree Duality involves only {\em single} cuts of the 
%one-loop Feynman diagrams.
%Both the FTT and the LTD can be derived by using the 
%Residue Theorem\footnote{
%	Within the context of loop integrals, the use of the residue 
%	theorem has been considered many times in textbooks and in the literature.
%}. 

In this chapter, we closely follow Refs. \cite{Catani:2008xa,Bierenbaum:2010cy} 
\footnote{
	Ref. \cite{Catani:2008xa} is a good introduction 
	to the Duality relation. It gives further insight into its nature 
	and its comparison to the FTT. We refer the interested reader 
	to that paper. For the purpose of this thesis we will follow it 
	(and \cite{Bierenbaum:2010cy}) only as far 
	as necessary to understand the later chapters.
}
to illustrate and derive the Loop--Tree Duality.
%Since the FTT has attracted a renewed interest 
%\cite{Brandhuber:2005kd} in the context of twistor-inspired methods
%\cite{Witten:2003nn,Cachazo:2004zb} to evaluate one-loop scattering 
%amplitudes~\cite{Bern:2007dw}, we also give a small example to
%discuss its correspondence (including similarities and differences)
%with the duality relation.

%The outline of the chapter is as follows. In Section~\ref{sec:not}, we introduce 
%our notation. In Section~\ref{sec:ft}, we briefly recall how the FTT relates 
%one-loop integrals with multiple-cut phase-space integrals. In 
%Section~\ref{sec:dt}, we present one of the main results of this publication: 
%we derive and illustrate the duality relation between one-loop integrals 
%and single-cut phase-space integrals. We also prove that the duality relation 
%requires to properly regularize propagators by a complex Lorentz-covariant
%prescription, which is different from the customary $+i0$ prescription of the 
%Feynman propagators. The duality is illustrated in Section~\ref{sec:2p} by
%considering the two-point function as the simplest example application. 

%----------------------------------------------------------------------------------------
%	Notation
%----------------------------------------------------------------------------------------

\newpage

\section{Notation}
\label{sec:intro}

The FTT and the LTD can be illustrated with no loss of generality
by considering their application to the basic ingredient of any one-loop
Feynman diagrams, namely a generic one-loop scalar integral $L^{(1)}$ with 
$N$ ($N \geq 2$) external legs. In the following and throughout this thesis, 
we assume the considered diagrams to be free from UV and IR divergencies.

\begin{figure}[h]
\centering
\includegraphics[scale=1]{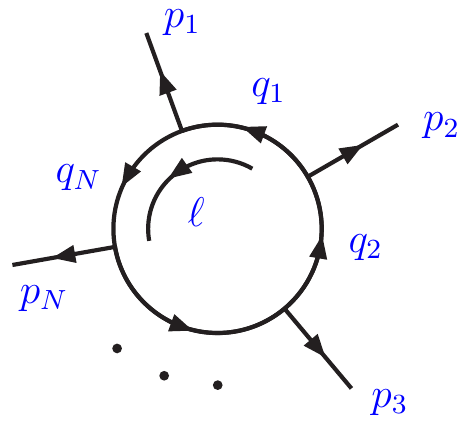}
%\vspace*{-6mm}
\caption{\label{f1loop} 
Momentum configuration of the one-loop $N$-point scalar integral.
}
\end{figure}

The momenta of the external legs are denoted by $p_1^\mu, p_2^\mu, \dots, p_N^\mu$
and are clockwise ordered (Fig.~\ref{f1loop}). All are taken as outgoing. To 
simplify the notation and the presentation, we also limit ourselves in the
beginning to considering massless internal lines only. Thus, the one-loop integral 
$L^{(1)}$ can in general be expressed as:  
\beq
\label{Ln}
L^{(1)}(p_1, p_2, \dots, p_N) = - i \, 
\int \frac{d^d \ell}{(2\pi)^d} \;
\prod_{i=1}^{N} \, \frac{1}{q_i^2+i 0} \;\;,
\eeq
where $\ell^\mu$ is the loop momentum (which flows anti-clockwise). The momenta
of the internal lines are denoted by $q_i^\mu$; they are given by
\beq
\label{defqi}
q_i = \ell + \sum_{k=1}^i p_k\;\;,
\eeq
and momentum conservation results in the constraint
\beq
\sum_{i=1}^N p_i = 0 \;\;.
\eeq
The value of the label $i$ of the external momenta is defined modulo 
$N$, i.e. $p_{N+i} \equiv p_{i}$.

The number of space-time dimensions is denoted by $d$ (the convention for the
Lorentz-indices adopted here is $\mu=0, 1, \dots, d-1$) with metric tensor 
$g^{\mu \nu} = {\rm diag}(+1,-1,\dots,-1)$. 
%Note that $d$ does not necessarily have 
%integer value, but it must satisfy $d \geq 1$ (as in the case of loop integrals in 
%dimensional regularization).
The space-time coordinates of any momentum $k_\mu$ are 
denoted as $k_\mu=(k_0, {\bf k})$, where $k_0$ is the energy (time component) of 
$k_\mu$. It is also convenient to introduce light-cone coordinates
$k_\mu=(k_+, \bf{k}_{\perp}, k_-)$, where $k_{\pm} = (k_0 \pm k_{d-1})/{\sqrt 2}$.
Throughout the chapter we consider loop integrals and phase-space integrals. If
the integrals are ultraviolet or infrared divergent, we always assume that they
are regularized by using analytic continuation in the number of space-time
dimensions (dimensional regularization). Therefore, $d$ is not fixed and 
does not necessarily have integer value.

%Throughout the paper we use 
We introduce 
the following shorthand notation:
\beq
- i \, \int \frac{d^d \ell}{(2\pi)^d} \;\;\bullet \equiv
\int_{\ell} \;\; \bullet \;\;.
\eeq
When we factorize off in a loop integral the integration over the momentum coordinate 
$\ell_0$ or $\ell_+$, we write
\beq
- i \, \int_{-\infty}^{+\infty} d\ell_0 \;\int \frac{d^{d-1}{\boldsymbol{\ell}}}{(2\pi)^d}
 \;\;\bullet \equiv \int d\ell_0 \;
\int_{\vec{\boldsymbol{\ell}}} \;\; \bullet \;\;,
\eeq
and
\beq
\label{pslc}
- i \, \int_{-\infty}^{+\infty} d\ell_+ \;\int_{-\infty}^{+\infty} d\ell_-
\int \frac{d^{d-2}{\boldsymbol{\ell}_{\perp}}}{(2\pi)^d}
 \;\;\bullet \equiv \int d\ell_+ \;
\int_{(\ell_-,\boldsymbol{\ell}_{\perp})} \;\; \bullet \;\;,
\eeq
respectively. The customary phase-space integral of a physical
particle with momentum $\ell$ (i.e. an on-shell particle with positive-definite 
energy: $\ell^2=0$, $\ell_0\geq 0$) reads
\beq
\label{psm}
\int \frac{d^d \ell}{(2\pi)^{d-1}} \;\theta(\ell_0) \;\delta(\ell^2) \;\;\bullet \equiv
\int_\ell \widetilde\delta (\ell) \;\;\bullet \;\;,
\eeq
where we have defined
\beq
\td(\ell) \equiv 2 \pi \, i \,\theta(\ell_0) \;\delta(\ell^2) 
= 2 \pi \, i \;\delta_+(\ell^2) \;\;.
\eeq

\begin{figure}[h]
\centering
\includegraphics[scale=1]{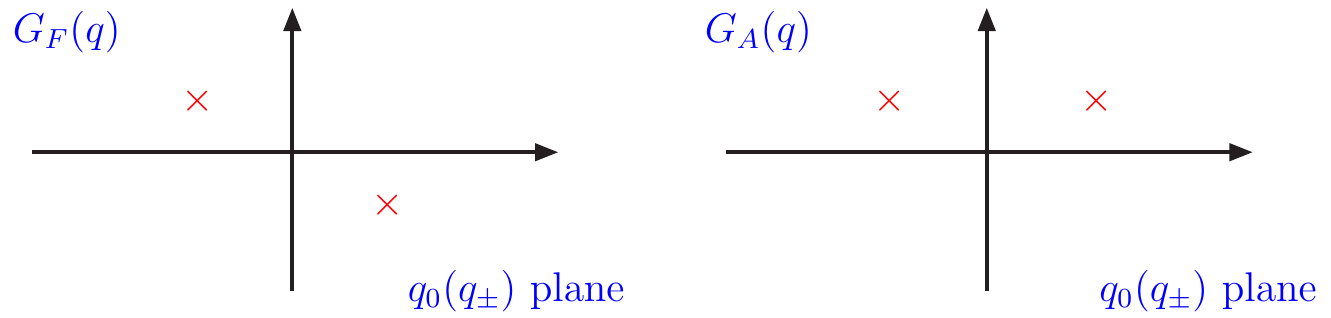}
\caption{\label{fvsa}
Location of the particle poles of the Feynman 
%$G(q)$ 
(left)
and advanced 
%$G_A(q)$ 
(right) propagators, $G_F(q)$ and $G_A(q)$,
in the complex plane of the variable 
$q_{0}$ or $q_{\pm}$.
}
\end{figure}

Using this shorthand notation, the one-loop integral $L^{(1)}$ in Eq.~(\ref{Ln}) 
can be cast into
\beq
\label{lng}
L^{(1)}(p_1, p_2, \dots, p_N) = \int_\ell \;\; \prod_{i=1}^{N} \,G_F(q_i) \;\;,
%G(q_1) \cdots G(q_N)~.
\eeq 
where $G_F(q)$ denotes the customary Feynman propagator,
\beq
G_F(q) \equiv \frac{1}{q^2+i0} \;\;.
\eeq
We also introduce the advanced propagator $G_A(q)$,
\beq
G_A(q) \equiv \frac{1}{q^2-i0\,q_{0}} \;\;.
\eeq
We recall that the Feynman and advanced propagators only differ in the 
position of the particle poles in the complex plane (Fig.~\ref{fvsa}). Using 
$q^2 = q_{0}^2 - {\bf q}^2= 2q_{+}q_{-} - {\bf q}_{\perp}^2$, we therefore have
\beq
\label{fpole}
\left[ G_F(q)\right]^{-1} = 0  \quad \Longrightarrow \quad
q_{0} = \pm  {\sqrt {{\bf q}^2 -i0}} \;\;, {\rm or} \;\;
q_{\pm} = \frac{ {\bf q}_{\perp}^2-i0}{2q_{\mp}} \;\;,
\eeq
and
\beq
\left[ G_A(q)\right]^{-1} = 0 \quad \Longrightarrow \quad
q_{0} \simeq \pm  {\sqrt {{\bf q}^2}} +i0 \;\;, {\rm or} \;\;
q_{\pm} \simeq \frac{ {\bf q}_{\perp}^2}{2q_{\mp}} +i0 \;\;.
\eeq
Thus, in the complex plane of the variable $q_0$ (or, equivalently\footnote{
	To be precise, each propagator leads to two poles in the plane $q_0$ 
	and to only one pole in the plane $q_+$ (or $q_-$).}, 
$q_{\pm}$), the pole with positive (negative) energy of the Feynman propagator is 
slightly displaced below (above) the real axis, while both poles (independently 
of the sign of the energy) of the advanced propagator are slightly 
displaced above the real axis.

%----------------------------------------------------------------------------------------
%	The Feynman Tree Theorem
%----------------------------------------------------------------------------------------

\section{The Feynman Tree Theorem}
\label{sec:ft}

In this section we briefly recall the FTT \cite{Feynman:1963ax,Feynman:1972mt}. 
To this end, we first introduce the advanced one-loop integral $L_A^{(1)}$, which 
is obtained from $L^{(1)}$ in Eq.~(\ref{lng}) by replacing the Feynman 
propagators $G_F(q_i)$ with the corresponding advanced propagators $G_A(q_i)$:
\beq
\label{lna}
L_A^{(1)}(p_1, p_2, \dots, p_N) = \int_{\ell} \;\; \prod_{i=1}^{N} \,G_A(q_i) \;\;.
\eeq 
Then, we note that 
\beq
\label{lna1}
L_A^{(1)}(p_1, p_2, \dots, p_N) = 0 \;\;.
\eeq 

\begin{figure}[h]
\centering
\includegraphics[scale=1]{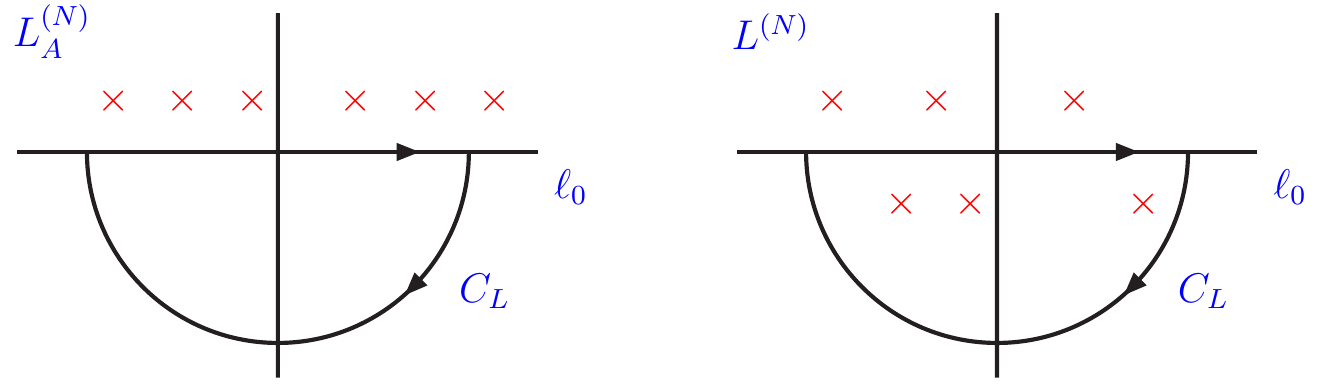}
\caption{\label{contour}
Location of poles and integration contour $C_L$ in 
	the complex $\ell_0$-plane for the advanced (left) and Feynman (right)
	one-loop integrals, $L_A^{(1)}$ and $L^{(1)}$.  
}
\end{figure}

The proof of Eq.~(\ref{lna1}) can be carried out in an elementary way by 
using the Cauchy Residue Theorem and choosing a suitable integration path
$C_L$. We have
\beeq
\label{lna2}
\lefteqn{L_A^{(1)}(p_1, p_2, \dots, p_N) = \int_{\vec{\boldsymbol{\ell}}} \;\;\; \int d\ell_0 \;
 \;\; \prod_{i=1}^{N} \,G_A(q_i) } \\
&=& \int_{\vec{\boldsymbol{\ell}}} \;\int_{C_L} d\ell_0 \;
 \;\; \prod_{i=1}^{N} \,G_A(q_i) 
= - \,2 \pi i \; \int_{\vec{\boldsymbol{\ell}}} \;\;\sum \; \res
 \;\left[ \;\prod_{i=1}^{N} \,G_A(q_i) \right] = 0 \nn\;\;. 
\eeeq 
The loop integral is evaluated by integrating first over the energy component 
$\ell_0$. Since the integrand is convergent when $\ell_0 \to \infty$, the $\ell_0$ 
integration can be performed along the contour $C_L$, which is
closed at $\infty$ in the lower half-plane of the 
complex variable $\ell_0$ (Fig.~\ref{contour}--{\em left}). The only 
singularities of the integrand with respect to the variable $\ell_0$ are 
the poles of the advanced propagators $G_A(q_i)$, which are located in the
upper half-plane. The integral along $C_L$ is then equal to 
the sum of the residues at the poles in the lower half-plane and therefore 
it vanishes. 

The advanced and Feynman propagators are related by
\beq
\label{gavsg}
G_A(q)=G_F(q)+\td(q) \;\;,
\eeq
which can straightforwardly be obtained by using the elementary identity 
\beq
\label{pvpid}
\frac{1}{x \pm i0} = {\rm PV}\left( \frac{1}{x} \right) \mp i \pi \, \delta(x) 
\;\;,
\eeq
where ${\rm PV}$ denotes the principal-value prescription. Inserting 
Eq.~(\ref{gavsg}) into the right-hand side of Eq.~(\ref{lna}) and collecting the 
contributions with an equal number of factors $G_F(q_i)$ and $\td(q_j)$, we obtain 
a relation between $L_A^{(1)}$ and the one-loop integral $L^{(1)}$:
\beeq
\label{lnavsln}
&& L_A^{(1)}(p_1, p_2, \dots, p_N) = \int_{\ell} \;\; \prod_{i=1}^{N} \,\left[
G_F(q_i)+\td(q_i) \right] \\
&& = L^{(1)}(p_1, p_2, \dots, p_N) + L_{\rm{1-cut}}^{(1)}(p_1, p_2, \dots, p_N)
+ \dots + L_{\rm{N-cut}}^{(1)}(p_1, p_2, \dots, p_N) \nn\;\;.
\eeeq
Here, the single-cut contribution is given by
\beq
\label{1cut}
L_{\rm{1-cut}}^{(1)}(p_1, p_2, \dots, p_N) = \int_{\ell} \;\; \sum_{i=1}^N
\; \td(q_i) \;
\prod_{\stackrel {j=1} {j\neq i}}^{N} 
\, G_F(q_j) \;\;.
\eeq
In general, the $m$-cut terms $L_{\rm{m-cut}}^{(1)}$ $(m \leq N)$ are the 
contributions with precisely $m$ delta functions $\td(q_i)$:
\beq
\label{lmcut}
L_{\rm{m-cut}}^{(1)}(p_1, p_2, \dots, p_N) = \int_{\ell} \;\; \left\{
\td(q_1) \dots \td(q_m) 
\; G_F(q_{m+1}) \dots G_F(q_{N}) 
+ {\rm uneq. \; perms.} \right\} \;\;,
\eeq
where the sum in the curly brackets includes all the permutations of
$q_1,\dots,q_N$ that give unequal terms in the integrand. 

Recalling that $L_A^{(1)}$ vanishes, cf.~Eq.~(\ref{lna1}), Eq.~(\ref{lnavsln})
results in:
\beq
\label{lftt}
L^{(1)}(p_1, p_2, \dots, p_N) = - \left[ \;
L_{\rm{1-cut}}^{(1)}(p_1, p_2, \dots, p_N)
+ \dots + L_{\rm{N-cut}}^{(1)}(p_1, p_2, \dots, p_N) \; \right] \;\;.
\eeq
This equation is the FTT in the specific case of the one-loop integral $L^{(1)}$. 
The FTT relates the one-loop integral $L^{(1)}$ to the multiple-cut\footnote{
	If the number of space-time dimensions is $d$, the right-hand side 
	of Eq.~(\ref{lftt}) receives contributions only from the terms with 
	$m \leq d$; the terms with larger values of $m$ vanish, since the 
	corresponding number of delta functions in the integrand is larger than 
	the number of integration variables.}
integrals $L_{\rm{m-cut}}^{(1)}$. Each delta function $\td(q_i)$ in 
$L_{\rm{m-cut}}^{(1)}$ replaces the corresponding Feynman propagator in $L^{(1)}$ 
by cutting the internal line with momentum $q_i$. This is synonymous to setting
the respective particle on shell. An $m$-particle cut decomposes the one-loop 
diagram in $m$ tree diagrams: in this sense, the FTT allows us to calculate 
loop-level diagrams from tree-level diagrams.

\begin{figure}[h]
\centering
\includegraphics[scale=1]{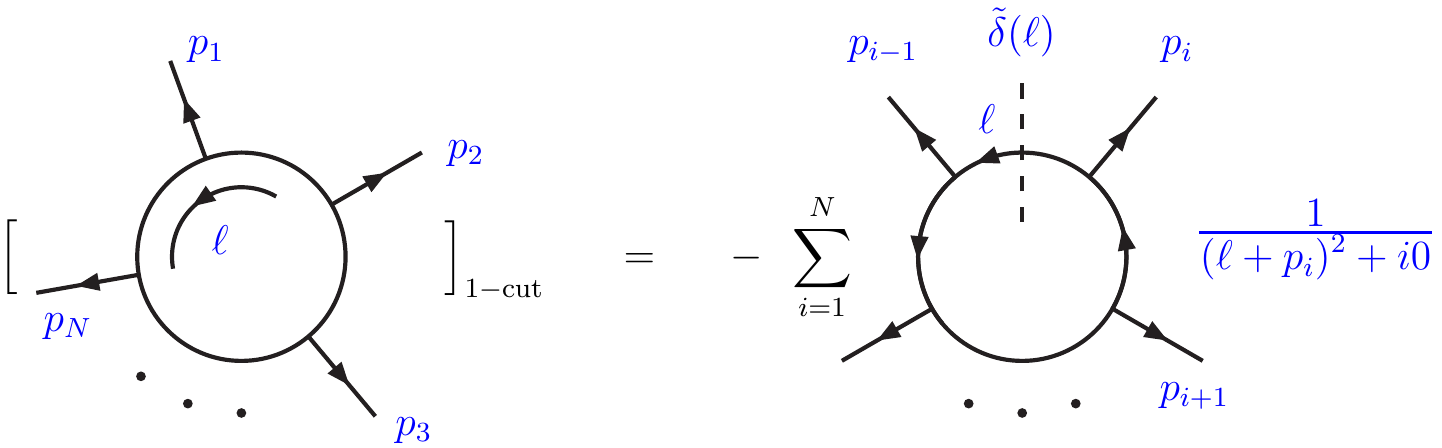}
\caption{\label{fig1cut}
The single-cut contribution of the Feynman Tree Theorem to the 
	one-loop $N$-point scalar integral. Graphical representation 
	as a sum of $N$ basic single-cut phase-space integrals.
}
\end{figure}

In view of the discussion in the following sections, it is useful 
to consider the single-cut contribution $L_{\rm{1-cut}}^{(1)}$ on the right-hand 
side of Eq.~(\ref{lftt}). In the case of single-cut contributions, the FTT 
replaces the one-loop integral with the customary one-particle phase-space 
integral, see Eqs.~(\ref{psm}) and (\ref{1cut}). Using the invariance of 
the loop-integration measure under translations of the loop momentum $\ell$,
we can perform the momentum shift $\ell \to \ell - \sum_{k=1}^i p_k$ in the term
proportional to $\td(q_i)$ on the right-hand side of Eq.~(\ref{1cut}). Thus,
cf.~Eq.~(\ref{defqi}), we have $q_i \to \ell$ and 
$q_j \to \ell + (p_{i+1} + p_{i+2} + \dots + p_{i+j})$, with $i \neq j$. We can 
repeat the same shift for each of the terms $(i=1,2,\dots,N)$ in the sum on the 
right-hand side of Eq.~(\ref{1cut}), and we can rewrite $L_{\rm{1-cut}}^{(1)}$ as 
a sum of $N$ basic phase-space integrals (Fig.~\ref{fig1cut}):
\beeq
\label{1cutsum}
L_{\rm{1-cut}}^{(N)}(p_1, p_2, \dots, p_N) &=& 
I_{\rm{1-cut}}^{(N-1)}(p_1, p_1+p_2, \dots, p_1+p_2+\dots+p_{N-1})
+ \,{\rm cyclic \;perms.} \nn \\
&=& \sum_{i=1}^N \;
I_{\rm{1-cut}}^{(N-1)}(p_i, p_i+p_{i+1}, \dots, p_i+p_{i+1}+\dots+p_{i+N-2})
\;\;.
\eeeq 
We denote the basic one-particle phase-space integrals with $n$ Feynman propagators
by $I_{\rm{1-cut}}^{(n)}$. They are defined as follows:
\beq
\label{i1cut}
I_{\rm{1-cut}}^{(n)}(k_1, k_2, \dots, k_n) = \int_{\ell} \td(\ell) \;
\prod_{j=1}^{n} G_F(\ell+k_j) = \int_{\ell} \td(\ell) \;\prod_{j=1}^{n} 
\frac{1}{2\ell\cdot k_j + k_j^2 + i0} \;\;.
\eeq

The extension of the FTT from the one-loop integrals $L^{(1)}$ to one-loop
scattering amplitudes ${\cal A}^{({\rm 1-loop})}$ (or Green's functions) in
perturbative field theories is straightforward, provided the corresponding
field theory is {\em unitary} and {\em local}. The generalization of 
Eq.~(\ref{lftt}) to arbitrary scattering amplitudes is \cite{Feynman:1963ax,Feynman:1972mt}:
\beq
\label{aftt}
{\cal A}^{({\rm 1-loop})} = - \left[ \;
{\cal A}^{({\rm 1-loop})}_{\rm{1-cut}} + {\cal A}^{({\rm 1-loop})}_{\rm{2-cut}}
+ \dots \; \right] \;\;,
\eeq
where ${\cal A}^{({\rm 1-loop})}_{\rm{m-cut}}$ is obtained in the same way as
$L_{\rm{m-cut}}^{(N)}$, i.e. by starting from ${\cal A}^{({\rm 1-loop})}$
and considering all possible replacements of $m$ Feynman propagators $G_F(q_i)$ of 
its loop internal lines with the `cut propagators' $\td(q_i)$.

The proof of Eq.~(\ref{aftt}) directly follows from Eq.~(\ref{lftt}):
${\cal A}^{({\rm 1-loop})}$ is a linear combination of one-loop integrals that
differ from $L^{(N)}$ only by the inclusion of interaction vertices. As briefly recalled below, this difference has 
harmless consequences on the derivation of the FTT.

Including particle masses in the advanced and Feynman propagators has an effect 
on the location of the poles produced by the internal lines in the loop. However, 
as long as the masses are {\em real}, as in the case of unitary theories, the 
position of the poles in the complex plane of the variable $q_0$ is affected only 
by a translation parallel to the real axis, with no effect on the imaginary part 
of the poles. This translation does not interfere with the proof of the FTT 
as given in Eqs.~(\ref{lna})--(\ref{lftt}). Therefore, the effect of a particle
mass $M_i$ in a loop internal line with momentum $q_i$ simply amounts to modifying 
the corresponding on-shell delta function $\td(q_i)$ when this line is cut to 
obtain ${\cal A}^{({\rm 1-loop})}_{\rm{m-cut}}$. This modification then leads to 
the obvious replacement:
\beq
\label{dmass}
\td(q_i) \to \td(q_i) = 2 \pi \, i \,\theta(q_{i, 0}) \;\delta(q_i^2-m_i^2) 
= 2 \pi \, i \;\delta_+(q_i^2-m_i^2) \;\;.
\eeq

Including interaction vertices has the effect of introducing numerator factors
in the integrand of the one-loop integrals. As long as the theory is local,
these numerator factors are at worst polynomials of the integration momentum $\ell$
\footnote{This statement is not completely true in the case of gauge theories and, 
	in particular, in the case of gauge-dependent quantities. For the discussion 
	of the additional issues that arise in gauge theories see Ref. \cite{Catani:2008xa} Section 9.} . 
In the  complex plane of the variable $\ell_0$, this polynomial behavior does not 
lead to additional singularities at any finite values of $\ell_0$. The only danger, 
when using the Cauchy theorem as in Eq.~(\ref{lna2}) to prove the FTT, stems from 
polynomials of high degree that can spoil the convergence of the $\ell_0$-integration 
at infinity. Nonetheless, if the field theory is unitary, these singularities at 
infinity never occur since the degree of the polynomials in the various integrands 
is always sufficiently limited by the unitarity constraint.

%----------------------------------------------------------------------------------------
%	The Loop--Tree Duality
%----------------------------------------------------------------------------------------

\section{The Loop--Tree Duality}
\label{sec:dt}

In this Section we derive and illustrate the Loop--Tree Duality relation between one-loop
integrals and single-cut phase-space integrals. This relation is the main
general result of this chapter.

Rather than starting from $L_A^{(1)}$, we directly apply the Residue Theorem 
to the computation of  $L^{(1)}$. We proceed exactly as in
Eq.~(\ref{lna2}), and obtain
\beeq
\label{ln4}
\lefteqn{L^{(1)}(p_1, p_2, \dots, p_N) = \int_{\vec{\boldsymbol{\ell}}} \;\;\; \int d\ell_0 \;
 \;\; \prod_{i=1}^{N} \,G_F(q_i) }\nn\\
&=& \int_{\vec{\boldsymbol{\ell}}} \;\int_{C_L} d\ell_0 \;
 \;\; \prod_{i=1}^{N} \,G_F(q_i)
= - \,2 \pi i \; \int_{\vec{\boldsymbol{\ell}}} 
 \;\;\sum \; \res
 \;\left[ \;\prod_{i=1}^{N} \,G_F(q_i) \right] \;\;. 
%\nn
\eeeq 
At variance with $G_A(q_i)$, each of the Feynman propagators $G_F(q_i)$ has single 
poles in both the upper and lower half-planes of the complex variable $\ell_0$ (see 
Fig.~\ref{contour}--{\em right}) and therefore the integral does not vanish as in 
the case of the advanced propagators. In contrast, here, the $N$ poles in the 
lower half-plane contribute to the residues in Eq.~(\ref{ln4}).

The calculation of these residues is elementary, but it involves several
subtleties. The detailed calculation, including a discussion of its subtle points, 
is presented in the following. 
%For the moment we limit ourselves to sketching the derivation of the result of this computation. 
 
The sum over residues in Eq.~(\ref{ln4}) receives contributions from $N$ terms,
namely the $N$ residues at the poles with negative imaginary part of each of the 
propagators $G_F(q_i)$, with $i=1,\dots,N$, see Eq.~(\ref{fpole}). 
Considering the residue at the $i$-th pole we write
\beq
\label{ipole}
{\rm Res}_{\{i{\rm-th \;pole}\}}
\;\left[ \;\prod_{j=1}^{N} \,G_F(q_j) \right] =
\left[ {\rm Res}_{\{i{\rm-th \;pole}\}} \;G_F(q_i) \right]
\;\left[ \;\prod_{\stackrel {j=1} {j\neq i}}^{N}
\,G_F(q_j) 
\right]_{\{i{\rm-th \;pole}\}} \;\;,
\eeq
where we have used the fact that the propagators $G_F(q_j)$, with $j\neq i$, are not 
singular at the value of the pole of $G_F(q_i)$. Therefore, they can be directly 
evaluated at this value.

%-----------------------------------
%	Evaluating the residues
%-----------------------------------

%\subsection{Evaluating the residues\label{sec:evres}}

%In Sect.~\ref{sec:dt} we have started the derivation of the Loop-Tree 
%duality relation in Eqs.~(\ref{ldt}) and (\ref{dcut})
%by using the Residue Theorem. The derivation is simple. However,
%it involves some subtle points. 
%in the evaluation of the residues.
%These points shall be discussed now.

Applying the Residue Theorem in the complex plane of the variable $\ell_0$, 
the computation of the one-loop integral $L^{(1)}$ reduces to the evaluation of
the residues at $N$ poles, according to Eqs.~(\ref{ln4}) and (\ref{ipole}).

The evaluation of 
%these residues 
the residues in Eq.~(\ref{ipole})
is a key 
%(and subtle) 
point in the derivation 
of the Loop--Tree Duality relation.
To make this point as clear as possible, we first introduce the notation
$q_{i,0}^{(+)}$ to explicitly denote the location of the $i$-th pole,
i.e. the location of the pole with negative imaginary part
(see Eq.~(\ref{fpole})) that is 
produced by the propagator $G_F(q_i)$.
We 
%prefer to 
further simplify our
notation with respect to the explicit dependence on the subscripts that label
the momenta. We write $G_F(q_j)= G_F(q_i + (q_j-q_i))$, where
$q_i$ depends on the loop momentum while $(q_j-q_i)=k_{ji}$ is a linear
combination of the external momenta (see Eq.~(\ref{defqi})). Therefore,
to carry out
the explicit computation of the $i$-th residue in Eq.~(\ref{ipole}),
%Eq.~(\ref{lnres1}), 
we re-label 
the momenta by $q_i \to q$ and $q_j \to q + k_j$, and we simply evaluate the 
term
\beq
\label{resid}
\left[ {\rm Res}_{\{q_0^{~}=q_{0}^{(+)}\}}
\,G_F(q) \right]
% \res
\;\left[ \;\prod_{j} \,G_F(q+ k_j) 
\right]_{q_0^{~}=q_{0}^{(+)}} \;\;,
\eeq
where (see Eq.~(\ref{fpole}))
\beq
q_0^{(+)} = {\sqrt {{\bf q}^2 -i0}} \;\;.
\eeq
In the next paragraphs, %we follow the steps of Sect.~\ref{sec:dt}
%(see Eqs.~(\ref{resGi}) and (\ref{respre})) and we
separately compute the residue of $G_F(q)$ and its
pre-factor -- the associated factor arising from the propagators $G_F(q+ k_j)$.

The computation of the residue of $G_F(q)$ gives
\beeq
{\rm Res}_{\{q_0^{~}=q_{0}^{(+)}\}} \,G_F(q) 
%&=& \lim_{q_0 \,\to \,q_{0}^{(+)}} 
%\; (q_0^{~} - q_{0}^{(+)}) \, G(q) = 
&=& \lim_{q_0^{~} \,\to \,q_{0}^{(+)}} 
\; \left\{ (q_0^{~} - q_{0}^{(+)}) \,\frac{1}{q_0^2 - {\bf q}^2 +i0 } \right\}= 
%\frac{1}{2 {\sqrt {{\bf q}^2 -i0}}} 
\frac{1}{2 \,q_{0}^{(+)}} 
\nn \\
\label{resg}
&=& \frac{1}{2 {\sqrt {{\bf q}^2}}} = \int dq_0 \;\delta_+(q^2) \;\;,
\eeeq
thus leading to the result in Eq.~(\ref{resGi}).
Note that the first equality in the second line of Eq.~(\ref{resg}) is 
obtained by removing the $i0$ prescription 
from the previous expression. This is fully justified. The term 
%$1/{\sqrt {{\bf q}^2 -i0}}$
$(q_{0}^{(+)})^{-1}=({\sqrt {{\bf q}^2 -i0}})^{-1}$ becomes singular 
when ${\bf q}^2 \to 0$,
and this corresponds to an end-point singularity in the integration over
${\bf q}$: therefore the $i0$ prescription  has no regularization effect on
such end-point singularity.
The second equality in the second line of 
Eq.~(\ref{resg}) simply follows from the definition of the on-shell delta
function $\delta_+(q^2)$, which we will later call the dual delta function.

%The calculation of the residue of $G_F(q_i)$ is straightforward and gives
Hence, the calculation of the residue of $G_F(q_i)$ gives 
\beq
\label{resGi}
\left[ {\rm Res}_{\{i{\rm-th \;pole}\}} \;G_F(q_i) \right] =
\left[ {\rm Res}_{\{i{\rm-th \;pole}\}} \;\frac{1}{q_i^2+i0} \right] 
= \int d \ell_0 \;\delta_+(q_i^2) \;\;.
\eeq
This result shows that considering the residue of the Feynman propagator of the 
internal line with momentum $q_i$ is equivalent to cutting that line by including the 
corresponding on-shell propagator $\delta_+(q_i^2)$. The subscript 
$+$ of $\delta_+$ refers to the on-shell mode with positive definite energy, 
$q_{i,0}=\sqrt{{\bf q}_i^2}$: the positive-energy mode is selected by the Feynman $i0$ 
prescription of the propagator $G_F(q_i)$. The insertion of Eq.~(\ref{resGi}) in 
Eq.~(\ref{ln4}) directly leads to a representation of the one-loop integral as a 
linear combination of $N$ single-cut phase-space integrals.

%-----------------------------------
%	Evaluating the product part
%-----------------------------------

%\subsection{Evaluating the product part \label{sec:prodpart}}

We now consider the evaluation of the residue pre-factor (the
second square-bracket factor in 
Eq.~(\ref{resid})). We first recall that the $i0$ prescription of the 
Feynman propagators has played an important role in the application 
(see Eqs.~(\ref{ln4}) and (\ref{resid}))
of the Residue Theorem to the computation of the loop integral: 
having 
selected the pole with negative imaginary part, $q_0^{~}=q_{0}^{(+)}$, 
the prescription eventually singled out 
the on-shell mode with positive definite energy, $q_0=\sqrt{{\bf q}^2}$ (see Eq.~(\ref{resg})). 
However, the evaluation
of the one-loop integrals by the direct application of the Residue Theorem
(as in Eq.~(\ref{ln4})) involves some subtleties. The subtleties mainly concern
the correct treatment of the Feynman $i0$ prescription in the calculation 
of the residue pre-factors. A consistent treatment
requires the {\em strict}
computation of the residue pre-factor in Eq.~(\ref{resid}): 
the $i0$ prescription in both $G_F(q+ k_j)$ and $q_{0}^{(+)}$ has to be dealt 
with by considering the imaginary part $i0$ as a {\em finite}, though
possibly small, quantity; the limit of infinitesimal values of $i0$ 
has to be taken only at the {\em very} end of the computation, 
thus leading to
%providing 
the interpretation of the
ensuing $i0$ prescription as mathematical distribution.
Applying this strict procedure, we obtain
\beeq
&&\left[ \;\prod_{j} \,G_F(q+ k_j) 
\right]_{q_0^{~}=q_{0}^{(+)}} =  
\left[ \;\prod_{j} \,\frac{1}{(q+ k_j)^2 + i0} 
\right]_{q_0^{~}=q_{0}^{(+)}} =
%\left[ \;\prod_{j} \,\frac{1}{2qk_j+ k_j^2} \right]_{q_0^{~}=q_{0}^{(+)}} \\
\prod_{j} 
\,\frac{1}{2q_{0}^{(+)}k_{j0} - 2{\bf q}\cdot {\bf k}_j+ k_j^2} \nn \\
\label{prefact}
&& \quad = \prod_{j} 
\,\frac{1}{2 |{\bf q}| k_{j0} - 2{\bf q\cdot k}_j+ k_j^2 - i0 k_{j0}/|{\bf q}|}
 = \left[ \prod_{j} 
\,\frac{1}{ 2qk_j+ k_j^2 - 
i0 k_{j0}/q_0} \right]_{q_0=|{\bf q}|}
\;\;.
\eeeq
The last equality on the first line of Eq.~(\ref{prefact}) simply follows from
setting $q_0^{~}=q_{0}^{(+)}$ in the expression on the square-bracket (note,
in particular, that $q^2=-i0$). The first equality on the second line 
%of Eq.~(\ref{prefact})
follows from  $2q_{0}^{(+)} \simeq 2|{\bf q}| - i0/|{\bf q}|$ (i.e. 
from expanding $q_{0}^{(+)}$ at small values of $i0$).

The result in Eq.~(\ref{prefact}) for the residue pre-factor is well-defined
and leads to a well-defined (i.e. non singular) expression once it is inserted 
in Eq.~(\ref{ln4}). The possible singularities from each of
the propagators $1/(q+k_j)^2$ are regularized by the displacement produced by 
the associated imaginary amount $i0 k_{j0}/q_0$. 
Performing the limit of infinitesimal values of $i0$, only the sign 
of the $i0$ prescription (and not its actual magnitude) is relevant. 
Therefore, since $q_0$ is positive,
in Eq.~(\ref{prefact}) we can
perform the replacement $i0 k_{j0}/q_0 \to i0 \,\eta k_j$, where $\eta^\mu$
is %the vector $\eta^\mu=(\eta_0, {\bf 0})$ with $\eta_0 > 0$; we finally obtain
a {\em future-like} vector with 
\beq
\label{etadef}
\eta_\mu = (\eta_0, {\bf \eta}) \;\;, \;\; \quad \eta_0 \geq 0, 
\; \eta^2 = \eta_\mu \eta^\mu \geq 0 \;\;,
\eeq
i.e.~a $d$-dimensional vector that can be either light-like $(\eta^2=0)$ or 
time-like $(\eta^2 > 0)$ with positive definite energy $\eta_0$. Hence, we finally obtain
\beeq
\label{resideta}
\left[ \;\prod_{j} \,G_F(q+ k_j) 
\right]_{q_0^{~}=q_{0}^{(+)}} =  \left[ \prod_{j} 
\,\frac{1}{ (q+ k_j)^2 - 
i0 \,\eta k_{j}}
\right]_{q_0=|{\bf q}|}
\;\;.
\eeeq
After reintroducing the original labels of the
momenta of the loop integral according to the replacements $q \to q_i$, $k_j 
\to q_j - q_i$, see the discussion above Eq.~(\ref{resid}), the calculation of the residue pre-factor on the r.h.s. of Eq.~({\ref{ipole}) yields
\beq
\label{respre}
\left[ \;\prod_{j\neq i} \,G_F(q_j) \right]_{\{i{\rm-th \;pole}\}} = 
\left[ \;\prod_{j\neq i} \,\frac{1}{q_j^2 + i0} \right]_{\{i{\rm-th \;pole}\}}
 = \prod_{j\neq i} \; \frac{1}{q_j^2 - i0 \,\eta (q_j-q_i)}
\;\;,
\eeq
%where $\eta$ is 
Note that the 
calculation of the residue at the pole of the internal line with momentum $q_i$ 
changes the propagators of the other lines in the loop integral. Although the 
propagator of the $j$-th internal line still has the customary form $1/q_j^2$, its 
singularity at $q_j^2=0$ is regularized by a different $i0$ prescription: the 
original Feynman prescription $q_j^2 +i0$ is modified in the new prescription 
$q_j^2 - i0 \,\eta (q_j-q_i)$, which we name the `dual' $i0$ prescription or, 
briefly, the $\eta$ prescription. The dual $i0$ prescription arises from the 
fact that the original Feynman propagator $1/(q_j^2 +i0)$ is evaluated at the 
{\em complex} value of the loop momentum $q$, which is determined by the location of 
the pole at $q_i^2+i0 = 0$. The $i0$ dependence from the pole has to be combined with the 
$i0$ dependence in the Feynman propagator to obtain the total dependence as given by 
the dual $i0$ prescription. The presence of the vector $\eta_\mu$ is a consequence of using 
the Residue Theorem. To apply it to the calculation of the $d$ dimensional loop 
integral, we have to specify a system of coordinates (e.g.~space-time or light-cone 
coordinates) and select one of them to be integrated over at fixed values of the 
remaining $d-1$ coordinates. Introducing the auxiliary vector $\eta_\mu$ with 
space-time coordinates $\eta_\mu =(\eta_0, {\bf 0_{\perp}}, \eta_{d-1})$, 
the selected system of coordinates can be denoted in a Lorentz-invariant form.
Applying the Residue Theorem in the complex plane of the variable $q_0$ at fixed 
(and {\em real}) values of the coordinates $\qt$ and $q_{d-1}^\prime =
q_{d-1} - q_0 \eta_{d-1}/\eta_0$ (to be precise,
in Eq.~(\ref{ln4}) we actually used $\eta_\mu = (1, {\bf 0})$), we obtain the
result in Eq.~(\ref{respre}). 

%-----------------------------------
%	Bringing the results together
%-----------------------------------

%\subsection{Bringing the results together}

The $\eta$ dependence of the ensuing $i0$ prescription is thus a consequence of the 
fact that the residues at each of the poles are not Lorentz-invariant quantities. 
The Lorentz-invariance of the loop integral is recovered only after summing over 
all the residues.

\begin{figure}[h]
\centering
\includegraphics[scale=1]{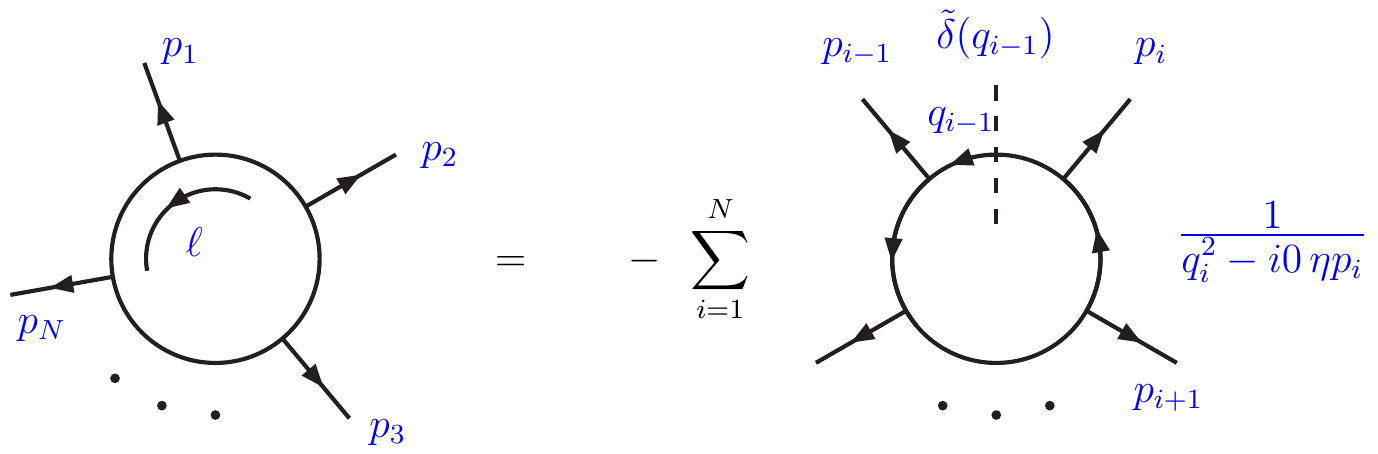}
\caption{\label{fig:dt}
The Duality relation for the one-loop $N$-point scalar integral.
Graphical representation as a sum of $N$ basic dual integrals.
}
\end{figure}
%%=============
%{\em The single-cut contribution of the Feynman Tree Theorem to the 
%one-loop $N$-point scalar integral.
%Graphical representation 
%as a sum of $N$ basic single-cut phase-space integrals.
%}}

Inserting the results of Eq.~(\ref{ipole})--(\ref{respre}) in Eq.~(\ref{ln4})
we directly obtain the Duality relation
between one-loop integrals and phase-space integrals:
\beq
\label{ldt}
L^{(1)}(p_1, p_2, \dots, p_N) = - \; {\widetilde L}^{(1)}(p_1, p_2, \dots, p_N)
\;\;,
\eeq
where the explicit expression of the phase-space integral ${\widetilde L}^{(1)}$ 
is (Fig.~\ref{fig:dt}) 
\bea
\label{dcut}
{\widetilde L}^{(1)}(p_1, p_2, \dots, p_N) = &\int_{\ell} \;\; \sum_{i=1}^N
\; \td(q_i) \;
\prod_{\stackrel {j=1} {j\neq i}}^{N} 
%\prod_{j=1 {\stackrel j\neq i}}^{N}
\;  G_D(q_i;q_j)  \;\;,\\
\label{eq:mlduprop}
\text{with}\quad &G_D(q_i;q_j)=\frac{1}{q_j^2 - i0 \,\eta k_{ji}} 
\quad \text{and} \quad k_{ji}=q_j-q_i,
\eea
where $\eta$ is the auxiliary vector defined in Eq.~(\ref{etadef}). Each of the $N-1$ 
propagators in the integrand is regularized by the dual $i0$ prescription and, 
thus, it is named `dual' propagator. Note that the momentum difference $q_j-q_i$ 
is independent of the integration momentum $\ell$: it only depends on the momenta of 
the external legs (see Eq.~(\ref{defqi})).

Using the invariance of the integration measure under translations of the momentum 
$\ell$, we can perform the same momentum shifts as described in Section \ref{sec:ft}.
In analogy to Eq.~(\ref{1cutsum}), we can rewrite Eq.~(\ref{dcut}) as a sum of $N$ 
basic phase-space integrals (Fig.~\ref{fig:dt}):
\beeq
\label{d1cutsum}
{\widetilde L}^{(1)}(p_1, p_2, \dots, p_N)
 &=& 
I^{(N-1)}(p_1, p_1+p_2, \dots, p_1+p_2+\dots+p_{N-1})
+ \,{\rm cyclic \;perms.} \nn \\
&=& \sum_{i=1}^N \;
I^{(N-1)}(p_i, p_i+p_{i+1}, \dots, p_i+p_{i+1}+\dots+p_{i+N-2})
\;\;.
\eeeq 
The basic one-particle phase-space integrals with $n$ dual propagators are denoted 
by $I^{(n)}$, and are defined as follows:
\beq
\label{idual}
I^{(n)}(k_1, k_2, \dots, k_n) =  \int_{\ell} \td(\ell) 
\;\; {\cal I}^{(n)}(\ell;k_1, k_2, \dots, k_n) =
\int_{\ell} \td(\ell) \;\prod_{j=1}^{n} \;
\frac{1}{2\ell k_j + k_j^2 - i0 \,\eta k_j} \,.
\eeq

We now comment on the comparison between the FTT (Eqs.~(\ref{1cut})--(\ref{i1cut})) 
and the Loop--Tree Duality (Eqs.~(\ref{ldt})--(\ref{idual})). The multiple-cut 
contributions $L_{\rm{m-cut}}^{(N)}$, with $m \geq 2$, of the FTT are completely 
absent from the Loop--Tree Duality relation, which only involves single-cut
contributions similar to those in $L_{\rm{1-cut}}^{(N)}$. However, 
the Feynman propagators present in $L_{\rm{1-cut}}^{(N)}$ are replaced by dual 
propagators in ${\widetilde L}^{(1)}$. This compensates for 
the absence of multiple-cut contributions in the Loop--Tree Duality. 

The $i0$ prescription of the dual propagator depends on the auxiliary vector 
$\eta$. The basic dual contributions $I^{(n)}$ are well defined for arbitrary values 
of $\eta$. However, when computing ${\widetilde L}^{(1)}$, the future-like vector 
$\eta$ has to be the {\em same} across all dual contributions 
(propagators): only then ${\widetilde L}^{(1)}$ does not depend on $\eta$.

In our derivation of the Loop--Tree Duality relation, the auxiliary vector $\eta$ originates 
from the use of the Residue Theorem. Independently of its origin, we can comment on 
the role of $\eta$ in the Duality relation. The one-loop integral 
$L^{(1)}(p_1, p_2, \dots, p_N)$ is a function of the Lorentz-invariants $(p_i p_j)$. 
This function has a complicated analytic structure, with pole and branch-cut 
singularities (scattering singularities), in the multidimensional space of the
complex variables $(p_i p_j)$. The $i0$ prescription of the Feynman propagators
selects a Riemann sheet in this multidimensional space and, thus, it unambiguously 
defines $L^{(1)}(p_1, p_2, \dots, p_N)$ as a single-valued function. Each 
single-cut contribution to ${\widetilde L}^{(1)}$ has additional (unphysical) 
singularities in the multidimensional complex space. The dual $i0$ prescription 
fixes the position of these singularities. The auxiliary vector $\eta$ 
{\em correlates} the various single-cut contributions in ${\widetilde L}^{(1)}$, 
so that they are evaluated on the same Riemann sheet: this leads to the 
cancellation of the unphysical single-cut singularities. In contrast, in the FTT, 
this cancellation is produced by the introduction of the multiple-cut 
contributions $L_{\rm{m-cut}}^{(N)}$.

\setcounter{footnote}{0}

We remark that the expression (\ref{d1cutsum}) of ${\widetilde L}^{(1)}$ as a sum 
of dual contributions is just a matter of notation: for massless internal 
particles ${\widetilde L}^{(1)}$ is actually a {\em single} phase-space integral 
whose integrand is the sum of the terms obtained by cutting each of the internal 
lines of the loop. In explicit form, we can write:
\beq
\label{dlsin}
{\widetilde L}^{(1)}(p_1, p_2, \dots, p_N) =
\int_{\ell} \td(\ell) \;\sum_{i=1}^N \;
{\cal I}^{(N-1)}(\ell;p_i, p_i+p_{i+1}, \dots, p_i+p_{i+1}+\dots+p_{i+N-2})
\;\;,
\eeq
%\beeq
%\label{dlsin}
%\!\!\!\!\!\!\!\!\!&&\!\!\!\!\!\! \!\!\!\!
% {\widetilde L}^{(N)}(p_1, p_2, \dots, p_N) \nn \\
%&& \quad = \int_q \td(q) \;\left\{ 
%{\cal I}^{(N-1)}(q;p_1, p_1+p_2, \dots, p_1+p_2+\dots+p_{N-1})
%+ \,{\rm cyclic \;perms.} \right\}  \,.
%\eeeq
where the function ${\cal I}^{(n)}$ is the integrand of the dual contribution in
Eq.~(\ref{idual}). Therefore, the Loop--Tree Duality relation (\ref{ldt}) directly expresses
the one-loop integral as the phase-space integral of a tree-level quantity. To 
name Eq.~(\ref{ldt}), we have introduced the term `duality' precisely to point out 
this direct relation\footnote{The word duality also suggests a stronger 
	(possibly one-to-one) correspondence between dual 
	integrals and loop integrals, which is further discussed in Ref. \cite{Rodrigo:2008fp}  Sect. 7.}
between the $d$-dimensional integral over the loop momentum and the
$(d-1)$-dimensional integral over the one-particle phase-space. For the 
FTT, the relation between loop-level and tree-level quantities is more involved, 
since the multiple-cut contributions $L_{\rm{m-cut}}^{(N)}$ (with $m \geq 2$) 
contain integrals of expressions that correspond to the product of $m$
tree-level diagrams over the phase-space for different number of particles.

The simpler correspondence between loops and trees in the context of the 
Loop--Tree Duality relation is further exploited in Ref. \cite{Rodrigo:2008fp} Sect. 10, where the 
Green's functions and scattering amplitudes are discussed.

%----------------------------------------------------------------------------------------
%	Explicit example: The scalar two-point function
%----------------------------------------------------------------------------------------

\section{Explicit example: The scalar two-point function}
\label{sec:2p}

In this Section we illustrate the application of the Loop--Tree Duality
relation to the evaluation of the one-loop two-point function $L^{(2)}$. A detailed 
discussion (including a comparison between FTT and Loop--Tree Duality as 
well as detailed results in analytic form and numerical results) 
of higher-point functions is found in \cite{Catani:2008xa} (see also 
Ref.~\cite{Gleisberg:2007zz}).

\begin{figure}[h]
\centering
\includegraphics[scale=1]{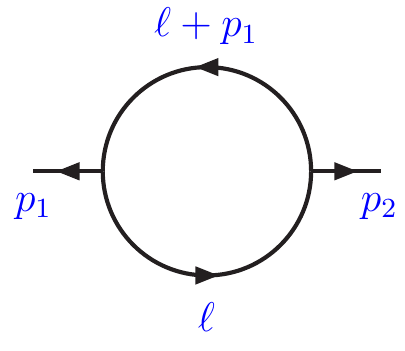}
\caption{\label{fig:bu}
The one-loop two-point scalar integral $L^{(2)}(p_1,p_2)$.
%function.
}
\end{figure}

The two-point function (Fig.~\ref{fig:bu}), also known as bubble function $\b$, 
is the simplest non-trivial one-loop integral with massless internal lines:
\beq
\label{l2}
\b (p_1^2) \equiv L^{(2)}(p_1,p_2) = - i \, 
\int \frac{d^d \ell}{(2\pi)^d} \;
%\frac{1}{q^2+i 0} \;\frac{1}{(q+p_1)^2+i 0} \;\;.
\frac{1}{\left[ \ell^2+i 0 \right] \left[(\ell+p_1)^2+i 0 \right]} \;\;.
\eeq
Here, we have visibly implemented momentum conservation $(p_1+p_2=0)$ and exploited
Lorentz invariance ($L^{(2)}(p_1,p_2)$ can only depend on $p_1^2$, which is the
sole available invariant). Since most of the one-loop calculations have been 
carried out in four-dimensional field theories (or in their 
dimensionally-regularized versions), we set $d=4-2\ep$. Note, however, that we 
present results for arbitrary values of $\ep$ or, equivalently, for any value 
$d$ of space-time dimensions.

The result of the one-loop integral in Eq.~(\ref{l2}) is well known:
\beq
\label{bub}
\b (p^2) = c_{\Gamma} 
\;\frac{1}{\ep (1-2\ep)} \; \left( -p^2 -i 0 \right)^{-\ep} \;\;,
\eeq
where $c_{\Gamma}$ is the customary $d$-dimensional volume factor that appears
from the calculation of one-loop integrals:
\beq
c_{\Gamma} \equiv \frac{\Gamma(1+\epsilon) \; 
\Gamma^2(1-\epsilon)}{\left(4\pi\right)^{2-\epsilon}\Gamma(1-2\epsilon)} \;\;.
\eeq

We recall that the $i0$ prescription in Eq.~(\ref{bub}) follows from the
corresponding prescription of the Feynman propagators in the integrand of
Eq.~(\ref{l2}). The $i 0$ prescription defines $\b (p^2)$ as a single-value
function of the real variable $p^2$. In particular, it gives $\b (p^2)$ an
imaginary part with an unambiguous value when $p^2 > 0$:
\beq
\label{bubri}
\b (p^2) = c_{\Gamma} 
\;\frac{1}{\ep (1-2\ep)} 
\; \left( |p^2| \right)^{-\ep}
\left[ \, \theta(-p^2) + \theta(p^2) \;e^{i\pi \ep} \,
\right] \;\;.
\eeq

%-----------------------------------
%	General form of single-cut integrals
%-----------------------------------

\subsection{General form of single-cut integrals}

To apply the Loop--Tree Duality relation, we have to compute the single-cut 
integrals $I_{\rm{1-cut}}^{(1)}$ and $I^{(1)}$, respectively. Since these integrals 
only differ because of their $i 0$ prescription, we introduce a more general 
regularized version, $I_{\rm reg}^{(1)}$, of the single-cut integral. We define: 
\beq
\label{i1reg}
I_{\rm reg}^{(1)}(k;c(k)) = \int_{\ell} \td(\ell) 
\;\frac{1}{2\ell k + k^2 + i0 \,c(k)} = \int \frac{d^d \ell}{(2\pi)^{d-1}} \; 
\delta_+(\ell^2) \;\frac{1}{2\ell k + k^2 + i0 \,c(k)} \;\;.
\eeq
Although $c(k)$ is an arbitrary function of $k$, $I_{\rm reg}^{(1)}$ only depends 
on the sign of the $i0$ prescription, i.e.~on the sign of the function
$c(k)$: setting $c(k)=+1$ we recover $I_{\rm{1-cut}}^{(1)}$, cf.~Eq.~(\ref{i1cut}), 
while setting $c(k)=- \eta k$ we recover $I^{(1)}$ (see Eq.~(\ref{idual})).

The calculation of the integral in Eq.~(\ref{i1reg}) is elementary, and the
result is
\beq
\label{i1regres}
I_{\rm reg}^{(1)}(k;c(k)) = -
\frac{c_{\Gamma}}{2 \cos (\pi \ep)} 
\;\frac{1}{\ep (1-2\ep)} 
\;\left[ \frac{k^2}{k_0} - i0 \,k^2\,c(k) \right]^{-\ep}
\;\left[ k_0 - i0 \,k^2\,c(k) \right]^{-\ep}
\;\;.
\eeq
%\beq
%\label{i1regres}
%I_{\rm reg}^{(1)}(k;c(k)) = -
%\frac{c_{\Gamma}}{2 \cos (\pi \ep)} 
%\;\frac{1}{\ep (1-2\ep)} 
%\;\left[ -k^2 -i 0 \,k_0 \,c(k) \right]^{\ep} 
%\;\left[ - k^2 - i0 \,c(k) \right]^{-\ep}
%\;\left[ k^2 + i0 \,c(k) \right]^{-\ep}
%\;\;.
%\eeq
Note that the typical volume factor, ${\widetilde c}_{\Gamma}$, of the
$d$-dimensional phase-space integral is 
\beq
{\widetilde c}_{\Gamma} = \frac{\Gamma(1-\epsilon) \; 
\Gamma(1+2\epsilon)}{\left(4\pi\right)^{2-\epsilon}} \;\;.
\eeq
The factor $\cos (\pi \ep)$ in Eq.~(\ref{i1regres}) originates from the difference 
between ${\widetilde c}_{\Gamma}$ and the volume factor $c_{\Gamma}$ of the loop 
integral:
\beq
\frac{{\widetilde c}_{\Gamma}}{c_{\Gamma}} = 
\frac{\Gamma(1+2\epsilon) \, 
\Gamma(1-2\epsilon)}{\Gamma(1+\epsilon) \;\Gamma(1-\epsilon)} =
\frac{1}{\cos (\pi \ep)} \;\;.
\eeq
We also note that the result in Eq.~(\ref{i1regres}) depends on the sign of
the energy $k_0$. This follows from the fact that the integration measure
in Eq.~(\ref{i1reg}) has support on the future light-cone, which is selected
by the positive-energy requirement of the on-shell constraint 
$\delta_+(\ell^2)$. 

The denominator contribution $(2\ell k + k^2)$ in the integrand of  Eq.~(\ref{i1reg}) 
is positive definite in the kinematical region where $k^2>0$ and $k_0>0$. In this 
region the $i0$ prescription is inconsequential, and  $I_{\rm reg}^{(1)}$ has no 
imaginary part. Outside this kinematical region, $(2\ell k + k^2)$ can vanish, leading
to a singularity of the integrand. The singularity is regularized by the $i0$ 
prescription, which also produces a non-vanishing imaginary part. The result in 
Eq.~(\ref{i1regres}) explicitly shows these expected features, since it can be 
rewritten as 
\beeq
%\label{i1regri}
I_{\rm reg}^{(1)}(k;c(k)) 
%&=& -
%\frac{c_{\Gamma}}{2 \cos (\pi \ep)} 
%\;\frac{\left( |k^2| \right)^{-\ep}}{\ep (1-2\ep)} 
%\left\{ \theta(-k^2) \left[ \theta(c(k)) \;e^{-i\pi \ep}
%+ \theta(-c(k)) \;e^{+i\pi \ep} \right] \right. \nn \\
%&+& \left. \theta(k^2) \left[ \theta(k_0) + \theta(-k_0)
%\left(\theta(c(k)) \;e^{+ 2 i\pi \ep}
%+ \theta(-c(k)) \;e^{- 2 i\pi \ep} \right) \right]
%\right\} \nn \\
&=& - 
\frac{c_{\Gamma}}{2 \cos (\pi \ep)} 
\;\frac{\left( |k^2| \right)^{-\ep}}{\ep (1-2\ep)} 
\left\{ \theta(-k^2) \left[ \cos(\pi \ep) - i \,\sin(\pi \ep) \;\sg(c(k))
\right] \right. \nn \\
\label{i1regri}
&+& \left. \theta(k^2) \left[ \theta(k_0) + \theta(-k_0)
\left( \cos(2\pi \ep) + i \,\sin(2\pi \ep) \;\sg(c(k))
\right) \right]
\right\}
\;\;.
\eeeq
%\beeq
%\label{i1regri}
%I_{\rm reg}^{(1)}(k;c(k)) &=& -
%\frac{c_{\Gamma}}{2 \cos (\pi \ep)} 
%\;\frac{\left( |k^2| \right)^{-\ep}}{\ep (1-2\ep)} 
%\left\{ \theta(-k^2) \left[ \theta(c(k)) \;e^{-i\pi \ep}
%+ \theta(-c(k)) \;e^{+i\pi \ep} \right] \right. \nn \\
%&+& \left. \theta(k^2) \left[ \theta(k_0) + \theta(-k_0)
%\left(\theta(c(k)) \;e^{+ 2 i\pi \ep}
%+ \theta(-c(k)) \;e^{- 2 i\pi \ep} \right) \right]
%\right\} \nn \\
%&=& - 
%\frac{c_{\Gamma}}{2 \cos (\pi \ep)} 
%\;\frac{\left( |k^2| \right)^{-\ep}}{\ep (1-2\ep)} 
%\left\{ \theta(-k^2) \left[ \cos(\pi \ep) - i \,\sin(\pi \ep) \;\sg(c(k))
%\right] \right. \nn \\
%&+& \left. \theta(k^2) \left[ \theta(k_0) + \theta(-k_0)
%\left( \cos(2\pi \ep) + i \,\sin(2\pi \ep) \;\sg(c(k))
%\right) \right]
%\right\}
%\;\;.
%\eeeq
We note that the functions $\b (k^2)$ and $I_{\rm reg}^{(1)}(k;c(k))$ have 
different analyticity properties in the complex $k^2$ plane. The bubble function 
has a branch-cut singularity along the positive real axis, $k^2 > 0$. The 
phase-space integral $I_{\rm reg}^{(1)}(k;c(k))$ has a branch-cut singularity 
along the entire real axis if $k_0 < 0$, while the branch-cut singularity is 
placed along the negative real axis if $k_0 > 0$.

%-----------------------------------
%	Duality relation for the two-point function
%-----------------------------------

\subsection{Duality relation for the two-point function}

\begin{figure}[h]
\centering
\includegraphics[scale=1]{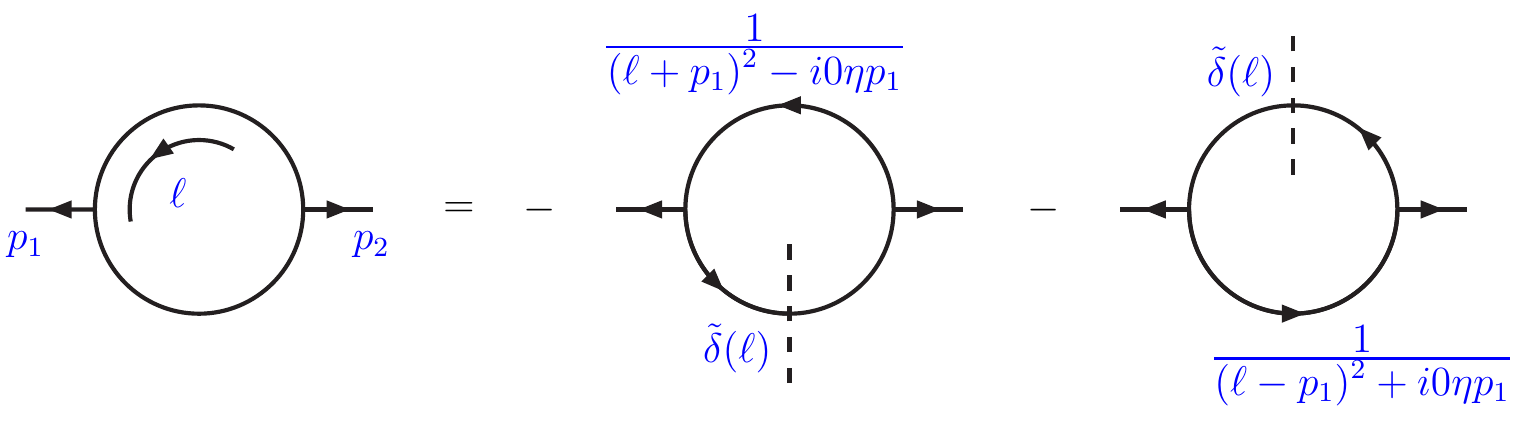}
\caption{\label{fig:bd}
One-loop two-point function: the Duality relation.
%{\em duality relation of the one-loop two-point function.}
}
\end{figure}

We now consider the Loop--Tree Duality (Fig.~\ref{fig:bd}) in the context of this
example. The dual representation of the one-loop two-point function is given by
\beq
\label{l2dual}
{\widetilde L}^{(2)}(p_1, p_2) = 
I^{(1)}(p_1) + \Bigl( p_1 \leftrightarrow - p_1 \Bigr) \;\;,
\eeq
cf.~Eqs.~(\ref{d1cutsum}) and (\ref{idual}). The dual contribution $I^{(1)}(k)$ 
is obtained by setting $c(k) = - \eta k$ in Eq.~(\ref{i1regres}). Since $\eta^\mu$ 
is a future-like vector, $c(k)$ has the following important property:
\beq
\label{etakey}
\sg(\eta k) = \sg(k_0) \;\;, \quad {\rm if} \;\; k^2 \geq 0 \;\;.
\eeq
Using this property, the result in Eq.~(\ref{i1regres}) can be written as 
\beq
\label{i1dual}
I^{(1)}(k) = - \;
\frac{c_{\Gamma}}{2} 
\;\frac{\left( - k^2 -i 0 \right)^{-\ep}}{\ep (1-2\ep)} 
\left[ 1 - i  \,\frac{\sin(\pi \ep)}{\cos(\pi \ep)} 
\;\sg(k^2 \eta k) \right] \;\;.
\eeq
%\beq
%I^{(1)}(k) = - \;
%\frac{c_{\Gamma}}{2 \cos (\pi \ep)} 
%\;\frac{\left( |k^2| \right)^{-\ep}}{\ep (1-2\ep)} 
%\left[ \theta(-k^2) + \theta(k^2) \;e^{+i\pi \ep} \right]
%\left[ \cos(\pi \ep) - i \;\sg(k^2 \eta k) \sin(\pi \ep)
%\right]
%\eeq
Comparing this expression with Eq.~(\ref{bub}), we see that the imaginary 
contribution in the square bracket is responsible for the difference with the 
two-point function. However, since $\sg(- \eta k) = - \sg(\eta k)$, this 
contribution is odd under the exchange $k \to -k$ and, therefore, it cancels 
when Eq.~(\ref{i1dual}) is inserted in Eq.~(\ref{l2dual}). Taken together,
\beq
\label{l2dualr}
{\widetilde L}^{(2)}(p_1, p_2) = 
I^{(1)}(p_1) + \Bigl( p_1 \leftrightarrow - p_1 \Bigr) =
- \;
c_{\Gamma} 
\;\frac{\left( - p_1^2 -i 0 \right)^{-\ep}}{\ep (1-2\ep)} 
\;\;,
\eeq
which fully agrees with the Duality relation 
${\widetilde L}^{(2)}(p_1, p_2) = - \,\b (p_1^2)$.

%----------------------------------------------------------------------------------------
%	Loop--Tree Duality with generic masses
%----------------------------------------------------------------------------------------

\section{Loop--Tree Duality with generic masses}
\label{sec:3sum}

This section serves to recapitulate the chapter and give the explicit formulae
 for generic masses as well as some explicit example calculations to illustrate 
 how the Loop--Tree Duality works in practice. 
 
 %-----------------------------------
%	Generic masses
%-----------------------------------

\subsection{Generic masses}
 
 The Feynman propagators $G_F(q_i)$ in \Eq{Ln} with real 
 internal masses $m_i$ read:
%============================================================
\beq
G_F(q_i) = \frac{1}{q_i^2-m_i^2+i0}~.
\eeq
%============================================================
The derivation of the Loop--Tree Duality Theorem is exactly the same regardless of the
internal lines being massive or massless ($m_i=0$), as long as the masses are
real. Non--vanishing internal real masses only account for a displacement of
the poles of the propagators along the real axis, which does not change the
derivation of the Duality Theorem, as will become obvious in the
following. Moreover, they do not alter the relationship between Feynman,
advanced, retarded and dual propagators, which is the basis of both, the
Duality Theorem as a duality to the FTT.\newline
Besides the customary Feynman propagators $G_F(q_i)$, we also encounter
advanced, $G_A(q_i)$, and retarded, $G_R(q_i)$, propagators, defined by:
%============================================================
\beq 
G_A(q_i) = \frac{1}{q_i^2-m_i^2-i0\,q_{i,0}}~,\qquad
G_R(q_i) = \frac{1}{q_i^2-m_i^2+i0\,q_{i,0}}~.
\eeq
%============================================================
The Feynman, advanced, and retarded propagators only differ in
the position of the particle poles in the complex plane. Using $q_i^2 =
q_{i,0}^2 - {\bf q}_i^2$, we therefore find the poles of the Feynman and
advanced propagators in the complex plane of the variable $q_{i,0}$ at:
%============================================================
\beq
\label{fpole2}
\left[ G_F(q_i)\right]^{-1} = 0  \:\:  \Longrightarrow \:\:
q_{i,0} = \pm  {\sqrt {{\bf q}_i^2 -m_i^2-i0}} \;\;
\:\:\: \mbox{and} \:\:\:
\left[ G_A(q_i)\right]^{-1} = 0 \:\: \Longrightarrow \:\:
q_{i,0} \simeq \pm  {\sqrt {{\bf q}_i^2 -m_i^2}} +i0~. 
\eeq
%============================================================
Thus, the pole with positive/negative energy of the Feynman propagator is
slightly displaced below/above the real axis, while both poles of the
advanced/retarded propagator, independently of the sign of the energy, are
slightly displaced above/below the real axis (cf. Fig.~\ref{fvsa}).  Similarly to the
 massless case, we further define
%============================================================
\beq
\label{tdp}
\td(q_i) \equiv 2 \pi \, i \, \theta(q_{i,0}) \, \delta(q_i^2-m_i^2) 
= 2 \pi \, i \, \delta_+(q_i^2-m_i^2)~,
\eeq
%============================================================
where again the subscript $+$ of $\delta_+$ refers to the on--shell mode with
positive definite energy, $q_{i,0}\geq 0$. Hence, the phase--space integral of
a physical particle with momentum $q_i$, i.e., an on--shell particle with
positive--definite energy, $q_i^2=m_i^2$, $q_{i,0}\geq 0$, reads:
%============================================================
\beq
%\label{psm}
\int \frac{d^d \ell}{(2\pi)^{d-1}} \, \theta(q_{i,0}) \, \delta (q_i^2-m_i^2) 
\; \cdots \equiv \int_{\ell} \td (q_i) \; \cdots~.
\eeq
In order to derive the Duality Theorem, one directly
applies the Residue Theorem to the computation of $L^{(1)}(p_1, p_2, \dots,
p_N)$ in \Eq{Ln}: Each of the Feynman propagators $G_F(q_i)$ has single poles
in both the upper and lower half--planes of the complex variable $\ell_{0}$.
Since the integrand is convergent when $\ell_{0}\to \infty$, by closing the
contour at $\infty$ in the lower half--plane and applying the Cauchy theorem,
the one--loop integral becomes the sum of $N$ contributions, each of them
obtained by evaluating the loop integral at the residues of the poles with
negative imaginary part belonging to the propagators $G_F(q_i)$.  The
calculation of the residue of $G_F(q_i)$ gives
%============================================================
\beq
\label{resGi35}
{\rm Res}  [   G_F(q_i) ]_{{\rm Im}(q_{i,0}) < 0} 
= \int d \ell_{0} \, \delta_+(q_i^2-m_i^2)~,
\eeq
%============================================================
with $\delta_+(q_i^2-m_i^2)$ defined in Eq.~(\ref{tdp}).  This result shows
that considering the residue of the Feynman propagator of the internal line
with momentum $q_i$ is equivalent to cutting that line by including the
corresponding on--shell propagator $\delta_+(q_i^2-m_i^2)$.  The propagators
$G_F(q_j)$, with $j\neq i$, are not singular at the value of the pole of
$G_F(q_i)$ and can therefore be directly evaluated at this point, yielding to
%============================================================
\beq
\label{respre35}
\prod_{j\neq i} \, G_F(q_j) 
\bigg|_{\substack{G_F(q_i)^{-1}=0\\ {\rm Im}(q_{i,0}) < 0}} = 
\prod_{j\neq i} \; G_D(q_i;q_j)~,
\eeq
%============================================================
where
%============================================================
\beq
G_D(q_i;q_j) = \frac{1}{q_j^2 -m_j^2 - i0 \,\eta\cdot k_{ji}}~,
\label{eq:dupro}
\eeq
%============================================================
is (massive) dual propagator %, as defined in Ref.~\cite{Catani:2008xa},
with $\eta$ a {\em future--like} vector, defined as in (\ref{etadef}).
%============================================================
%\beq
%\label{etadef}
%\eta_\mu = (\eta_0, {\bf \eta}) \;\;, \;\; \quad \eta_0 \geq 0, 
%\; \eta^2 = \eta_\mu \eta^\mu \geq 0 \;\;,
%\eeq
%============================================================
i.e.,~a $d$--dimensional vector that can be either light--like $(\eta^2=0)$ or
time--like $(\eta^2 > 0)$ with positive definite energy $\eta_0$ and $k_{ji}=q_j-q_i$. 
Collecting the 
results from \Eq{resGi35} and \Eq{respre35}, the Loop--Tree Duality
Theorem at one--loop takes the final form
%============================================================
\bea
\label{oneloopduality2}
L^{(1)}(p_1, p_2, \dots, p_N) \quad
= \quad - \sum\limits_i \, \int_{\ell} \; \td(q_i) \,
\prod_{\substack{j=1 \\ j\neq i}}^{N} \,G_D(q_i;q_j)~.
\eea 
%============================================================
%Contrary to the FTT, \cite{Feynman:1963ax}, \Eq{oneloopduality2} contains
%only single--cut integrals. Multiple--cut integrals, like those that appear in
%the FTT, are absent thanks to modifying the original $+i0$ prescription of the
%uncut Feynman propagators in \Eq{respre35} by the new prescription $- i0 \,\eta
%\cdot k_{ji}$, which is named the `dual' $i0$ prescription or, briefly, the
%$\eta$ prescription. 
%The dual $i0$ prescription arises from the fact that the original Feynman
%propagator $G_F(q_j)$ is evaluated at the {\em complex} value of the loop
%momentum $\ell$, which is determined by the location of the pole at
%$q_i^2-m_i^2+i0 = 0$.  The $i0$ dependence of the pole of $G_F(q_i)$ modifies
%the $i0$ dependence in the Feynman propagator $G_F(q_j)$, leading to the total
%dependence as given by the dual $i0$ prescription. The presence of the vector
%$\eta_\mu$ is a consequence of using the Residue Theorem and the fact that the
%residues at each of the poles are not Lorentz--invariant quantities.  The
%Lorentz--invariance of the loop integral is recovered after summing over all
%the residues. Furthermore, in the one--loop case, the momentum difference
%$\eta\cdot k_{ji}$ is independent of the integration momentum $\ell$, and only
%depends on the momenta of the external legs (cf. Eq.~(\ref{defqi})).

%-----------------------------------
%	Explicit example
%-----------------------------------

\subsection{Explicit example \label{sec:masslexmp}}

Here we present a very simple yet fully featured example of a triangle graph 
with generic internal masses. The integral to calculate is 
\bea
L^{(1)}(p_1,p_2,p_3)=\int_{\ell} G_{F}(q_1)G_{F}(q_2)G_{F}(q_3)~,
\eea
with the three Feynman propagators
\bea
G_{F}(q_1)=\frac{1}{q_1^2-m_1^2+i0}~,\quad
G_{F}(q_2)=\frac{1}{q_2^2-m_2^2+i0}~,
  \quad\text{and}\quad G_{F}(q_3)=\frac{1}{q_3^2-m_3^2+i0}
\eea
where $q_1=\ell+p_1$, $q_2=\ell+p_1+p_2=\ell$ and $q_3 = \ell$ 
according to the conventions 
established in Section \ref{sec:intro}. Applying Duality means using \Eq{dcut} 
to rewrite the integral $I$. A triangle has three internal lines, thus $N=3$. Integral $I$ takes 
the form:
\bea
L^{(1)}(p_1,p_2,p_3)=&\int_{\ell}\td(q_1)G_D(q_1;q_2)G_D(q_1;q_3) & &\text{first contribution}\nn\\
+&\int_{\ell}G_D(q_2;q_1)\td(q_2)G_D(q_2;q_3) & &\text{second contribution}\nn\\
+&\int_{\ell}G_D(q_3;q_1)G_D(q_3;q_2)\td(q_3) & &\text{third contribution}
\eea
Each summand of \Eq{dcut} is a dual contribution. 
In the first dual contribution the line carrying $q_1$ gets cut, i.e. it becomes a 
dual delta function, i.e. $\td(q_1)=2\pi i\; \theta(q_{1,0}) \; \delta(q_1^2)$. 
The Feynman propagators that correspond to the other internal 
line get promoted to a dual propagators as in \Eq{eq:mlduprop}. The first argument 
of the dual propagator indicates the cut, the second one the momentum of the 
internal line it is assigned to.\\
In order to produce dual contribution two, the next internal line 
(the one associated with $q_2$) gets cut, and all 
of the other lines (= lines carrying $q_1$ and $q_3$) converted to dual propagators, similarly to the first 
dual contribution.\\
To evaluate the dual delta functions we take 
advantage of 
\bea
\delta(g(x))=\sum\limits_{i=1}^n\frac{\delta(x-x_i)}{|g'(x_i)|}~,
\eea
where $x_i$ are the zeros of $g(x)$. Hence the dual deltas 
yield:
\bea
\td(q_1)&=\frac{\delta(\ell_0-(-p_{1,0}+\sqrt{(\boldsymbol{\ell}+\mathbf{p}_1)^2+m_1^2}))}{2\sqrt{(\boldsymbol{\ell}+\mathbf{p}_1)^2+m_1^2)}}~,\nn\\
\td(q_2)&=\frac{\delta(\ell_0-(-p_{1,0}-p_{2,0}+\sqrt{(\boldsymbol{\ell}+\mathbf{p}_1+\mathbf{p}_2)^2+m_2^2}))}{2\sqrt{(\boldsymbol{\ell}+\mathbf{p}_1+\mathbf{p}_2)^2+m_2^2)}}~,\nn\\
\td(q_3)&=\frac{\delta(\ell_0-\sqrt{\boldsymbol{\ell}^2+m_3^2})}{2\sqrt{\boldsymbol{\ell}^2+m_3^2}}~.
\eea
There is a crucial difference to note: The dual delta functions 
$\td(q_1)$, $\td(q_2)$ and $\td(q_3)$ force the zero component of the loop 
integration to different values. For example, in contribution one we 
have $q_{1,0}^{(+)}=\sqrt{\mathbf{q}_1^2}\Rightarrow \ell_0=-p_{1,0}
+\sqrt{(\boldsymbol{\ell}+\mathbf{p}_1)^2+m_1^2}$ whereas in 
contribution three the zero component is fixed to 
$q_{3,0}^{(+)}=\sqrt{\mathbf{q}_3^2+m_3^2}\Rightarrow \ell_0=
\sqrt{\boldsymbol{\ell}^2+m_3^2}$. This will heavily affect the structure 
of the dual contributions. Now we can insert for the dual 
propagators and apply the dual delta functions. To give the reader 
a better idea of how the outcome looks we write down contribution three, 
which we will call $I_3$, explicitly:
\bea
\label{eq:explmassexample}
I_3=-\int_{\boldsymbol{\ell}}&\frac{1}{2p_{1,0}\sqrt{\boldsymbol{\ell}^2+m_3^2}
+2\boldsymbol{\ell}\cdot \mathbf{p}_1-m_1^2+m_3^2
+p_1^2-i0\eta k_{13}}\cdot
\frac{1}{2\sqrt{\boldsymbol{\ell}^2+m_3^2}}\cdot\\
&\frac{1}{2(p_{1,0}+p_{2,0})\sqrt{\boldsymbol{\ell}^2+m_3^2}
+2\boldsymbol{\ell}\cdot (\mathbf{p}_1+\mathbf{p}_2)
+(p_1+p_2)^2-m_2^2+m_3^2-i0\eta k_{23}}\nn
\eea
The other two dual contributions look very similar. 
This is the final result that we obtain from using the Loop-Tree 
Duality. We see that it is a phase-space integration which 
runs only over the loop three momenta. 
We could now put numbers for $p_1$ and $p_2$ and perform an 
actual calculation. Therefore we would pass the integral 
to a numerical integrator as we will do in Chapter 
\ref{Chapter7}. In fact, Section \ref{sec:restriangle} 
shows the results for this exact triangle that we have 
discussed here.
% Chapter Template

\chapter{Loop--Tree Duality beyond One--Loop} % Main chapter title

\label{Chapter4} % Change X to a consecutive number; for referencing this chapter elsewhere, use \ref{ChapterX}

In the previous chapter, the Loop--Tree Duality method was 
introduced for the one-loop case. It has the appealing property of recasting the virtual 
corrections in a from which is very similar to the real ones, thus giving 
rise to the idea of directly combining the two. In the introduction, this 
Duality relation has been derived for the one--loop case. The purpose 
of this chapter is twofold.\\
First we want to establish a better-suited 
notation which then in turn allows the generalization of the Duality Theorem 
to situations involving two-loops. Therefore, a systematic procedure is 
presented which can be employed repeatedly to cover diagrams with even 
more loops. For the scope of this thesis it is sufficient to illustrate 
the technique at the two--loop level. Readers who are interested 
in applying it beyond two-lops are advised to check Ref. 
\cite{Bierenbaum:2010cy}.\\
Second, this chapter helps to prepare to following one, in which we 
will be dealing with higher order poles. At the one--loop level, these can 
be avoided through an adequate choice of gauge, however, at two--loops 
this no longer the case.

\newpage

%----------------------------------------------------------------------------------------
%	Duality relation at one--loop
%----------------------------------------------------------------------------------------

\section{Duality relation at one--loop}
\label{sec:one-loop}

It was shown in Chapter \ref{Chapter3} that using the Cauchy residue theorem
the one--loop integral can be written in the form:
\bea
\label{oneloopduality4}
L^{(1)}(p_1, p_2, \dots, p_N) 
\quad=\quad - \sum_{i} \, \int_{\ell_1} \; \td(q_i) \,
\prod_{\substack{j=1 \\ j\neq i}}^{N} \,G_D(q_i;q_j)~,
\eea 
where
%============================================================
\beq
G_D(q_i;q_j) = \frac{1}{q_j^2 -m_j^2 - i0 \,\eta (q_j-q_i)}~,
\eeq
%============================================================
is the so--called dual propagator, as defined in Ref.~\cite{Catani:2008xa},
with $\eta$ a {\em future--like} vector,
%============================================================
\beq
\label{etadef4}
 \eta_0 \geq 0, 
\; \eta^2 = \eta_\mu \eta^\mu \geq 0 \;\;,
\eeq
%%============================================
%\begin{figure}[t]
%\centering
%\includegraphics[scale=1]{Figures/OneLoopAmplitude.pdf}
%\caption{\label{f1loop} 
%{\em Momentum configuration of the one--loop $N$--point scalar integral.}}
%\end{figure}
%%============================================

% %%============================
% \begin{figure}[t]
% \begin{center}
% \begin{picture}(350,110)(0,0)
% \SetWidth{1.2}
% %
% \Line(0,50)(100,50)
% \Line(200,50)(300,50)
% \LongArrow(100,50)(150,50)
% \LongArrow(300,50)(350,50)
% \LongArrow(75,10)(75,90)
% \LongArrow(275,10)(275,90)
% %
% \Text(10,85)[]{$G_F(q_i)$}
% \Text(210,85)[]{$G_A(q_i)$}
% \Text(140,10)[]{$q_{i,0}$ plane}
% \Text(340,10)[]{$q_{i,0}$ plane}
% \Text(110,35)[]{$\times$}
% \Text(40,65)[]{$\times$}
% \Text(315,65)[]{$\times$}
% \Text(240,65)[]{$\times$}
% \end{picture}
% \end{center}
% \caption{\label{fvsa}
% {\em Location of the particle poles of the Feynman  (left)
% and advanced (right) propagators $G_F(q_i)$ and $G_A(q_i)$
% in the complex plane of the variable $q_{i,0}$.}}
% \end{figure}
% %%============================

The extension of the Duality Theorem to two loops has been discussed in detail in 
\cite{Bierenbaum:2010cy,Bierenbaum:2010dk}. Here we recall the basic points.
We extend the definition of propagators 
of single momenta to combinations of propagators of
sets of internal momenta. Let $\alpha_k$ be any set of internal momenta
$q_i,q_j$ with $i,j \in \alpha_k$.  We then define Feynman and dual
propagator functions of this set $\alpha_k$ in the following way:
%============================================================
\beq
\label{eq:multi}
G_{F} ( \alpha_k) = \prod_{i \in \alpha_k} G_{F}( q_i)~, \qquad
G_D( \alpha_k) = \sum_{i \in \alpha_k} \, \td(q_{i}) \, 
\prod_{\substack{j \in \alpha_k \\ j \neq i }} \, G_D( q_i; q_j)~.
\eeq
%============================================================
By definition, $G_D(\alpha_k)=\td(q_i)$, when $\alpha_k = \{i\}$ and thus
consists of a single four momentum. At one--loop order, $\alpha_k$ is
naturally given by all internal momenta of the diagram which depend on the
single integration loop momentum $\ell_1$, $\alpha_k=\{1,2,\ldots, N\}$.
However, let us stress that $\alpha_k$ can in principle be any set of
internal momenta. At higher order loops, e.g., several integration loop momenta
are needed, and we can define \emph{several loop lines}  $\alpha_k$ to label all the
internal momenta (cf. \Eq{lines}) where \Eq{eq:multi} will be used for these
loop lines or unifications of these. We also define:
%============================================================
\beq
\label{eq:multiminus}
G_D(-\alpha_k) = \sum_{i \in \alpha_k} \, \td(-q_{i}) \, 
\prod_{\substack{j \in \alpha_k \\ j \neq i }} \, G_D(-q_i;-q_j)~, 
%\qquad
%G_{A} (-\alpha_k) = \prod_{i \in \alpha_k} G_{A}(-q_i) = G_{R} (\alpha_k)~,
\eeq
%============================================================
where the sign in front of $\alpha_k$ indicates that we have reversed the
momentum flow of all the internal lines in $\alpha_k$. For Feynman
propagators, moreover, $G_F(-\alpha_k)=G_F(\alpha_k)$.
Using this notation the following relation holds for any set of
internal momenta $\alpha_k$:
%============================================================
\beq
G_A(\alpha_k) = G_F(\alpha_k) + G_D(\alpha_k)~,
\label{eq:relevant}
\eeq
%============================================================
where $G_A(q_i)$ is the advanced propagator:
\beq
G_A(q_i) = \frac{1}{q_i^2-m_i^2-i0\,q_{i,0}}~,
\eeq
and 
\beq
G_{A} ( \alpha_k) = \prod_{i \in \alpha_k} G_{A}( q_i)~.
\eeq
The proof of \Eq{eq:relevant} can be found in Ref. \cite{Bierenbaum:2010cy}.
Note that individual
terms in $G_D(\alpha_k)$ depend on the dual vector $\eta$, but the sum over
all terms contributing to $G_D(\alpha_k)$ is independent of it.  
Another crucial relation for the following is given by a formula that
allows to express the dual function of a set of momenta in terms of chosen
subsets. Considering the following set $\beta_N \equiv \alpha_1 \cup ... \cup
\alpha_N$, where $\beta_N$ is the unification of various subsets
$\alpha_i$, we can obtain the relation: 
\bea
\label{eq:GAinGDGeneralN}
G_D(\alpha_1 \cup \alpha_2 \cup ... \cup \alpha_N) 
% &=& \prod_{i=1}^N G_A(\alpha_i) - \prod_{i=1}^N G_F(\alpha_i)\nn \\ 
% &=& \prod_{i=1}^N \left[G_F(\alpha_i)+G_D(\alpha_i)\right] - \prod_{i=1}^N
% G_F(\alpha_i)\nn \\ 
\quad= \sum_{\substack{\beta_N^{(1)} \cup     \beta_N^{(2)}  = \beta_N}} \,
\prod_{\substack{i_1\in \beta_N^{(1)}}} \, G_D(\alpha_{i_1}) \,
\prod_{\substack{i_2\in \beta_N^{(2)}}} \,G_F(\alpha_{i_2})\,.  
\eea 
%============================================================
The sum runs over all partitions of $\beta_N$ into exactly two blocks
$\beta_N^{(1)}$ and $\beta_N^{(2)}$ with elements $\alpha_i,\, i\in
\{1,...,N\}$, where, contrary to the usual definition, we include the case:
$\beta_N^{(1)} \equiv \beta_N$, $\beta_N^{(2)} \equiv \emptyset$.
For the case of $N=2$, e.g.,
where $\beta_2 \equiv \alpha_1 \cup \alpha_2$, we have:
%============================================================
\beq
\label{eq:twoGD}
G_D(\alpha_1 \cup \alpha_2) 
= G_D(\alpha_1) \, G_D(\alpha_2)
+ G_D(\alpha_1) \, G_F(\alpha_2)
+ G_F(\alpha_1) \, G_D(\alpha_2)~.  
\eeq
%============================================================
Naturally it holds that:
\beq
G_F(\alpha_1 \cup \alpha_2 \cup ... \cup \alpha_N)= 
\prod_{i=1}^N G_F(\alpha_i)~.
\eeq
Since in general relation (\ref{eq:GAinGDGeneralN}) holds for any
set of basic elements $\alpha_i$ which are sets of internal momenta,
one can look at these expressions in different ways, depending on the given
sets and subsets considered. If we define, for example, the basic subsets
$\alpha_i$ to be given by single momenta $q_i$, and since in that case
$G_D(q_i) = \td(q_i)$, Eq.~(\ref{eq:GAinGDGeneralN}) then denotes a sum over
all possible differing m--tuple cuts for the momenta in the set $\beta_N$,
while the uncut propagators are Feynman propagators. These cuts start from
single cuts up to the maximal number of cuts given by the term where all the
propagators of the considered set are cut.
Using this notation, the Duality Theorem at one--loop can be written in
the compact form:
\beq
\label{eq:simpledual}
L^{(1)}(p_1, p_2, \dots, p_N) = - \int_{\ell_1} G_D(\alpha_1)~, 
\eeq
%============================================================
where $\alpha_1$ as in \Eq{defqi} labels {\it all} internal momenta $q_i$.
In this way, we directly obtain the Duality relation between one--loop
integrals and single--cut phase--space integrals and hence \Eq{eq:simpledual}
can also be interpreted as the application of the Duality Theorem to the given set
of momenta $\alpha_1$. It obviously agrees, at one loop, with
\Eq{oneloopduality4}. 

\begin{figure}[t]
\centering
\includegraphics[scale=1]{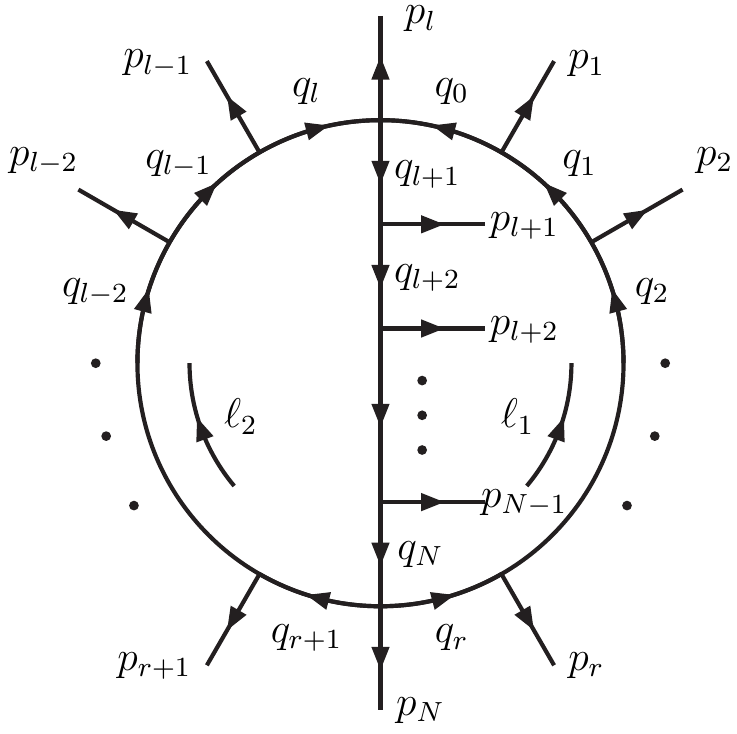}
\caption{\label{f2loop}
Momentum configuration of the two--loop $N$--point scalar integral.}
\end{figure}
%%============================================

%----------------------------------------------------------------------------------------
%	Duality relation at one-- and two--loops
%----------------------------------------------------------------------------------------

\section{Duality relation at two--loops}
\label{sec:two-loops}

We now turn to the general two--loop master diagram, as presented in
Figure~\ref{f2loop}.  Again, all external momenta $p_i$ are taken as outgoing,
and we have $p_{i,j}=p_i+p_{i+1}+\ldots +p_j$, with momentum conservation
$p_{1,N} = 0$. The label $i$ of the external momenta is defined modulo $N$,
i.e., $p_{N+i} \equiv p_{i}$. In the two--loop case, unlike at the one--loop order, the number of
external momenta might differ from the number of internal momenta. The loop momenta
are $\ell_1$ and $\ell_2$, which flow anti--clockwise and clockwise
respectively.  The momenta of the internal lines are denoted by $q_i$ and are
explicitly given by
%============================================================
\beq
\label{defqi2l}
q_i = \left\{
\begin{tabular}{ll}
$\ell_1+p_{1,i}$ & , $i \in \alpha_1$ \\
$\ell_2+p_{i,l-1}$ & , $i \in \alpha_2$ \\
$\ell_1+\ell_2+p_{i,l-1}$ &  , $i\in \alpha_3$ ~,
\end{tabular}
\right.  \eeq 
%============================================================
where $\alpha_k$, with $k=1,2,3$, are defined as the set of lines, propagators
respectively, related to the momenta $q_i$, for the following ranges of $i$:
%============================================================
\beq
\label{lines}
\alpha_1\equiv \{0,1,...,r\}~,\qquad \alpha_2\equiv \{r+1,r+2,...,l\}~,
\qquad \alpha_3\equiv \{l+1,l+2,...,N\}~.
\eeq
%============================================================
In the following, we will use $\alpha_k$
for denoting a set of indices or the set of the corresponding internal momenta
synonymously. Furthermore, we will refer to these lines often simply as the
``loop lines''.

We shall now extend the Duality theorem to the two--loop case, 
by applying \Eq{eq:simpledual} iteratively. We consider first, in 
the most general form, a set of several loop lines $\alpha_1$ to $\alpha_N$
depending on the same integration momentum $\ell_i$, and find
%============================================================
\beq
\label{eq:Applydual}
\int_{\ell_i} \; G_F(\alpha_1 \cup \alpha_2 \cup ... \cup \alpha_N) 
= 
- \int_{\ell_i} \; G_D(\alpha_1 \cup \alpha_2 \cup ... \cup \alpha_N)~,
\eeq
%============================================================
which states the application of the Duality Theorem, \Eq{eq:simpledual}, to the set
of loop lines belonging to the same loop. \Eq{eq:Applydual} is the
generalization of the Duality Theorem found at one--loop to a single loop of a
multi--loop diagram. Each subsequent application of the Duality Theorem to
another loop of the same diagram will introduce an extra single cut, and by
applying the Duality Theorem as many times as the number of loops, a given
multi--loop diagram will be opened to a tree--level diagram. The Duality
Theorem, \Eq{eq:Applydual}, however, applies only to Feynman propagators, and
a subset of the loop lines whose propagators are transformed into dual
propagators by the application of the Duality Theorem to the first loop might
also be part of another loop (cf., e.g., the ``middle'' line belonging to
$\alpha_3$ in Fig.~\ref{f2loop}). The dual function of the unification of
several subsets can be expressed in terms of dual and Feynman functions of the
individual subsets by using \Eq{eq:GAinGDGeneralN} (or \Eq{eq:twoGD}),
and we will use these expressions to transform part of the dual propagators
into Feynman propagators, in order to apply the Duality Theorem to the second
loop. Therefore, applying \Eq{eq:Applydual} to the loop with loop momentum $\ell_1$, 
reexpressing the result via \Eq{eq:twoGD} 
in terms of dual and Feynman propagators and applying \Eq{eq:Applydual} to the second
loop with momentum $\ell_2$, we obtain the Duality relation at two loops in the form:
%============================================================
\bea
\label{Advdual}
 \!\!\!\!\!\!\!\!\!\!
&L^{(2)}(p_1, p_2, \dots, p_N)  \\
& =   \int_{\ell_1} \int_{\ell_2} \, \left\{
- G_D(\alpha_1) \, G_F(\alpha_2) \, G_D(\alpha_3) 
+ G_D(\alpha_1) \, G_D(\alpha_2\cup \alpha_3)
+ G_D(\alpha_3) \, G_D(-\alpha_1\cup \alpha_2) \right\}~.\nn
 \!\!\!\!\!\!\!\!\!\!
\eea 
%============================================================
This is the dual representation of the two--loop scalar integral as a function
of double--cut integrals only, since all the terms of the integrand in
\Eq{Advdual} contain exactly two dual functions as defined in \Eq{eq:multi}.
The integrand in \Eq{Advdual} can then be reinterpreted as the sum over
tree--level diagrams integrated over a two--body phase--space.

The integrand in \Eq{Advdual}, however, contains several dual functions of two
different loop lines, and hence dual propagators whose dual $i0$ prescription
might still depend on the integration momenta. This is the case for dual
propagators $G_D(q_i;q_j)$ where each of the momenta $q_i$ and $q_j$ belong to
different loop lines. If both momenta belong to the same loop line the
dependence on the integration momenta in $\eta(q_j-q_i)$ obviously cancels,
and the complex dual prescription is determined by external momenta only. The
dual prescription $\eta(q_j-q_i)$ can thus, in some cases, change sign within
the integration volume, therefore moving up or down the position of the poles
in the complex plane. To avoid this, we should reexpress the dual
representation of the two--loop scalar integral in \Eq{Advdual} in terms of
dual functions of single loop lines. This transformation was unnecessary at
one--loop because at the lowest order all the internal momenta depend on the
same integration loop momenta; in other words, there is only a single loop
line.

Inserting \Eq{eq:twoGD} in \Eq{Advdual} and reordering some terms, we arrive
at the following representation of the two--loop scalar integral
%============================================================
\bea
\label{Advdualstar}
% \!\!\!\!\!\!\!\!\!\!
&L^{(2)}(p_1, p_2, \dots, p_N)  \\
&=   \int_{\ell_1} \int_{\ell_2} \, \left\{
  G_D(\alpha_1)  \, G_D(\alpha_2) \, G_F(\alpha_3) 
+ G_D(-\alpha_1) \, G_F(\alpha_2) \, G_D(\alpha_3)
+ G^*(\alpha_1)  \, G_D(\alpha_2) \, G_D(\alpha_3) \right\}~, \nn 
\eea 
%============================================================
where
%============================================================
\beq
\label{Gstar1}
G^*(\alpha_1) \equiv G_F(\alpha_1) + G_D(\alpha_1) + G_D(-\alpha_1)~.
\eeq
%============================================================
In \Eq{Advdualstar}, the
$i0$ prescription of all the dual propagators depends on external momenta
only. Through \Eq{Gstar1}, however, \Eq{Advdualstar} contains also triple
cuts, given by the contributions with three $G_D(\alpha_k)$. The triple cuts
are such that they split the two--loop diagram into two disconnected
tree--level diagrams. By definition, however, the triple cuts
are such that there is no more than one cut per loop line $\alpha_k$. Since
there is only one loop line at one--loop, it is also clear why we did not
generate disconnected graphs at this loop order.
For a higher number of loops, we expect to
find at least the same number of cuts as the number of loops, and topology-dependent 
disconnected tree diagrams built by cutting up to all the loop lines
$\alpha_k$.
These results can be generalized at three--loops and beyond without any additional effort. The 
reader is referred to \cite{Bierenbaum:2010cy} for further details. 
% Chapter Template

\chapter{Loop--Tree Duality beyond Simple Poles} % Main chapter title

\label{Chapter5} % Change X to a consecutive number; for referencing this chapter elsewhere, use \ref{ChapterX}

%\lhead{Chapter 4. \emph{Loop--Tree Duality beyond One--Loop}} % Change X to a consecutive number; this is for the header on each page - perhaps a shortened title

In the previous chapters, the derivation of the Duality relied on having only simple poles in the loop integral, 
i.e. it was only applicable to Feynman graphs that do not feature identical propagators. At one--loop this situation 
can always be avoided by a convenient choice of gauge \cite{Catani:2008xa}, but for two--loop and higher order 
corrections this isn't the case anymore. Hence, in this chapter we first want to address the issue of higher order poles. We will see that a straightforward application of the Residue Theorem similar to the one--loop case is feasible 
but leads to complex expressions. Therefore a second, more elegant, way is shown which involves the use of 
Integration By Parts (IBP) relations.

\newpage

%----------------------------------------------------------------------------------------
%	Duality relation for multiple poles
%----------------------------------------------------------------------------------------

\section{Duality relation for multiple poles}
\label{sec:double-poles}

% %%============================================
\begin{figure}[t]
\centering
\includegraphics[scale=1]{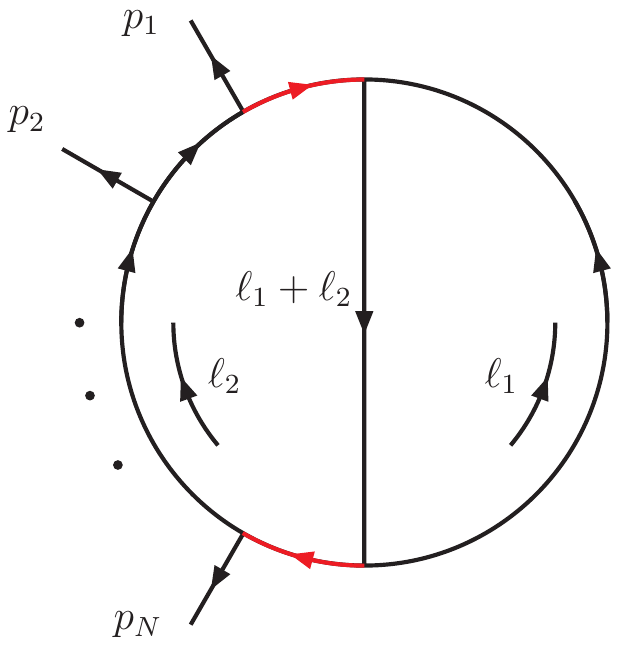}
\caption{\label{dpoleloop}
The two--loop $N$--point scalar integral with a double pole marked in red.}
\end{figure}

%============================================================
In the previous chapter we applied the Residue Theorem to one-- and two--loop graphs that
contain only single poles, i.e. no identical propagators. At one--loop this is always the case for a suitable 
choice of gauge \cite{Catani:2008xa}. However,
at higher loops there exists the possibility of identical propagators, i.e. higher order
poles \cite{Bierenbaum:2012th}. Obviously, we need to generalize the Duality Theorem to accommodate for such graphs.
The first occurrence of higher order poles is at the two--loop
level, with the sole double pole generic graph shown in Fig.~\ref{dpoleloop}. The Duality Theorem
can be derived for such graphs as well, using the Residue Theorem for multiple poles
\begin{align}
\text{Res}_{\{z=z_0\}}f(z)=\left.\frac{1}{(k-1)!}\left(\frac{d^{k-1}}{dz^{k-1}}(z-z_0)^{k}f(z)\right)\right|_{z=z_0}~.
\end{align}
The derivation follows similar steps as with the single pole case and is independent of any particular
coordinate system. We will derive an expression both in Cartesian and light--cone coordinates, to demonstrate
this independence. We start with the cartesian system.
We write the Feynman propagator in a form that makes the poles explicit, i.e,
\begin{align}
G_F(q_i)=\frac{1}{(q_{i0}-q_{i0}^{(+)})(q_{i0}+q_{i0}^{(+)})}~,
\end{align}
where $q_{i0}^{(+)}=\sqrt{\mathbf{q}_i^2+m_i^2-i0}$ is the position of the pole. Then, applying the Residue Theorem 
by selecting poles with negative imaginary part, we have
\begin{align}
\text{Res}_{\{\textrm{Im} \,  q_{i0}<0\}}G_F^2(q_i)=-\frac{2}{(2q_{i0}^{(+)})^3}=
-\int dq_0\frac{1}{2(q_{i0}^{(+)})^2}\delta_+(q_i^2-m_i^2).
\end{align}
The imaginary component of the new denominator $1/2(q_{i0}^{(+)})^2$ is irrelevant, because it is always a 
positive quantity. We refer the reader to \cite{Catani:2008xa} where the calculation for the case of simple poles 
is explained in more detail. Then, we assume the following Lorentz-invariant prescription of the residue
\begin{align}
\text{Res}_{\{\text{Im} \, q_{i0}<0\}}G_F^2(q_i)=-\int dq_0\frac{\eta^2}{2(\eta q_{i})^2}\delta_+(q_i^2-m_i^2)~,
\label{ResCart}
\end{align}
where $\eta^{\mu}=(\eta_0,\mathbf{0})$ is a future-like vector, $\eta_0>0$, in Cartesian coordinates.
Contrary to the one--loop case, where numerators depending on the loop momentum do not modify the Duality prescription, 
in the two--loop and higher orders cases the derivative in the residue calculation introduced by the higher order poles 
act on every single term in the numerator and also on the remaining propagators. 
Let $N(\alpha_k)$ be a function of a set of momenta $q_l$, with $l \in \alpha_k$. 
Then the residue of a double pole is given by
\begin{align}
\text{Res}_{\{\text{Im} \, q_{i0}<0\}} \left\lbrace G_F^2(q_i)\left(\prod\limits_{j\neq i}G_F(q_j)\right)N(\alpha_k) 
\right\rbrace =
\left.\frac{\partial}{\partial q_0}\frac{1}{(q_{i0}+q_{i0}^{(+)})^2}
\left(\prod\limits_{j\neq i}G_F(q_j)\right)N(\alpha_k)\right|_{q_{i0}=q_{i0}^{(+)}}\nonumber\\
=\left(\prod\limits_{j\neq i}G_D(q_i;q_j)\right)\frac{1}{(2q_{i0}^{(+)})^2}
\left[-\frac{1}{q_{i0}^{(+)}}-\sum\limits_{j\neq i}(2q_{j0})G_D(q_i;q_j)+\frac{\partial}{\partial q_0}\right]N(\alpha_k)~,
\end{align}
which can be written as
\begin{align}
\label{eq:doublepolecart}
\text{Res}_{\{\text{Im} \, q_{i0}<0\}} 
\left\lbrace G_F^2(q_i)\left(\prod\limits_{j\neq i}G_F(q_j)\right)N(\alpha_k) \right\rbrace =
\int dq_0\delta_+(q_i^2-m_i^2)\left(\prod\limits_{j\neq i}G_D(q_i;q_j)\right)\nonumber\\
\times\left[-\frac{\eta^2}{2(\eta q_i)^2}-\sum\limits_{j\neq i}\frac{\eta q_j}{\eta q_i}G_D(q_i;q_j)
+\frac{1}{2\eta q_i}\frac{\partial}{ \partial \eta q_i}\right]N(\alpha_k)~.
\end{align}

In light--cone coordinates we choose our coordinates such that in the plus component complex plane the poles 
with negative imaginary part are located at:
\begin{align}
q_{i+}^{(+)}=\frac{q_{i\perp}^2+m_i^2-i0}{2q_{i-}},\qquad\text{with}\quad q_{i-}>0~.
\end{align}
In these light--cone coordinates the Feynman propagator reads:
\begin{align}
G_F(q_i)=\frac{1}{2q_{i-}(q_{i+}-q_{i+}^{(+)})}~,
\end{align}
and thus
\begin{align}
\text{Res}_{\left\lbrace \text{Im\,}q_{i0}<0 \right\rbrace }\theta(q_{i-})G_F^2(q_i)=0~,
\end{align}
which, at first sight, seems to contradict equation \Eq{ResCart}. This contradiction can be resolved by taking into 
account the fact that in light cone coordinates, the dual vector $\eta$ is light-like and therefore $\eta^2=0$. 
Hence equation \Eq{ResCart} remains valid. Now, we are ready to calculate the residue of a double pole in light cone 
coordinates:
\begin{align}
\label{eq:double-lightcone}
&\text{Res}_{\{\text{Im\,}q_{i0}<0 \}} 
\left\lbrace \theta(q_{i-})G_F^2(q_i)\left(\prod\limits_{j\neq i}G_F(q_j)\right)N(\alpha_k) \right\rbrace =
\left.\frac{\theta(q_{i-})}{(2q_{i-})^2}\frac{\partial}{\partial q_+}
\left(\prod\limits_{j\neq i}G_F(q_j)\right)N(\alpha_k)\right|_{q_{i+}=q_{i+}^{(+)}}\nonumber\\
&=\int dq_+\delta_+(q_i^2-m_i^2)\left(\prod\limits_{j\neq i}G_D(q_i;q_j)\right)
\left[-\sum\limits_{j\neq i}\frac{\eta q_j}{\eta q_i}G_D(q_i;q_j)+\frac{1}{2\eta q_i}
\frac{\partial}{\partial\eta q_i}\right]N(\alpha_k)~,
\end{align}
where now $\eta^{\mu}=(\eta_+,\eta_-=0,\mathbf{0}_{\perp})$. \Eq{eq:double-lightcone} has the same
functional form as in \Eq{eq:doublepolecart}, although with a different dual vector $\eta$. Thus we can
generalize \Eq{eq:doublepolecart} and \Eq{eq:double-lightcone} to an arbitrary coordinate system and
combining simple and double poles in a single formula we get in Cartesian coordinates:
\begin{align}
\int_q G_F^2(q_i)&\left(\prod\limits_{j\neq i}G_F(q_j)\right)N(\alpha_k)=\nonumber\\
&-\int_q\Bigg\{\tilde{\delta}(q_i)\left(\prod\limits_{j\neq i}G_D(q_i;q_j)\right)
\left[-\frac{\eta^2}{2(\eta q_i)^2}-\sum\limits_{n\neq i}\frac{\eta q_j}{\eta q_i}G_D(q_i;q_j)
+\frac{1}{2\eta q_i}\frac{\partial}{\partial \eta q_i}\right]\nonumber\\
&+\sum\limits_{j\neq i}\tilde{\delta}(q_j)G_D^2(q_j;q_i)\left(\prod\limits_{k\neq i,j}G_D(q_j;q_k)\right)\Bigg\}
N(\alpha_k).
\label{doublepoleformula}
\end{align}

Equation (\ref{doublepoleformula}), is the main result of this section. It extends the Duality Theorem
to integrals with identical propagators or, to put it differently, with double poles in the complex plane.
For the case of the generic two--loop graph in Fig.~\ref{dpoleloop}, this result can be seen as an extension
of \Eq{eq:twoGD}. If we have two groups of momenta, $\alpha_k , \alpha_2$, one of which contains the 
double propagator, i.e. $\alpha_k=\left\lbrace q_n=\ell_1+\ell_2 \right\rbrace $ and 
$\alpha_2= \left\lbrace q_2=\ell_2,q_3=\ell_2+p_1,\ldots,q_{n-1}=\ell_2+p_{1,N-1},q_2=\ell_2  \right\rbrace $, 
and we denote by $\alpha_2^{\prime}$ a group that contains all the momenta of $\alpha_2$ leading to single poles, namely
\newline
$\alpha_2^{\prime}=\left\lbrace q_2=\ell_2,q_3=\ell_2+p_1,\ldots,q_{n-1}=\ell_2+p_{1,N-1} \right\rbrace$,
then we can write:
\bea
G_D(\alpha_k \cup \alpha_2)&= \tilde{\delta}(q_2)\left(\prod_{j \in \alpha_2^{\prime},\alpha_k }^{n}G_D(q_2;q_j)\right)
\left[-\frac{\eta^2}{2(\eta q_2)^2}-\sum_{j \in \alpha_2^{\prime},\alpha_k}^{n}\frac{\eta q_j}{\eta q_2}G_D(q_2;q_j)
\right] \nn \\
&+\quad\sum_{i \in \alpha_2^{\prime},\alpha_k}^{n}\tilde{\delta}(q_i)G_D^2(q_i;q_2)
\left(\prod\limits_{j \neq i}^{n} G_D(q_i;q_j)\right).
\eea
This result states that for the case of a double pole, one follows the usual procedure of cutting every propagator
line once, including the double propagator, and transforming the rest of the propagators to dual propagators.
A similar formula can be derived for the case of multiple (triple and higher) poles. 
The calculation of the residue of a multiple pole introduces, however, contributions with powers of dual propagators.
In absence of a general transformation formula analogous
to \Eq{eq:GAinGDGeneralN}, it is not possible to rewrite \Eq{doublepoleformula} in terms
of dual propagators whose dual $+i0$ prescription depends
on the external momenta only.  For that reason, we will present in the next section a different strategy 
for dealing with higher order poles based on the reduction of the integral using Integration By Parts.

%----------------------------------------------------------------------------------------
%	Reducing to single poles with IBPs
%----------------------------------------------------------------------------------------

\section{Reducing to single poles with IBPs}

In this section, we discuss a different approach to the generalization of the Duality Theorem to higher order
poles. We will use Integration By Parts (IBP) \cite{Chetyrkin:1981qh,Smirnov:2006ry}
to reduce the integrals with multiple poles to ones with simple poles. We emphasize the fact 
the \emph{we do not need to reduce the integrals to a particular integral basis}. 
We just need to reduce them "enough", so that the higher order poles disappear.

To give a short introduction to the method and establish our notation, let us consider a general $m$--loop scalar integral
in $d$ dimensions, with $n$ denominators $D_1,\ldots,D_n$ raised to exponents $a_1,\ldots,a_n$ 
and external momenta $p_1,\ldots,p_N$:
\beq
\int_{\ell_1}  \cdots \int_{\ell_m}  \frac{1}{D_1^{a_1} \cdots D_n^{a_n}} \; .
\eeq
If we notice that
\beq
\int_{\ell_1}  \cdots \int_{\ell_m}   
\frac{\partial}{\partial s^\mu}\frac{t^\mu}{D_1^{a_1} \cdots D_n^{a_n}} = 0 \; ,
\label{eq:IBPdef}
\eeq
where $s^\mu=\ell_1^\mu, \ell_2^\mu, \ldots, \ell_m^\mu$, the integrand being a total derivative with
respect to the loop momenta, we can find relations between scalar integrals with different exponents
$a_i$. This will allow us to express integrals with exponents larger than one, in terms of simpler ones. In effect,
we will be able to write integrals with multiple poles in terms of sums of integrals with simple poles.
In the numerator of the integrand of \Eq{eq:IBPdef} we can use 
$t^\mu=\ell_1^\mu, \ldots, \ell_m^\mu,p_1^\mu,\ldots,p_{N}^\mu$, to obtain a
system of equations that relate the various integrals. For simplicity, when referring to an IBP we will
use the shorthand notation:
\beq
\frac{\partial}{\partial s} \cdot t
\eeq
to denote Eq.~(\ref{eq:IBPdef}).
The differentiation will raise or leave an exponent
unchanged, while, contractions with the loop and external momenta in the numerator of the integrand, 
can be expressed in terms
of the propagators to lower an exponent. Often times though, this is not possible, leaving scalar products
of momenta, which cannot be expressed in terms of denominators. These are called Irreducible Scalar Products (ISP).
We will consider ISPs as additional denominators, $D_{ij}=\ell_i \cdot p_j$ \, , with a negative index $a_i$. 
We use the notation:
\beq
F(a_1 a_2 \cdots a_n)= \int_{\ell_1} \int_{\ell_2} \frac{1}{D_1^{a_1}D_2^{a_2} \cdots D_n^{a_n}}
\eeq
to denote a generic two--loop integral with $n$ propagators raised to an arbitrary integer power, with 
$D_i=G_F^{-1}(q_i)=q_i^2-m_i^2+i0$ and $q_i$ denotes any combination of external and loop momenta. In the following the 
prescription $+i0$ for the propagators is understood. 
We will use the symbol ${\bf a_i^+}$ to denote the raising of the index $a_i$ by one i.e. 
${\bf 1^+}F(a_1, a_2, \cdots a_n)=F(a_1+1, a_2, \ldots ,a_n)$ and the symbol ${\bf a_i^-}$ 
to denote the lowering of the index $a_i$ by one i.e. 
${\bf 2^-}F(a_1, a_2, \cdots a_n)=F(a_1, a_2-1, \ldots ,a_n)$. A combination of the two means that the operators
apply at the same time i.e. ${\bf 1^+ 2^-}F(a_1, a_2, \cdots a_n)=F(a_1+1, a_2-1, \ldots ,a_n)$.
In the following we will use two automated codes, for the reduction, {\tt FIRE} \cite{Smirnov:2008iw}, 
a {\tt MATHEMATICA} package for the reduction of integrals and {\tt REDUZE 2} \cite{vonManteuffel:2012np}
\footnote{Since the most obvious first approach seems to be to try to express the
  integrals with multiple poles in terms of the same integrals with
  only single poles, c.f. \Eq{eq:firstexample}, we used, in addition
  to the ``usual'' version of {\tt REDUZE 2}, in some cases a special
  patch for {\tt REDUZE 2} which provides a modification of its integral
  ordering in the final result. This modified version of {\tt REDUZE 2}
  delivered the results for the integrals in this desired form stated
  in the subsequent sections, while we used the normal version of the
  integral ordering for the remaining cases. Note
  that we also calculated explicitly the relations obtained from the modified version, in the easiest cases of the
  massless two-- and three--loop integrals which can be built by
  insertion of the the massless one--loop two--point function, and
  found agreement.} ,
a package written in C++, using {\tt GiNaC} \cite{Bauer:2000cp}.

%-----------------------------------
%	The case for two--loop diagrams
%-----------------------------------

\subsection{The case for two--loop diagrams}

The only generic two--loop scalar graph with $N$-legs and a double propagator is shown in Fig.~\ref{dpoleloop}. 
The simplest case is the two--point function with massless internal lines. 
The denominators are:
\beq
D_1=\ell_1^2 \; \; , \; \;  D_2=\ell_2^2 \; \; , \; \; D_3=(\ell_2+p)^2 \; \; , \; \; D_4=(\ell_1+\ell_2)^2 
\; \; , \; \; D_5 = \ell_1 \cdot p \, , \nn
\eeq
where we have introduced an ISP as an additional denominator 
\footnote{For {\tt REDUZE 2} the corresponding propagator is added and used as input instead.}. 
For the rest of this section the prescription $+i0$ for the propagators is understood. 
In our notation, the integral we want to reduce is $F(12110)$ and to this end we use the six total derivatives 
\beq
\frac{\partial}{\partial \ell_i} \cdot \ell_j \; , \; \frac{\partial}{\partial \ell_i} \cdot p
\; \; , \;\; i,j=1,2.
\label{eq:IBP2loop2p}
\eeq
Applying these IBPs on $F(a_1a_2a_3a_4a_5)$ we get a system of recursive equations. Using specific 
values for the exponents $a_i$ we can solve this system and obtain $F(12110)$. For this particular case, 
we solve the system explicitly and the reader is referred
to the Appendix \ref{app:systemsolve} for details. Finally we arrive at:
\beq
\label{eq:firstexample}
F(12110)=\frac{-1+3\epsilon}{(1+\epsilon)s} \; F(11110) \, ,
\eeq
where $s=p^2+i0$, a result which contains only single poles and can be treated using the Duality Theorem 
\cite{Bierenbaum:2010cy}. 
For the rest of the 
cases below and in the three--loop case in the next section, we have used {\tt FIRE} and {\tt REDUZE 2}
to perform the reductions and check our results.
For three external legs $p_1^2=p_2^2=0,p_3^2=(p_1+p_2)^2$ and massless internal lines, we have the denominators: 
\beq
D_1=\ell_1^2\; \; , \; \;D_2=\ell_2^2 \; \; , \; \; D_3=(\ell_2+p_1)^2 \; \; , \; \; D_4=(\ell_2+p_1+p_2)^2 \; \; , \; \;
D_5=(\ell_1+\ell_2)^2  \; \; , \; \; \nn
\eeq
\beq
D_6= \ell_1 \cdot p_1 \; \; , \; \; D_7=\ell_1 \cdot p_2  \; , \nn
\eeq
where the last two are the ISPs that appear in this case. The integral we want to reduce is $F(1211100)$.
We use eight IBPs:
\beq
\frac{\partial}{\partial \ell_i} \cdot \ell_j \; , \; \frac{\partial}{\partial \ell_i} \cdot p_j
\; \; , \;\; i,j=1,2.
\label{eq:IBP2loop3p}
\eeq
A similar analysis to the one above, gives:
\beq
 F(1211100)=\frac{3\epsilon}{(1+\epsilon)s} F(1111100)=
-\frac{3(1-3\epsilon)(2-3\epsilon)}{\epsilon (1+\epsilon)s^3} F(1001100) \; ,
\eeq
where $s=(p_1+p_2)^2+i0$,
which, again contains only single poles and can be treated with the Duality Theorem. 

The inclusion of masses does not affect the general picture of the reduction. It solely
introduces numerators in some integrals after the reduction is done. But, as
we have stressed already, the application of the Duality Theorem is not affected by numerators since
it only operates on denominators \cite{Bierenbaum:2010cy}. As an illustrative example, let us consider the two--loop
graph with two external legs and one massive loop (see Fig.~\ref{dpoleloop}). For the case of the left
loop being massive (related to $\ell_2$), with mass $m$, the denominators involved are
\beq
D_1=\ell_1^2 \; \; , \; \; D_2=\ell_2^2-m^2 \; \; , \; \; D_3=(\ell_2+p)^2-m^2 \; \; , \; \; 
D_4= (\ell_1+\ell_2)^2-m^2 \; ,\nn
\eeq
with the addition of the irreducible scalar product
\beq
D_5= \ell_1 \cdot p \; ,
\eeq
needed to perform the reduction. Using the same IBPs of \Eq{eq:IBP2loop2p}, the result of the reduction, with
{\tt FIRE} is:
\bea
F(12110)\quad=&\quad\frac{(\epsilon-1) \left[  -\epsilon s^2 +2 m^2 (9 \epsilon -2 \epsilon^2 -3) s +
   4 m^4 (-3+2\epsilon)(-1+2\epsilon)  \right]  }{2 \epsilon (2
   \epsilon-1) m^4 s \left(4 m^2-s\right)^2} F(00110) \nn \\
+&\quad\frac{A }{2 \epsilon (2 \epsilon-1) m^4 s \left(4m^2-s\right)^2}F(10110) 
-\frac{(\epsilon-1) }{2 (2 \epsilon-1) m^4 s} F(1-1110)  \nn \\
-&\quad\frac{(\epsilon-1)^2 \left(2 m^2-s\right)}{(2 \epsilon-1) m^4 s \left(4 m^2-s\right)} F(01010) 
-\frac{(\epsilon-1) \left(4 \epsilon m^2+2 m^2-s\right)}{2 (2 \epsilon-1) m^4 \left(4 m^2-s\right)} F(01110) \nn \\
+&\quad\frac{2 (\epsilon-1) \left(m^2-s\right) \left(10 \epsilon m^2-\epsilon s-3
   m^2\right) }{\epsilon (2 \epsilon-1) m^4 s \left(4 m^2-s\right)^2} F(1011-1) ,
\eea
with $s=p^2+i0$ and 
\beq
\label{eq:A}
A=(1-\epsilon) \left[ \epsilon s +2 (3-8\epsilon)m^2  \right]s^2 
+ 2 (1-2\epsilon) m^4 \left[2(3-4\epsilon)m^2-(6-5\epsilon)s \right] \; .
\eeq
The reduction generates two integrals with a numerator, namely
\bea
F(1-1110)\quad=&\quad\int_{\ell_1}\int_{\ell_2} \frac{\ell_2^2-m^2}{D_1D_3D_4} \; , \nn \\
F(1011-1)\quad=&\quad\int_{\ell_1}\int_{\ell_2} \frac{\ell_1 \cdot p}{D_1D_3D_4} \; , \nn
\eea
but the double poles have now disappeared. The result with {\tt REDUZE 2} reads:
\bea
  F(1 2 1 1 0)\quad=&\quad
    -
    \frac{(\epsilon-1) \left(4 \epsilon m^2+2
   m^2-s\right)}{2 (2 \epsilon-1) m^4 \left(4
   m^2-s\right)}F(0 1 1 1 0) \nn \\
    +&\quad
     \frac{3 (\epsilon-1) \left(8 \epsilon m^4-12 \epsilon
   m^2 s+\epsilon s^2-4 m^4+4 m^2 s\right)}{2
   \epsilon (2 \epsilon-1) m^4 s \left(4 m^2-s\right)^2}F(1 -1 1 1 0) \nn \\
    +&\quad
     \frac{A}{2 \epsilon (2
   \epsilon-1) m^4 s \left(4 m^2-s\right)^2}F(1 0 1 1 0) \nn \\
    +&\quad
     \frac{(\epsilon-1) \left(8 \epsilon^2 m^2 s-2 \epsilon^2 s^2-16
   \epsilon m^4+6 \epsilon m^2 s+\epsilon s^2+12
   m^4-6 m^2 s\right)}{2 \epsilon (2 \epsilon-1)
   m^4 s \left(4 m^2-s\right)^2}F(0 1 0 1 0) \; , \nn \\
\eea
where $A$ is given by \Eq{eq:A}. Despite the appearance of
different integrals the two results are of course equivalent. This is because, the integrals $F(00110)$ and
$F(01010)$, in the result obtained with {\tt FIRE}, are identical (as can be seen by shifting the loop momenta), 
so the sum of their coefficients gives exactly the coefficient of the result obtained with  {\tt REDUZE 2}. 
The same argument applies for the integrals $F(1011-1)$ and $F(1-1110)$.
The appearance of the numerators does not affect the application
of the Duality Theorem for integrals with single poles as was detailed in \cite{Bierenbaum:2010cy}.
For the case of the right loop in Fig.~\ref{dpoleloop} being massive (related to $\ell_1$), we have the denominators:
\beq
D_1=\ell_1^2-m^2 \; \; , \; \; D_2=\ell_2^2 \; \; , \; \; D_3=(\ell_2+p)^2 \; \; , \; \; 
D_4= (\ell_1+\ell_2)^2-m^2 \; \; , \; \; D_5= \ell_1 \cdot p \; . \nn
\eeq
Using the IBPs from \Eq{eq:IBP2loop2p}, we get with {\tt FIRE}:
\bea
F(12110)\quad=&\quad\frac{\left(32 \epsilon^2 m^4+8 \epsilon^2 m^2 s+\epsilon^2 s^2-32 \epsilon m^4-11 \epsilon m^2 s-\epsilon s^2+6 m^4+3 m^2 s\right) }{6 m^4
   s^2} F(10110) \nn \\ 
-&\quad\frac{(\epsilon-1) \left(16 \epsilon^3 m^2+4 \epsilon^3 s-20 \epsilon^2 m^2-8 \epsilon m^2-7 \epsilon s+3 m^2+3 s\right) }{6 (2 \epsilon-1)
   (2 \epsilon+1) m^4 s^2} F(10010) \nn \\
-&\quad\frac{(\epsilon-1) \epsilon }{3 m^4 s} F(1011-1) -\frac{(\epsilon-1) (2 \epsilon-1) }{2 m^2
   s} F(01110) \nn \\
-&\quad\frac{(\epsilon-1) \left(12 \epsilon m^2+\epsilon s-3 m^2\right) }{6 m^4 s^2} F(1-1110)\nn \\
-&\quad\frac{(\epsilon-1) \left(6 \epsilon m^2+\epsilon s-3
   m^2\right) }{6 m^4 s} F(11100) \; , 
\eea
and with {\tt REDUZE 2}:
\bea
  F(1 2 1 1 0)\quad=&\quad
     -\frac{(\epsilon-1) \left(8 \epsilon m^2+\epsilon s-2
   m^2\right)}{4 m^4 s^2}
     F(1 -1 1 1 0) \nn \\
    +&\quad
    \frac{64 \epsilon^2 m^4+16 \epsilon^2 m^2 s+\epsilon^2
   s^2-64 \epsilon m^4-22 \epsilon m^2 s-\epsilon
   s^2+12 m^4+6 m^2 s}{12 m^4 s^2}F(1 0 1 1 0)\nn \\
    -&\quad\frac{(\epsilon-1) \left(12 \epsilon m^2+\epsilon s-6
   m^2\right)}{6 m^4 s}F(1 1 1 0 0)\nn \\
   -&\quad\frac{(\epsilon-1) (2 \epsilon-3) \left(16 \epsilon^2 m^2+2
   \epsilon^2 s+4 \epsilon m^2+3 \epsilon s-2 m^2-2
   s\right)}{12 (2 \epsilon-1) (2 \epsilon+1) m^4 s^2}F(1 0 0 1 0) \; .
\eea
For the case of the double pole, two--loop graph, with three external legs and one massive loop,
we have the denominators:
\beq
D_1=\ell_1^2 \; \; , \; \; D_2=\ell_2^2-m^2 \; \; , \; \; D_3=(\ell_2+p_1)^2-m^2 \; \; , \; \; 
D_4=(\ell_2+p_1+p_2)^2-m^2 \; \; , \; \; D_5=(\ell_1+\ell_2)^2-m^2 \; ,  \nn 
\eeq 
\beq
D_6= \ell_1 \cdot p_1 \; \; , \; \; D_7= \ell_1 \cdot p_2 \; . \nn
\eeq
Using the IBPs from \Eq{eq:IBP2loop3p}  we get with {\tt FIRE}:
\bea
F(1211100)\quad=&\quad\frac{(\epsilon-1) 
\left\lbrace  (1+4\epsilon) s^2 -4 \epsilon m^2 s (11-2\epsilon) -8m^4 (4\epsilon^2-8\epsilon-1) \right\rbrace  }
{8 \epsilon (2 \epsilon-1) m^6 s \left(4 m^2-s\right)^2} F(0001100)\nn \\
-&\quad\frac{A_1}{2 \epsilon (2 \epsilon-1) m^6 s \left(4
   m^2-s\right)^2} F(1001100) \nn \\
+&\quad \frac{\left(8 \epsilon^3-12 \epsilon^2+4 \epsilon-1\right) (\epsilon-1) }
{8 \epsilon (2 \epsilon-1)^2 m^6 s} F(0010100)  \nn \\
-&\quad\frac{(\epsilon-1) }{2 (2 \epsilon-1) m^4} \left[ F(0111100) - F(1011100) \right]  
+\frac{2(\epsilon-1) }{m^2 s \left(4 m^2-s\right)} F(0101100) \nn \\
+&\quad\frac{(\epsilon-1) \left(2 \epsilon m^2-\epsilon s-m^2\right)
   }{2 \epsilon (2 \epsilon-1) m^6 s^2} F(1-101100) \nn \\
+&\quad\frac{(\epsilon-1)^2 \left(8 \epsilon m^2-2 \epsilon s-6 m^2+s\right)}{2 (2 \epsilon-1) m^6 s \left(4 m^2-s\right)} F(0100100) \nn \\
-&\quad\frac{2 (\epsilon-1) \left(m^2-s\right) 
\left\lbrace  -\epsilon s^2 +m^2 s (6\epsilon-1)+8m^4 (2\epsilon-1)  \right\rbrace  }
{\epsilon (2 \epsilon-1) m^6 s^2 \left(4 m^2-s\right)^2} F(10011-10) \; ,
\eea
where $s=(p_1+p_2)^2+i0$,  and:
\beq
A_1=\epsilon(1-\epsilon) s^3 + m^2 (-1+\epsilon)(-1+9\epsilon)s^2 +
  m^4 (1-\epsilon)(5+6\epsilon)s +2 m^6 (1+2\epsilon)(-3+4\epsilon) \; ,
\eeq
while, with {\tt REDUZE 2}, we get:
\bea
 F(1 2 1 1 1 0 0)
   \quad=&\quad
    -\frac{\epsilon-1}{2 (2 \epsilon-1) m^4}
    F(0 1 1 1 1 0 0) \nn \\
    +&\quad\frac{\epsilon-1}{2 (2 \epsilon-1) m^4}
    F(1 0 1 1 1 0 0) \nn \\
    +&\quad\frac{2 (\epsilon-1)}{m^2 s \left(4 m^2-s\right)}
    F(0 1 0 1 1 0 0) \nn \\
    -&\quad\frac{3 (\epsilon-1) \left(4 \epsilon m^4-8 \epsilon m^2 s+\epsilon
   s^2+2 m^4+m^2 s\right)}{2 \epsilon (2 \epsilon-1) m^6 s \left(4
   m^2-s\right)^2}
    F(1 -1 0 1 1 0 0) \nn \\
     -&\quad\frac{A_2}{2 \epsilon (2 \epsilon-1)
   m^6 s \left(4 m^2-s\right)^2}
    F(1 0 0 1 1 0 0) \nn \\
     +&\quad\frac{A_3}{4 \epsilon (2 \epsilon-1)^2 m^6 s \left(4
   m^2-s\right)^2}
    F(0 1 0 0 1 0 0) \; ,
\eea
where 
\bea
A_2\quad=&\quad16 \epsilon^2 m^6-6 \epsilon^2 m^4 s+9 \epsilon^2 m^2
   s^2-\epsilon^2 s^3 \nn \\
-&\quad4 \epsilon m^6+\epsilon m^4 s-10 \epsilon m^2
   s^2+\epsilon s^3-6 m^6+5 m^4 s+m^2 s^2 \; ,
\eea
 and
\bea
A_3\quad=&\quad(\epsilon-1) \left(128 \epsilon^4 m^4-64 \epsilon^4 m^2 s+8
   \epsilon^4 s^2-256 \epsilon^3 m^4+112 \epsilon^3 m^2 s- 12\epsilon^3 s^2+192 \epsilon^2 m^4 \right. \nn \\
-&\quad \left. 92 \epsilon^2 m^2 s+8 \epsilon^2
   s^2-40 \epsilon m^4+26 \epsilon m^2 s-\epsilon s^2-12 m^4+4 m^2
   s-s^2\right) \; .
\eea

The cases with additional external legs can be treated in a similar manner. It can always be reduced to sums of 
integrals with single propagators at the expense of introducing numerators. Although no formal proof exists,
in all cases studied so far it has been possible to reduce to integrals where only single poles appear. The generality
of this result seems plausible \cite{Gluza:2010ws}.

Our strategy is now clear. For a two--loop calculation, first we reduce all double pole graphs using
IBPs or any other method. The remaining integrals all contain single poles and can be treated using
the Duality Theorem at two--loops. The appearance of vector or tensor integrals does not spoil this strategy
since the Duality Theorem for single poles, affects only the denominators of the integrands.

%-----------------------------------
%	The case for three--loop diagrams
%-----------------------------------

\subsection{The case for three--loop diagrams}

\begin{figure}[!htb]
%\begin{center}
\centering
\includegraphics[scale=1]{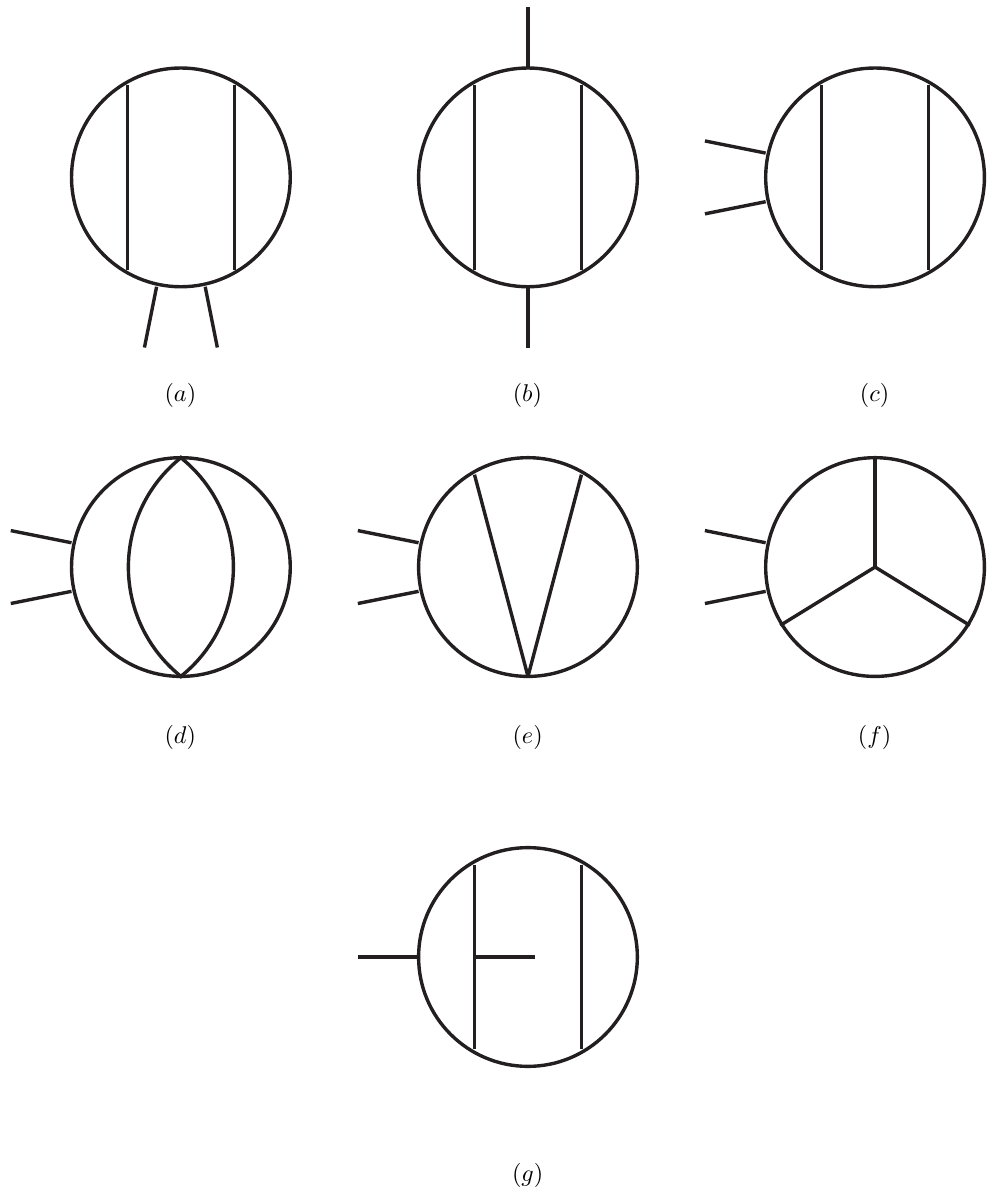}
\caption{Master topologies of three--loop scalar
    integrals with multiple powers of internal propagators.  
    Each internal line already dressed with some external 
    leg can be dressed with additional external lines.}
\label{3LoopTop}
%\end{center}
\end{figure}

For three--loop graphs there exists one topology with a triple propagator and a number of topologies
with a double propagator. All topologies are shown in Fig.~\ref{3LoopTop}. The arguments for the two--loop
case are valid here as well. We first reduce the multiple pole integrands by using IBPs until we have integrals with 
only single poles (possibly with numerators) and then we can then apply single-pole Duality Theorem as it was described
for the three--loop case in Ref. \cite{Bierenbaum:2010cy}. In the following,  we show explicitly the reduction 
of the two-point function for the different topologies and for massless internal lines.
We use the notation:
\beq
F(a_1 a_2 \cdots a_n)= \int_{\ell_1} \int_{\ell_2} \int_{\ell_3} \frac{1}{D_1^{a_1}D_2^{a_2} \cdots D_n^{a_n}}
\eeq
to denote a generic three--loop integral with $n$ propagators raised to an arbitrary integer power, with 
$D_i=G_F^{-1}(q_i)=q_i^2-m_i^2+i0$ or $D_i$ equal to any ISP and $q_i$  any combination of external and loop momenta. 
We also have $s=p^2+i0$.
The IBPs to be used for the reduction are:
\beq
\frac{\partial}{\partial \ell_i} \cdot \ell_j \; , \; \frac{\partial}{\partial \ell_i} \cdot p
\; \; , \;\; i,j=1,2,3
\label{eq:IBP3loop2p}
\eeq
In the following, we present first the result obtained with {\tt REDUZE 2} and then with {\tt FIRE}.
For the single triple pole graph (a) in Fig. (\ref{3LoopTop}), we have the following expressions:
\begin{description}
 \item[(a)] The denominators used are:
     \beq
       D_1=\ell_1^2 \; , \; D_2=\ell_2^2 \; , \; D_3=\ell_3^2 \; , \; D_4=(\ell_2-p)^2 \; , \; 
       D_5=(\ell_1-\ell_2)^2 \; , \; D_6=(\ell_3-\ell_2)^2 \; , \nn
     \eeq
     \beq
      D_7=\ell_1 \cdot p \; , \; D_8=\ell_3 \cdot p \; , \; D_9=\ell_1 \cdot \ell_3 \; , \nn
     \eeq
     with the result:
     \beq
      F(131111000)=\frac{2\epsilon(-1+4\epsilon)}{(1+\epsilon)(1+2\epsilon) s^2} F(111111000)=
       \frac{2(-1+2\epsilon)(-1+4\epsilon)}{ (1+\epsilon) (1+2\epsilon)s^3} \; F(101111000) \; .
     \eeq
 \end{description}
For the graphs with doubles poles, (b)-(g), Fig. (\ref{3LoopTop}), we find:
\begin{description}
 \item[(b)] The denominators are:
     \beq
       D_1=\ell_1^2 \; , \; D_2=\ell_2^2 \; , \; D_3=\ell_3^2 \; , \; D_4=(\ell_2-p)^2 \; , \; 
       D_5=(\ell_1-\ell_2)^2 \; , \; D_6=(\ell_3-\ell_2+p)^2 \; , \nn
     \eeq
     \beq
      D_7=\ell_1 \cdot p \; , \; D_8=\ell_1 \cdot \ell_3 \; , \; D_9=\ell_3 \cdot p \; , \nn
     \eeq
     with the result:
     \bea
      F(121211000)\quad=&\quad
     \frac{3(-1+4\epsilon)(1+3\epsilon)}{(1+\epsilon)^2 s^2}F(111111000) \nn \\
      =&\quad \; 
      \frac{6(-2+3\epsilon)(-1+3\epsilon)(1+3\epsilon)(-3+4\epsilon)(-1+4\epsilon)}
       {\epsilon^2 (1+\epsilon)^2 (-1+2\epsilon) s^4}
      \; F(101011000)\; . \nn \\
     \eea
 \item[(c)] The denominators are:
     \beq
       D_1=\ell_1^2 \; , \; D_2=\ell_2^2 \; , \; D_3=\ell_3^2 \; , \; D_4=(\ell_3+p)^2 \; , \; 
       D_5=(\ell_3-\ell_2)^2 \; , \; D_6=(\ell_1-\ell_2)^2 \; , \nn
     \eeq
     \beq
      D_7=\ell_1 \cdot p \; , \; D_8=\ell_2 \cdot p \; , \; D_9=\ell_1 \cdot \ell_3 \; , \nn
     \eeq
     with the result:
     \bea
      F(122111000)\quad=&\quad\frac{2\epsilon (-1+4\epsilon)(-1+3\epsilon)}
     {(1+2\epsilon)(1+\epsilon)^2 s^2}F(111111000) \nn \\
      =&\quad 
      \frac{2(-2+3\epsilon)(-1+3\epsilon)(-3+4\epsilon)(-1+4\epsilon)}{\epsilon (1+\epsilon)^2 (1+2\epsilon) s^4}
       \; F(100111000)\; .
     \eea
 \item[(d)] The denominators are:
     \beq
       D_1=\ell_1^2 \; , \; D_2=\ell_2^2 \; , \; D_3=\ell_3^2 \; , \; D_4=(\ell_3-p)^2 \; , \; 
       D_5= (\ell_2+\ell_3-\ell_1)^2  \; , \nn
     \eeq
     \beq
      D_6=\ell_1 \cdot p \; , \; D_7=\ell_2 \cdot p \; , \; D_8=\ell_1 \cdot \ell_2\; , \; D_9=\ell_1 \cdot \ell_3 \; , \nn
     \eeq
     with the result:
     \beq
      F(112110000)=\frac{(-1+2\epsilon)}{\epsilon s} \; F(111110000)
              =\frac{(-3+4 \epsilon)}{\epsilon s^2}  \; F(110110000)\; .
     \eeq
 \item[(e)] The denominators are:
     \beq
       D_1=\ell_1^2 \; , \; D_2=\ell_2^2 \; , \; D_3=\ell_3^2 \; , \; D_4=(\ell_3-p)^2 \; , \; 
       D_5=(\ell_1+\ell_2)^2 \; , \; D_6=(\ell_2+\ell_3)^2 \; , \nn
     \eeq
     \beq
      D_7=\ell_1 \cdot p \; , \; D_8=\ell_1 \cdot \ell_3 \; , \; D_9=\ell_2 \cdot p \; , \nn
     \eeq
     with the result:
     \beq
      F(112111000)=
       \frac{(-1+4\epsilon)}{(1+2\epsilon) s}F(111111000)
        = \frac{(-2+3\epsilon)(-3+4\epsilon)(-1+4\epsilon)}{\epsilon^2 (1+2\epsilon) s^3}
       \; F(100111000)\; .
     \eeq
 \item[(f)] The denominators are:
     \beq
       D_1=\ell_1^2 \; , \; D_2=\ell_2^2 \; , \; D_3=\ell_3^2 \; , \; D_4=(\ell_2+p)^2\; , \; 
       D_5=(\ell_1+\ell_2)^2 \; , \; D_6=(\ell_1+\ell_3)^2 \; , \nn
     \eeq
     \beq
     D_7=(\ell_3-\ell_2)^2\; , \;  D_8=\ell_1 \cdot p \; , \; D_9=\ell_3 \cdot p \; , \nn
     \eeq
     with the result:
     \bea
      F(121111100)=&\quad\frac{2\epsilon}{(1+\epsilon)s} F(111111100) \nn \\
      =&\quad
      \frac{2(-2+3\epsilon)(-1+3\epsilon)(-3+4\epsilon)(-1+4\epsilon)}{\epsilon^2 (1+\epsilon) (1+2\epsilon) s^4}
       \; \left[  F(001111000)+F(100101100) \right]  \nn \\
      +&\quad \frac{2(-1+2\epsilon)^2 (-1+4\epsilon)}{\epsilon (1+\epsilon) (1+2\epsilon)s^3} \; F(101110100)\; .
     \eea
 \item[(g)] The denominators are:
     \beq
       D_1=\ell_1^2 \; , \; D_2=\ell_2^2 \; , \; D_3=\ell_3^2 \; , \; D_4=(\ell_3-p)^2 \; , \nn
     \eeq
     \beq
       D_5=(\ell_1-\ell_2)^2 \; , \; D_6=(\ell_3-\ell_2)^2 \; , \; D_7=(\ell_3-\ell_2-p)^2 \nn
     \eeq
     \beq
      D_8=(\ell_1-\ell_3)^2\; , \;
      D_9=(\ell_1-p)^2
     \eeq
     with the result:
     \bea
     F(1 2 1 1 1 1 1 00 )
     \quad=&\quad
    \frac{(-1+3 \epsilon)^2(1+5 \epsilon)}{(1+\epsilon)(1+2 \epsilon)^2 s^2}
     F(1 1 1 1 1 1 1 -1 0)\nn \\
    +&\quad\frac{\epsilon (9 \epsilon^2-11 \epsilon-4)}{(1+\epsilon) (1+2 \epsilon)^2  s^2}
     F(1 1 1 1 1 1 1 00)\; .
     \eea

\end{description}
The difference between the results of {\tt FIRE} on the one hand and {\tt REDUZE 2} 
on the other is due to the fact that the second is expressed in terms of
basis integrals while the first is expressed in terms of integrals with single poles of the 
same type as the multiple pole integral (in effect the
first result can be further reduced to the second). Since we do not seek a particular basis for our
reduction, as was stressed earlier, both results are equally useful as far as application of the Duality 
Theorem is concerned.

%----------------------------------------------------------------------------------------
%	Conclusions
%----------------------------------------------------------------------------------------

\section{Conclusions}
\label{sec:conclusions}

We have extended the Duality Theorem to two-- and
three--loop integrals with multiple poles. A Lorentz--invariant expression
for the residues of double poles has been derived, which can
be extended straightforwardly to triple and, in general,
multiple poles. In the absence of a systematic procedure to
%express the dual $+i0$ prescription in terms of external
%momenta exclusively, as in the case of single poles,
reexpress dual propagators in terms of Feynman propagators 
(cf. \Eq{eq:GAinGDGeneralN} for the case of simple poles)
we have explored an alternative approach.
We use IBP identities to reduce the integrals with identical propagators to ones with only single poles. 
Therefore, the essential features of the Loop--Tree Duality now remain intact.  We reiterate that our goal 
is not to reduce everything to some set of master integrals. Rather, we reduce the integrals until 
there are no multiple poles left. Then, we can use the Duality Theorem in its original form for single pole
propagators, to rewrite them as integrals of a tree--level object over a modified
phase-space. The appearance of additional tensor integrals, due to the reduction, does
not affect our procedure, since applying the Duality Theorem in its single-pole version, only cuts propagators, leaving
the numerators of the integrals unaffected.

% Chapter Template

\chapter{On the Cancellation of Singularities} % Main chapter title

\label{Chapter6} % Change X to a consecutive number; for referencing this chapter elsewhere, use \ref{ChapterX}

%\lhead{Chapter 5. \emph{Cancelling Singularities}} % Change X to a consecutive number; this is for the header on each page - perhaps a shortened title

In the previous chapters the Loop--Tree Duality was extended to cover Feynman diagrams which 
involve multiple loops as well as higher order poles \cite{Bierenbaum:2010cy,Bierenbaum:2012th}.\\
Keeping in mind that the final aim of the Loop--Tree Duality is to treat virtual and real corrections 
at the same time. In this chapter we analyse the singular behaviour of one-loop integrals and 
scattering amplitudes in the framework of the Loop--Tree Duality method. 
We begin with a discussion of the cancellation of singularities among dual 
contributions at the integrand level. After that, we present a phase space mapping 
between virtual corrections in the dual representation and the real corrections 
for the local cancellation of infrared divergencies.

\newpage

%----------------------------------------------------------------------------------------
%	The singular behaviour of the loop integrand
%----------------------------------------------------------------------------------------

\section{The singular behaviour of the loop integrand}
\label{sec:loop}

We consider a general one-loop $N$-leg scalar integral
\beq
\label{Ln5}
L^{(1)}(p_1, p_2, \dots, p_N) =
\int_{\ell} \, \prod_{i \in \alpha_1} \, G_F(q_i)~, \qquad
\int_{\ell} \bullet =-i \int \frac{d^d \ell}{(2\pi)^{d}} \; \bullet~,
\eeq
where 
\beq
G_F(q_i)=\frac{1}{q_i^2-m_i^2+i0}
\label{eq:feynman}
\eeq
are Feynman propagators that depend on the 
loop momentum $\ell$, which flows anti-clockwise, 
and the four-momenta of the external legs $p_{i}$, 
$i \in \alpha_1 = \{1,2,\ldots N\}$, which are taken as outgoing and 
are ordered clockwise. 
We use dimensional regularization with $d$  
the number of space-time dimensions. 
The momenta of the internal lines $q_{i,\mu} = (q_{i,0},\mathbf{q}_i)$, 
where $q_{i,0}$ is the energy (time component) and $\mathbf{q}_{i}$ are 
the spacial components, are defined as $q_{i} = \ell + k_i$ with 
$k_{i} = p_{1} + \ldots + p_{i}$, and $k_{N} = 0$ by momentum conservation. 
We also define $k_{ji} = q_j - q_i$.

The loop integrand becomes singular in regions of the 
loop momentum space in which subsets of internal lines go on-shell,
although the existence of singular points of the integrand 
is not enough to ensure the emergence in the loop integral
of divergences in the dimensional regularization parameter. 
Nevertheless, numerical integration over integrable singularities 
still requires a contour deformation~\cite{Gong:2008ww,Nagy:2006xy,Kramer:2002cd,Soper:2001hu,Soper:1999xk,Soper:1998ye,Becker:2012nk,Becker:2012bi}, 
namely, to promote the loop momentum to the complex 
plane in order to smoothen the loop matrix elements in the singular 
regions of the loop integrand. Hence, the relevance to 
identify accurately all the integrand singularities. 

In Cartesian coordinates, the Feynman propagator in~\Eq{eq:feynman}  
becomes singular at hyperboloids with origin in $-k_{i}$, 
where the minimal distance between each hyperboloid and 
its origin is determined by the internal mass $m_i$.
This is illustrated in Fig.~\ref{fig:cartesean}, where for simplicity
we work in $d=2$ space-time dimensions. Figure~\ref{fig:cartesean}~(left)
shows a typical kinematical situation where two 
momenta, $k_1$ and $k_2$, are separated by a time-like distance, 
$k_{21}^2 > 0$, and a third momentum, $k_3$, is space-like separated  
with respect to the other two, 
$k_{31}^2 <0$ and $k_{32}^2 <0$. The on-shell forward hyperboloids
($q_{i,0}>0$) are represented in Fig.~\ref{fig:cartesean} by solid lines, 
and the backward hyperboloids ($q_{i,0}<0$) by dashed lines. 
For the discussion that will follow it is important to stress that Feynman 
propagators become positive inside the respective 
hyperboloid and negative outside. 
Two or more Feynman propagators become simultaneously singular 
where their respective hyperboloids intersect. 
In most cases, these singularities, due to normal or anomalous 
thresholds~\cite{Mandelstam:1960zz,Rechenberg:1972rq}
of intermediate states, are integrable.
However, if two massless propagators are separated by a 
light-like distance, $k_{ji}^2=0$, then the overlap 
of the respective light-cones is tangential, 
as illustrated in Fig.~\ref{fig:cartesean}~(right),
and leads to non-integrable collinear singularities. 
In addition, massless propagators can generate soft singularities 
at $q_{i}=0$. 

%%%%%%%%%%%%%%%
\begin{figure}[h]
\centering
\includegraphics[scale=1]{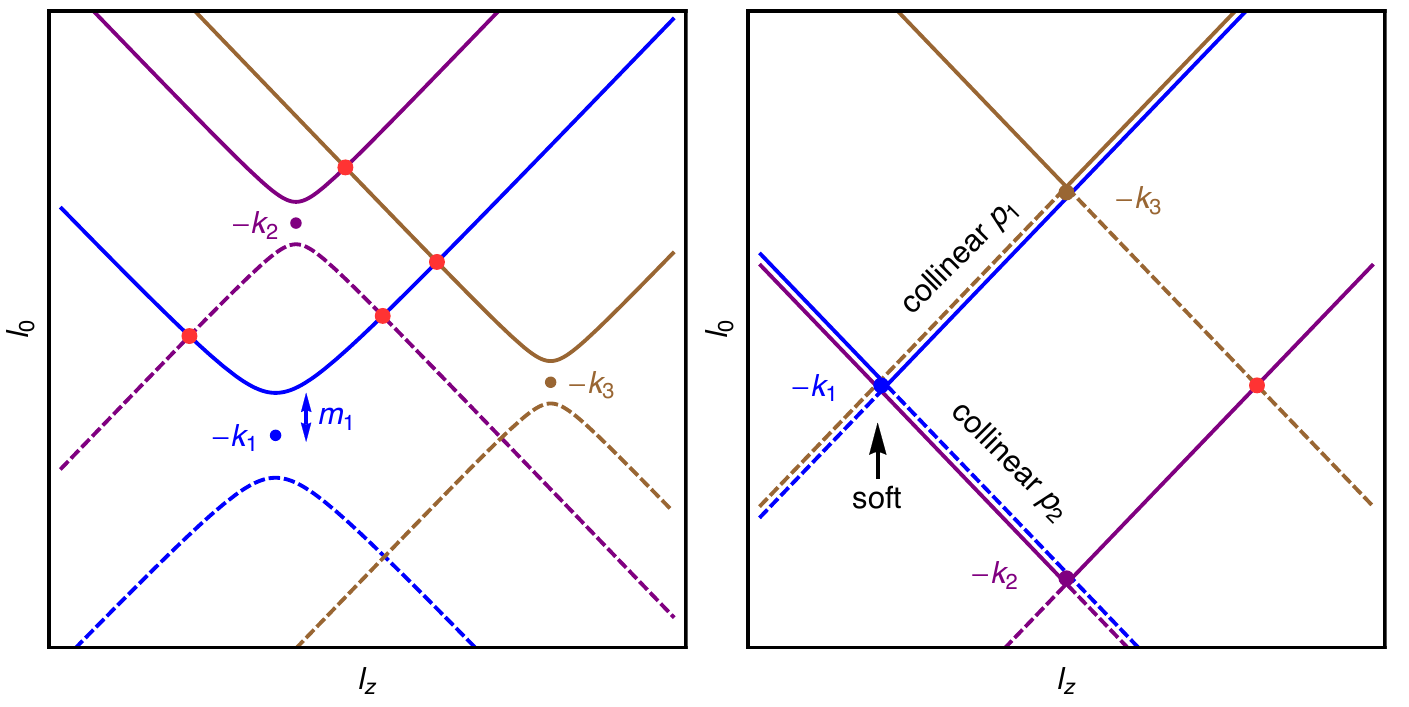}
\caption{On-shell hyperboloids for three arbitrary propagators 
in Cartesian coordinates in the ($\ell_0$,$\ell_z$) space (left). 
Kinematical configuration with infrared singularities (right).
In the latter case, the on-shell hyperboloids degenerate to light-cones. 
\label{fig:cartesean}}
\end{figure}
%%%%%%%%%%%%%%%

As we have seen in Chapter \ref{Chapter3}, 
the dual representation of the scalar one-loop integral in \Eq{Ln}
is the sum of $N$ dual integrals~\cite{Catani:2008xa,Bierenbaum:2010cy}:
\bea
\label{oneloopduality}
L^{(1)}(p_1, p_2, \dots, p_N) 
\quad=\quad - \sum_{i \in \alpha_1} \, \int_{\ell} \; \td(q_i) \,
\prod_{\substack{j \in \alpha_1 \\ j\neq i}} \,G_D(q_i;q_j)~,
\eea 
where
\beq
G_D(q_i;q_j) = \frac{1}{q_j^2 -m_j^2 - i0 \, \eta \cdot k_{ji}}
\eeq
are the so-called dual propagators, as defined in Ref.~\cite{Catani:2008xa},
with $\eta$ a {\em future-like} vector, $\eta^2 \ge 0$, 
with positive definite energy $\eta_0 > 0$.
The delta function 
$\td(q_i) \equiv 2 \pi \, i \, \theta(q_{i,0}) \, \delta(q_i^2-m_i^2)$
sets the internal lines on-shell by selecting the pole of the propagators
with positive energy $q_{i,0}$ and negative imaginary part. 
In the following we take $\eta_\mu = (1,\mathbf{0})$, and thus 
$- i0 \, \eta \cdot k_{ji} = -i0 \, k_{ji,0}$. This is equivalent 
to performing the loop integration along the on-shell forward hyperboloids.
Let us mention that in the light-cone coordinates 
($\ell_+$, $\ell_-$, $\mathbf{l}_\perp$), 
where $\ell_\pm=(\ell_0 \pm \ell_{d-1})/\sqrt{2}$,
Feynman propagators vanish at hyperboloids in the plane ($\ell_+$,$\ell_-$) 
which are similar to those depicted in Fig.~\ref{fig:cartesean} but 
rotated by 45 degrees. Consequently, by selecting the forward hyperboloids 
the integration limits of either $\ell_+$ or $\ell_-$ are restricted 
and the restrictions are different for each dual contribution.
For this reason, although~\Eq{oneloopduality}
is valid for any system of coordinates, we will stick 
for the rest of the thesis to Cartesian coordinates where 
all the dual contributions share the same integration limits
for the loop three-momentum. 

A crucial point of our discussion is the observation that 
dual propagators can be rewritten as 
\beq
\td(q_i) \, G_D(q_i;q_j) = 
i\, 2 \pi \, 
\frac{\delta(q_{i,0}-q_{i,0}^{(+)})}{2 q_{i,0}^{(+)}} \, 
\frac{1}{(q_{i,0}^{(+)} + k_{ji,0})^2-(q_{j,0}^{(+)})^2}~,
\label{eq:newdual}
\eeq
where
\beq
q_{i,0}^{(+)} = \sqrt{\mathbf{q}_i^2 + m_i^2-i0}
\label{qi0distance}
\eeq
is the loop energy measured along the on-shell hyperboloid 
with origin at $-k_i$. By definition we have ${\rm Re}(q_{i,0}^{(+)}) \ge 0$. 
The factor $1/q_{i,0}^{(+)}$ can become singular for $m_i=0$, but the integral 
$\int_\ell \delta(q_{i,0}-q_{i,0}^{(+)})/q_{i,0}^{(+)}$ is still convergent by two 
powers in the infrared. 
Soft singularities require two dual propagators, where each of the two
dual propagators contributes with one power in the infrared.
From \Eq{eq:newdual} it is obvious that dual propagators become 
singular, $G_D^{-1}(q_i;q_j)=0$, if one of the following conditions is fulfilled:
\bea
& q_{i,0}^{(+)}+q_{j,0}^{(+)}+k_{ji,0}=0~, \label{ellipsoid} \\
& q_{i,0}^{(+)}-q_{j,0}^{(+)}+k_{ji,0}=0~. \label{hyperboloid}
\eea
The first condition, \Eq{ellipsoid}, is satisfied if the 
forward hyperboloid of $-k_i$ intersects 
with the backward hyperboloid of $-k_j$. 
The second condition, \Eq{hyperboloid}, is true when 
the two forward hyperboloids intersect each other. 

In the massless case, \Eq{ellipsoid} and \Eq{hyperboloid} are the equations
of conic sections in the loop three-momentum space; 
$q_{i,0}^{(+)}$ and $q_{j,0}^{(+)}$ are the distance to the {\it foci} 
located at $-\mathbf{k}_i$ and $-\mathbf{k}_j$, respectively, 
and the distance between the foci is $\sqrt{\mathbf{k}_{ji}^2}$.
If internal masses are non-vanishing, \Eq{qi0distance} can be reinterpreted 
as the distance associated to a four-dimensional space with 
one ``massive'' dimension and the foci now located 
at $(-\mathbf{k}_i,-m_i)$ and $(-\mathbf{k}_j,-m_j)$, respectively. 
Then, the singularity arises at the intersection of 
the conic sections given by \Eq{ellipsoid} or \Eq{hyperboloid} 
in this generalized space with the zero mass plane. This picture 
is useful to identify the singular regions of the loop integrand 
in the loop three-momentum space.

The solution to \Eq{ellipsoid} is an ellipsoid and clearly requires $k_{ji,0}<0$.
Moreover, since it is the result of the intersection of a forward 
with a backward hyperboloid the distance between the two 
propagators has to be future-like, $k_{ji}^2 \ge 0$. 
Actually, internal masses restrict this condition. Bearing in mind 
the image of the conic sections in the generalized massive space so 
we can deduce intuitively that \Eq{ellipsoid} has solution for 
\beq
k_{ji}^2-(m_j+m_i)^2 \ge 0~, \qquad k_{ji,0}<0~, \qquad 
\rm{forward~with~backward~hyperboloids}~.
\label{eq:generalizedtimelike}
\eeq
The second equation, \Eq{hyperboloid}, leads to a hyperboloid in the 
generalized space, and there are solutions for $k_{ji,0}$ either 
positive or negative, namely when either of the two momenta are 
set on-shell. However, by interpreting the result in the generalized 
space it is clear that the intersection with the zero mass plane 
does not always exist, and if it exists, it can be either an ellipsoid 
or a hyperboloid in the loop three-momentum space. 
Here, the distance between the momenta of the propagators 
has to be space-like, although also time-like configurations 
can fulfill~\Eq{hyperboloid} as far as the time-like 
distance is small or close to light-like. 
The following condition is necessary:
\beq
k_{ji}^2-(m_j-m_i)^2 \le 0~, \qquad \rm{two~forward~hyperboloids}~.
\label{eq:generalizedspacelike}
\eeq
In any other configuration, the singularity appears for loop 
three-momenta with imaginary components.

%----------------------------------------------------------------------------------------
%	Cancellation of singularities among dual integrands
%----------------------------------------------------------------------------------------

\section{Cancellation of singularities among dual integrands}
\label{sec:cancel}

In this section we prove one of the main properties of the Loop--Tree 
Duality method, namely the partial cancellation of 
singularities among different dual integrands.
This represents a significant advantage with respect to the integration 
of regular loop integrals in the $d$-dimensional space, where one 
single integrand cannot obviously lead to such cancellation. 

Let us consider first two Feynman propagators separated by 
a space-like distance, $k_{ji}^2 < 0$ (or more generally 
fulfilling~\Eq{eq:generalizedspacelike}). 
In the corresponding dual representation one of these
propagators is set on-shell and the other becomes dual, 
and the integration occurs along the respective on-shell forward 
hyperboloids. See again Fig.~\ref{fig:cartesean}~(left) for a graphical 
representation of this set-up.
There, the two forward hyperboloids of $-k_1$ and $-k_3$ intersect 
at a single point. Integrating over $\ell_z$ along the forward 
hyperboloid of $-k_1$ we find that the dual propagator 
$G_D(q_1;q_3)$, which is negative below the intersection point where 
the integrand becomes singular, changes sign above this point 
as we move from outside to inside the on-shell hyperboloid of $-k_3$. 
The opposite occurs if we set $q_3$ on-shell;
$G_D(q_3;q_1)$ is positive below the intersection point, and 
negative above. The change of sign leads to the cancellation of the 
common singularity. Notice that also the dual $i0$ prescription 
changes sign. In order to prove 
analytically this cancellation, we define 
$x = q_{i,0}^{(+)} - q_{j,0}^{(+)} + k_{ji,0}$. 
In the limit $x\to 0$:
\beq
\lim_{x \to 0} \, \left( 
\td(q_i) \, G_D(q_i;q_j) + (i \leftrightarrow j) 
\right)
= \left( \frac{1}{x}-\frac{1}{x} \right) \, 
\frac{1}{2 q_{j,0}^{(+)}} \, \td(q_i) + {\cal O}(x^0)~,
\label{eq:spacecancel1}
\eeq
and thus the leading singular behaviour cancels
among the two dual contributions. 
The cancellation of these singularities is not altered by 
the presence of other non-vanishing dual propagators 
(neither by numerators) because
\beq
\lim_{x \to 0} \, G_D(q_j;q_k) = 
\lim_{x \to 0} \, 
\frac{1}{(q_{j,0}^{(+)} + k_{ki,0} - k_{ji,0})^2-(q_{k,0}^{(+)})^2} =
\lim_{x \to 0} \, G_D(q_i;q_k)~,
\label{eq:spacecancel2}
\eeq
where we have used the identity $k_{kj,0} = k_{ki,0} - k_{ji,0}$. 
If instead, the separation is time-like (in the sense 
of~\Eq{eq:generalizedtimelike}), 
we define $x = q_{i,0}^{(+)} + q_{j,0}^{(+)} + k_{ji,0}$, and find
\beq
\lim_{x \to 0} \, \left( 
\td(q_i) \, G_D(q_i;q_j) + (i \leftrightarrow j) 
\right)
= - \theta(-k_{ji,0}) \, 
\frac{1}{x} \, \frac{1}{2 q_{j,0}^{(+)}} \, 
\td(q_i) + (i \leftrightarrow j) + {\cal O}(x^0)~.
\label{eq:timecancel}
\eeq
In this case the singularity of the integrand remains because of the 
Heaviside step function.

We should consider also the case in which more than two 
propagators become simultaneously singular. 
To analyse the intersection of three forward 
hyperboloids, we define 
\beq
\lambda \, x = q_{i,0}^{(+)} - q_{j,0}^{(+)} + k_{ji,0}~, \qquad
\lambda \, y = q_{i,0}^{(+)} - q_{k,0}^{(+)} + k_{ki,0}~.
\eeq
As before, we use the identity $k_{kj,0} = k_{ki,0} - k_{ji,0}$, 
and thus $q_{j,0}^{(+)} - q_{k,0}^{(+)} + k_{kj,0} = \lambda \, (y-x)$.
In the limit in which the three propagators become simultaneously singular:
\bea
& \lim_{\lambda \to 0}
\left( \td(q_i) \, G_D(q_i;q_j) \, G_D(q_i;q_k) + {\rm perm.} \right) 
= \nn \\ 
&\frac{1}{\lambda^2} \,
\left( \frac{1}{x\, y} + \frac{1}{x\, (x-y)} + \frac{1}{y\, (y-x)} \right) \, 
\frac{1}{2 q_{j,0}^{(+)}} \, \frac{1}{2 q_{k,0}^{(+)}} \, 
\td(q_i) + {\cal O}(\lambda^{-1})~,
\label{eq:threespacecancel}
\eea
and again the leading singular behaviour cancels in the sum. 
Although not shown for simplicity in \Eq{eq:threespacecancel}, 
also the ${\cal O}(\lambda^{-1})$ terms cancel in the sum, thus 
rendering the integrand finite in the limit $\lambda \to 0$.
For three propagators there are also more possibilities:
two forward hyperboloids might intersect simultaneously with a backward 
hyperboloid, or two backward hyperboloids might intersect with a forward 
hyperboloid. In the former case, we define 
$\lambda \, x = q_{i,0}^{(+)} + q_{k,0}^{(+)} + k_{ki,0}$, and
$\lambda \, y = q_{j,0}^{(+)} + q_{k,0}^{(+)} + k_{kj,0}$,
with $k_{ki,0}<0$ and $k_{kj,0}<0$, and hence 
$q_{i,0}^{(+)} - q_{j,0}^{(+)} + k_{ji,0} = \lambda (x-y)$. 
In the $\lambda \to 0$ limit 
\bea
& \lim_{\lambda \to 0}
\left( \td(q_i) \, G_D(q_i;q_j) \, G_D(q_i;q_k) + {\rm perm.}
\right) = \nn \\ 
&\theta(-k_{ki,0}) \, \theta(-k_{kj,0}) \, 
\frac{1}{\lambda^2} \, \left( \frac{1}{x\, (y-x)} + \frac{1}{y\, (x-y)} \right) \, 
\frac{1}{2 q_{j,0}^{(+)}} \, \frac{1}{2 q_{k,0}^{(+)}} \, 
\td(q_i) + {\cal O}(\lambda^{-1})~.
\label{eq:spacecancel3}
\eea
Notice that the singularity in $1/(x-y)$ cancels in \Eq{eq:spacecancel3} 
(also at ${\cal O}(\lambda^{-1})$). 
In the latter case, we set as before $\lambda \, x = q_{i,0}^{(+)} + q_{k,0}^{(+)} + k_{ki,0}$, 
and define $\lambda \, z = q_{i,0}^{(+)} + q_{j,0}^{(+)} + k_{ji,0}$, then 
\bea
& \lim_{\lambda \to 0}
\left( \td(q_i) \, G_D(q_i;q_j) \, G_D(q_i;q_k) + {\rm perm.}
\right) = 
- \theta(-k_{ki,0}) \, \nn \\ & \times \theta(-k_{ji,0}) \, 
\frac{1}{\lambda^2} \, \left( \frac{1}{x\, z} \right) \,
\frac{1}{2 q_{j,0}^{(+)}} \, \frac{1}{2 q_{k,0}^{(+)}} \, 
\td(q_i) + {\cal O}(\lambda^{-1})~. 
\label{eq:spacecancel3b}
\eea
Similarly, it is straightforward to 
prove that four forward hyperboloids
do not lead to any common singularity and more generally  
that the remaining multiple singularities are only driven 
by propagators that are time-like connected and less energetic 
than the propagator which is set on-shell. 

Thus, we conclude that singularities 
of space-like separated propagators~\footnote{
Including light-like and time-like configurations such that 
\Eq{eq:generalizedspacelike} is fulfilled.}, 
occurring in the intersection of on-shell forward hyperboloids,
are absent in the dual representation of the loop integrand. 
The cancellation of these singularities 
at the integrand level already represents a big advantage of 
the Loop--Tree Duality with respect to the direct integration in the 
four-dimensional loop space; it makes unnecessary the use of contour 
deformation to deal numerically with the integrable singularities of 
these configurations. 
This conclusion is also valid for loop scattering amplitudes. 
Moreover, this property can be extended in a straightforward manner 
to prove the partial cancellation of infrared singularities.

Collinear singularities occur when two massless propagators are 
separated by a light-like distance, $k_{ji}^2=0$. In that case, the 
corresponding light-cones overlap tangentially along 
an infinite interval. Assuming $k_{i,0} > k_{j,0}$, however,  
the collinear singularity for $\ell_0 > -k_{j,0}$ appears 
at the intersection of the two forward light-cones, 
with the forward light-cone of $-k_{j}$ located
inside the forward light-cone of $-k_{i}$, 
or equivalently, with the forward light-cone 
of $-k_{i}$ located outside the forward light-cone of $-k_{j}$, 
Thus, the singular behaviour of the two dual components 
cancel against each other, following the same qualitative 
arguments given before. 
For $-k_{i,0} < \ell_0 < -k_{j,0}$, instead, it is the forward 
light-cone of $-k_i$ that intersects tangentially 
with the backward light-cone of $-k_j$ according to~\Eq{ellipsoid}. 
The collinear divergences survive in this energy strip, 
which indeed also limits the range of the loop three-momentum 
where infrared divergences can arise. 
If there are several reference momenta separated by light-like distances
the infrared strip is limited by the minimal and maximal energies
of the external momenta. The soft singularity of the integrand
at $q_{i,0}^{(+)}=0$ leads to soft divergences only if two other 
propagators, each one contributing with one power in the infrared, 
are light-like separated from $-k_i$. In Fig.~\ref{fig:cartesean}~(right)
this condition is fulfilled only at $q_{1,0}^{(+)}=0$, 
but not at $q_{2,0}^{(+)}=0$ neither at $q_{3,0}^{(+)}=0$.

In summary, both threshold and infrared singularities are 
constrained in the dual representation of the loop integrand 
to a finite region where the loop three-momentum is of the order 
of the external momenta. Singularities outside this region, 
occurring in the intersection of on-shell forward hyperboloids
or light-cones, cancel in the sum of all the dual contributions.

%----------------------------------------------------------------------------------------
%	Cancellation of infrared singularities with real corrections
%----------------------------------------------------------------------------------------

\section{Cancellation of infrared singularities with real corrections}
\label{sec:real}

Having constrained the loop singularities to a finite region of the 
loop momentum space, we discuss now how to map this region into the 
finite-size phase-space of the real corrections for the cancellation 
of the remaining infrared singularities.
The use of collinear factorization and splitting matrices, 
encoding the collinear singular behaviour of scattering amplitudes
as introduced in Ref.~\cite{Catani:2003vu,Sborlini:2013jba}, 
is suitable for this discussion.

%%%%%%%%%%%%%%%
\begin{figure}[h]
\centering
\includegraphics[width=15cm]{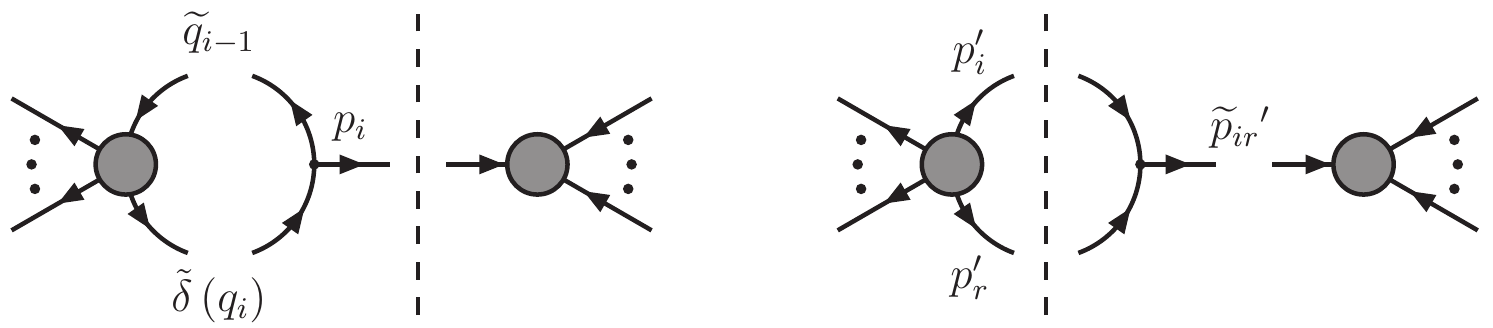}
\caption{\label{fig:collinear} 
Factorization of the dual one-loop and tree-level squared amplitudes 
in the collinear limit. The dashed line represents the momentum conservation
cut.}
\end{figure}
%%%%%%%%%%%%%%%

We consider the interference of the one-loop scattering amplitude
${\cal M}^{(1)}_N$ with the corresponding $N$-parton 
tree-level scattering amplitude ${\cal M}^{(0)}_N$, 
which is integrated with the appropriate phase-space factor 
\beq
\int d\Phi_N(p_1; p_2,\ldots, p_N) = 
\left(\prod_{i=2}^N \int_{p_i} \td(p_i) \right) \, 
(2\pi)^d \, \delta^{(d)} (\sum_{i=1}^{N} p_i)~,
\eeq
where we assume that only the external momentum $p_1$ is incoming ($p_{1,0}<0$).
Then, we select the corresponding dual contribution with the internal massless 
line $q_i$ on-shell
\bea
I^{(1)}_{i} \quad &=\quad 
2 {\rm Re} \, \int d\Phi_N(p_1; p_2,\ldots, p_N) 
\int_{\ell}  \,  \td(q_i) \, \theta(p_{i,0}-q_{i,0}^{(+)}) \nn \\ &\times\quad  
\bra{{\cal M}^{(0)}_N (p_1, \ldots, p_{N})} \, 
{\cal M}^{(0)}_{N+2} (\ldots, p_i, -q_i, q_i, p_{i+1}, \ldots) \ra~, 
\label{eq:I1ith}
\eea
where the loop energy in~\Eq{eq:I1ith} is restricted 
by the energy of the adjacent external massless particle $p_{i,0}$
to select the infrared sector, according to the discussion of the previous sections.
We also consider the $N+1$-parton tree-level scattering amplitude 
\beq
\ket{{\cal M}^{(0),\, ir}_{N+1}(p_1, p_2', \ldots)} = 
\ket{{\cal M}^{(0),\, ir}_{N+1} (\ldots, p_{ir}' \to p_i'+p_r', \ldots)}~, 
\label{eq:trees}
\eeq
where an extra particle is radiated from parton $i$, 
with $p_{ir}' = p_i' + p_r'$, and the complementary 
scattering amplitude ${\cal M}^{(0)}_{N+1}$
that contains all the tree-level contributions with the exception 
of those already included in ${\cal M}^{(0),\, ir}_{N+1}$.
The corresponding interference, integrated over the 
phase-space of the final-state particles, is 
\bea
&& I^{(0)}_{ir} = 2 {\rm Re} \, 
\int d\Phi_{N+1}(p_1; p_2', \ldots) \,
\bra{{\cal M}^{(0),\, ir}_{N+1} (p_1, p_2', \ldots)} \, 
{\cal M}^{(0)}_{N+1} (p_1, p_2', \ldots) \ra~.
\label{eq:I0ab}
\eea
For the simplicity of the presentation, 
we do not consider explicitly in this section 
the square of ${\cal M}^{(0),\, ir}_{N+1}$, which is related with 
a self-energy insertion in an external leg and whose infrared 
divergences are removed by wave-function 
renormalization~\cite{Catani:2008xa}.  
The final-state external momenta of the loop 
and tree amplitudes in \Eq{eq:I1ith} and \Eq{eq:I0ab}, 
although labelled with the same indices, are constrained by  
different phase-space momentum conservation delta functions. 
A mapping between the primed (real amplitudes) and unprimed 
(virtual amplitudes) momenta is necessary to show the 
cancellation of collinear divergences. 

In the limit where $\mathbf{p}_i$ and $\mathbf{q}_i$ become collinear the
dual one-loop matrix element ${\cal M}^{(0)}_{N+2}$ 
in \Eq{eq:I1ith} factorizes as 
\beq
\ket{{\cal M}^{(0)}_{N+2} (\ldots, p_i, -q_i, q_i, \ldots)} =
\sp^{(0)}(p_i, -q_i; -\widetilde{q}_{i-1}) \, 
\ket{\overline{\cal M}^{(0)}_{N+1} 
(\ldots, -\widetilde{q}_{i-1}, q_i, \ldots)} 
+ {\cal O}(\sqrt{q_{i-1}^2})~,
\label{eq:piqicoll}
\eeq
where the reduced matrix element $\overline{\cal M}^{(0)}_{N+1}$
is obtained by replacing the two collinear partons of 
${\cal M}^{(0)}_{N+2}$ by a single parent parton with light-like 
momentum 
\beq
\widetilde{q}_{i-1}^\mu = q_{i-1}^\mu - \frac{q_{i-1}^2 \, n^\mu}{2  \, n q_{i-1}}~, 
\eeq
with $n^\mu$ a light-like vector, $n^2=0$. 
Similarly, in the limit where $\mathbf{p}_i'$ and $\mathbf{p}_r'$
become collinear the tree-level matrix element 
${\cal M}^{(0),\, ir}_{N+1}$ factorizes as 
\beq
\bra{{\cal M}^{(0),\, ir}_{N+1} (p_1, p_2', \ldots, p_{N+1}')} =
\bra{\overline{\cal M}^{(0)}_N 
(\ldots, p_{i-1}', \widetilde{p}_{ir}', p_{i+1}', \ldots)} \, 
\sp^{(0) \dagger}(p_i', p_r'; \widetilde{p}_{ir}') 
+ {\cal O}(\sqrt{s_{ir}'})~,
\label{eq:piprimecoll}
\eeq
where $s_{ir}' = p_{ir}'^2$, and
\beq
\widetilde{p}_{ir}'^{\mu} = p_{ir}'^{\mu} 
- \frac{s_{ir}' \, n^\mu}{2\, n  p_{ir}'}
\eeq
is the light-like momentum of the parent parton. 
A graphical representation of the collinear limit of both virtual 
and real corrections is illustrated in Fig.~\ref{fig:collinear}.  
This graph suggests that in the collinear limit the mapping 
between the four-momenta of the virtual and real matrix elements 
should be such that 
$p_i = \widetilde{p}_{ir}'$,
$p_{j} = p_{j}' (j \ne i)$,
$-\widetilde{q}_{i-1} = p_i'$ and 
$q_i = p_r'$ in the collinear limit. 
Notice that $p_r'$ is restricted by momentum conservation 
but $q_i$ is not. However, the relevant infrared region is 
bound by $q_{i,0}^{(+)} \le p_{i,0}$  in \Eq{eq:I1ith}. This restriction 
allows to map $q_i$ to $p_r'$. The mapping, nevertheless, 
is not as obvious as can be induced from Fig.~\ref{fig:collinear}
as the propagators that become singular in the collinear limit 
in the virtual and real matrix elements are different. 
Reconsidering $p_i'$ as the parent parton momentum of the 
collinear splitting, we find the following relation 
between splitting matrices entering the real matrix elements
\beq
\sp^{(0) \dagger}(p_i', p_r'; \widetilde{p}_{ir}') = 
\frac{(\widetilde{p}_{ir}' - p_r')^2}{s_{ir}'} \, 
\sp^{(0)}(\widetilde{p}_{ir}', - p_r'; p_i')~, 
\label{antisplittings}
\eeq
where $(\widetilde{p}_{ir}' - p_r')^2/s_{ir}' = - n p_i'/n p_{ir}'$.
We show now that the factor $- n p_i'/n p_{ir}'$ is compensated 
by the phase-space. By introducing the following identity in 
the phase-space of the real corrections
\beq
1 = \int \, d^d p_{ir}' \, 
\delta^{(d)}\left(p_{ir}' - p_i' - p_r' \right)~,
\eeq
and performing the integration over the three-momentum ${\bf p}_i'$
and the energy component of $p_{ir}'$, the real 
phase-space becomes 
\beq
\int d\Phi_{N+1}(p_1; p_2', \ldots) = 
\int d\Phi_N(p_1; \ldots,p_{ir}',\ldots) \, 
\int_{p_r'} \, \td(p_r') \, 
\frac{E_{ir}'}{E_i'}~,
\label{pscollinear}
\eeq
where the factor $(n p_i'/n p_{ir}')(E_{ir}'/E_i')$
equals unity in the collinear limit.
Inserting~\Eq{eq:piqicoll} in~\Eq{eq:I1ith}, 
and~\Eq{eq:piprimecoll}, \Eq{antisplittings} and \Eq{pscollinear} 
in \Eq{eq:I0ab} the loop and tree contributions show to have a 
very similar structure with opposite sign and match each other 
at the integrand level in the collinear limit. Correspondingly, 
soft singularities at $p_r'\to 0$ can be treated consistently as 
the endpoint limit of the collinear mapping.

%----------------------------------------------------------------------------------------
%	Conclusions and outlook
%----------------------------------------------------------------------------------------

\section{Conclusions and outlook}
\label{sec:conclusions}

The Loop--Tree Duality method exhibits attractive theoretical
aspects and nice properties which are manifested by a direct physical 
interpretation of the singular behaviour of the loop integrand. 
Integrand singularities occurring in the intersection 
of on-shell forward hyperboloids or light-cones cancel among dual contributions. 
The remaining singularities, excluding UV divergences,
are found in the intersection of forward with backward on-shell 
hyperboloids or light-cones and are 
produced by dual propagators that are light-like or time-like separated 
and less energetic than the internal propagator that is set on-shell. 
Therefore, these singularities can be interpreted in terms of causality and 
are restricted to a finite region of the loop three-momentum space, 
which is of the size of the external momenta. As a result, 
a local mapping at the integrand level is possible between one-loop 
and tree-level matrix elements to cancel soft and collinear divergences. 
One can anticipate that a similar analysis at higher orders of the 
Loop--Tree Duality relation is expected to provide equally interesting results. 
We leave this analysis for future work.

% Chapter Template

\chapter{Multi-leg Scalar Integrals} % Main chapter title

\label{Chapter7} % Change X to a consecutive number; for referencing this chapter elsewhere, use \ref{ChapterX}

%\lhead{Chapter 6. \emph{Multi-leg Scalar Integrals}} % Change X to a consecutive number; this is for the header on each page - perhaps a shortened title

After all these 
theoretical foundations haven been established in the previous chapters, let us turn to an actual numerical implementation. 
We have written a computer program in C++ to give a practical 
proof-of-concept of the theoretical framework, and to be used for the calculation 
of one-loop diagrams. The program at its current state will serve as the basis of any 
further developments. The specifics of the program are given in Section \ref{sec:program}. 
For the moment let us add here that from the point of view of a dual integral, calculating a 
phase--space point in the absence of threshold singularities 
is comparatively easy. Things become 
more challenging for points with complex results, because these feature 
singularities that do not cancel among dual contributions which means 
they have to be dealt with by contour deformation. Since the energy 
component is already fixed due the prior application of the Residue 
Theorem, a dual integral and consequently the contour deformation 
lives only in three spatial dimensions.

%we then compare to analytical results when availiable.

\newpage

%----------------------------------------------------------------------------------------
%	Preparation of the Numerical Implementation
%----------------------------------------------------------------------------------------

\section{Preparation of the Numerical Implementation}

The goal is to calculate a one-loop scalar integral in its dual representation: 
\bea
%\label{oneloopduality}
L^{(1)}(p_1, p_2, \dots, p_N) 
\quad=\quad - \sum_{i \in \alpha_1} \, \int_{\ell} \; \td(q_i) \,
\prod_{\substack{j \in \alpha_1 \\ j\neq i}} \,G_D(q_i;q_j)
\eea 
Therefore, the program relies on a set of fundamental equations. These shall be 
presented here. Since the program shall work for generic masses, 
we need, as a first ingredient, the massive dual propagator
\bea
G_D(q_i;q_j) = \frac{1}{q_j^2 -m_j^2 - i0 \, \eta \, k_{ji}}~.
\label{eq:dualprop}
\eea
However, in practice we will make use of its rewritten form as in \Eq{eq:newdual}
\beq
\td(q_i) \, G_D(q_i;q_j) = 
i\, 2 \pi \, 
\frac{\delta(q_{i,0}-q_{i,0}^{(+)})}{2 q_{i,0}^{(+)}} \, 
\frac{1}{(q_{i,0}^{(+)} + k_{ji,0})^2-(q_{j,0}^{(+)})^2}~,
%\label{eq:newdual}
\eeq
where
\beq
q_{i,0}^{(+)} = \sqrt{\mathbf{q}_i^2 + m_i^2-i0} \quad \text{and} \quad k_{ji} = k_j-k_i~.
\label{qi0distance2}
\eeq
This means actually, that the integration to perform is a three-dimensional one, 
because $\mathbf{q}_i=\boldsymbol{\ell}+\mathbf{k}_i$ where $\boldsymbol{\ell}$ 
is the loop three-momentum and $\mathbf{k}_i = \sum_{j=1}^i \mathbf{p}_j$ 
the sum of the external three-momenta up to the i-th. 
The zero-component has already been integrated out as explained in 
Chapter \ref{Chapter3} and as a result the $q_{i,0}$-component is fixed according 
to \Eq{qi0distance2}. This is indicated by the presence of the delta function $\td(q_i)$.\\
The next ingredient is given by
\bea
k_{ji}^2-(m_j+m_i)^2 &\geq 0,\quad k_{ji,0} < 0 & &\text{ellipsoid singularity}\\
k_{ji}^2-(m_j-m_i)^2 &\leq 0 & &\text{hyperboloid singularity}
\eea
These two equations are used to identify the type of singularity that we are dealing with. 
In Section \ref{sec:dualsing} we will see how they come into play.

%----------------------------------------------------------------------------------------
%	The Mapping
%----------------------------------------------------------------------------------------

\section{The Mapping}

We use Cuhre and VEGAS from the Cuba library \cite{Hahn:2004fe} as numerical integrators. Cuhre is 
a deterministic integrator that uses cubature rules and 
performs well as long as the dimensionality of the integral 
is not too high. As stated before, dual integrals are only three-dimensional making Cuhre 
an excellent choice for the task. VEGAS on the other hand, is a Monte Carlo-integrator that 
uses importance sampling for variance reduction. While it is much slower, it is useful to 
crosscheck results and most probably will be used in later stages when real and virtual corrections are 
combined.\\
Both integrators have in common that they assume the integration region to be the unit cube. 
Hence, in order to perform an integral over the entire phase-space, we must find a mapping 
$(-\infty,\infty)^3\to[0,1]^3$.\\
The first and easiest option that comes to mind is simply rescaling every dimension. One 
possible mapping to do so is
\bea
\ell_i = \tan\left(\pi\left(x_i-\frac{1}{2}\right)\right)~,\quad i=1,2,3
\eea
When $x_i \to 0$, the argument of the tangent goes to 
$ -\pi/2$ and the tangent of it to $-\infty$, when $x_i\to 1$, the argument goes to 
$+\pi/2$ and the tangent of it to $+\infty$.\\
However, with regard to program stability it is advantageous to choose a mapping that
relies on the use of spherical coordinates. In fact, it is a two-step process: In the first step 
the Cartesian coordinates $\ell_x, \ell_y, \ell_z$ running from $-\infty$ to $+\infty$ get mapped 
to $r,\cos(\theta)$ and $\phi$, with $r$ being the radius, $\phi$ 
azimuth and $\theta$ the polar angle, respectively.
\bea
\ell_x &= r\cos(\phi)\sqrt{1-\cos^2(\theta)}\nn\\
\ell_y &= r\sin(\phi){\sqrt{1-\cos^2(\theta)}}\nn\\
\ell_z &= r\cos(\theta)
\eea
In the second step these get 
mapped to $x,y,z\in [0,1]$.
\bea
r &= \frac{x}{x-1}\nn\\
\cos(\theta) &= 1-2y\nn\\
\phi &= 2\pi z
\eea
%$x,y,z\in [0,1]$ get mapped to $r,\cos(\theta)$ and $\phi$, with $r$ being the radius, $\phi$ 
%azimuth and $\theta$ the polar angle, respectively.
%\bea
%r &= \frac{x}{x-1}\nn\\
%\cos(\theta) &= 1-2y\nn\\
%\phi &= 2\pi z
%\eea
%In the second step these get 
%mapped to Cartesian coordinates $\ell_x, \ell_y, \ell_z$ running from $-\infty$ to $+\infty$.
%\bea
%\ell_x &= r\cos(\phi)\sqrt{1-\cos^2(\theta)}\nn\\
%\ell_y &= r\sin(\phi){\sqrt{1-\cos^2(\theta)}}\nn\\
%\ell_z &= r\cos(\theta)
%\eea
Despite seeming more complex at first glance, it is actually slightly faster. 
 %\begin{table}[H]
%\setlength{\tabcolsep}{0.3pc}
% \caption{Pentagon example results}
   % {\small
  % \centering
%\begin{tabular}{lccc}
%&\\
%\hline
%\hline
%\\
% & Runtime Triangle & Runtime Box & Runtime Pentagon\\
%\\
%\hline
%\hline
%\\
%Cartesian coordinates & 64.3\text{s} & 107.8s & 163.9\\
%
%Spherical coordinates & 63.2\text{s} & 106.5s & 163.6\\
%
%\hline
%\hline
%\end{tabular}
%}
%\caption{The choice of mapping consistently affects the runtime.}
%\label{tab:runtime}
%\end{table}
%The values of Table \ref{tab:runtime} have been generated by 
%averaging over 100 runs of the same momentum configuration. 
%To magnify the differences due to the choice of mapping, VEGAS 
%has been used as numerical integrator. For this reason, the displayed 
%program runtimes are much longer than the ones given in Section \ref{sec:results}.\\
Of course, every change of variables demands the inclusion of 
the corresponding Jacobian.

\newpage

%----------------------------------------------------------------------------------------
%	Singular behaviour of dual contributions
%----------------------------------------------------------------------------------------

\section{Singular behaviour of dual contributions \label{sec:dualsing}}

Dual integrals feature certain types of singularities. This has already been thoroughly 
discussed in Section \ref{sec:cancel}, thus, in this section, we want to quickly 
recapitulate the parts that are going to be relevant to a numerical implementation. 
%as well as point out the consequences.\\
For generic masses, the loop integrand becomes singular at on-shell 
hyperboloids with $q_{i,0}^{(+)} = \sqrt{\mathbf{q}_i^2 + m_i^2-i0}$ (forward-hyperboloids, 
solid lines of Fig. \ref{fig:cartesean}) and $q_{i,0}^{(-)} = -\sqrt{\mathbf{q}_i^2 + m_i^2-i0}$ 
(backward-hyperboloids, dashed lines of Fig. \ref{fig:cartesean}). The origins of the corresponding
on-shell hyperboloids are at $-\mathbf{k}_i$. %This situation is drawn once again 
%in two dimensions in Figure \ref{fig:oshmassive} for a three propagator setup.
%\begin{SCfigure}[][h]
%\centering
%\includegraphics[scale=1]{Figures/LCHyperboloidsMassive.pdf}
%\caption{On-shell hyperboloids of three propagators in the ($\ell_0,\ell_z$)-plane. 
%The red dots mark the locations of the singularities. 
%Loop--Tree Duality is equivalent to integrating along the forward 
%hyperboloids. \label{fig:oshmassive}}
%\end{SCfigure}
There are two main types of singularities to distinguish: 
\begin{itemize}
\item Forward-foward intersection. In Fig. \ref{fig:cartesean}, 
these are the singularities that correspond to the intersection of two solid
lines. They cancel among dual contributions. In short, the reason is the following 
(for more details, see Section \ref{sec:cancel}):  
Propagators are positive inside and negative outside. When integrating along 
the forward-hyperboloids, every singularity is passed twice. One 
time going from the inside to the outside  (or vice versa) and the second time 
from the outside to the inside (or vice versa). The crucial point is that therefore, 
the contributions coming from the two integrations have opposite sign and thus 
cancel out.\\
The situation for this type of singularity is drawn in Fig. \ref{fig:singhyperboloid}, once in 
loop three--momentum space (\ref{fig:singhyperbolid3d}) and for easier 
legibility in two dimensions (\ref{fig:singhyperbola2d}). The surface (curve) represents 
all points for which the dual propagator in consideration has a pole. The 
dots indicate the positions of the foci ($=-\mathbf{k}_i$) as introduced in
Chapter \ref{Chapter6}.
%Since Figure \ref{fig:oshmassive} is only a two-dimensional slice, these 
%singularities appear as points. In the full four-dimensional world they take 
%the form of hyperboloids. Therefore we will refer to them as 
%hyperboloid singularities from now on.
\item Forward-backward intersection. These singularities originate from 
the intersection of a solid with a dashed line in Fig. \ref{fig:cartesean}. 
They remain and require to 
be dealt with by contour deformation. In Section \ref{sec:contourdef} we 
will see how this is done.\\
In Fig. \ref{fig:singellipsoid}, the shapes of the surfaces at which 
propagators with such a singularity become infinite are illustrated. 
Again we show the actual loop three--momentum space plot 
(Figure \ref{fig:singellipsoid3d}) alongside a 
simpler two-dimensional version (Fig. \ref{fig:singellipse2d}).
%Due to Figure \ref{fig:oshmassive} only being a two-dimensional plot 
%these singularities show up as mere points. In reality, i.e. in four 
%dimensions, these singularities have the shape of ellipsoids. Therefore, 
%we will call them ellipsoid singularities.
\end{itemize}
Because Fig. \ref{fig:cartesean} is only a two-dimensional plot 
these singularities show up as mere points. In reality, i.e. in $1+3$ 
dimensions (with the expression $1+n, n\in \mathbb{N}$, dimensions, 
we refer to a Minkowski space with 1 time and $n$ spacial dimensions), 
these singularities have the shape of hyperboloids or ellipsoids, 
respectively, as has been shown in Fig. \ref{fig:singhyperbolid3d} and 
\ref{fig:singellipsoid3d}. 
Therefore, we will call them hyperboloid or ellipsoid singularities 
from now on.
 \begin{figure}[h]
\centering
\begin{subfigure}[b]{0.45\textwidth}
\includegraphics[width=\textwidth]{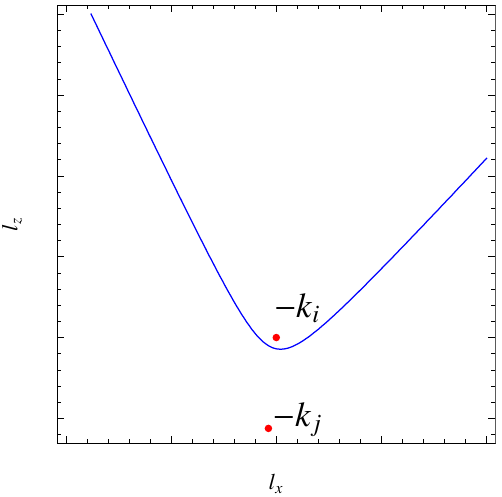}
\caption{Two-dimensional hyperbola}
\label{fig:singhyperbola2d}
\end{subfigure}
\quad
\begin{subfigure}[b]{0.45\textwidth}
\includegraphics[width=\textwidth]{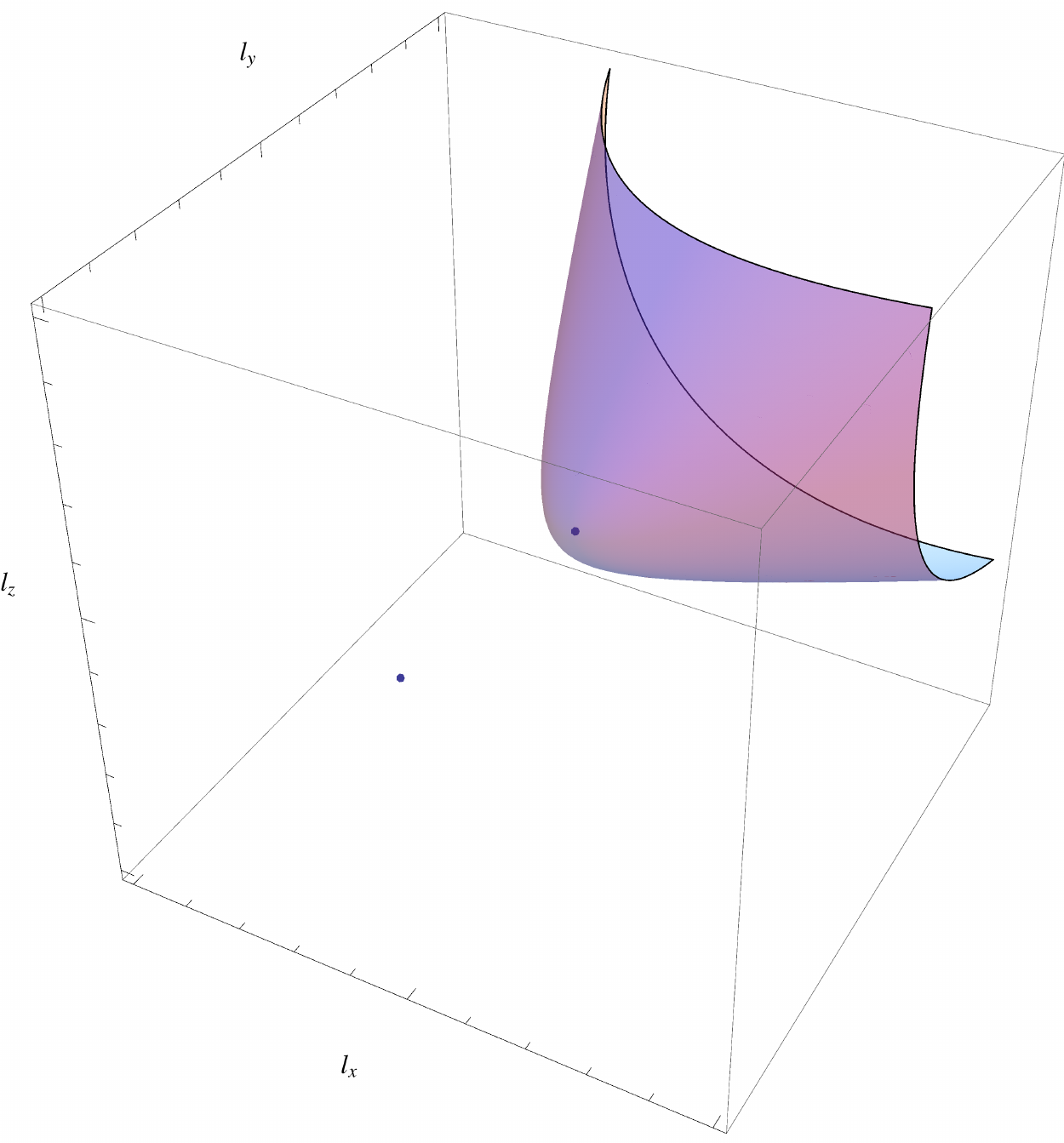}
\caption{Three-dimensional hyperboloid}
\label{fig:singhyperbolid3d}
\end{subfigure}
\caption{Example of a hyperboloid singularity. In 1+2 dimensions the hyperboloid 
degenerates to a hyperbola, as displayed on the left.}
\label{fig:singhyperboloid}
\end{figure}
 \begin{figure}[h]
\centering
\begin{subfigure}[b]{0.45\textwidth}
\includegraphics[width=\textwidth]{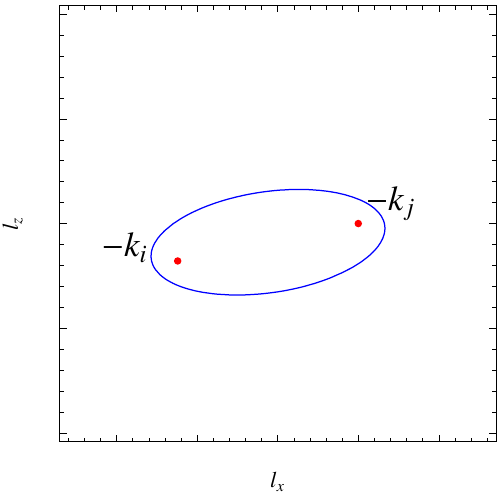}
\caption{Two-dimensional ellipse}
\label{fig:singellipse2d}
\end{subfigure}
\quad
\begin{subfigure}[b]{0.45\textwidth}
\includegraphics[width=\textwidth]{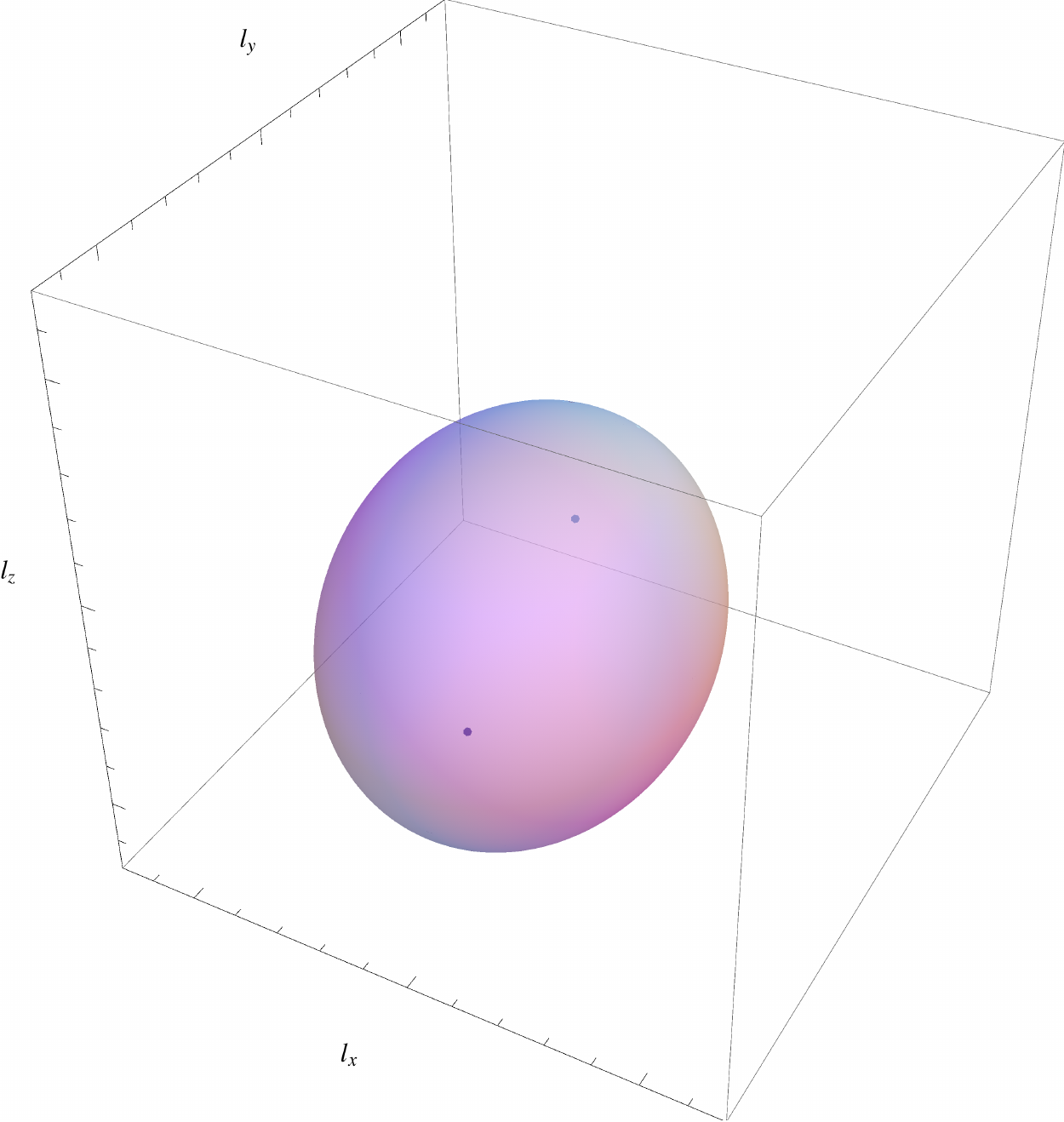}
\caption{Three-dimensional ellipsoid}
\label{fig:singellipsoid3d}
\end{subfigure}
\caption{Example of an ellipsoid singularity. In 1+2 dimensions the ellipsoid 
degenerates to an ellipse, as displayed on the left.}
\label{fig:singellipsoid}
\end{figure}

%These two points combined have a major on impact on how to employ 
%contour deformation when using duality. This will be discussed in Section
 %\ref{sec:interplay}.
 \newpage

%----------------------------------------------------------------------------------------
%	Interplay of dual contributions
%----------------------------------------------------------------------------------------

\section{Interplay of dual contributions\label{sec:interplay}}

Loop--Tree Duality transforms one-loop integrals into so-called dual 
integrals:
\beq
L^{(N)}(p_1, p_2, \dots, p_N) = \int_\ell \;\; \prod_{i=1}^{N} \,G_F(q_i) 
\longrightarrow
- \sum_{i \in \alpha_1} \, \int_{\ell} \; \td(q_i) \,
\prod_{\substack{j \in \alpha_1 \\ j\neq i}} \,G_D(q_i;q_j)
\label{eq:trafodual}
\eeq 
with $\alpha_1$ a set of internal momenta belonging to the same 
loop. %From now on, we will refer to the summands of the sum of 
%\Eq{eq:trafodual} as \textit{dual contributions}.\\
\\
Symbolically speaking, we could express Duality with the following 
scheme:
\begin{figure}[H]
\centering
\bea
\begin{ropmatrix}
G_F\cdot G_F \cdots G_F\quad\xrightarrow{\hspace*{1.0cm}\text{\normalsize Duality}\hspace*{1.0cm}}\qquad\\
\\
\\
\\
\\
\end{ropmatrix}
\begin{matrix} 
\delta & G_D & G_D & \cdots & G_D\\
 G_D & \delta & G_D & \cdots & G_D\\
 G_D & G_D & \delta & \cdots & G_D\\
 \vdots & \vdots & \vdots & \ddots & \vdots\\
 G_D & G_D & G_D & \cdots & \delta
 \end{matrix}\nn
 \eea
 \caption{Symbolical representation of the Loop--Tree Duality. 
 $G_F=$ Feynman propagator, $G_D=$ dual propagator, 
 $\delta=\td(q_i)$. \label{fig:ltdsymb}}
 \end{figure}
Each line in the matrix-like structure on the right side of the arrow of Fig. \ref{fig:ltdsymb} 
represents a dual contribution whereas the columns give the number of the 
corresponding external leg. This scheme can now be used to indicate 
the positions of the different singularities in a dual integral. Consider the 
following small example, in which zero means no singularity, 
H stands for hyperboloid singularity and 
E for ellipsoid singularity:
\bea
\begin{matrix}
0 & \text{\bf{H}} & 0 & \text{\bf{H}}\\
\text{\bf{H}} & 0 & \text{\bf{H}} & \text{\bf{E}}\\
\text{\bf{E}} & \text{\bf{H}} & 0 & \text{\bf{E}}\\
\text{\bf{H}} & 0 & 0 & 0
\end{matrix}
\label{eq:notationexample}
\eea
The correct interpretation would be: 
There is a one-loop box integral, hence we have four dual contributions. 
On the main diagonal we have only zeros because the dual delta functions 
cannot produce hyperboloid or ellipsoid singularities. The first dual 
contribution has two hyperboloid singularities, namely at positions two and four, 
the second dual contribution has hyperboloid singularities at 
positions one and three and an ellipsoid singularity at position 
four. The third dual contribution has two ellipsoid singularities at positions 
one and four and one hyperboloid singularity at position two and the fourth 
dual contribution has only one hyperboloid singularity at position 
one. This way of denoting singularities will become useful momentarily.\\
There is one further observation to make: 
Apparently the hyperboloid singularities are distributed symmetrically 
around the main diagonal. This is not by accident. Going back to Section 
\ref{sec:loop} we have \Eq{hyperboloid} which is the 
defining equation for hyperboloid singularities. Due to its symmetry 
under the exchange of $i$ ($i$ counts dual contributions) and 
$j$ ($j$ counts leg positions) the hyperboloid singularities 
always appear in pairs and are distributed symmetrically around 
the main diagonal of (\ref{eq:notationexample}). 
Inspecting \Eq{ellipsoid}, which is the defining equation for 
ellipsoid singularities, we see that this equation is not 
symmetric under the exchange of indices. Thus for every 
ellipsoid singularity in \ref{eq:notationexample} we have a 
zero as its counterpart.\\
In Section \ref{sec:dualsing}, and with more detail in Section \ref{sec:cancel},
we established that hyperboloid 
singularities cancel among dual contributions and therefore do 
not need to be treated via contour deformation. Yet they impact the 
way we have to deform. In order to preserve the cancellation of 
hyperboloid singularities, dual contributions featuring the same 
hyperboloid singularity (pair) must receive the same deformation. To 
further illustrate this point, let us look at the following pentagon example:
\begin{figure}[H]
\scalebox{1.1}{\parbox{.5\linewidth}{%
\bea
\begin{matrix*}[l]
0 & \text{\bf{H}} & 0 & 0 & 0 & \rdelim\}{3}{10mm}[] & \text{Contributions are coupled:}\\
\text{\bf{H}} & 0 & \text{\bf{H}} & \text{\bf{E}} & 0 & & \text{Every contribution receives all deformations}\\
\text{\bf{E}} & \text{\bf{H}} & 0 & \text{\bf{E}} & 0 & & \text{that occur within the coupling.}\\
\text{\bf{E}} & \text{\bf{E}} & \text{\bf{E}} & 0 & \text{\bf{E}} &  \; \to\; & \text{Deform with ellipsoids that itself contains.}\\
0 & 0 & 0 & 0 & 0 &  \;\to\; & \text{No deformation needed here.}
\end{matrix*}
\nn
\eea
}}
\caption{Pentagon with dual contributions coupled by hyperboloid
singularities.\label{fig:pentacouple}}
\end{figure}
In Fig. \ref{fig:pentacouple}, contributions one, 
two and three are coupled 
via their common hyperboloid singularities. 
Thus, they need to receive the very same deformation that accounts 
for \textit{all} ellipsoid singularities occurring within those contributions. 
%In the example these 
These are found at position four of the second contribution 
and positions one and four of the third contribution. The fourth 
dual contribution is not coupled to any other contribution and can be 
deformed as standalone. The fifth contribution does not require any 
treatment.\\
As a general strategy, one organizes the dual contributions into groups. 
A group is a set of pairwisely coupled contributions. Each of the groups is 
deformed independently from the others. Within a group every contribution 
receives the \textit{same} deformation that accounts for \textit{all} the 
ellipsoids of the group.\\
Turning back to the example of Fig. \ref{fig:pentacouple}, we would have three groups: the first 
group involves contributions one to three, the second group constitutes of 
contribution four and the third group constitutes of contribution five.

%----------------------------------------------------------------------------------------
%	Contour deformation
%----------------------------------------------------------------------------------------

\section{Contour deformation \label{sec:contourdef}}

The ellipsoid singularities (forward-backward type) lie on the real axis. To avoid 
them, we need to deform the integration path into the imaginary space.

%-----------------------------------
%	A one-dimensional example
%-----------------------------------

\subsection{A one-dimensional example\label{sec:cdef1d}}

To motivate the basic concept of contour deformation, let us have a look 
at the following simple example. The function
\bea
f(\ell_x) = \frac{1}{\ell_x^2-E^2+i0}
\eea
has poles at $\ell_{x\pm}=\pm(E-i0)$. Simply integrating along the 
real axis would lead to infinities. Therefore, an integration path that 
goes \textit{around} the singular points is needed. One possible 
way to achieve this would be:
\bea
\ell_x \to \ell_x' = \ell_x + i\lambda\ell_x\exp\left(-\frac{\ell_x^2-E^2}{2E^2}\right)
\label{eq:defo2d}
\eea
The parameter $\lambda$ serves to scale the deformation along the imaginary 
axis. At the position of the pole, the exponent becomes $0$ and thus the 
exponential function hits its maximum, which is $1$. Far away from the poles 
the exponent is a large negative number, hence exponentiating it 
suppresses the deformation.
\begin{figure}[H]
\centering
\includegraphics[scale=1]{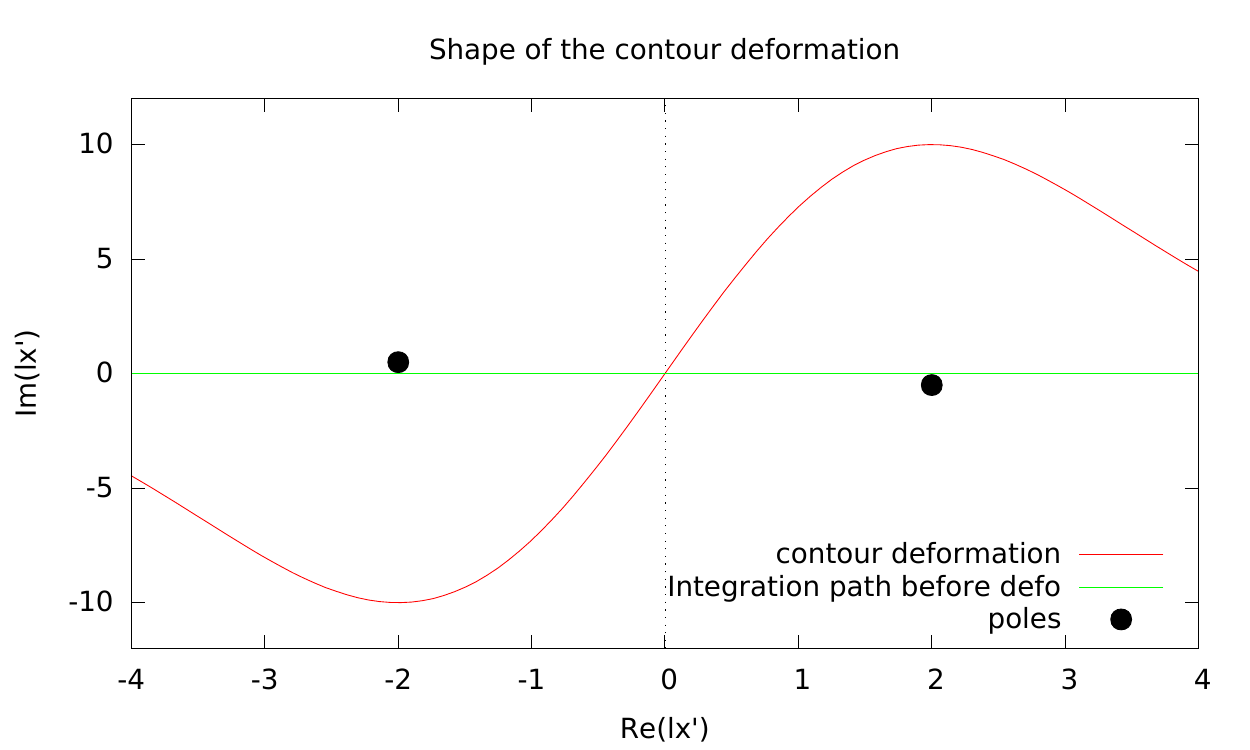}
\caption{Contour deformation as in \Eq{eq:defo2d} for $\lambda=5$ and $E=2$.\label{fig:defo2d}}
\end{figure}

%-----------------------------------
%	Deformation in three dimensions
%-----------------------------------

\subsection{Deformation in three dimensions\label{sec:defo3d}}

Every valid deformation must satisfy a certain set of 
requirements \cite{Soper:1999xk}:
\begin{enumerate}
\item The deformation has to respect the $i0$-prescription of the propagator:\\
In general, a ($d$-dimensional) contour deformation has the form:
\bea
\ell\to\ell' = \ell + i\kappa
\label{eq:gendef}
\eea
where $\kappa$ usually is a function of the loop momentum $\ell$. In our case, we want 
to perform the integration over a product of dual propagators. Plugging the deformation 
of \Eq{eq:gendef} into the on-shell energy (\Eq{qi0distance2}), we obtain
\bea
q_{i,0}^{(+)}=\sqrt{-\boldsymbol{\kappa}^2+
2i\boldsymbol{\kappa}\cdot\mathbf{q}_i+\mathbf{q}_i^2+m_i^2-i0}~.
\eea
The Feynman prescription $-i0$ tells us in which direction to deform when coming 
close to a singularity. Hence, any valid deformation must match this prescription. 
Consequently, we need to have
\bea
\boldsymbol{\kappa}\cdot\mathbf{q}_i<0~.
\eea
%\item 
\item The deformation should vanish at infinity:\\
We are looking for a deformation that does not change the actual value of the integral. 
Therefore, we do not want $|\kappa|$ to grow for $|\ell|\to\infty$. An easy 
way to satisfy this condition is to choose $\kappa$ such that $|\kappa|\to 0$ 
as $|\ell |\to \infty$.\footnote{
Strictly speaking, there is a third condition:\\
The deformation must vanish at the position of soft or collinear singularities:\\
This point is of importance for the matching of soft and collinear singularities 
between real and virtual corrections. If the deformation shifts those singularities, 
alongside everything else, the cancellation will be spoilt. 
However, in the scope of this thesis, we are only dealing with finite diagrams.
}
\end{enumerate}
With these conditions in mind, we construct the deformation in the following way:\\
As explained in Section \ref{sec:interplay}, we first organize the dual contributions 
into groups. For every ellipsoid singularity of the group we include a factor:
\bea
\lambda_{ij}\left(\frac{\mathbf{q}_i}{\sqrt{\mathbf{q}_i^2}}+
\frac{\mathbf{q}_j}{\sqrt{\mathbf{q}_j^2}}\right)\exp\left(
\frac{-G_D^{-2}(q_j;q_i)}{A_{ij}}\right)~,
\eea
with $\mathbf{q}_i=\boldsymbol{\ell}+\mathbf{k}_i$ and $\boldsymbol{\ell}$ 
the loop three-momentum. It is made up of two main components: \\
The \textit{vector-part} 
$\left(\mathbf{q}_i/\sqrt{\mathbf{q}_i^2}+\mathbf{q}_j/\sqrt{\mathbf{q}_j^2}\right)$
is designed to always point to the outside of the singularity
ellipsoid, see Fig. \ref{fig:vectorpart}. $\mathbf{q}_i$ is a vector 
that points from focus $i$ to a particular point in loop three-momentum 
space. This is indicated by the dashed lines. Keep in mind that, 
despite its name, a `focus' is not a focal point of the ellipse/ellipsoid, but rather 
refers to the foci introduced in Section \ref{sec:loop}, i.e. focus $i = -\mathbf{k}_i$. 
Dividing by $\sqrt{\mathbf{q}_i^2}$ rescales the vector to length 1. The same 
procedure is repeated with $\mathbf{q}_j$. Adding the two together 
results in the angle between 
$\mathbf{q}_i$ and the resulting vector part always being $\leq \pi/2$. 
By choosing all the scaling parameters 
$\lambda_{ij}<0$ for all possible combinations $\{ij\}$ we satisfy 
the first condition.
\begin{figure}[H]
\centering
\includegraphics[scale=0.7]{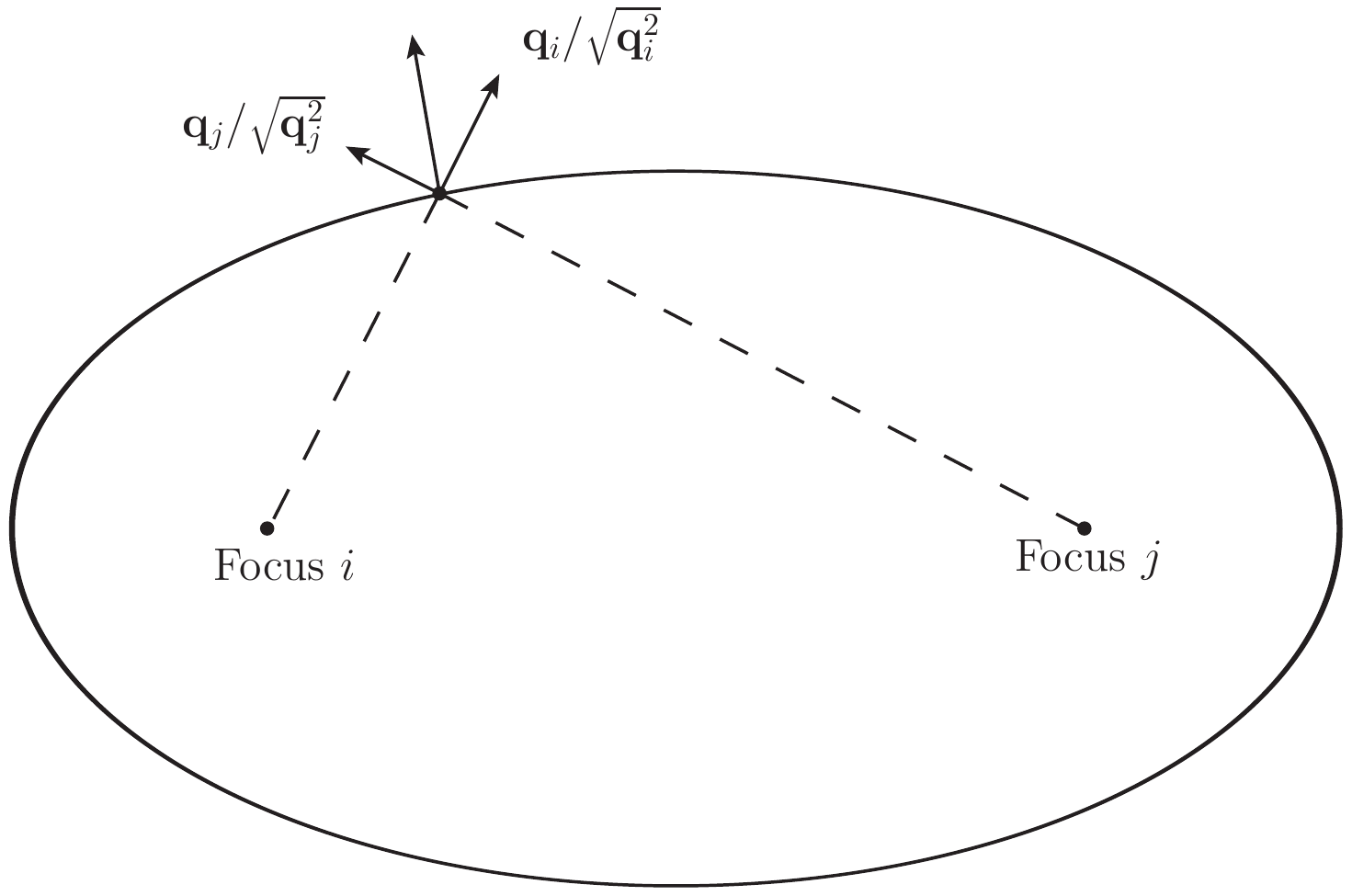}
\caption{Two-dimensional slice of 
the singularity ellipsoid of dual contribution $i$ 
at position $j$. The resulting vector gives 
the orientation of the vector part.\label{fig:vectorpart}}
\end{figure}
Inside the singularity ellipsoid  
the two vectors $\mathbf{q}_i/\sqrt{\mathbf{q}_i^2}$ and 
$\mathbf{q}_j/\sqrt{\mathbf{q}_j^2}$
cancel totally along the 
major axis of the ellipsoid. This cancellation in the vicinity of the 
center of the singularity ellipsoid is shown in Fig. \ref{fig:vecpart}. 
In order to be able to plot the situation, the graphs are restricted to 1+2 
dimensions. Keep in mind that the imaginary part of the plotted deformation 
is a two component vector where each component is a 
function of $\ell_x$ \textit{and} $\ell_y$. Therefore, the graphs 
\ref{fig:vecpartx} and \ref{fig:vecparty} indicate how far the deformation 
goes in the direction of the imaginary part of the $\ell_x$- and $\ell_y$-axis.
\begin{figure}[H]
\centering
\begin{subfigure}[b]{0.48\textwidth}
\includegraphics[width=\textwidth]{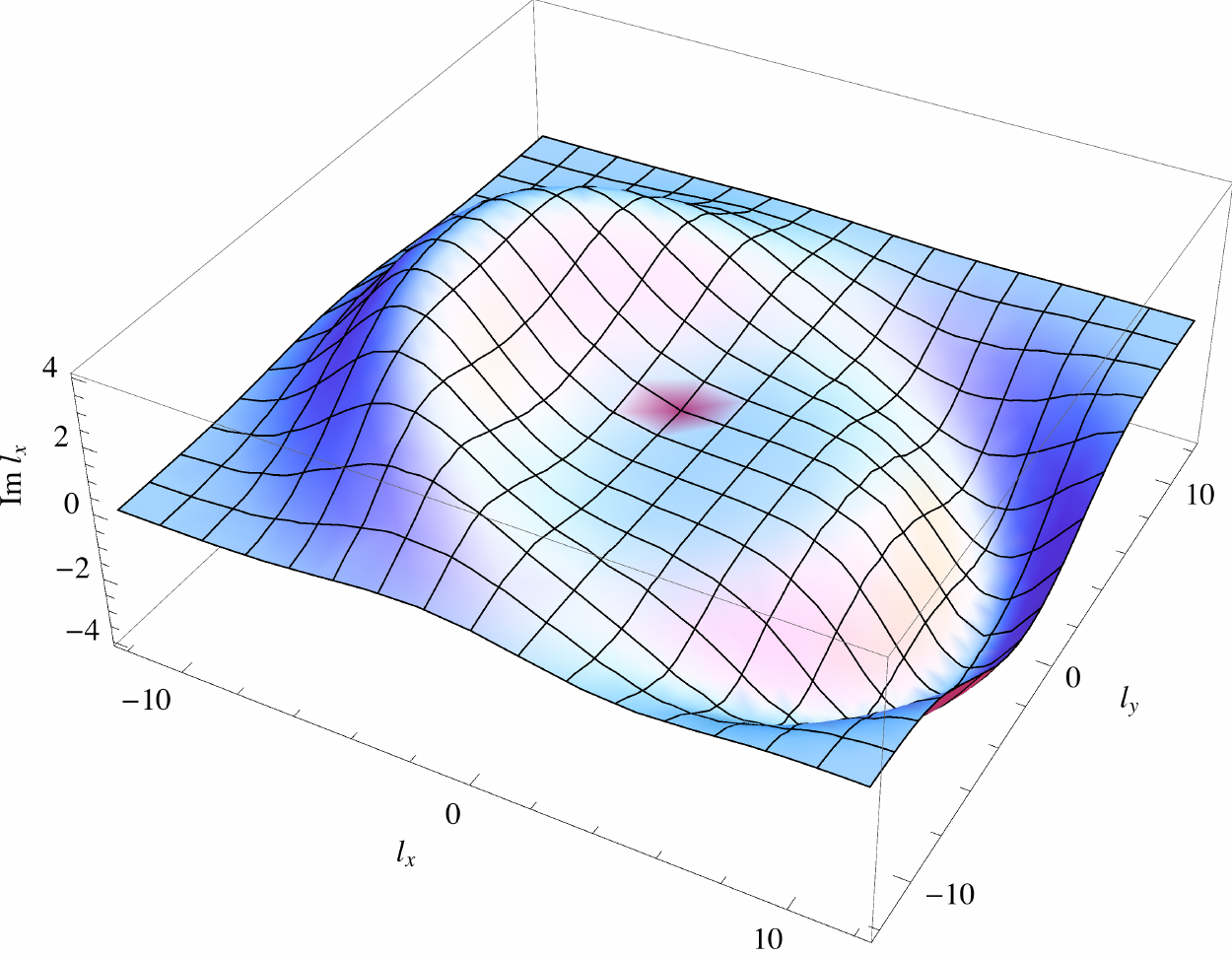}
\caption{x-component}
\label{fig:vecpartx}
\end{subfigure}
\quad
\begin{subfigure}[b]{0.48\textwidth}
\includegraphics[width=\textwidth]{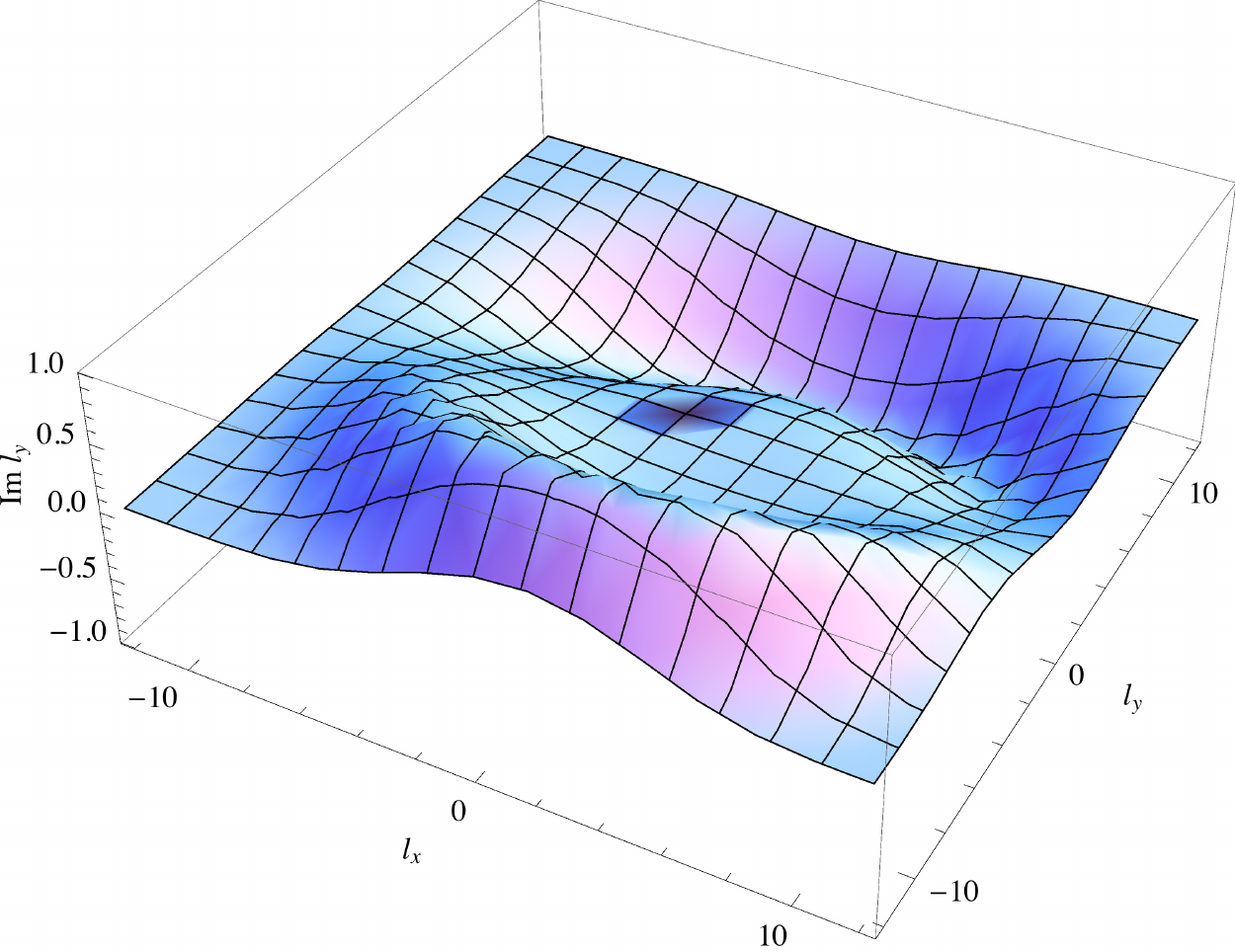}
\caption{y-component}
\label{fig:vecparty}
\end{subfigure}
\caption{Imaginary part of the deformation in 1+2 dimensions. 
The vector part causes the deformation to flatten inside the 
singularity ellipsoid.}
\label{fig:vecpart}
\end{figure}

\newpage

The \textit{suppression factor} $\exp(-G_D^{-2}(q_j;q_i)/A_{ij})$ fulfills the 
second condition. At the position of the singularity, $G_D^{-2}(q_j;q_i)$ is 
0 and thus the suppression factor reaches its maximum. Far away from 
the singularity, $-G_D^{-2}(q_j;q_i)$ is a large negative value and thus the 
exponential is close to zero. This behaviour is demonstrated in Fig. 
\ref{fig:volcano}.
\begin{figure}[H]
\centering
\includegraphics[scale=1.1]{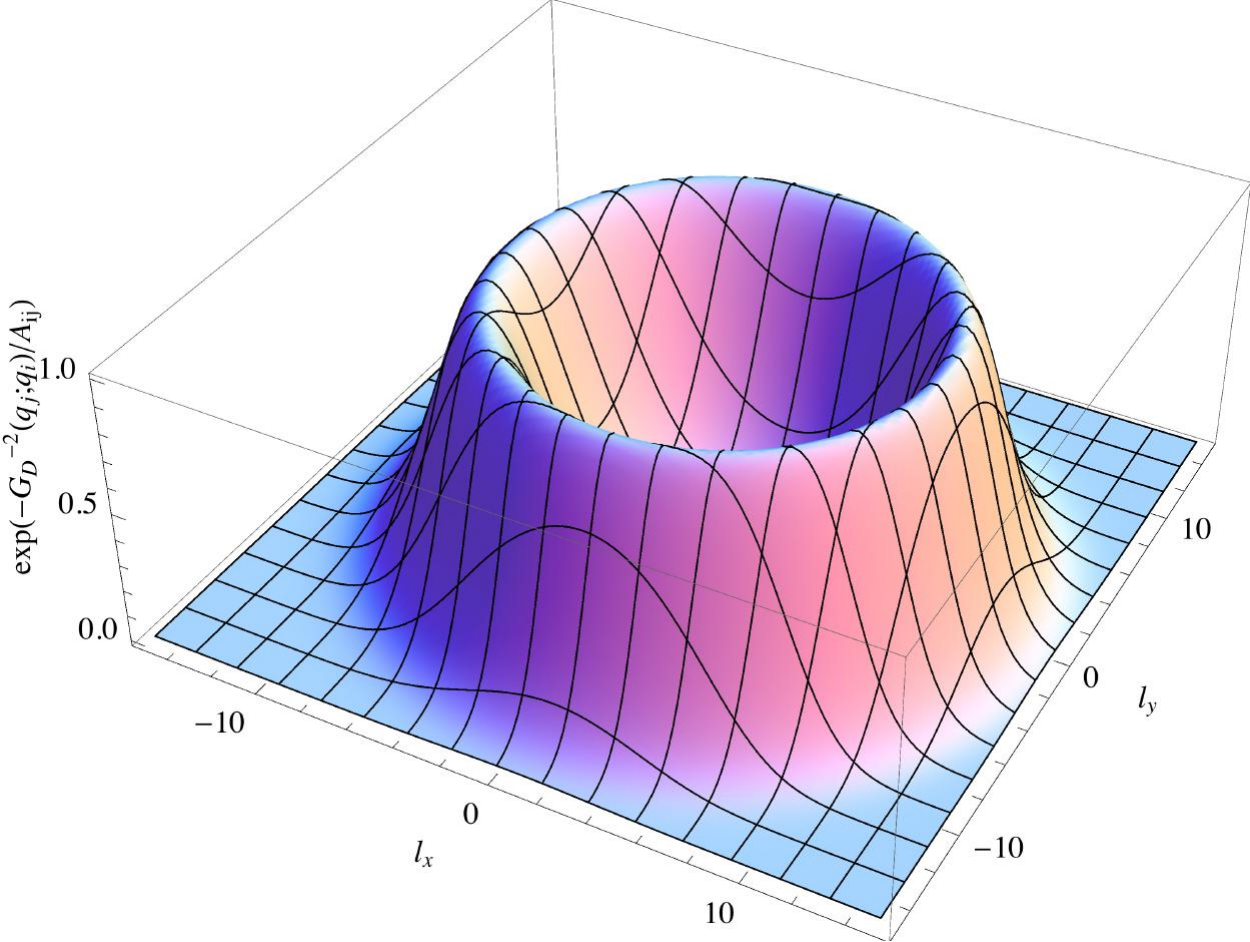}
\caption{Volcano-shaped suppression factor in 1+2 dimensions. 
The crater line is exactly the singularity ellipse.\label{fig:volcano}}
\end{figure}
$\lambda_{ij}$ is a scaling factor analogous 
to $\lambda$ in Section \ref{sec:cdef1d}. $A_{ij}$ is the width of the 
deformation. The indices $ij$ in $\lambda_{ij}$ and $A_{ij}$ indicate that those 
parameters can be chosen individually for each contribution to the deformation 
for optimization purposes. 
Then we sum over the entire group and arrive at:
\bea
i\boldsymbol{\kappa} = i\sum\limits_{i,j\in \text{group}}
\lambda_{ij}\left(\frac{\mathbf{q}_i}{\sqrt{\mathbf{q}_i^2}}+
\frac{\mathbf{q}_j}{\sqrt{\mathbf{q}_j^2}}\right)\exp\left(
\frac{-G_D^{-2}(q_j;q_i)}{A_{ij}}\right)
\label{eq:defogroup}
\eea
where, of course, $i$ is the imaginary unit. Finally, we add the imaginary 
contribution to the loop momentum to make the deformation complete.
\bea
\boldsymbol{\ell}\to\boldsymbol{\ell}'=\boldsymbol{\ell}+i\boldsymbol{\kappa}
\eea
The corresponding Jacobian can be calculated analytically.

%----------------------------------------------------------------------------------------
%	Choosing the parameters
%----------------------------------------------------------------------------------------

\section{Choosing the parameters}

As mentioned before, the parameters $\lambda_{ij}$ and $A_{ij}$ are 
the scaling and the width of the deformation. Typically, $\lambda_{ij}$ 
is a dimensionless negative number and $A_{ij}$ a positive number 
of dimension $q^4$; setting all $\lambda_{ij}=0$ 
completely switches off the deformation.\\
Although they can all be chosen differently,
selecting $\lambda_{ij}=\lambda=-0.5$ and 
$A_{ij}=A=10^6$ produces very good results for most of the examples 
that we are presenting within this thesis. As a rule of thumb, 
$A_{ij}\approx 10^2m^4$ where $m^4$ 
is a measure for the $\text{energy}^4$ of the process in consideration.

%-----------------------------------
%	The scaling parameter $\lambda$
%-----------------------------------

\subsection{The scaling parameter $\lambda$}

The parameter $\lambda_{ij}$ determines how far an 
individual contribution to the deformation 
goes around a pole. This is illustrated in Fig. \ref{fig:lambda124}, 
in which for the sake of clarity we restricted ourselves to $1+1$ dimensions 
and a deformation that consists of only one contribution.
\begin{figure}[H]
\centering
\includegraphics[scale=1.5]{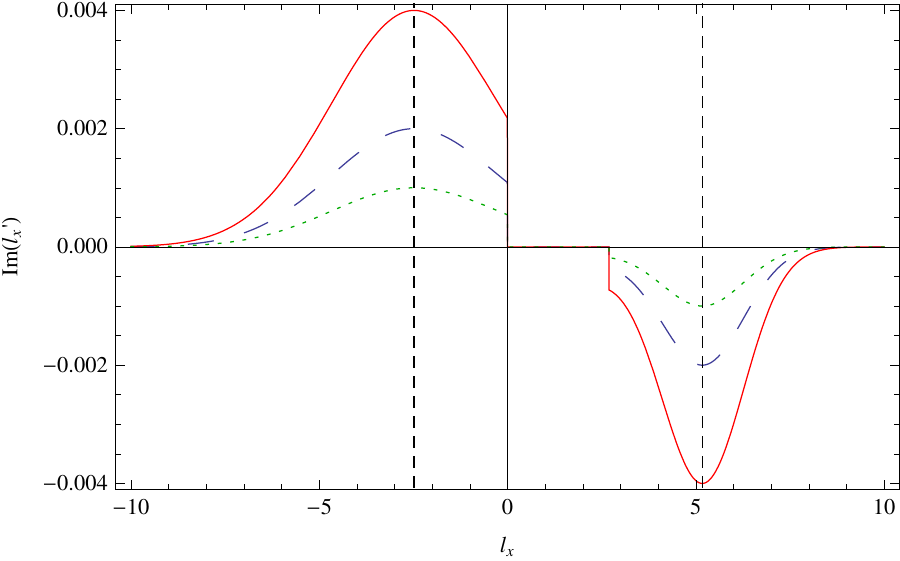}
\caption{The deformation at different values of $\lambda$\label{fig:lambda124}. 
The dashed vertical lines mark the positions of the singularities.}
\end{figure}
From the dotted to the dashed and finally to the solid line, $\lambda_{ij}=\lambda$ 
has been doubled every time while keeping the width constant. 
Therefore, the peak of the solid curve is four 
times higher (lower) than the dotted one. Between $l_x=0$ and $l_x=3$ there 
is a region where the deformation vanishes. This is the cancelling effect 
of the vector part mentioned in Section \ref{sec:defo3d}. In 1+1 dimensions 
this leads to a total cancellation. It is equivalent to slicing Fig. \ref{fig:vecpart} 
along the axis that connects the two foci of the singularity ellipse.\\
There is one further aspect to it which is inspired by \cite{Becker:2012nk}. 
On the one hand, we would like to choose $\lambda$ as large as possible to stay away 
from the singularity. One the other hand, we would like to make $\lambda$ as small 
as possible in order to ensure that we do not enclose other (unwanted) singularities. 
To escape from this dilemma, consider the on shell energy $q_{i,0}^{(+)}$ with 
the deformation $\boldsymbol{\ell}+i\bar{\lambda}\boldsymbol{\kappa}$ plugged in:
\bea
\mathbf{q}_{i,0}^{(+)}=\sqrt{-\bar{\lambda}^2\boldsymbol{\kappa}^2+
2i\bar{\lambda}\boldsymbol{\kappa}\cdot\mathbf{q}_i+\mathbf{q}_i^2+m_i^2-i0}~.
\eea
Here we have introduced an overall factor $\bar{\lambda}$. It scales the entire 
deformation as opposed to the $\lambda_{ij}$ which scale individual contributions 
to the deformation (see \Eq{eq:defogroup}).\\
Since $q_{i,0}^{(+)}\geq 0$, this puts a constraint on $\bar{\lambda}$. We set the 
square root to zero and solve the quadratic equation in $\bar{\lambda}$:
\bea
\bar{\lambda}_{\pm}=iX_i\pm\sqrt{Y_i-X_i^2}
\eea
with
\bea
X_i&=\frac{\mathbf{q}_i\cdot\boldsymbol{\kappa}}{\boldsymbol{\kappa}^2}, & Y_i&=\frac{\mathbf{q}_i^2+m_i^2}{\boldsymbol{\kappa}^2}
\eea
where $X_i\leq 0$ and $Y_i\geq 0$. If $X_i^2>Y_i$, the poles of 
$q_{i,0}^{(+)}$ lie on the imaginary $\bar{\lambda}$-axis and the real value of 
$\bar{\lambda}$ may take any value. If however, $X_i^2<Y_i$ 
and $X_i\to 0$ we have 
a pole at $\bar{\lambda}=\sqrt{Y_i}$. Thus we define $\bar{\lambda}$ 
in the following way:
\bea
\bar{\lambda} = \begin{cases} \frac{1}{2}\sqrt{Y_i} & \text{if} \quad X_i^2 < Y_i/2,\\
\sqrt{X_i^2-\frac{Y_i}{4}} & \text{if} \quad Y_i/2 < X_i^2 < Y_i,\\
1 & \text{if} \quad Y_i<X_i^2
\end{cases}
\eea
After performing this check for every dual contribution in the deformation 
group, we end up with a set of ``safety $\bar{\lambda}$s'' out of which we pick the 
smallest. Using this procedure, we make sure that our deformation does not come too 
close to other singularities originating from the dual delta function.

%-----------------------------------
%	The width of the deformation
%-----------------------------------

\subsection{The width of the deformation}

The parameter $A_{ij}$ determines how broad or narrow the peaks 
of the deformation contributions are. This is illustrated in Fig. \ref{fig:A148}. 
Again, to improve clarity, the plot is restricted to $1+1$ dimensions and 
the deformation used has only a single contribution to it.
\begin{figure}[h]
\centering
\includegraphics[scale=1.5]{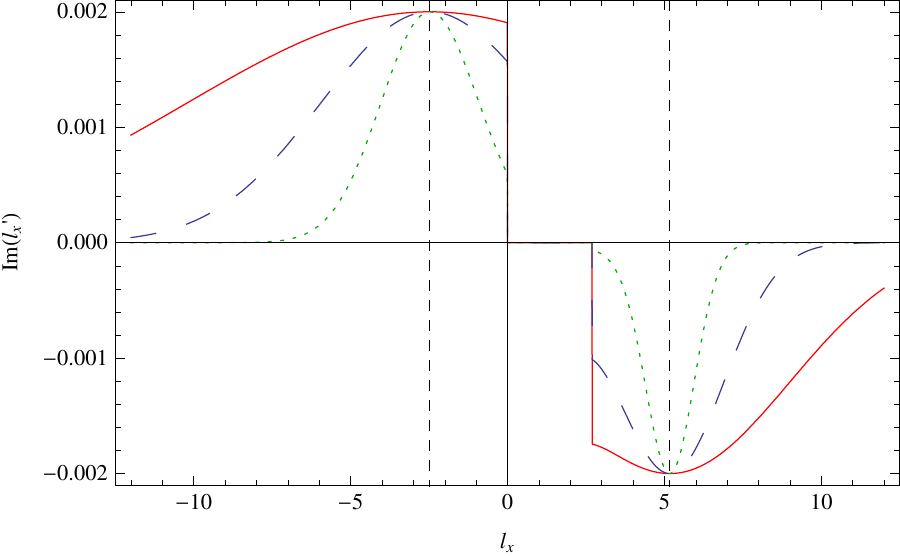}
\caption{The deformation at different values of $A$\label{fig:A148}. 
The dashed vertical lines mark the positions of the singularities.}
\end{figure}
This time the width has been increased by a factor of five from the dotted 
to the dashed curve and another factor five from the dashed to the solid curve. The scaling 
parameter $\lambda$ has been kept constant.\\
As mentioned, setting $A_{ij}=A=10^6$ works fine for the majority of 
momentum configurations that we tested. Nonetheless, there are some cases where 
a bigger (or smaller) choice of width is advised to get good precision. 
To understand why, let us have a look at the dual propagator in its 
rewritten form:
\bea
G_D(q_i;q_j)=\frac{1}{(q_{i,0}^{(+)} + k_{ji,0})^2-(q_{j,0}^{(+)})^2}~.
\eea
We see that the inverse of the dual propagator is proportional to 
momentum $G_D^{-1} \sim q$. In the 
suppression factor the inverse of the dual propagator gets squared. 
Hence the exponent of the suppression factor is $\sim q^2/A$. 
In order to avoid the deformation 
getting to wide or too narrow, it is necessary choose the 
width proportional to the physical energy scale of the external momenta.

%----------------------------------------------------------------------------------------
%	Numerical Results
%----------------------------------------------------------------------------------------

\section{Numerical Results\label{sec:results}}

Now we are in possession of all the necessary tools to make the numerical 
integration work. We have a code which runs on a desktop machine with an 
Intel i7 (3.4GHz) processor with 8 cores and 16 GB of RAM. 
Memory consumption is negligible.

%-----------------------------------
%	The program
%-----------------------------------

\subsection{Details on the implementation\label{sec:program}}

The program is written entirely in C++ and uses the Cuba library \cite{Hahn:2004fe} 
as a numerical integrator. This means it can be run on any 
machine on which C++ and the Cuba library  are available. 
As input the user has to specify the number of 
external legs, the external momenta themselves, the internal masses and, 
if necessary, the parameters of the contour deformation. The momenta and 
masses can be read in from a text file. With regard to the numerical integrator 
the user can choose between Cuhre \cite{Cuhre1, Cuhre2} and VEGAS 
and give the desired number 
of evaluations. At this stage, the program is ready to be compiled and executed. 
Schematically speaking, it performs the following steps:
\begin{enumerate}
\item Read in momenta and masses.
\item Check where ellipsoid singularities occur.
\item Check where hyperboloid singularities occur and group the dual contributions 
accordingly.
\item Call the integrator, taking into account the number of external legs as 
well as the results from the previous steps.
\end{enumerate}
We use {\tt MATHEMATICA 8.0} \cite{Mathematica} 
to generate randomized momenta and masses 
and {\tt LoopTools 2.10} \cite{Hahn:1999wr} to produce reference values for comparison. 
The goal is to scan 
as much of the phase-space as possible to make sure the program works 
properly in all regions. 
The momenta of all the example points and scans of the following sections 
are collected in Appendix \ref{AppendixB}.\\
We see two main paths along which the complexity of the calculation grows.
\begin{enumerate}
\item Increasing the number of external legs. In general, 
going from $n \to n+1$ legs means for the Duality:
\begin{itemize}
\item One extra dual contribution. %(compared to the triangle case).
\item Since each dual contribution does consist of one more dual propagator, 
there are much more possibilities for ellipsoid singularities to occur. Thus the 
deformation picks up more contributions.
\end{itemize}
Therefore, we investigate multi-leg scalar integrals up
to the pentagon level in this chapter.
\item The presence of tensor numerators. Having tensors of increasing 
rank in the numerator will render the function to be integrated more complex. We 
address tensor integrals in Chapter \ref{Chapter8} although it does not affect 
the singular behaviour of the integral.
\end{enumerate}
Our setup is such that we run points asking for a fixed number of evaluations. Then 
we modify the parameters of the deformation for best results.\footnote{Part of the 
results which we are going to show in the following will be published 
in a forthcoming paper \cite{paper}.}

%-----------------------------------
%	Scalar Triangles
%-----------------------------------

\subsection{Scalar Triangle\label{sec:restriangle}}

We consider first infrared finite triangle (three external legs) scalar integrals. 
Momentum configurations that do not need deformation 
(i.e. whose loop integral is purely real) are integrated in about $0.15$ 
 seconds with a precision of at least 4 digits. $5\cdot 10^4$ evaluations 
 are sufficient to achieve this result.
 
 \begin{table}[H]
%\setlength{\tabcolsep}{0.3pc}
% \caption{Pentagon example results}
   % {\small
   \centering
\begin{tabular}{lllll}
&\\
\hline
%\hline
%\\
 & Point \plabel{point1}{1}  & Point 1 Error &  Point \plabel{point2}{2} & Point 2 Error\\
%\\
\hline
%\hline
%\\
LoopTools &-5.85694E-5 & 0 & -3.39656E-7 & 0\\
Loop--Tree Duality & -5.85685E-5 & 2.4E-9 & -3.39688E-7 & 5.3E-11\\
\hline
%\hline
\end{tabular}
%}
\caption{Example for two non-deformation phase-space points. 
Since there is no deformation, 
all values are purely real.}
\label{tab:trinodef}
\end{table}
Point \ref{point1} of Table \ref{tab:trinodef} has all internal masses equal 
while Point \ref{point2} has three different internal masses. Momenta 
and masses 
were chosen randomly between $-100$ and $+100$. This 
even allows for unphysical momentum configurations, but at this 
stage we want to test stability and precision regardless of whether 
we are in the physical region or not. 
For example, although Triangle \ref{point2} 
represents a situation that is not realized in nature, it is computed 
without any problems. From the program's point of view there is no difference; the internal 
masses are mere parameters.\\
For momentum configurations that require deformation the function to integrate 
is more complicated, hence we have to evaluate it more often. With $10^6$ 
iterations, the calculation time increases to around $2.5$ seconds. 
Four digits of precision are achievable by optimizing the parameters of the 
deformation.
 \begin{table}[H]
%\setlength{\tabcolsep}{0.3pc}
% \caption{Pentagon example results}
   % {\small
   \centering
\begin{tabular}{lllll}
&\\
\hline
%\hline
%\\
 & Real Part  & Real Error &  Imaginary Part & Imaginary Error\\
%\\
\hline
%\hline
%\\
LoopTools P.\plabel{point3}{3} & 5.37305E-4 & 0 & -6.68103E-4 & 0\\
Loop--Tree Duality P.3 & 5.37307E-4 & 8.6E-9 & -6.68103E-4 & 8.6E-9\\
\hline
LoopTools P.\plabel{point4}{4} & -5.61370E-7 & 0 & -1.01665E-6 & 0\\
Loop--Tree Duality P.4 & -5.61371E-7 & 7.2E-10 & -1.01666E-6 & 7.2E-10\\
\hline
%\hline
\end{tabular}
%}
\caption{Example for two phase-space point that need deformation. With a 
deformation applied, the results are complex.}
\label{tab:tridef}
\end{table}
Similar to the non-deformation points before, Point \ref{point3} of Table \ref{tab:tridef} 
has all internal masses equal whereas in Point \ref{point4} all three of them 
have different values. Regardless of kinematics, the Loop--Tree Duality 
is capable of producing accurate results in both situations.\\

\begin{figure}[h]
\centering
   \begin{subfigure}[b]{1\textwidth}
   \includegraphics[width=1\linewidth]{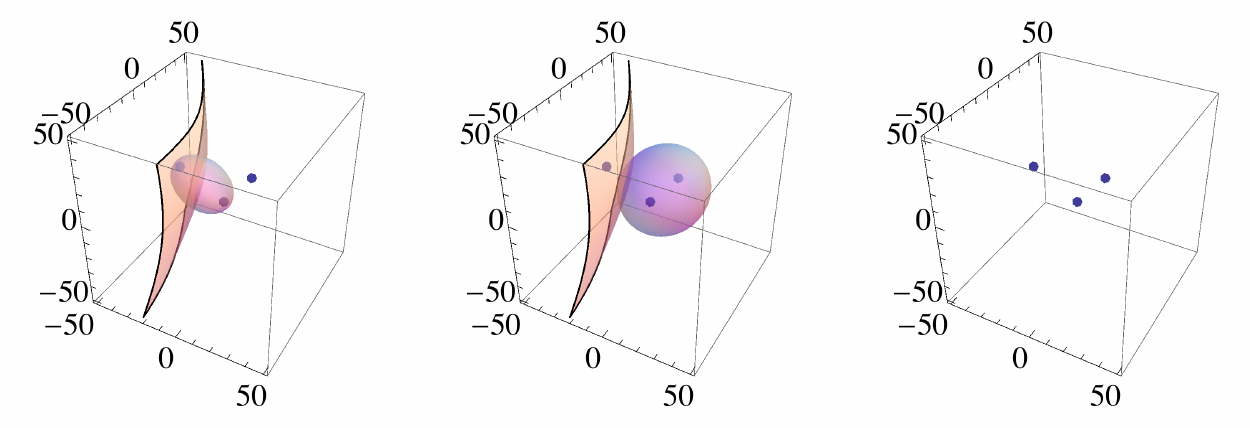}
   %\caption{Real Part}
   \label{fig:tri1lms} 
\end{subfigure}

\begin{subfigure}[b]{1\textwidth}
   %\includegraphics[width=1\linewidth]{Figures/triangleim.pdf}
   %\caption{Imaginary Part}
   \bea
   \begin{matrix}
   0 & \text{\bf{H}} &  \text{\bf{E}} \\
    \text{\bf{H}} & 0 &  \text{\bf{E}} \\
   0 & 0 & 0
   \end{matrix}\nn
   \eea
   \label{fig:tri1sch}
\end{subfigure}
\caption{Singularities of a triangle graph in loop-momentum 
space and singularity scheme for comparison.\label{fig:trilms}}
\end{figure}

Fig. \ref{fig:trilms} gives an additional view on a triangle example 
phase-space point: Each of the cubes in the first line represents 
one dual contribution. Drawn are the ellipsoid (violet surfaces) 
and hyperboloid (orange surfaces) 
singularities in loop-momentum space. The blue dots are the 
positions of the foci of the on-shell hyperboloids, i.e. 
$-\mathbf{k}_i,\,\,\, i\in \{1,2,3\}$. 
In the second line we have the singularity 
scheme. This means each line of the singularity scheme corresponds 
to one box.\\
\\
Another important check is the scan around threshold. This 
region is usually numerically unstable. The Loop--Tree Duality excels here, because the 
algorithm does not have to do extra work. 
In Fig. \ref{fig:trithreshold} all internal masses are equal, i.e.
$m_i = m$, $i\in\{1,2,3\}$, and 
the center-of-mass energy $s$ was kept 
constant while the mass $m$ was varied. As you can see from the plots, 
there is no drop in precision around threshold. Calculation time remains 
constant, as well.\\
%To put things into perspective, we ran Point \ref{point4} through SecDec \cite{Borowka:2012yc}, asking for the 
%same amount of precision, which in this case was around four digits. To calculate that 
%Point it took SecDec 33 seconds (LTD 2 seconds) from start to finish. However, taking 
%a point just below threshold from Figure \ref{fig:trithreshold}, calculation 
%time increases to 3 minutes.
\begin{figure}[H]
\centering
   \begin{subfigure}[b]{1\textwidth}
   \includegraphics[width=1\linewidth]{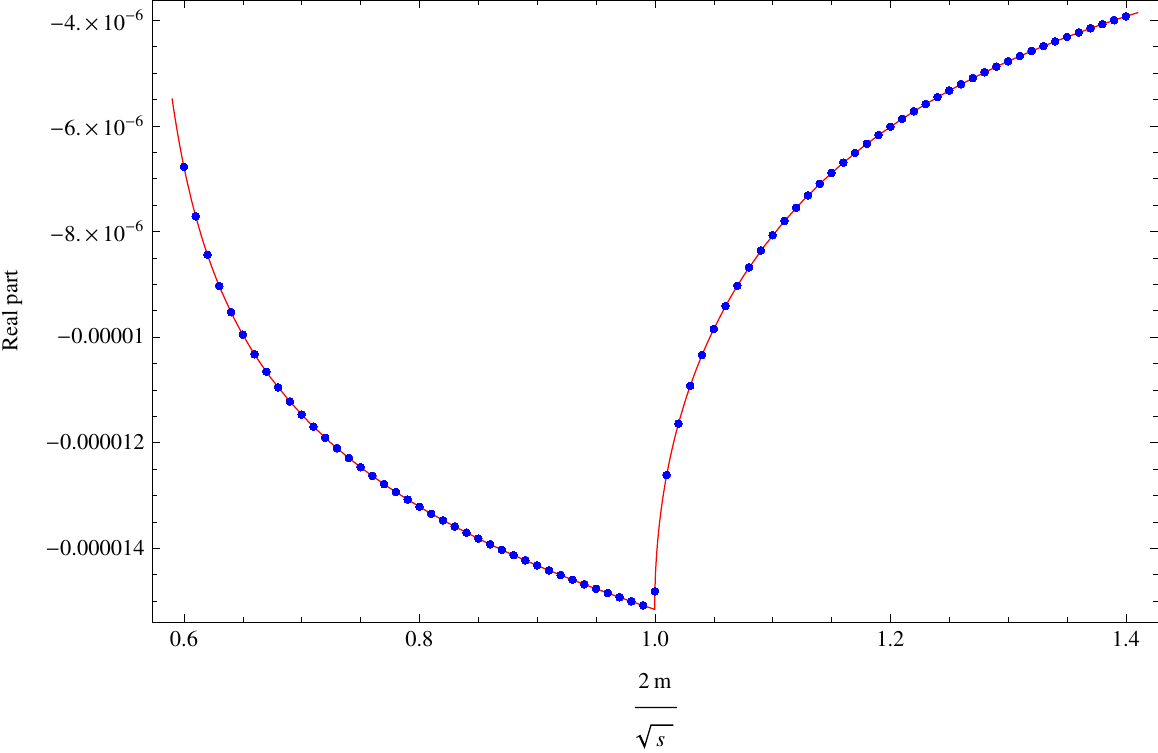}
   %\caption{Real Part}
   \label{fig:trire} 
\end{subfigure}

\begin{subfigure}[b]{1\textwidth}
   \includegraphics[width=1\linewidth]{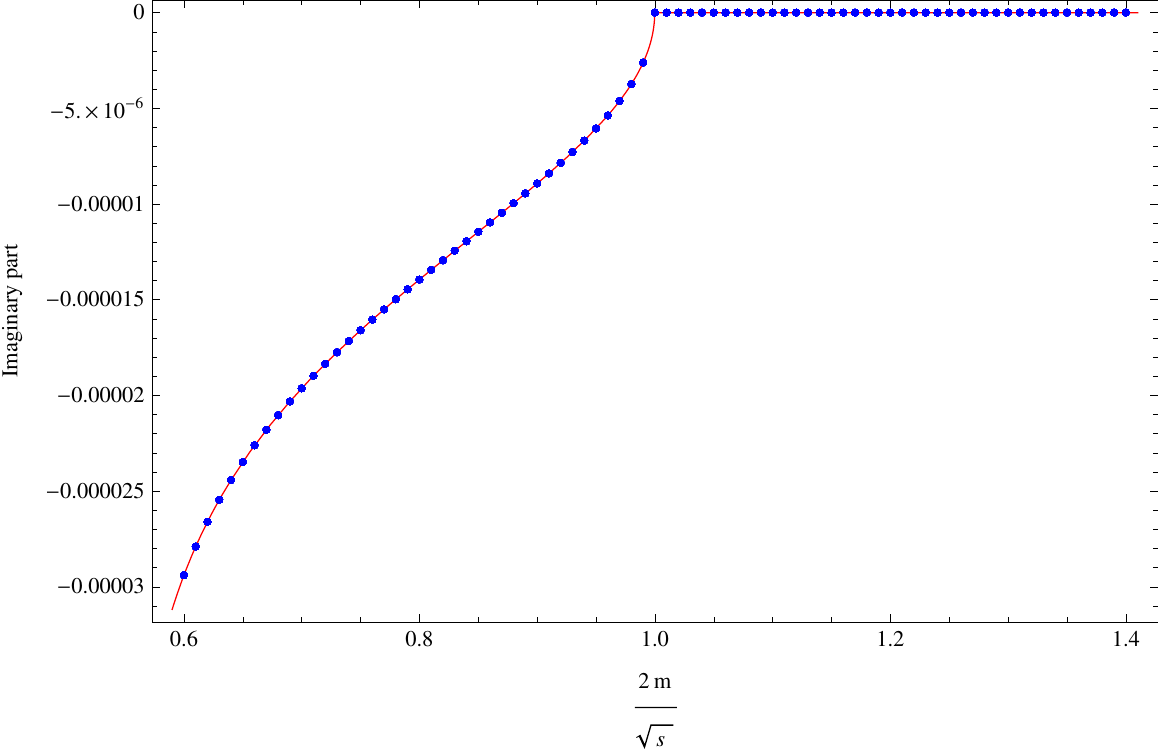}
   %\caption{Imaginary Part}
   \label{fig:triim}
\end{subfigure}
\caption{Scan of the region around threshold. The red curve 
is LoopTools, the blue points are the Loop--Tree Duality.\label{fig:trithreshold}}
\end{figure}

%\newpage

%-----------------------------------
%	Scalar Boxes
%-----------------------------------

\subsection{Scalar Boxes}

We repeated the analysis for the box (four external legs) case. The main difference 
is that now there are four dual contributions with three dual propagators 
each.\\ 
To get good precision, we chose to evaluate boxes that need deformation for 
$4\cdot 10^6$ times, while the number of evaluations for non-deformation phase-space 
points was kept at $5\cdot 10^4$ evaluations, the same as in the triangle case. 
This is reflected in the program runtimes. Deformation points take 16 seconds, 
non-deformation points take about 0.25 seconds, respectively.\\
%While non-deformation phase-space 
%points are still integrated in less than $0.5$ seconds (still $5\cdot 10^4$ evaluations), 
%the points that need 
%deformation have to be evaluated 4 million times to consistently
%get good precision. 
%As a direct consequence, calculation time increases to 
%around 10 seconds.\\
%Generally speaking, it is much harder to get the phase-space 
%points with deformation. The corresponding dual integrals have 
%to be evaluated many more times to reach the same level of accuracy. 
While it is practically guaranteed to get the non-deformation points 
with good 
precision, the quality of results depends on the proper choice of 
parameters for points with deformation. Therefore, we focus our 
attention towards such points by giving many more examples 
for them.\\
A very good method to do so is to perform a mass-scan as we already 
did in the triangle case. 
%Since boxes are more complex, the are more 
%variables to keep track of. 
From Fig. \ref{fig:boxalleq} to Fig. 
\ref{fig:boxvart} we present 
scans for the following momentum configurations:
\begin{itemize}
\item All internal masses equal. The center-of-mass energy $s$ is kept constant 
while the mass is varied. This is shown in Fig. \ref{fig:boxalleq}.
\item Two adjacent internal masses equal and the two opposing masses 
equal, i.e. $m_1=m_2$ and $m_3=m_4$. $m_1$ and therefore 
also $m_2$ is varied, the other 
momenta and masses are kept constant. This is shown in Fig. \ref{fig:box2adjm}.
%\item All four internal masses different. Mass $m_1$ gets varied, rest is kept 
%constant. This is shown in Figure \ref{fig:box4mvarm1}.
%\item Opposing masses are chosen equal, i.e. $m_1=m_3$ and $mh2=m_4$. 
%Mass $m_1 (=m_3)$ is varied, the rest is kept constant. This is shown in 
%Figure \ref{fig:boxoppo}.
%\item One mass different from the others which are chosen to be equal, i.e. 
%$m_1\neq m_2=m_3=m_4$. This is shown in Figure \ref{fig:box4m31}.
\item Scan in which the Mandelstam variable $t$ is varied. This is realized by 
varying $p_3$ while keeping $p_3^2$ constant. Of course, due to 
momentum conservation, this involves 
$p_4^2=(p_1+p_2+p_3)^2$ not being constant. This is shown in Fig. \ref{fig:boxvart}.
\end{itemize}
Concrete values for each of those situations are found in Table \ref{tab:boxes}.\\
The scans show how the program performs in certain slices of the phase space. This 
is important because we want the program to be generically applicable to any momentum 
configuration. From the plots and tables, you can see that the program deals well
with all kinds of boxes, even when many different kinematical scales are involved.\\
\\
Point \ref{point5} and \ref{point7} of Table \ref{tab:boxes} correspond to a 
momentum 
configuration, in which all four internal masses are equal. In 
Point \ref{point6} and \ref{point8} all masses are different. In Point \ref{point9} two 
adjacent 
internal lines have equal masses as well as the two opposing 
ones. Point \ref{point10} represents a situation in which opposite 
lines have equal masses.%; in Point \#11 (vm) one internal mass 
%is different form the other three which are equal.
 \begin{table}[H]
%\setlength{\tabcolsep}{0.3pc}
% \caption{Pentagon example results}
   % {\small
   \centering
\begin{subtable}{1\textwidth}
\begin{tabular}{lllll}
&\\
\hline
%\hline
%\\
 & Point \plabel{point5}{5}  & Point 5 Error &  Point \plabel{point6}{6} & Point 6 Error\\
%\\
\hline
%\hline
%\\
LoopTools & 2.15339E-13 & 0 & 1.39199E-11 & 0\\
Loop--Tree Duality & 2.15319E-13 & 5.2E-17 & 1.39199E-11 & 6.3E-16\\
\hline
%\hline
\end{tabular}
\caption{Two non-deformation phase-space points.\label{tab:boxnodefo}}
\end{subtable}
\begin{subtable}{1\textwidth}
\begin{tabular}{lllll}
&\\
\hline
%\hline
%\\
 & Real Part  & Real Error &  Imaginary Part & Imaginary Error\\
%\\
\hline
%\hline
%\\
LoopTools P.\plabel{point7}{7} & -2.38766E-10 & 0 & -3.03080E-10 & 0\\
Loop--Tree Duality P.7 & -2.38798E-10 & 8.2E-13 & -3.03084E-10 & 8.2E-13\\
\hline
LoopTools P.\plabel{point8}{8} & -4.27118E-11 & 0 & 4.49304E-11 & 0\\
Loop--Tree Duality P.8 & -4.27127E-11 & 5.3E-14 & 4.49301E-11 & 5.3E-14\\
\hline
LoopTools P.\plabel{point9}{9} & 6.43041E-11 & 0 & 1.61607E-10 & 0\\
Loop--Tree Duality P.9 & 6.43045E-11 & 8.4E-15 & 1.61607E-10 & 8.4E-15\\
\hline
LoopTools P.\plabel{point10}{10} & -4.34528E-11 & 0 & 3.99020E-11 & 0\\
Loop--Tree Duality P.10 & -4.34526E-11 & 3.5E-14 & 3.99014E-11 & 3.5E-14\\
\hline
%LoopTools P.\#11 & -4.27118E-11 & 0 & 4.49304E-11 & 0\\
%
%Loop--Tree Duality P.\#11 & -4.27127E-11 & 5.3E-14 & 4.49301E-11 & 5.3E-14\\
%
%\hline
%\hline
\end{tabular}
\caption{Four phase-space points with different kinematics that need deformation.}
\end{subtable}
%}
\caption{Examples for boxes with and without deformation.}
\label{tab:boxes}
\end{table}

Similar to the triangle case, Fig. \ref{fig:boxlms} shows the singularities 
in loop three-momentum space. It becomes immediatly obvious that, only by going 
to box graphs, the level of complexity has grown substantially. Because 
the hyperboloid singularities couple dual contributions together, an 
individual contribution receives a much more complicated deformation. 
%This is why we chose to evaluate the integrand four times more often, 
%as mentioned earlier.

\begin{figure}[H]
\centering
   \begin{subfigure}[b]{1\textwidth}
   \includegraphics[width=1\linewidth]{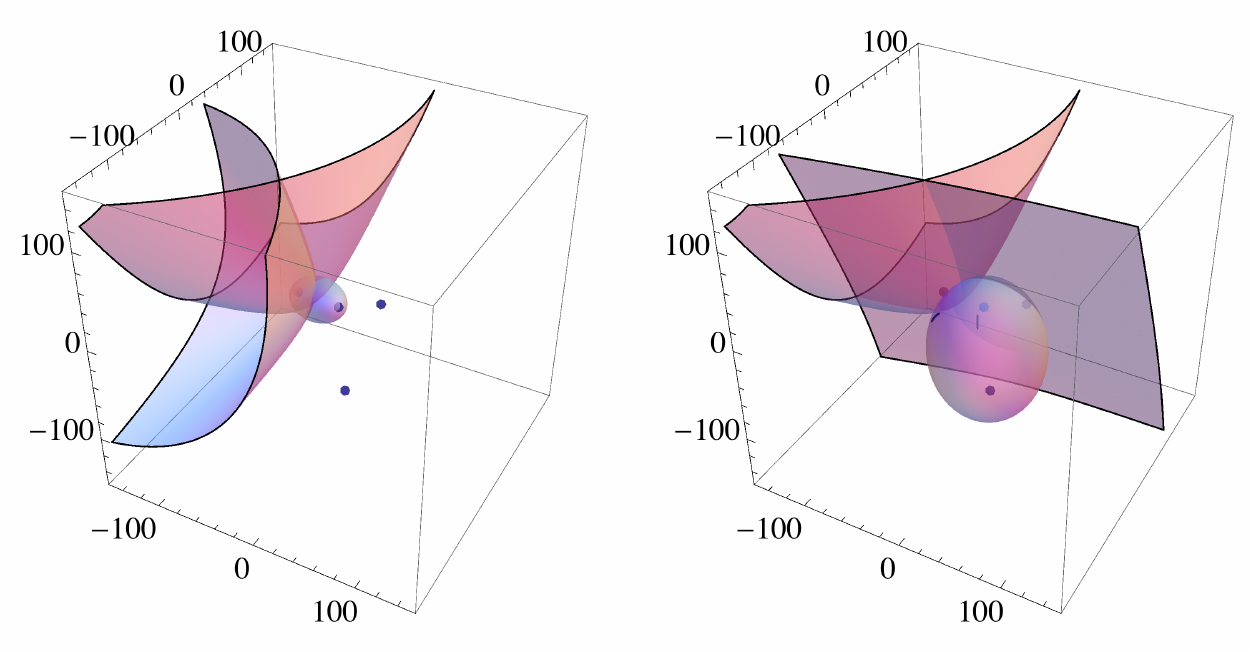}
   %\caption{Real Part}
   \label{fig:boxlms1} 
\end{subfigure}

   \begin{subfigure}[b]{1\textwidth}
   \includegraphics[width=1\linewidth]{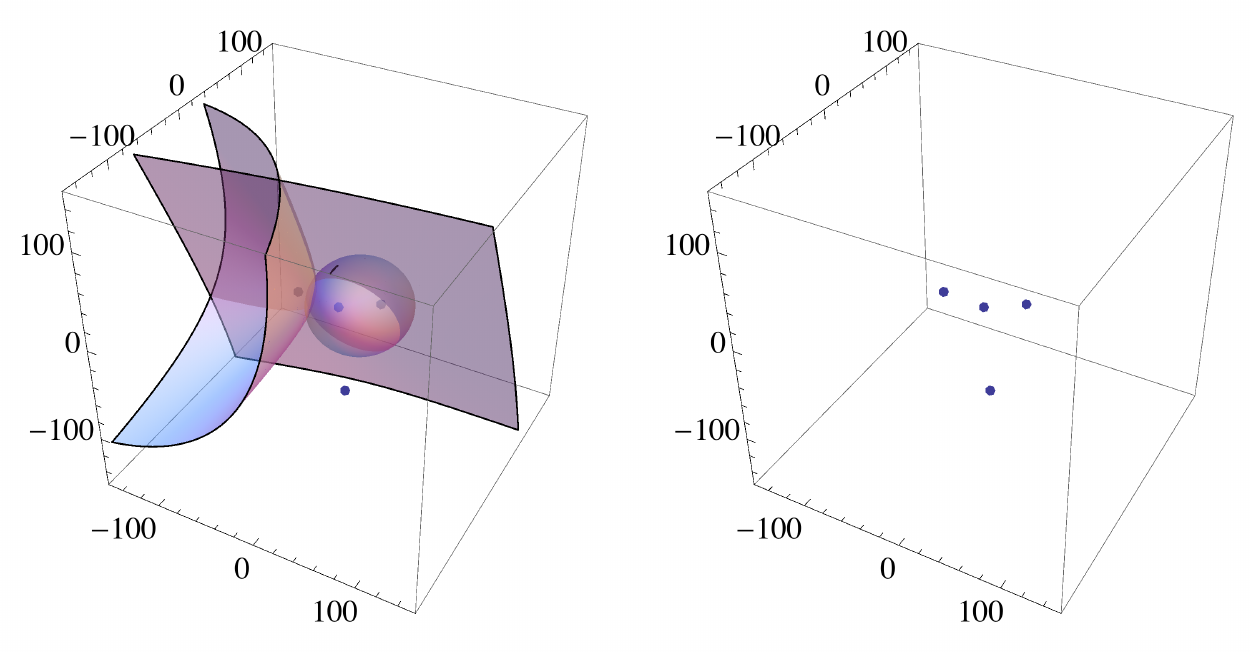}
   %\caption{Real Part}
   \label{fig:boxlms2} 
\end{subfigure}

\begin{subfigure}[b]{1\textwidth}
   %\includegraphics[width=1\linewidth]{Figures/triangleim.pdf}
   %\caption{Imaginary Part}
   \bea
   \begin{matrix}
   0 & \text{\bf{H}} &  \text{\bf{H}} &  \text{\bf{E}} \\
    \text{\bf{H}} & 0 &  \text{\bf{H}} & \text{\bf{E}} \\
    \text{\bf{H}} & \text{\bf{H}} & 0 & \text{\bf{E}} \\
   0 & 0 & 0 & 0
   \end{matrix}\nn
   \eea
   \label{fig:tboxsch}
\end{subfigure}
\caption{Singularities of a box graph in loop-momentum 
space and singularity scheme for comparison.\label{fig:boxlms}}
\end{figure}

In Fig. \ref{fig:boxlms}, the dots give the locations of the foci 
$-\mathbf{k}_i$, the surfaces are ellipsoid and 
hyperboloid singularities. Note how the 
hyperboloids always appear pairwisely across the dual 
contributions.

\begin{figure}[H]
\centering
   \begin{subfigure}[b]{1\textwidth}
   \includegraphics[width=1\linewidth]{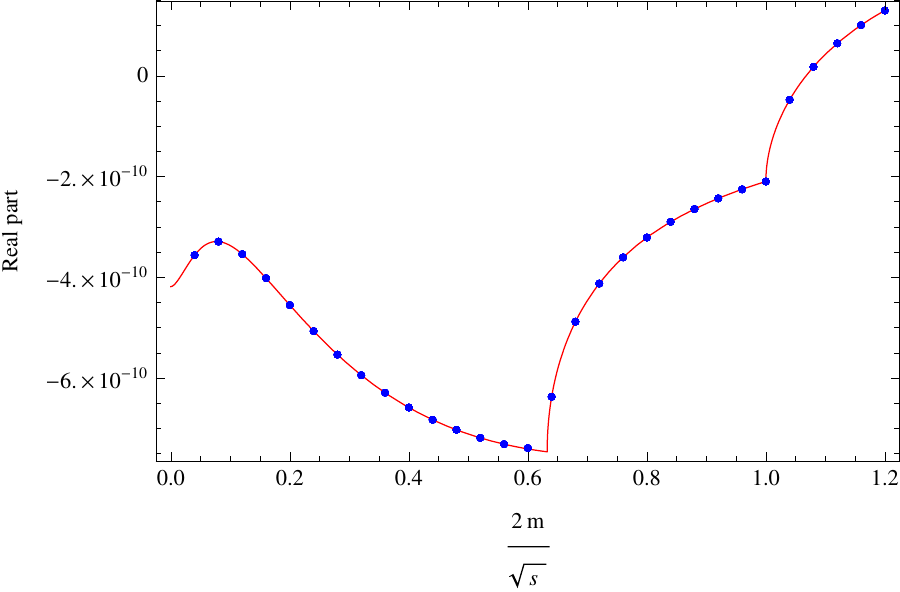}
   %\caption{Real Part}
   \label{fig:boxrealleq} 
\end{subfigure}

\begin{subfigure}[b]{1\textwidth}
   \includegraphics[width=1\linewidth]{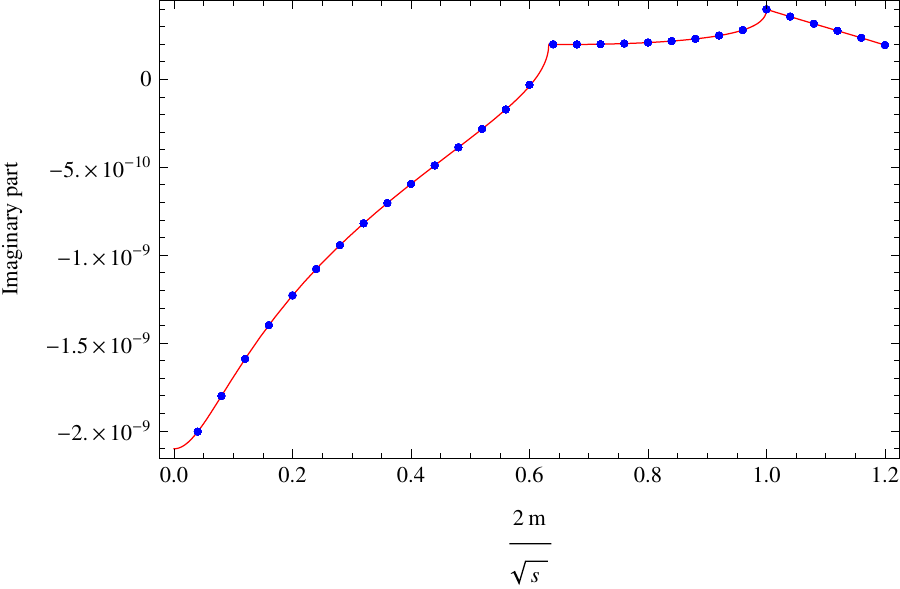}
   %\caption{Imaginary Part}
   \label{fig:boximalleq}
\end{subfigure}
\caption{Mass-scan of a box integral. The red curve 
is LoopTools, the blue points are the Loop--Tree Duality.\label{fig:boxalleq}}
\end{figure}

In Fig. \ref{fig:boxalleq}, two thresholds are passed at $2m/\sqrt{s}=0.65$ 
and $1$. From right to left, the number of ellipsoid singularities grows by 
one after each threshold from one to three.

\begin{figure}[H]
\centering
   \begin{subfigure}[b]{1\textwidth}
   \includegraphics[width=1\linewidth]{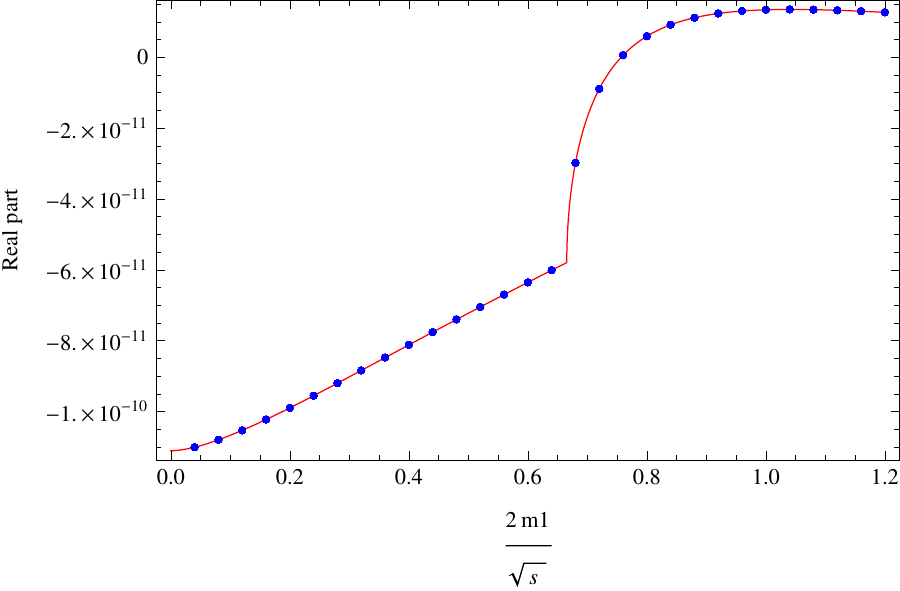}
   %\caption{Real Part}
   \label{fig:boxre2adjm} 
\end{subfigure}

\begin{subfigure}[b]{1\textwidth}
   \includegraphics[width=1\linewidth]{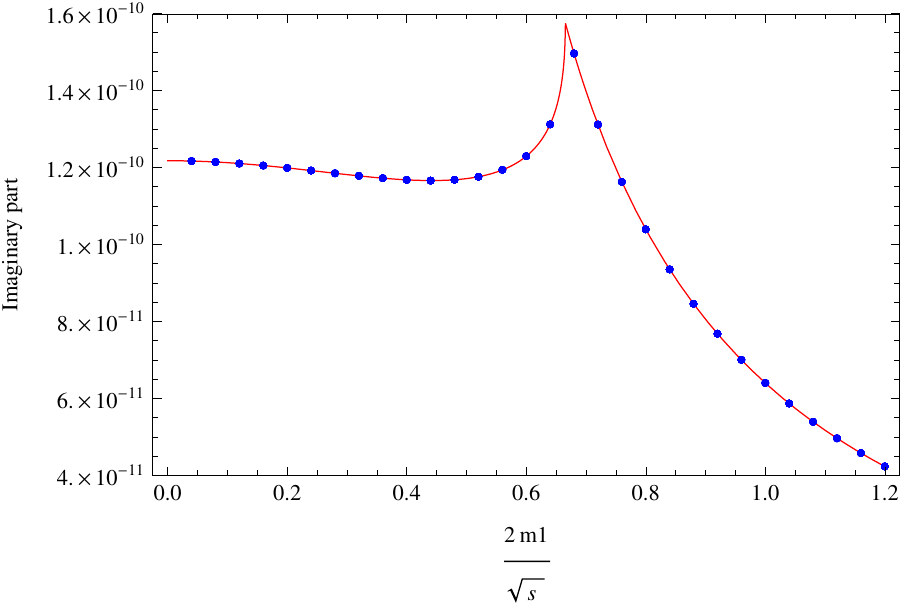}
   %\caption{Imaginary Part}
   \label{fig:boxim2adjm}
\end{subfigure}
\caption{Mass-scan of a box integral. The red curve 
is LoopTools, the blue points are the Loop--Tree Duality.\label{fig:box2adjm}}
\end{figure}

In Fig. \ref{fig:box2adjm}, one threshold is crossed at $2m_1/\sqrt{s}\approx 0.65$. 
From right to left the number of ellipsoid singularities goes from one to two.

\begin{figure}[H]
\centering
   \begin{subfigure}[b]{1\textwidth}
   \includegraphics[width=1\linewidth]{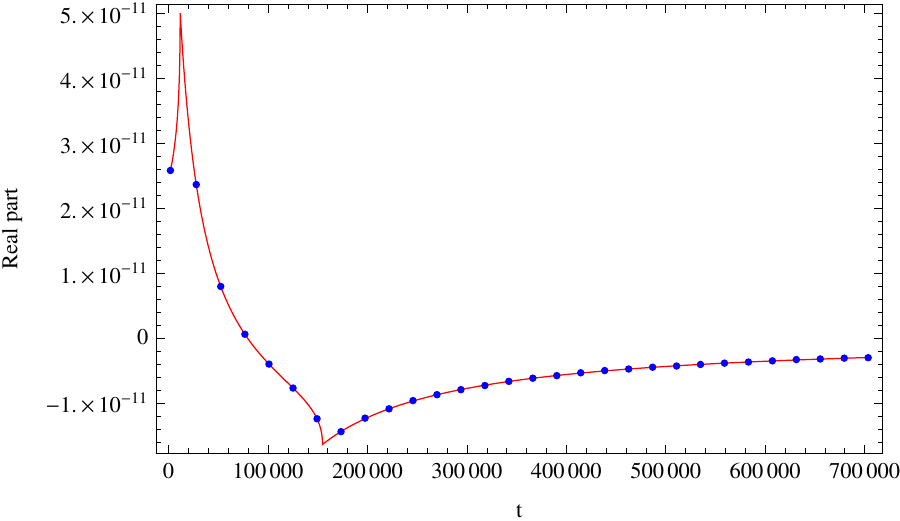}
   %\caption{Real Part}
   \label{fig:boxrevart} 
\end{subfigure}

\begin{subfigure}[b]{1\textwidth}
   \includegraphics[width=1\linewidth]{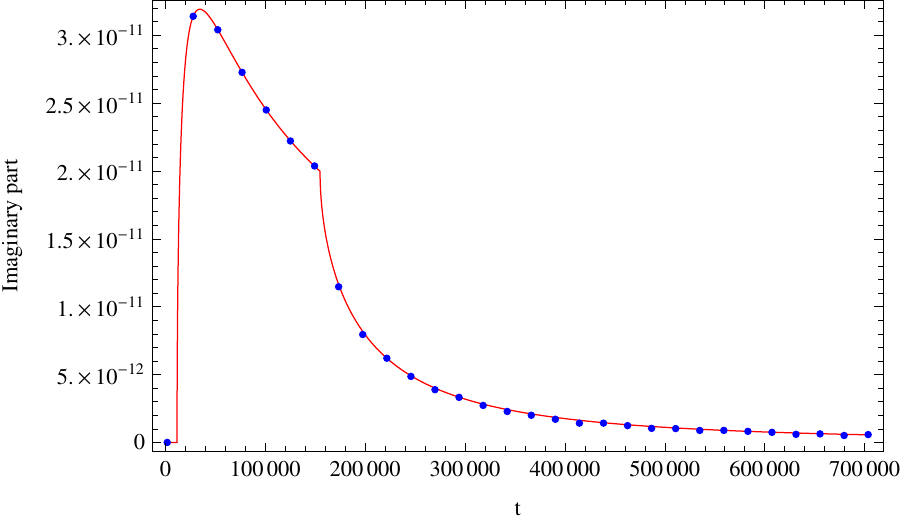}
   %\caption{Imaginary Part}
   \label{fig:boximvart}
\end{subfigure}
\caption{$t$-scan of a box integral. The red curve 
is LoopTools, the blue points are the Loop--Tree Duality.\label{fig:boxvart}}
\end{figure}

In Fig. \ref{fig:boxvart}, the Mandelstam variable $t$ is being varied. 
Two thresholds are passed, one at $t\approx 10^4$ and one at $t\approx 
150\cdot 10^4$. From left to right, the number of ellipsoid singularities 
increases by one after each crossing of a threshold from zero to two.

\newpage

%-----------------------------------
%	Scalar Pentagons
%-----------------------------------

\subsection{Scalar Pentagons}

After the evaluation of triangles and boxes, the next step is to check 
pentagon (five external legs) graphs. Complexity grows once more, which means:
\begin{itemize}
\item A dual integral now consists of five dual contributions.
\item Each dual contribution contains one additional 
($=$ four in total) dual propagator.
\end{itemize}
Therefore, we had to increase the number of evaluations: 
Non-deformation points are evaluated $10^5$ times 
which takes approximately $0.5$ seconds. Points with deformation 
demand $5\cdot 10^6$ evaluations to maintain the level of precision 
of the triangles and boxes. This results in an average calculation 
time of 28 seconds. This enhanced complexity is especially 
well-illustrated by the following Figure in which the singularities 
of the different dual contributions are plotted in loop-momentum 
space:

\begin{figure}[H]
\centering
   \begin{subfigure}[b]{1\textwidth}
   \includegraphics[width=1\linewidth]{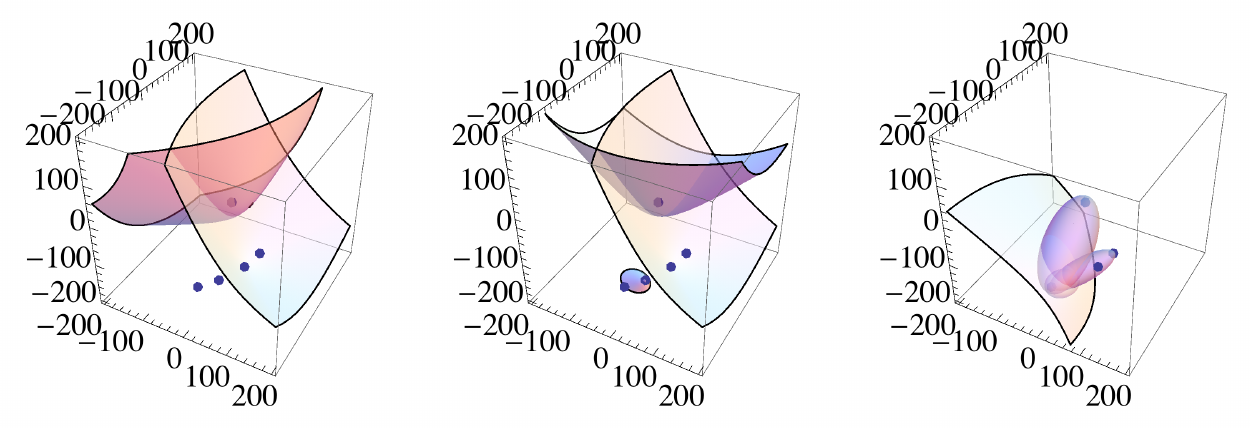}
   %\caption{Real Part}
   \label{fig:pentagonlms1} 
\end{subfigure}

   \begin{subfigure}[b]{0.66\textwidth}
   \includegraphics[width=1\linewidth]{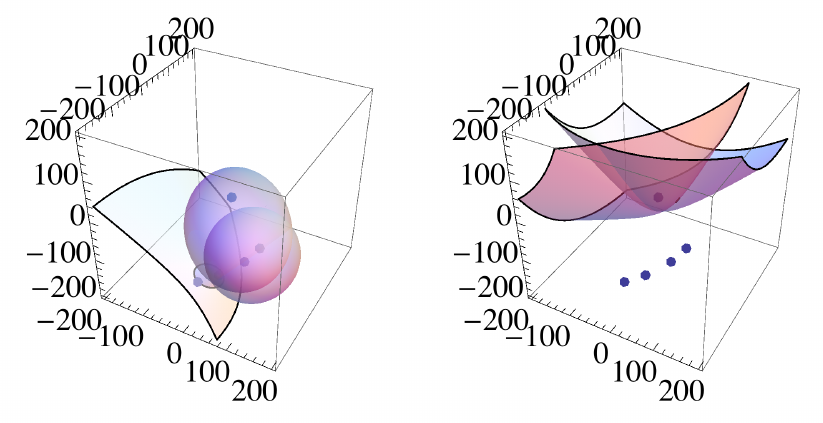}
   %\caption{Real Part}
   \label{fig:pentagonlms2} 
\end{subfigure}
\begin{subfigure}[b]{0.33\textwidth}
   %\includegraphics[width=1\linewidth]{Figures/triangleim.pdf}
   %\caption{Imaginary Part}
   \centering
   \bea
   \begin{matrix}
   0 & \text{\bf{H}} & 0 & 0 & \text{\bf{H}} \\
    \text{\bf{H}} & 0 & 0 & \text{\bf{H}} & \text{\bf{H}} \\
    \text{\bf{E}} & \text{\bf{E}} & 0 & \text{\bf{H}} & \text{\bf{E}} \\
   \text{\bf{E}} & \text{\bf{H}} & \text{\bf{H}} & 0 & \text{\bf{E}} \\
   \text{\bf{H}} & \text{\bf{H}} & 0 & 0 & 0
   \end{matrix}\nn
   \eea
   \vspace{8mm}
   \label{fig:tpentagonsch}
\end{subfigure}
\caption{Singularities of a box graph in loop-momentum 
space and singularity scheme for comparison.\label{fig:pentagonlms}}
\end{figure}

Again, dots are foci of the on-shell hyperboloids, surfaces are ellipsoid 
and hyperboloid singularities. From the singularity scheme we 
can read off that all five dual contributions are coupled together due to 
the interaction of the hyperboloids.

 \begin{table}[H]
%\setlength{\tabcolsep}{0.3pc}
% \caption{Pentagon example results}
   % {\small
   \centering
\begin{subtable}{1\textwidth}
\begin{tabular}{lllll}
&\\
\hline
%\hline
%\\
 & Point \plabel{point11}{11}  & Point 11 Error &  Point \plabel{point12}{12} & Point 12 Error\\
%\\
\hline
%\hline
%\\
LoopTools & -1.24025E-13 & 0 & -1.48356E-14 & 0\\
Loop--Tree Duality & -1.24027E-13 & 1.6E-17 & -1.48345E-14 & 1.1E-17\\
\hline
%\hline
\end{tabular}
\caption{Two non-deformation phase-space points.\label{pentex-non}}
\end{subtable}
\begin{subtable}{1\textwidth}
\begin{tabular}{lllll}
&\\
\hline
%\hline
%\\
 & Real Part  & Real Error &  Imaginary Part & Imaginary Error\\
%\\
\hline
%\hline
%\\
LoopTools P.\plabel{point13}{13} & 1.02350E-11 & 0 & 1.40382E-11 & 0\\
Loop--Tree Duality P.13 & 1.02353E-11 & 1.0E-16 & 1.40385E-11 & 1.0E-16\\
\hline
LoopTools P.\plabel{point14}{14} & 7.46345E-15 & 0 & -9.13484E-15 & 0\\
Loop--Tree Duality P.14 & 7.46309E-15 & 6.1E-18 & -9.13444E-15 & 6.1E-18\\
\hline
LoopTools P.\plabel{point15}{15} & 6.89836E-15 & 0 & 2.14893E-15 & 0\\
Loop--Tree Duality P.15 & 6.89848E-15 & 6.5E-18 & 2.14894E-15 & 6.5E-18\\
\hline
%\hline
\end{tabular}
\caption{Three phase-space points that need deformation.\label{pentex-defo}}
\end{subtable}
%}
\caption{Examples for pentagons with and without deformation.}
\label{tab:pentex}
\end{table}

Table \ref{tab:pentex} displays a collection of pentagon sample 
results for different kinematical configurations. In Points \ref{point11} and 
\ref{point13} all internal masses are equal; in Point \ref{point14} they are all distinct from 
each other and in Point \ref{point15} we have $m_1=m_2=m_3\neq m_4=m_5$. 
Again, the Loop-Tree Duality shows its robustness by producing accurate 
results regardless of the kinematical situation. This statement is further 
supported by various scans we performed:
\begin{itemize}
\item All internal masses equal. The center-of-mass energy $s$ is kept constant 
while the mass gets varied. This is shown in Fig. \ref{fig:pentalleq}.
\item All five internal masses different. Mass $m_1$ gets varied, the rest is kept 
constant. This is shown in Fig. \ref{fig:pentalldiff}.
\item Scan in which the center-of-mass energy $s$ is varied. This is realized by 
varying $p_1$. Of course, due to 
momentum conservation, this involves 
$p_4^2=(p_1+p_2+p_3)^2$ not being constant. This is shown in Fig. 
\ref{fig:pentachu}.
\end{itemize}

From the plots, you can see that the Loop-Tree 
Duality is able 
to achieve good precision for pentagons as well. 
Even situations in which many ellipsoid 
singularities are involved, are handled well by the program. For 
example, the number of ellipsoid singularities in Figure \ref{fig:pentalleq}  
increases from two to five when going from right to left.

\begin{figure}[H]
\centering
   \begin{subfigure}[b]{1\textwidth}
   \includegraphics[width=1\linewidth]{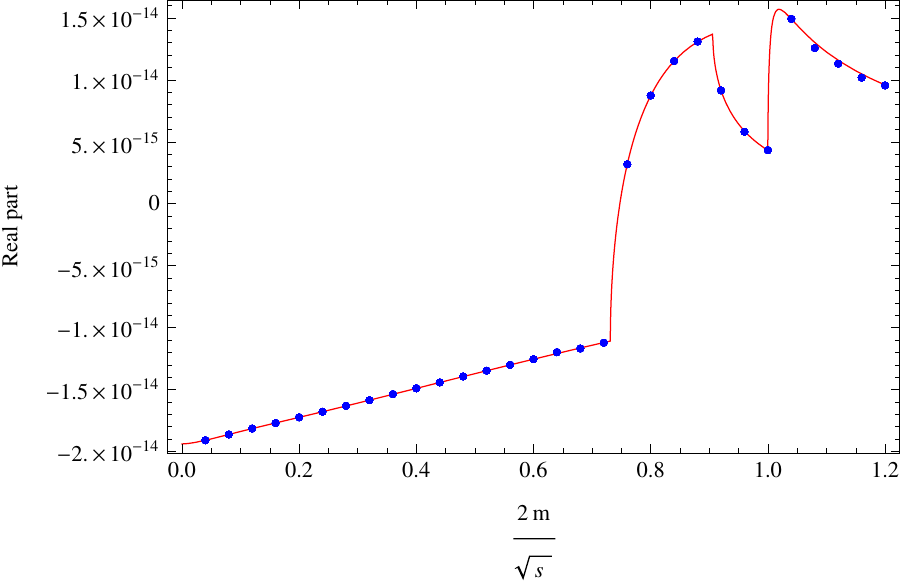}
   %\caption{Real Part}
   \label{fig:pentalleqre} 
\end{subfigure}

\begin{subfigure}[b]{1\textwidth}
   \includegraphics[width=1\linewidth]{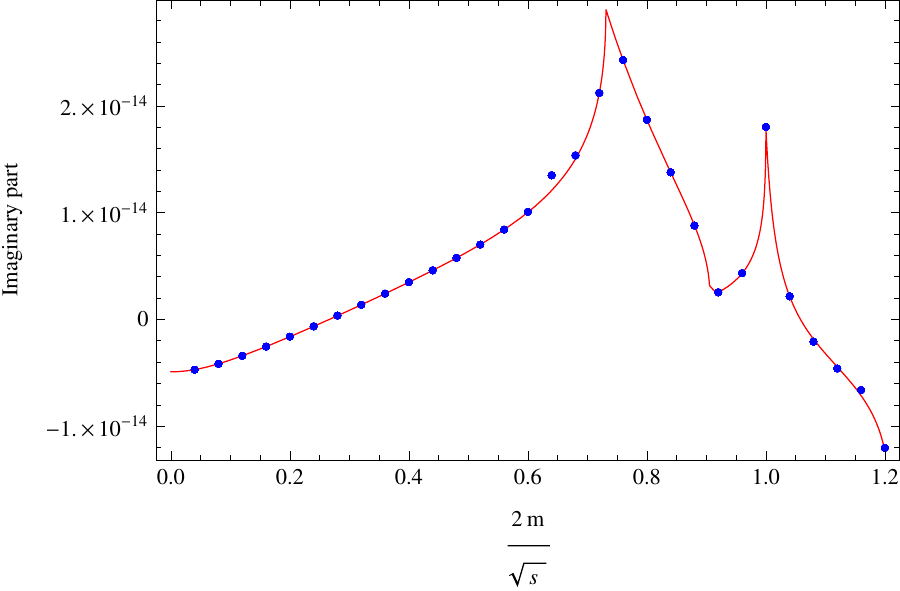}
   %\caption{Imaginary Part}
   \label{fig:pentalleqim}
\end{subfigure}
\caption{Mass-scan of a pentagon integral. The red curve 
is LoopTools, the blue points are the Loop--Tree Duality.\label{fig:pentalleq}}
\end{figure}

Three thresholds are passed in this scan. From right to left, we start at two 
ellipsoid singularities and arrive at five.

\begin{figure}[H]
\centering
   \begin{subfigure}[b]{1\textwidth}
   \includegraphics[width=1\linewidth]{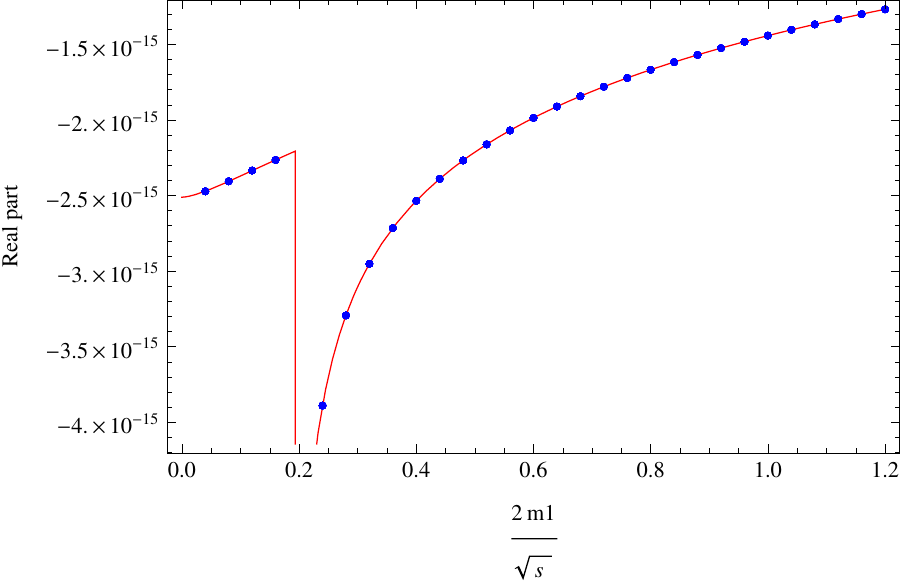}
   %\caption{Real Part}
   \label{fig:pentalldiffre} 
\end{subfigure}

\begin{subfigure}[b]{1\textwidth}
   \includegraphics[width=1\linewidth]{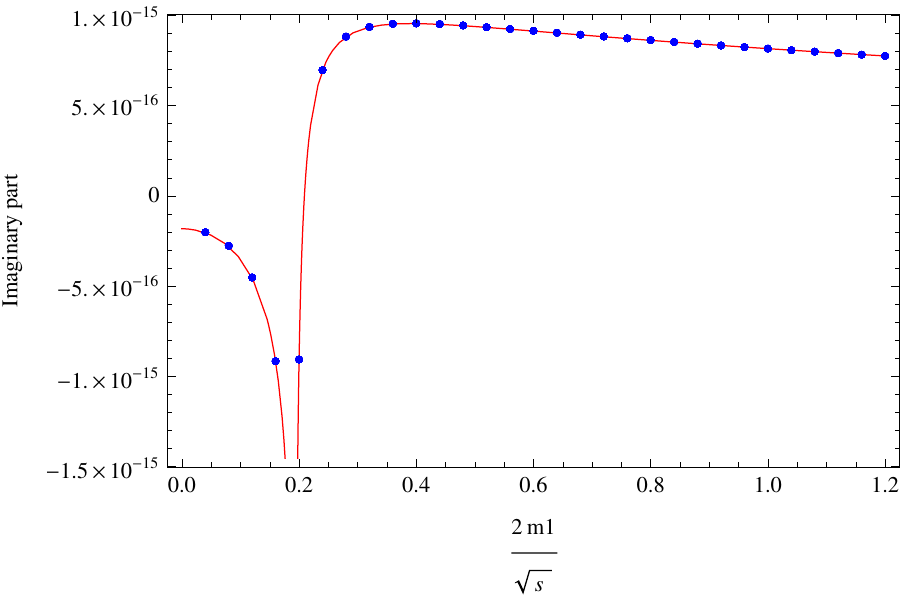}
   %\caption{Imaginary Part}
   \label{fig:pentalldiffim}
\end{subfigure}
\caption{Mass-scan of a pentagon integral. The red curve 
is LoopTools, the blue points are the Loop--Tree Duality.\label{fig:pentalldiff}}
\end{figure}

One threshold is crossed at $2m_1/\sqrt{s}\approx$ 0.2. From right to 
left, the number of ellipsoid singularities increases from five to six.

\begin{figure}[H]
\centering
   \begin{subfigure}[b]{1\textwidth}
   \includegraphics[width=1\linewidth]{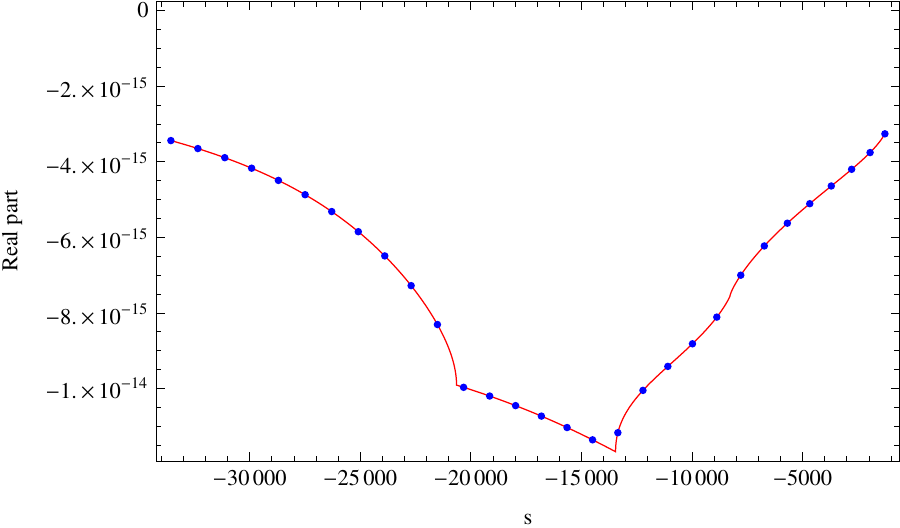}
   %\caption{Real Part}
   \label{fig:pentachure} 
\end{subfigure}

\begin{subfigure}[b]{1\textwidth}
   \includegraphics[width=1\linewidth]{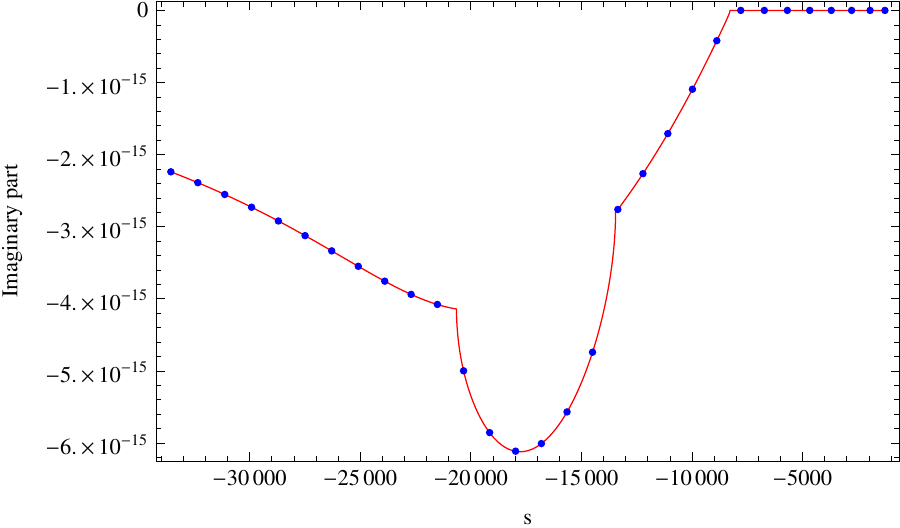}
   %\caption{Imaginary Part}
   \label{fig:pentachuim}
\end{subfigure}
\caption{Energy-scan of a pentagon integral. The red curve 
is LoopTools, the blue points are the Loop--Tree Duality.\label{fig:pentachu}}
\end{figure}

In this scan, we pass three thresholds at $s\approx -8.5\cdot 10^3, -13.5\cdot 
10^3$ and $-21\cdot 10^3$ which divide the graph into four zones. From 
right to left, we start with zero ellipsoid singularities in the first zone, then we have 
one in the second zone, two in the third zone and finally one in the last zone.
 
% Chapter Template

\chapter{Tensor Integrals} % Main chapter title

\label{Chapter8} % Change X to a consecutive number; for referencing this chapter elsewhere, use \ref{ChapterX}

%\lhead{Chapter 7. \emph{Numerators and more legs}} % Change X to a consecutive number; this is for the header on each page - perhaps a shortened title

Up to this point, we have shown that the 
Loop--Tree Duality is capable of dealing with all kinds of 
kinematical configurations. The only constraint that we have made 
was to limit ourselves to scalar integrals. In Chapter 
\ref{Chapter3}, we argued that the inclusion of numerators 
different from 1 
should not be an issue for the Loop--Tree Duality because it acts only 
on the denominator of the loop integral. Hence, in this chapter, 
we are going to relieve this restriction and repeat some of the 
analysis of the previous chapter with tensor integrals. 
Additionally, we are going one step further by calculating even 
hexagons, which are usually tough to compute, but from the 
perspective of LTD are a straightforward generalization.

\newpage

%----------------------------------------------------------------------------------------
%	The influence of numerators
%----------------------------------------------------------------------------------------

\section{The influence of numerators}

To start off, we restate the Duality Theorem for scalar integrals (cf. \Eq{oneloopduality}):
\bea
%\label{oneloopduality}
L^{(1)}(p_1, p_2, \dots, p_N) 
\quad=\quad - \sum_{i\in\alpha_1} \, \int_{\ell} \; \td(q_i) \,
\prod_{\substack{j\in\alpha_1 \\ j\neq i}} \,G_D(q_i;q_j)~,
\eea 
where
\beq
G_D(q_i;q_j) = \frac{1}{q_j^2 -m_j^2 - i0 \,\eta (q_j-q_i)}~.
\eeq
If our one-loop integral features a non-trivial numerator 
$\mathcal{N}(\ell,\{p_i\})$. Then, the 
Loop--Tree Duality Theorem takes the form
\bea
%\label{oneloopduality}
L^{(1)}(p_1, p_2, \dots, p_N) 
\quad=\quad - \sum_{i\in\alpha_1} \, \int_{\ell} \; \td(q_i) \,
\prod_{\substack{j\in\alpha1 \\ j\neq i}} \,G_D(q_i;q_j)\mathcal{N}(\ell,\{p_i\})~.
\eea 
While the numerator is formally left unchanged, there actually is a 
practical impact. The presence of the dual delta function demands 
$q_{i,0}^{(+)}=\sqrt{\mathbf{q}_i^2+m_i^2}$ which is equivalent to 
having the 
energy component of the loop-momentum forced to
\bea
\ell_0=-k_{i,0}+\sqrt{\mathbf{q}_i^2+m_i^2}~.
\label{eq:l0num}
\eea
In other words, whenever we perform a single cut of a Feynman graph, 
the numerator has to be evaluated at the position of the cut which is 
fixed by the dual delta function. As a direct consequence, the numerator 
takes a different form in each dual contribution.

Another important aspect to take into consideration is the cancellation of 
singularities among dual contributions. In Section \ref{sec:cancel}, we 
argued that numerators do not spoil the cancellation of the hyperboloid 
singularities. Here, we would like to make explicitly show why. 
A typical numerator is a product of 
scalar products of the form ``loop-momentum contracted with external 
momentum'': $\ell\cdot p_k$. Let us see what happens to a single factor 
when it hits the singularity. Note first, that the hyperboloid singularity is 
given by \Eq{hyperboloid} which we rewrite in the more suitable form
\bea
q_{i,0}^{(+)}-k_{i,0} = q_{j,0}^{(+)}-k_{j,0}~.
\label{eq:hyperboloidnewform}
\eea
Using \Eq{eq:l0num}, the loop-momentum $\ell$ contracted with some 
external momentum $p_k$ is:
\bea
\ell\cdot p_k\,|_{\text{i-th cut}} &= (-k_i+q_{i,0}^{(+)})p_{k,0}-\boldsymbol{\ell}\cdot \mathbf{p}_{k}\nn\\
&= (-k_j+q_{j,0}^{(+)})p_{k,0}-\boldsymbol{\ell}\cdot \mathbf{p}_{k} = \ell\cdot p_k\,|_{\text{j-th cut}}
\label{eq:cancelnum}
\eea
where we have used \Eq{eq:hyperboloidnewform} for the first equality on 
the second line of \Eq{eq:cancelnum}. It means that the numerators of two 
dual contributions $i$ and $j$ take the same value at their common pole, 
thus leaving the cancellation of hyperboloid singularities intact. 
This is an important property to take advantage of, because 
it allows us to straightforwardly apply the Loop--Tree-Duality to such diagrams 
without any additional effort.

%-----------------------------------
%	An illustrative example
%-----------------------------------

\subsection{An illustrative example}

To make things even more explicit, consider 
the following simple example. We take the triangle integral of Section  
\ref {sec:masslexmp} and give it the numerator 
$\mathcal{N}(\ell,p_1,p_2) = \ell\cdot p_1$:
\bea
I=\int_{\ell}\, \ell\cdot p_1\, \prod\limits_{i=1}^3 G_{F}(q_i)
 \eea 
$p_1, p_2$ and $p_3=-p_1-p_2$ are the external momenta. 
Since we are dealing with a triangle, obviously $N=3$. 
Consequently, the Loop--Tree Duality translates the Feynman 
integral into three dual contributions.
\bea
I=&\int_{\ell}\td(q_1)G_D(q_1;q_2)G_D(q_1;q_3)\,\ell\cdot p_1\nn\\
+&\int_{\ell}G_D(q_2;q_1)\td(q_2)G_D(q_2;q_3)\,\ell\cdot p_1\nn\\
+&\int_{\ell}G_D(q_3;q_1)G_D(q_3;q_2)\td(q_3)\,\ell\cdot p_1
\eea
This time, we investigate how 
the numerator affects the calculus. The dual delta 
functions of 
the different contributions fix the energy 
component of the numerators to different values. 
Hence the scalar product becomes
\bea
\ell\cdot p_1&\to
(-p_{1,0}
+\sqrt{(\boldsymbol{\ell}+\mathbf{p}_1)^2+m_1^2})\,p_{1,0}-
\boldsymbol{\ell}\cdot \mathbf{p}_1 & &\text{first contribution}\nn\\
\ell\cdot p_1&\to
(-p_{1,0}-p_{2,0}
+\sqrt{(\boldsymbol{\ell}+\mathbf{p}_1+\mathbf{p}_2)^2+m_2^2})\,p_{1,0}-
\boldsymbol{\ell}\cdot \mathbf{p}_1 & &\text{second contribution}\nn\\
\ell\cdot p_1&\to
\sqrt{\boldsymbol{\ell}^2+m_3^2}\,p_{1,0}-
\boldsymbol{\ell}\cdot \mathbf{p}_1 & &\text{third contribution}
\eea
This means, that finally we will arrive at with three dual contributions 
similar to the ones of \Eq{eq:explmassexample}. %To give the reader 
%a better idea of how these look like we write down contribution 3, 
%which we will call $I_3$, explicitly:
%\bea
%I_3=-\int_{\boldsymbol{\ell}}&\frac{\sqrt{\boldsymbol{\ell}^2+m_3^2}\,p_{1,0}-
%\boldsymbol{\ell}\cdot \mathbf{p}_1}{2p_{1,0}\sqrt{\boldsymbol{\ell}^2+m_3^2}
%+2\boldsymbol{\ell}\cdot \mathbf{p}_1-m_1^2+m_3^2
%+p_1^2-i0\eta k_{13}}\cdot
%\frac{1}{2\sqrt{\boldsymbol{\ell}^2+m_3^2}}\cdot\\
%&\frac{1}{2(p_{1,0}+p_{2,0})\sqrt{\boldsymbol{\ell}^2+m_3^2}
%+2\boldsymbol{\ell}\cdot (\mathbf{p}_1+\mathbf{p}_2)
%+(p_1+p_2)^2-m_2^2+m_3^2-i0\eta k_{23}}\nn
%\eea

In the next sections, we present the results for multi-leg tensor integrals. 
We only consider IR- and UV-finite diagrams.
The explicit momenta of all the points and scans presented within this chapter 
are found in Appendix \ref{AppendixB}.

\newpage

%----------------------------------------------------------------------------------------
%	Tensor Triangles
%----------------------------------------------------------------------------------------

\section{Tensor Triangles}

In Chapter \ref{Chapter7}, we convinced ourselves that the Loop--Tree 
Duality can produce accurate results for scalar one--loop integrals 
independent of the kinematical 
configuration. Now, we focus on checking various different 
numerator functions within a similar analysis.\\
We start with infrared-finite one--loop triangle graphs. 
The simplest non-trivial numerator function possible is the 
loop-momentum $\ell$ contracted with one of the external momenta 
$p_1$ or $p_2$, i.e. $\ell\cdot p_1$ or $\ell\cdot p_2$. Since we are 
limiting our discussion to UV-finite graphs, these are the only possible 
numerators at the triangle level.\\
The number of evaluations per phase-space point is the same as in the scalar case;
this means $5\cdot 10^4$ evaluations for non-deformation points and 
$10^6$ evaluations for phase-space points with deformation. 
Due to the presence of numerators, the integrand function is a 
bit more complex. Hence the calculation time, compared to scalar triangles 
increases from 2.5 seconds to 3.5 seconds for deformation-points. 
For non-deformation points there is hardly any measurable difference, 
i.e. they stay at around 0.15 seconds\\
A compilation examples is shown in Table \ref{tab:trilp}.
Points \ref{point16} and \ref{point18} correspond to kinematical situations 
in which all internal masses are equal and the numerator is $\ell\cdot p_2$, 
whereas Points \ref{point17} and \ref{point19} have all internal masses are 
different from each other and numerator $\ell\cdot p_1$.

 \begin{table}[H]
%\setlength{\tabcolsep}{0.3pc}
% \caption{Pentagon example results}
   % {\small
   \centering
\begin{subtable}{1\textwidth}
\begin{tabular}{lllll}
&\\
\hline
%\hline
%\\
 & Point \plabel{point16}{16}  & Point 16 Error &  Point \plabel{point17}{17} & Point 17 Error\\
%\\
\hline
%\hline
%\\
LoopTools & -1.07284E-2 & 0 & -1.59964E-3 & 0\\
Loop--Tree Duality & -1.07281E-2 & 5.4E-6 & -1.59985E-3 & 6.6E-7\\
\hline
%\hline
\end{tabular}
\caption{Two non-deformation phase-space points.\label{trilp-non}}
\end{subtable}
\begin{subtable}{1\textwidth}
\begin{tabular}{lllll}
&\\
\hline
%\hline
%\\
 & Real Part  & Real Error &  Imaginary Part & Imaginary Error\\
%\\
\hline
%\hline
%\\
LoopTools P.\plabel{point18}{18} & -2.64773E-3 & 0 & 1.37469E-2 & 0\\
Loop--Tree Duality P.18 & -2.64726E-3 & 5.5E-6 & 1.37448E-2 & 5.5E-6\\
\hline
LoopTools P.\plabel{point19}{19} & -1.19501E-2 & 0 & 1.35834E-3 & 0\\
Loop--Tree Duality P.19 & -1.19511E-2 & 3.6E-5 & 1.35859E-3 & 3.6E-5\\
\hline
%\hline
\end{tabular}
\caption{Two phase-space points that need deformation.\label{trilp-defo}}
\end{subtable}
%}
\caption{Examples for triangles involving numerators, with and without deformation.}
\label{tab:trilp}
\end{table}

In the triangle mass-scan of Fig. \ref{fig:trinum}, we varied the mass in 
the same way as we did in the previous chapter. Three thresholds are passed 
at $2m/\sqrt{s}\approx 0.15$, 0.45 and 1.

\begin{figure}[H]
\centering
   \begin{subfigure}[b]{1\textwidth}
   \includegraphics[width=1\linewidth]{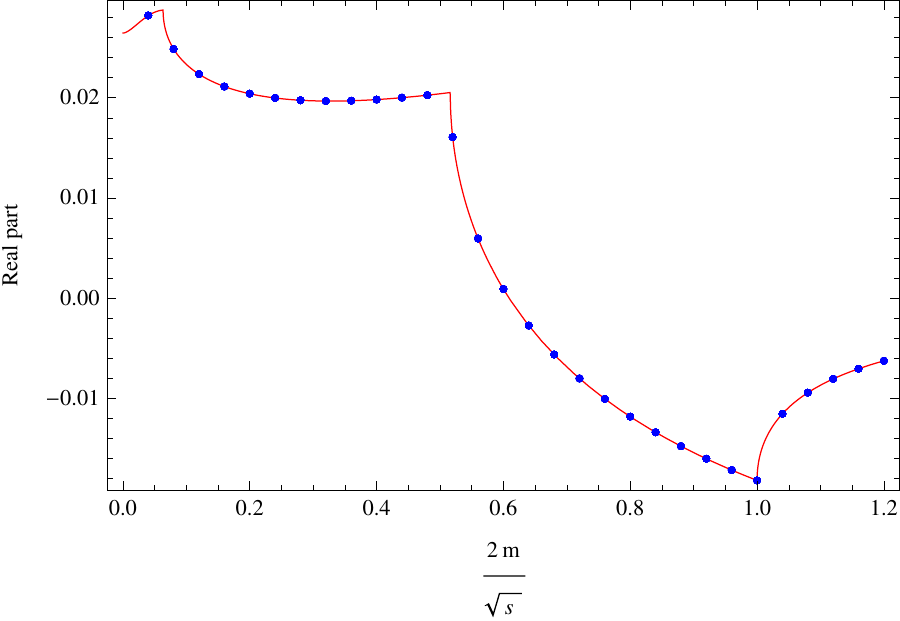}
   %\caption{Real Part}
   \label{fig:trirenum} 
\end{subfigure}

\begin{subfigure}[b]{1\textwidth}
   \includegraphics[width=1\linewidth]{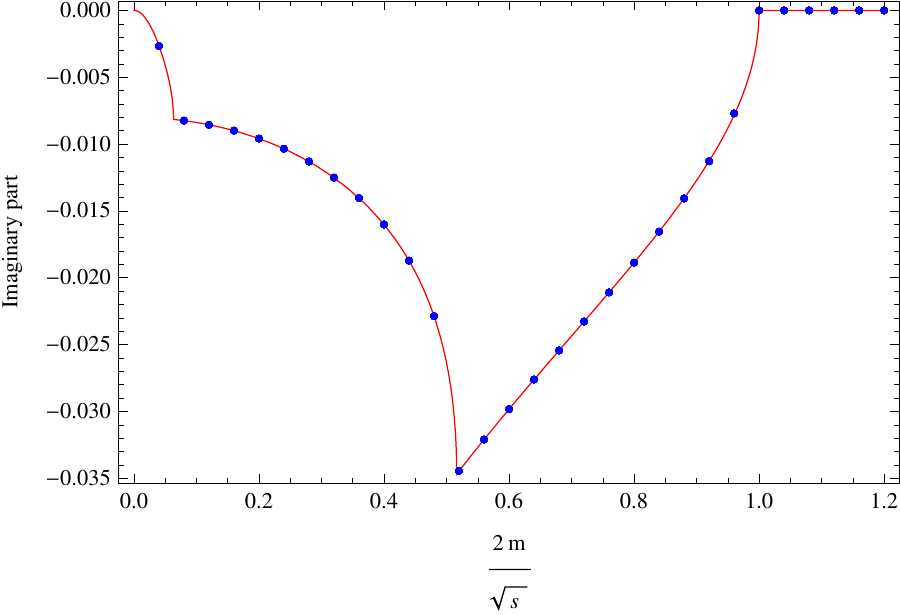}
   %\caption{Imaginary Part}
   \label{fig:triimnum}
\end{subfigure}
\caption{Mass-scan of a triangle with numerator $\ell\cdot p_3$. The red curve 
is LoopTools, the blue points are the Loop--Tree Duality.\label{fig:trinum}}
\end{figure}

%----------------------------------------------------------------------------------------
%	Tensor Boxes
%----------------------------------------------------------------------------------------

\section{Tensor Boxes}

Again, we repeat the analysis of the triangles at the box-level. We only want to 
deal with UV-finite diagrams, thus tensors up to rank three are allowed in the numerator. 
Since having a rank-one tensor is quite similar to the triangle case, we will focus on
rank two and rank three situations in order to have a more challenging 
test for the program.\\
We use the same number of evaluations as we did for scalars, this means $4\cdot 10^6$ 
evaluations for deformation points and $5\cdot 10^4$ evaluations for points that do 
not need deformation.
As far as deformation points are concerned, calculation times differ according to the 
rank of the tensor function in the numerator. Their general order of magnitude is 
around 20 seconds, the span between a rank zero and a rank three tensor amounts 
to 8 seconds. Non-deformation points take about 0.25 seconds, almost independently of the 
rank of the numerator function.

 \begin{table}[H]
%\setlength{\tabcolsep}{0.3pc}
% \caption{Pentagon example results}
   % {\small
   \centering
\begin{subtable}{1\textwidth}
\begin{tabular}{lllll}
&\\
\hline
%\hline
%\\
 & Point \plabel{point20}{20}  & Point 20 Error &  Point \plabel{point21}{21} & Point 21 Error\\
%\\
\hline
%\hline
%\\
LoopTools & -3.42913E-4 & 0 & 9.64909E+0 & 0\\
Loop--Tree Duality & -3.42905E-4 & 5.9E-8 & 9.64924E+0 & 4.6E-3\\
\hline
%\hline
\end{tabular}
\caption{Two non-deformation phase-space points.\label{boxlp-non}}
\end{subtable}
\begin{subtable}{1\textwidth}
\begin{tabular}{lllll}
&\\
\hline
%\hline
%\\
 & Real Part  & Real Error &  Imaginary Part & Imaginary Error\\
%\\
\hline
%\hline
%\\
LoopTools P.\plabel{point22}{22} & 1.33032E-3 & 0 & -1.07780E-3 & 0\\
Loop--Tree Duality P.22 & 1.33033E-3 & 5.6E-7 & -1.07779E-3 & 5.6E-7\\
\hline
LoopTools P.\plabel{point23}{23} & -2.15448E+2 & 0 & -1.10792E+2 & 0\\
Loop--Tree Duality P.23 & -2.15451E+2 & 8.7E-2 & -1.10789E+2 & 8.7E-2\\
\hline
%\hline
\end{tabular}
\caption{Two phase-space points that need deformation.\label{boxlp-defo}}
\end{subtable}
%}
\caption{Examples for boxes involving numerators of tensor rank two and three, with and without deformation.}
\label{tab:boxlp}
\end{table}

In Table \ref{tab:boxlp}, Points \ref{point20} and \ref{point22} are boxes with 
all internal masses equal and numerator $\ell\cdot p_1 \times \ell\cdot p_2$. Points 
\ref{point21} and \ref{point23} represent situations in which all internal masses are 
different and the numerator has the form $\ell\cdot p_1\times\ell\cdot p_3\times\ell\cdot p_4$.

We also did a scan of the Mandelstam variable $t$, see Fig. \ref{fig:boxnum}, for which we took a box with 
three scalar products in the numerator, $\ell\cdot p_1\times\ell\cdot p_2\times\ell\cdot p_3$, and varied 
$p_3$ while keeping $p_3^2$ constant. Since $p_3$ also appears in the numerator, 
both numerator and denominator are affected by this scan. 

\begin{figure}[H]
\centering
   \begin{subfigure}[b]{1\textwidth}
   \includegraphics[width=1\linewidth]{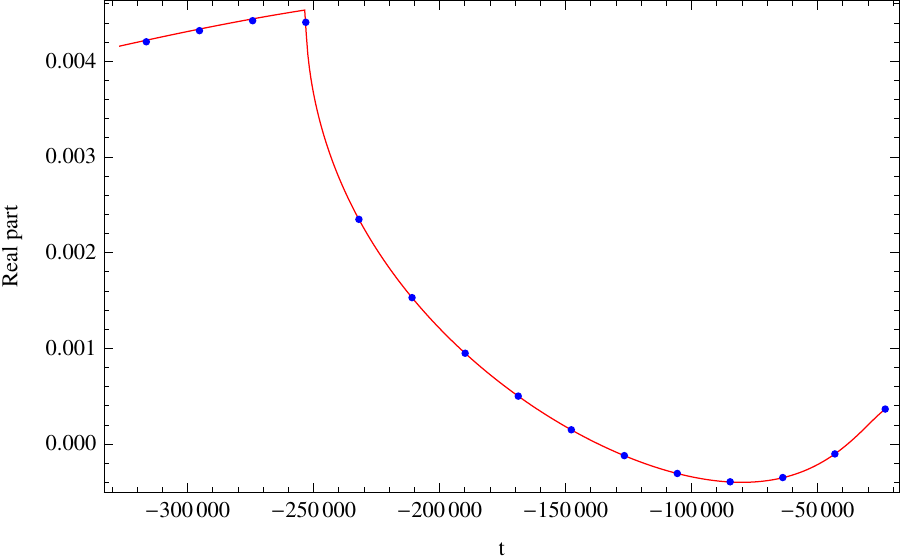}
   %\caption{Real Part}
   \label{fig:boxnumre} 
\end{subfigure}

\begin{subfigure}[b]{1\textwidth}
   \includegraphics[width=1\linewidth]{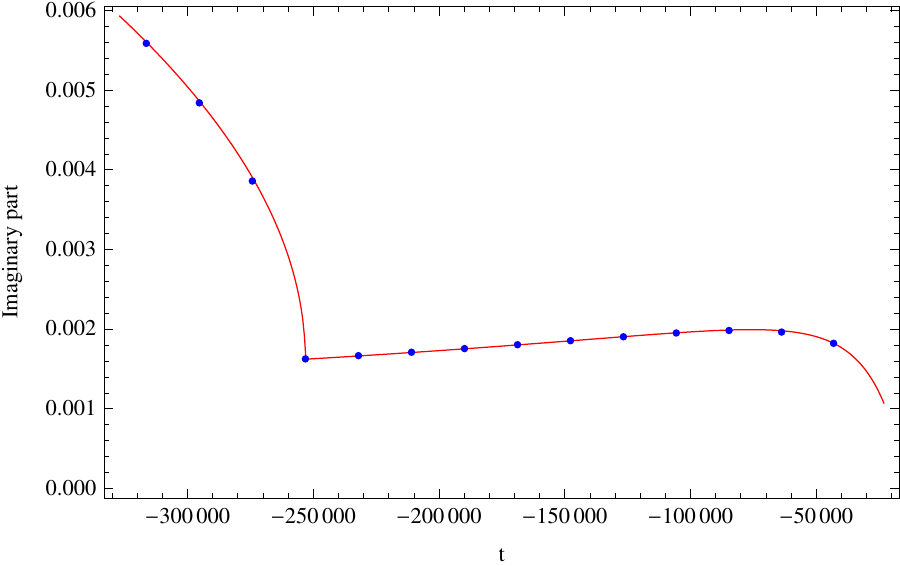}
   %\caption{Imaginary Part}
   \label{fig:boxnumim}
\end{subfigure}
\caption{Scan of the region around threshold. The red curve 
is LoopTools the blue points are the Loop--Tree Duality.\label{fig:boxnum}}
\end{figure}

%----------------------------------------------------------------------------------------
%	Tensor Pentagons
%----------------------------------------------------------------------------------------

\section{Tensor Pentagons}

Next, we investigate pentagon tensor integrals at the one--loop level with 
numerators up to rank three. The number of evaluations is chosen to be the same as in the 
scalar case, i.e. $10^5$ times for non-deformation and $5\cdot 10^6$  times for 
phase-space points that require deformation. This results in calculation times 
of 0.7 seconds and 33 seconds, respectively. The runtime difference between a point 
with a numerator of rank zero and rank three is negligible for non-deformation 
points and about 10 seconds for deformation points.\\
Table \ref{tab:pentagonlp} shows a selection of sample points.
Point \ref{point24} and \ref{point26} feature the rank two numerator 
$\ell\cdot p_3\times\ell\cdot p_4$ while Point \ref{point25} 
and \ref{point27} have the numerator $\ell\cdot p_3\times\ell\cdot p_4\times\ell\cdot p_5$. 
In all points all internal masses are equal. At first glance, Point \ref{point27} 
seems to be weaker but it actually contains six ellipsoid singularities whereas the other points 
usually have two to three. We include this point to demonstrate that 
the program holds well together even under such challenging circumstances.

 \begin{table}[H]
%\setlength{\tabcolsep}{0.3pc}
% \caption{Pentagon example results}
   % {\small
   \centering
\begin{subtable}{1\textwidth}
\begin{tabular}{lllll}
&\\
\hline
%\hline
%\\
 & Point \plabel{point24}{24}  & Point 24 Error &  Point \plabel{point25}{25} & Point 25 Error\\
%\\
\hline
%\hline
%\\
LoopTools & -1.86472E-8 & 0 & 1.74828E-3 & 0\\
Loop--Tree Duality & -1.86462E-8 & 2.6E-12 & 1.74808E-3 & 2.8E-7\\
\hline
%\hline
\end{tabular}
\caption{Two non-deformation phase-space points.\label{pentalp-non}}
\end{subtable}
\begin{subtable}{1\textwidth}
\begin{tabular}{lllll}
&\\
\hline
%\hline
%\\
 & Real Part  & Real Error &  Imaginary Part & Imaginary Error\\
%\\
\hline
%\hline
%\\
LoopTools P. \plabel{point26}{26} & -1.68298E-6 & 0 & 1.98303E-6 & 0\\
Loop--Tree Duality P. 26 & -1.68298E-6 & 7.4E-10 & 1.98299E-6 & 7.4E-10\\
\hline
LoopTools P. \plabel{point27}{27} & -8.34718E-2 & 0 & 1.10217E-2 & 0\\
Loop--Tree Duality P. 27 & -8.34829E-2 & 7.5E-5 & 1.10119E-2 & 7.5E-5\\
\hline
%\hline
\end{tabular}
\caption{Two phase-space points which need deformation.\label{pentalp-defo}}
\end{subtable}
%}
\caption{Examples for pentagons involving numerators of tensor rank two and three, with and without deformation.}
\label{tab:pentagonlp}
\end{table}

We also performed several scans; a sample is presented in Figure \ref{fig:pentanum}. 
In that scan, similarly to what we have done with scalar pentagons, 
we varied $p_1$ and thus the center-of-mass energy $s=(p_1+p_2)$. 
The corresponding numerator function 
is $\ell\cdot p_1\times\ell\cdot p_2 \times\ell\cdot p_3$, 
which means that both numerator and denominator take part in the scan. From the plot, 
we can see that the Loop-Tree Duality is able to pass this challenging test.

\begin{figure}[H]
\centering
   \begin{subfigure}[b]{1\textwidth}
   \includegraphics[width=1\linewidth]{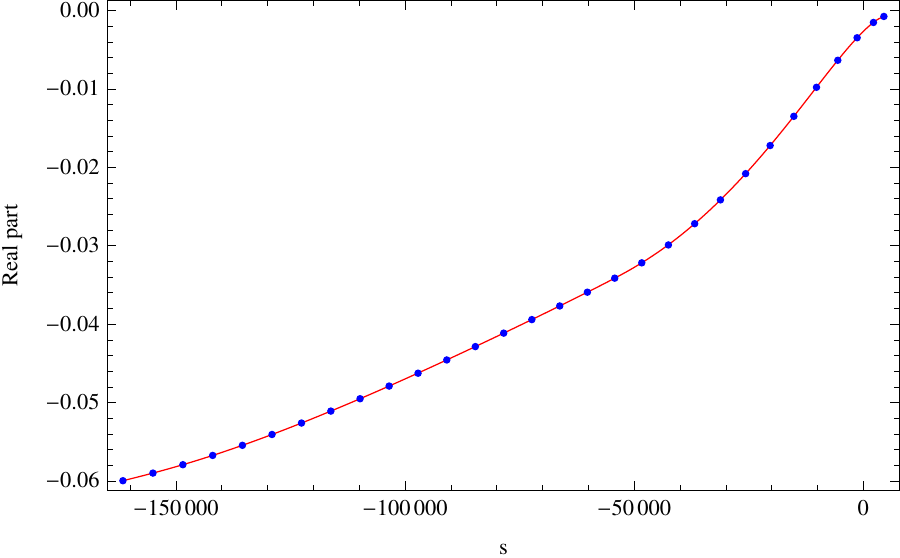}
   %\caption{Real Part}
   \label{fig:trire} 
\end{subfigure}

\begin{subfigure}[b]{1\textwidth}
   \includegraphics[width=1\linewidth]{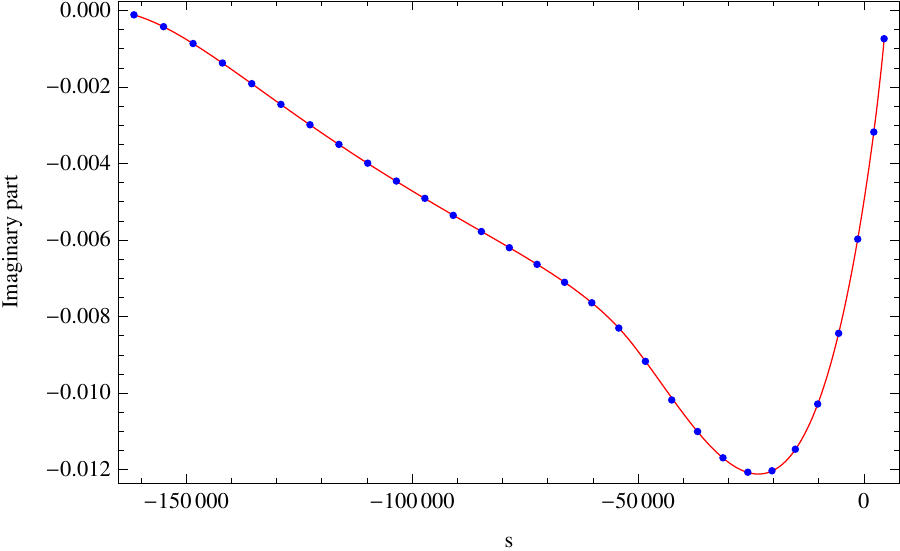}
   %\caption{Imaginary Part}
   \label{fig:triim}
\end{subfigure}
\caption{Scan of the region around threshold. The red curve 
is LoopTools, the blue points are the Loop--Tree Duality.\label{fig:pentanum}}
\end{figure}

%----------------------------------------------------------------------------------------
%	Tensor Hexagons
%----------------------------------------------------------------------------------------

\section{Tensor Hexagons}

We also analysed hexagon (six external legs) tensor integrals, for which we raised the 
number of evaluations for non-deformation points to $10^6$ and 
for deformation points to $8\cdot 10^6$. 
The corresponding calculation times are 8 and 75 seconds, respectively. 
Since we evaluate non-deformation 
points ten times more compared to pentagons, the different numerators have a 
measurable impact on the calculation time. The difference between rank zero and 
rank three is 6 seconds and 15 seconds for deformation points.\\
LoopTools can provide reference values only up to pentagons. Instead, we 
used the program SecDec version 3 \cite{Borowka:2015mxa} for comparison.\\
Again, we present a selection of sample points in Table \ref{tab:hexagonlp}. 
Point \ref{point28} and \ref{point30} feature the rank-one numerator 
$\ell\cdot p_1$, in the former all internal masses are different and in the latter 
they are all equal. Point \ref{point29} has six distinct internal masses and 
the numerator function  $\ell\cdot p_2\times\ell\cdot p_4\times\ell\cdot p_6$,
Point \ref{point31} possesses the numerator $\ell\cdot p_2\times \ell\cdot p_5$ and six 
different masses, as well. Finally, in Point \ref{point32}, which exhibits the numerator 
$\ell\cdot p_4\times\ell\cdot p_5\times \ell\cdot p_6$, all momenta are distinct 
form each other.

 \begin{table}[H]
%\setlength{\tabcolsep}{0.3pc}
% \caption{Pentagon example results}
   % {\small
   \centering
\begin{subtable}{1\textwidth}
\begin{tabular}{lllll}
&\\
\hline
%\hline
%\\
 & Point \plabel{point28}{28}  & Point 28 Error &  Point \plabel{point29}{29} & Point 29 Error\\
%\\
\hline
%\hline
%\\
SecDec & -1.21585E-15 & 1.2E-19 & 4.46117E-9 & 3.7E-13\\
Loop--Tree Duality & -1.21552E-15 & 3.5E-18 & 4.46136E-9 & 2.6E-15\\
\hline
%\hline
\end{tabular}
\caption{Two non-deformation phase-space points.\label{hexalp-non}}
\end{subtable}
\begin{subtable}{1\textwidth}
\begin{tabular}{lllll}
&\\
\hline
%\hline
%\\
 & Real Part  & Real Error &  Imaginary Part & Imaginary Error\\
%\\
\hline
%\hline
%\\
SecDec P. \plabel{point30}{30} & 1.01359E-15 & 2.3E-19 & 2.68657E-15 & 2.6E-19\\
Loop--Tree Duality P. 30 & 1.01345E-15 & 1.3E-18 & 2.68633E-15 & 1.3E-18\\
\hline
SecDec P. \plabel{point31}{31} & 2.45315E-12 & 2.4E-16 & -2.06087E-12 & 2.0E-16\\
Loop--Tree Duality P. 31 & 2.45273E-12 & 7.3E-15 & -2.06202E-12 & 7.3E-15\\
\hline
%\hline
SecDec  P. \plabel{point32}{32} & -2.07531E-6 & 1.9E-10 & 6.97158E-7 & 5.6E-11\\
Loop--Tree Duality P. 32 & -2.07526E-6 & 7.5E-13 & 6.97192E-7 & 7.5E-13\\
\hline
\end{tabular}
\caption{Two phase-space points that need deformation.\label{hexalp-defo}}
\end{subtable}
%}
\caption{Examples for pentagons involving numerators of tensor rank two and three, with and without deformation.}
\label{tab:hexagonlp}
\end{table}
%\pagestyle{lastpage}
% Chapter Template

\chapter{Summary and Future Work} % Main chapter title
%\addcontentsline{toc}{chapter}{Conclusion and Outlook} 
\label{Chapter9} % Change X to a consecutive number; for referencing this chapter elsewhere, use \ref{ChapterX}

%\lhead{\emph{Conclusion and outlook}} % Change X to a consecutive number; this is for the header on each page - perhaps a shortened title

%----------------------------------------------------------------------------------------
%	Conclusion
%----------------------------------------------------------------------------------------

\section{Summary}
%\addcontentsline{toc}{section}{Conclusion} 
One of the main difficulties in calculating scattering amplitudes 
at NLO and higher is the cancellation of infrared singularities among 
real and virtual corrections. The Loop--Tree Duality aims to solve that 
problem by recasting the virtual corrections in a form which closely 
resembles the real ones. This appealing property motivates the idea 
of combining the two, thus treating them simultaneously in a common 
Monte Carlo event generator.\\
Initially the LTD was introduced for scalar one--loop 
integrals \cite{Catani:2008xa} and in later works its applicability has been expanded 
by a large margin via a systematic procedure has been established to 
calculate Feynman graphs with an arbitrary number of loops 
\cite{Bierenbaum:2010cy}. In Chapter \ref{Chapter5} and in \cite{Bierenbaum:2012th} 
we have shown how to deal with loop integrals involving higher order poles. 
Instead of applying the Residue Theorem 
for higher order poles, using Integration By Parts techniques reduced the 
integrals featuring such poles to integrals with simple poles. 
After that we could go ahead and straightforwardly 
apply the known formulae. At that point, the major theoretical questions 
were answered and the important was to check the efficiency of the method. To that direction, as a
preparatory step (Chapter \ref{Chapter6} and \cite{Buchta:2014dfa}), 
we investigated the singularities of the dual contributions 
and found a partial cancellation of singularities among contributions 
happening at the \textit{integrand level}. In the case 
of massless diagrams, we were able to show that the infrared singularities 
are confined to a finite region of loop-momentum space which can then be 
mapped to the real corrections. With this information at hand, we did a 
numerical implementation of the Loop--Tree Duality in the form of a 
computer program written in C++. Being a first step towards the combined 
treatment, the program calculated one-loop integrals. Due to the partial 
cancellation of singularities, we only had to deal with singularities of 
the ellipsoid type. We dealt with the ellipsoid singularities by contour deformation. 
Still, the hyperboloid singularities affected the way we deformed:
In order to keep the cancellations intact, we grouped dual contributions featuring 
the same hyperboloid singularities together. Then, we employed a contour deformation 
that accounted for all the ellipsoid singularities of the entire group. This contour 
deformation was constructed to respect the $i0$-prescription of the propagators and 
fall rapidly to 0 where it was not needed so that the value of 
the overall integral would not be altered. Since dual 
integrals are three--dimensional, we only had to deform in loop three-momentum 
space. As integrators we used Cuhre and VEGAS from the Cuba library. This setup 
has proven to be successful. We were able to calculate (UV- and IR-finite) scalar 
with good precision. 
This has been illustrated by showing individual points as well as several scans 
through the phase space, in which we picked one mass or momentum to be varied 
while the other ones were kept constant. 
In Chapter \ref{Chapter8} we also demonstrated that the Loop-Tree Duality deals equally 
well with tensor integrals (numerators different from 1) up to rank three.\\
The numerical implementation is still in an early state. Neverthelss, the wide 
range of checks we have performed proves the potential of the method. 
In particular, the implementation works well independently from the number 
of kinematical scales (bilinears, trilinears, different masses) involved or whether 
a phase space point is in the vicinity of a threshold or not.

%----------------------------------------------------------------------------------------
%	Outlook
%----------------------------------------------------------------------------------------

\section{Future Work}
%\addcontentsline{toc}{section}{Outlook} 
Having shown that the Loop--Tree Duality method works and is efficient, 
the next step will be to address any issue the code might run into. 
Furthermore we would like to provide a 
systematical way to choose the parameters $\lambda$ and the width for 
optimal results. We also aim to try the LTD for a real, physical process 
(i.e. not a toy model), for example the six- or eight-photon amplitude. 
To that end, we will move towards treating a case where we combine 
the real and virtual corrections. The goal is to have 
a highly automated program which calculates the entire cross-section at the 
one-loop level. Finally, the next step in our list is to implement the LTD 
for two-loop amplitudes. This is quite an extensive list, but with the presented work 
we have laid a good basis. 
\pagestyle{intro}
\pagestyle{fancy}
%\include{Chapters/ChapterE}

%----------------------------------------------------------------------------------------
%	THESIS CONTENT - APPENDICES
%----------------------------------------------------------------------------------------

%\addtocontents{toc}{\vspace{2em}} % Add a gap in the Contents, for aesthetics

\appendix % Cue to tell LaTeX that the following 'chapters' are Appendices

% Include the appendices of the thesis as separate files from the Appendices folder
% Uncomment the lines as you write the Appendices

% Appendix A

\addtocontents{toc}{\vspace{2em}}

\chapter{Duality Nomenclature} % Main appendix title

\label{AppendixA} % For referencing this appendix elsewhere, use \ref{AppendixA}

%\lhead{Appendix A. \emph{duality Nomenclature}} % This is for the header on each page - perhaps a shortened title

\begin{itemize}
\item \textit{Loop--Tree Duality}: \\
The Loop--Tree Duality (in short: Duality) 
is a Feynman integral transformation 
that maps loop integrals to a sum of phase-space (tree--level)
 integrals. Within the conventions of Chapter \ref{Chapter3}, 
the Loop--Tree Duality at one-loop level looks like:
\bea
\int_{\ell}\,\prod\limits_{i=1}^N \,G_{F}(q_i)
\quad \longrightarrow
  \quad - \sum\limits_{i=1}^N \, \int_{\ell} \; \td(q_i) \,
\prod_{\substack{j=1 \\ j\neq i}}^{N} \,G_D(q_i;q_j)
\label{ap:ltd}
\eea 
%where $G_{F_i}$ are the Feynman and $G_D(q_i;q_j)$ 
%the \textit{dual propagators}. 
This transformation is achieved by performing the 
integration over the energy component of the loop 
integral. The resulting integration runs only over the 
loop three-momentum and is very similar to the real 
radiation corrections. Thus, it encourages the idea of 
combining the two, treating them simultaneously 
in a common Monte Carlo event generator.\\
Each summand is called \textit{dual contribution} and 
is constructed according to a special pattern: One of 
the internal lines gets on-shell, i.e. it is replaced by a 
\textit{dual delta function}, while all the other 
(non-cut) Feynman propagators are promoted to 
\textit{dual propagators}. This procedure is repeated for 
every internal line once and the results are added together 
to yield the \textit{dual integral}.
\item \textit{Dual delta function}:\\
The dual delta function $\td(q_i)$ is the delta function that 
appears after applying the Loop--Tree Duality:
\beq
\td(q_i) \equiv 2 \pi \, i \, \theta(q_{i,0}) \, \delta(q_i^2-m_i^2) 
= 2 \pi \, i \, \delta_+(q_i^2-m_i^2)\nn
\eeq
It serves several purposes. Formally it is there 
(cf. \Eq{ap:ltd}) to indicate 
that the energy integration has already been carried out, 
leaving an integral over the loop three-momentum only. 
It fixes the $\ell_0$-component to a certain value 
determined by the argument of the delta function. The theta function, 
which is included in the dual delta function, ensures that the 
positive energy solution gets picked from the argument of 
the dual delta function.
\item \textit{Dual propagator}: \\
Applying the Loop--Tree Duality involves converting all 
Feynman propagators to dual propagators, except the cut one. 
There are two representations of the dual propagator,
\beq
\td(q_i)G_D(q_i;q_j) = \frac{\td(q_i)}{q_j^2 -m_j^2 - i0 \,\eta\cdot k_{ji}}
 \quad \text{and} \quad 
G_D(q_i;q_j) = 
\frac{1}{(q_{i,0}^{(+)} + k_{ji,0})^2-(q_{j,0}^{(+)})^2}~,
\label{ap:duprop}
\eeq
where
\beq
q_{i,0}^{(+)} = \sqrt{\mathbf{q}_i^2 + m_i^2-i0}~.
\eeq
The main difference between a dual propagator and 
a Feynman propagator is the $i0$-prescription. The 
Feynman propagator depends on the loop four-momentum 
as opposed to the dual propagator which depends on the loop 
three-momentum.\\
The $i0$-prescription indicates 
which way to go around the singularities when integrating. Since 
the integration of the energy component has already been carried 
out, the dual propagator can only depend on the loop 
three-momentum and the $i0$-prescription gets modified (for 
details on how and why this happens, see Section \ref{sec:dt}). 
As a direct consequence, the new dual prescription is merely 
a way of bookkeeping on how to go around singularities in 
loop three-momentum space.\\
The form on the left of \Eq{ap:duprop} is a natural result of the 
derivation of the Loop--Tree Duality, the form on the right is a 
little more suited for numerical implementations.
\item \textit{Dual prescription}: \\
The modified $i0$-prescription of the dual propagator is called 
dual prescription. The dual prescription of the dual propagator 
$G_D(q_i;q_j)$ has the form
\bea
-i0\eta\cdot k_{ji}~,
\eea
where $\eta$ is a future-like four-vector, $\eta^2\geq 0, \eta_0>0$. 
Typically, $\eta$ is chosen to be $(1,0,0,0)$. In general, $\eta$ 
depends on the choice of the coordinate frame. Nonetheless, once all 
dual contributions have been added together, $\eta$ cancels out. 
In other words dual integrals are independent from the choice of 
$\eta$ whereas individual dual contributions do depend on it. At one--loop, the vector 
$k_{ji}=q_j-q_i$ 
does not depend on the loop momentum and determines the 
sign of the dual prescription.
\item \textit{Dual contribution}: \\
A dual contribution is one individual summand of the sum in 
\Eq{ap:ltd},
\bea
\int_{\ell} \; \td(q_i) \,
\prod_{\substack{j=1 \\ j\neq i}}^{N} \,G_D(q_i;q_j)~.
\eea
It represents one single cut of a one-loop diagram. The cut 
line gets replaced by a dual delta function, the Feynman 
propagators of all other internal lines are promoted to 
dual propagators. This is demonstrated explicitly in Section 
\ref{sec:masslexmp}.\\
Loop--Tree Duality converts a one-loop integral with $N$ 
external legs into a sum of $N$ dual contributions. While 
the original one-loop integral runs over the loop 
four-momentum, the integrations of the dual contributions are 
only three-dimensional thus simplifying the calculation.\\
Individual dual contributions can feature singularities 
(hyperboloid singularities, cf. Section \ref{sec:dualsing}) 
that partially disappear once all dual contributions are combined. 
\item \textit{Dual integral}: \\
The dual integral is the end result obtained by applying the 
Loop--Tree Duality to a one-loop integral:
\bea
- \sum\limits_{i=1}^N \, \int_{\ell} \; \td(q_i) \,
\prod_{\substack{j=1 \\ j\neq i}}^{N} \,G_D(q_i;q_j)
\eea
It features a couple of appealing properties: Since the 
$\ell_0$-integration has already been performed, the 
resulting dual integral only involves a three-dimensional 
integration. In fact, it resembles the form of the real 
corrections. Therefore, the Loop--Tree Duality aims 
towards a combined treatment of loop and tree 
contributions.
%\item \textit{Ellispoid Singularity}:\\
%\item \textit{Hyperboloid Singularity}:\\
%\item \textit{On-shell hyperboloid}:\\
\end{itemize}
% Appendix Template

\chapter{Proof of the Reduction of Eq. (\ref{eq:firstexample})} % Main appendix title

%\label{AppendixC} % Change X to a consecutive letter; for referencing this appendix elsewhere, use \ref{AppendixX}

%\lhead{Appendix X. \emph{Appendix Title Here}} % Change X to a consecutive letter; this is for the header on each page - perhaps a shortened title
\label{app:systemsolve}

%----------------------------------------------------------------------------------------
%	Proof of the reduction of Eq. (\ref{eq:firstexample})
%----------------------------------------------------------------------------------------

Here we solve the system of equations, explicitly, to arrive at Eq. (\ref{eq:firstexample}). We note that we are
not aiming for a full reduction to a set of master integrals but rather to reduce the multiple poles to single
poles. Therefore, any integral which has single propagators is to be considered known. 

Using the IBPs, \Eq{eq:IBP2loop2p}, on the generic integral $F(a_1a_2a_3a_4a_5)$, we get the system of equations:
%\begin{flushleft}
\beq
\label{eq:IBPsystem1}
d-2a_1-a_4-a_5 - a_4 {\bf 4^+} {\bf 1^-}+a_4 {\bf 4^+}{\bf 2^-}=0~,
\eeq
\beq
\label{eq:IBPsystem2}
a_1-a_4+ \frac{1}{2}s a_5 {\bf 5^+}+a_4 {\bf 4^+} ({\bf 1^-}-{\bf 2^-})+a_1 {\bf 1^+} ({\bf 2^-}-{\bf 4^-})
    +\frac{1}{2} a_5 {\bf 5^+} ({\bf 2^-}-{\bf 3^-})=0~, 
\eeq
\beq
\label{eq:IBPsystem3}
a_4 {\bf 4^+} (s+{\bf 2^-}-{\bf 3^-}-2 \; {\bf 5^-})-s a_5 {\bf 5^+}-2a_1 {\bf 1^+}{\bf 5^-}=0~, 
\eeq
\beq
\label{eq:IBPsystem4}
a_2-a_4+a_2 {\bf 2^+} ({\bf 1^-}-{\bf 4^-})+a_3 {\bf 3^+} ({\bf 1^-}+{\bf 2^-}-{\bf 4^-}-2 \, {\bf 5^-})+
   a_4 {\bf 4^+} ({\bf 2^-}-{\bf 1^-})=0~, 
\eeq
\beq
\label{eq:IBPsystem5}
d-2a_2-a_3-a_4 +a_3{\bf 3^+} (s-{\bf 2^-})+a_4 {\bf 4^+} ({\bf 1^-}-{\bf 2^-})=0~,
\eeq
\beq
\label{eq:IBPsystem6}
a_2-a_3+s  (a_2{\bf 2^+}-a_3{\bf 3^+}+a_4{\bf 4^+})+a_4 {\bf 4^+} ({\bf 2^-}-{\bf 3^-}-2 \, {\bf 5^-})
   +a_3{\bf 3^+ 2^-}-a_2{\bf 2^+ 3^-}=0~,
\eeq
%\end{flushleft}
where $s=p^2+i0$. The appearance of the operator ${\bf 5^-}$ signals that we have 
the ISP $\ell_1 \cdot p$ in the numerator of an integral. As long as these integrals possess single propagators,
we will not reduce them further but consider them known. We also note that a lot of the integrals that appear 
after setting particular values to the parameters $a_i$ in this system, are 
zero in dimensional regularization (in the massless case). 
Let us start by setting $a_2=2,a_1=a_3=a_4=1,a_5=-1$ in (\ref{eq:IBPsystem2}). We get:
\beq
\label{System1}
F(2111-1)-F(1112-1)-\frac{1}{2}sF(12110)-\frac{1}{2}F(11110)=0~.
\eeq
Taking the sum of (\ref{eq:IBPsystem5}) and (\ref{eq:IBPsystem6}) and setting $a_1=1,a_2=a_3=a_4=1,a_5=0$ we get:
\beq
\label{System2}
(d-4)F(11110)+sF(12110)+sF(11120)-2F(1112-1)=0~.
\eeq
Taking the difference between (\ref{eq:IBPsystem5}) and (\ref{eq:IBPsystem6}) 
and setting $a_1=2,a_2=a_3=a_4=1,a_5=0$ we get:
\beq
\label{System3}
-2F(2111-1)-sF(12110)+sF(11210)-F(10210)=0~.
\eeq
Finally, setting $a_1=2,a_2=a_3=a_4=1,a_5=-1$ in (\ref{eq:IBPsystem1}), we get:
\beq
\label{System4}
(d-4)F(2111-1)-F(1112-1)+F(2012-1)=0~.
\eeq
The integrals $F(10120)$ and $F(11210)$ and $F(2012-1)$, in this system of equations, can be computed simply by taking further, 
appropriate combinations of Eqs. (\ref{eq:IBPsystem1})-(\ref{eq:IBPsystem6}). Setting $a_1=a_2=a_3=a_4=1,a_5=0$ in
(\ref{eq:IBPsystem1}) we get $F(10120)=(3-d)F(11110)$. It also holds that $F(10120)=F(10210)$, as can be
seen by making the shifts in the loop momenta $\ell_2\rightarrow-\ell_1-\ell_2-p$.
From Eq. (\ref{eq:IBPsystem5}), by setting $a_1=a_2=a_3=a_4=1,a_5=0$ and using the value of  $F(10120)$, we get:
$s F(11210)=(10-3d)F(11110)$. Finally, adding Eqs. (\ref{eq:IBPsystem1}) and (\ref{eq:IBPsystem2}) and 
setting $a_1=a_2=1,a_3=2,a_4=1,a_5=-1$ we get: $F(2012-1)=(d-3)(d-4)F(11110)$. The rest of the system of equations
(\ref{System1})-(\ref{System4}) can now be solved sequencially, arriving at:
\beq
F(12110)= \frac{(3d-10)}{(d-6)s} F(11110) \, ,
\eeq
a result which, after putting $d=4-2\epsilon$, agrees with Eq. (\ref{eq:firstexample}).
% Appendix Template

\chapter{Momenta of the Example Points} % Main appendix title

\label{AppendixB} % Change X to a consecutive letter; for referencing this appendix elsewhere, use \ref{AppendixX}

%\lhead{Appendix X. \emph{Appendix Title Here}} % Change X to a consecutive letter; this is for the header on each page - perhaps a shortened title

Here we give the momenta and masses of the different phase-space points and scans shown in 
Chapter \ref{Chapter7} and \ref{Chapter8}. Due to momentum conservation $p_N=-\sum_{i=1}^{N-1}p_i$, 
therefore it is sufficient to give only the momenta $p_1$ to $p_{N-1}$.

\section{Individual points}

\begin{enumerate}[leftmargin=2.5cm]
  \item[{\bf Point \ref{point1}}] $p_1=\{5.23923,-4.18858,0.74966,-3.05669\}$\\
$p_2=\{6.99881,-2.93659,5.03338,3.87619\}$\\
$m_1=m_2=m_3=7.73358$
  \item[{\bf Point \ref{point2}}] $p_1=\{13.42254,58.79478,-73.11858,-91.95015\}$\\
$p_2=\{81.65928,-68.52173,8.75578,-95.05353\}$\\
$m_1= 49.97454, m_2=86.92490, m_3=80.22567$
  \item[{\bf Point \ref{point3}}] $p_1=\{10.51284,6.89159,-7.40660,-2.85795\}$\\
$p_2=\{6.45709,2.46635,5.84093,1.22257\}$\\
$m_1= m_2 = m_3 = 0.52559$
  \item[{\bf Point \ref{point4}}] $p_1=\{95.77004,31.32025,-34.08106,-9.38565\}$\\
$p_2=\{94.54738,-53.84229,67.11107,45.56763\}$\\
$m_1= 83.02643, m_2=76.12873, m_3=55.00359$
  \item[{\bf Figure \ref{fig:trilms}}] $p_1=\{44.38942,17.84418,12.70440,-23.67441\}$\\
$p_2=\{11.62982,-35.11756,-9.52573,1.27635\}$\\
$m_1= m_2 = m_3 = 7.89824$
\end{enumerate}
\newpage
\begin{enumerate}[resume*]%[label=\bfseries Point \#\arabic*, leftmargin=2.5cm]
  \item[{\bf Point \ref{point5}}] $p_1=\{31.54872,-322.40325,300.53015,-385.58013\}$\\
$p_2=\{103.90430,202.00974,-451.27794,-435.12848\}$\\
$p_3=\{294.76653,252.88958,447.09194,311.71630\}$\\
$m_1= m_2 = m_3 = m_4 = 4.68481$
  \item[{\bf Point \ref{point6}}] $p_1=\{50.85428,-55.74613,11.69987,94.92591\}$\\
$p_2=\{0.69914,67.19262,-5.78627,91.52776\}$\\
$p_3=\{52.35768,76.32258,43.82222,13.05874\}$\\
$m_1= 54.29650, m_2 = 53.54058, m_3 = 55.96814, m_4 = 51.74438$
  \item[{\bf Point \ref{point7}}] $p_1=\{62.80274,-49.71968,-5.53340,-79.44048\}$\\
$p_2=\{48.59375,-1.65847,34.91140,71.89564\}$\\
$p_3=\{76.75934,-19.14334,-17.10279,30.22959\}$\\
$m_1= m_2 = m_3 = m_4 = 9.82998$
  \item[{\bf Point \ref{point8}}] $p_1=\{98.04093, 77.37405, 30.53434,-81.88155\}$\\
$p_2=\{73.67657,-53.78754,13.69987,14.20439\}$\\
$p_3=\{68.14197,-36.48119,59.89499,-81.79030\}$\\
$m_1= 81.44869, m_2 = 94.39003, m_3 = 57.53145, m_4 = 0.40190$
  \item[{\bf Point \ref{point9}}] $p_1=\{90.15393, -60.44028, -18.19041,42.34210\}$\\
$p_2=\{75.27949,86.12082,19.15087,-95.80345\}$\\
$p_3=\{14.34134,2.00088,87.56698,39.80553\}$\\
$m_1= m_2 = 21.23407, m_3 = m_4 = 81.40164$
  \item[{\bf Point \ref{point10}}] $p_1=\{56.88939, 87.04163,-34.62173,-42.86104\}$\\
$p_2=\{92.86718,-91.88334,59.75945,38.70047\}$\\
$p_3=\{55.98527,-35.20008,9.02722,82.97219\}$\\
$m_1= m_3 = 67.88777, m_2 = m_4 = 40.77317$
  \item[{\bf Figure \ref{fig:boxlms}}] $p_1=\{95.95213, 65.25140,-40.62468,30.93648\}$\\
$p_2=\{68.47023,-60.09584,18.23998,84.29507\}$\\
$p_3=\{12.99839,12.08603,-99.08246,-34.58997\}$\\
$m_1= m_2 = m_3 = m_4 = 11.50163$
  \item[{\bf Figure \ref{fig:pentagonlms}}] $p_1=\{62.72107, -90.39721,46.19119,67.90895\}$\\
$p_2=\{57.27070,84.69403,21.86256,93.85136\}$\\
$p_3=\{97.74567,11.18811,68.32347,-31.80762\}$\\
$p_4=\{51.13803,-98.84980,-16.22424,-84.44255\}$\\
$m_1= m_2 = m_3 = m_4 = m_5 = 5.12753$
\end{enumerate}
\newpage
\begin{enumerate}[resume*]%[label=\bfseries Point \#\arabic*, leftmargin=2.5cm]
  \item[{\bf Point \ref{point11}}] $p_1=\{33.74515, 45.72730,31.15254,-7.47943\}$\\
$p_2=\{31.36435,-41.50734,46.47897,2.04203\}$\\
$p_3=\{4.59005,17.07010,32.65403,41.93628\}$\\
$p_4=\{29.51054,-28.25963,46.17333,-35.08918\}$\\
$m_1= m_2 = m_3 = m_4 = m_5 = 5.01213$
  \item[{\bf Point \ref{point12}}] $p_1=\{33.76482, 45.44063,-10.68084,16.41925\}$\\
$p_2=\{72.93498,67.49170,-11.81485,-36.28455\}$\\
$p_3=\{8.01673,-49.40112,-66.09200,-0.11414\}$\\
$p_4=\{-86.54188,-97.01228,68.12494,32.94875\}$\\
$m_1= 98.42704, m_2 = 28.89059, m_3 = 40.51436$\\
$m_4 = 75.45643, m_5 = 11.08327$
  \item[{\bf Point \ref{point13}}] $p_1=\{1.58374, 6.86200,-15.06805,-10.63574\}$\\
$p_2=\{7.54800,-3.36539,34.57385,27.52676\}$\\
$p_3=\{43.36396,-49.27646,-25.35062,-17.68709\}$\\
$p_4=\{22.58103,38.31530,-14.67581,-3.08209\}$\\
$m_1= m_2 = m_3 = m_4 = m_5 = 2.76340$
  \item[{\bf Point \ref{point14}}] $p_1=\{-93.06712,-36.37997,-27.71460,38.42206\}$\\
$p_2=\{-46.33465,-11.90909,32.33395,46.42742\}$\\
$p_3=\{8.41724,-83.92296,56.21715,34.04937\}$\\
$p_4=\{-15.23696,71.33931,48.68306,-53.67870\}$\\
$m_1= 59.10425, m_2 = 60.25099, m_3 = 76.79109$\\
$m_4 = 65.27606, m_5 = 5.99925$
  \item[{\bf Point \ref{point15}}] $p_1=\{-32.14401,-64.50445,46.04455,-75.56462\}$\\
$p_2=\{-96.90340,-27.60002,-71.50486,86.25541\}$\\
$p_3=\{-37.95135,46.18586,25.67520,-71.38501\}$\\
$p_4=\{-87.67870,66.66463,-36.20151,-27.37362\}$\\
$m_1= m_2 = m_3 = 79.63229, m_4 = m_5 = 51.70237$
  \item[{\bf Point \ref{point16}}] $p_1=\{4.38045,-34.94383,6.34750,-24.71187\}$\\
$p_2=\{37.32326,-11.93504,-29.29012,31.65489\}$\\
$m_1= m_2 = m_3 = 7.27906$
  \item[{\bf Point \ref{point17}}] $p_1=\{-42.91300,62.46030,42.97296,-18.69577\}$\\
$p_2=\{0.56662,67.82929,16.29586,-8.45274\}$\\
$m_1= 85.86724, m_2 = 81.52322, m_3 = 76.12179$
  \item[{\bf Point \ref{point18}}] $p_1=\{31.22641,-11.44458,11.90191,5.51688\}$\\
$p_2=\{39.14661,48.24103,-22.72685,18.32313\}$\\
$m_1= m_2 = m_3 = 6.89867$
\end{enumerate}
\newpage
\begin{enumerate}[resume*]%[label=\bfseries Point \#\arabic*, leftmargin=2.5cm]
  \item[{\bf Point \ref{point19}}] $p_1=\{91.31175,59.18576,-1.99192,-24.84234\}$\\
$p_2=\{71.50353,-58.97396,-27.79064,-4.80365\}$\\
$m_1= 2.21844, m_2 = 36.02116, m_3 = 19.04975$
  \item[{\bf Point \ref{point20}}] $p_1=\{-2.85715,23.32734,73.44393,-19.93952\}$\\
$p_2=\{-78.71545,23.81905,-94.38603,7.22745\}$\\
$p_3=\{75.48933,-0.49095,-37.15293,-43.56893\}$\\
$m_1 = m_2 =  m_3 = m_4 = 99.24288$
  \item[{\bf Point \ref{point21}}] $p_1=\{96.38513,72.23517,25.81581,23.30879\}$\\
$p_2=\{-7.41341,-96.03161,-79.52827,-28.56493\}$\\
$p_3=\{-5.92534,98.48338,97.62395,-72.38643\}$\\
$m_1 = 32.69503, m_2 = 95.90407, m_3 = 53.60699, m_4 = 42.59971$
  \item[{\bf Point \ref{point22}}] $p_1=\{18.47170,12.68290,4.34693,-11.26827\}$\\
$p_2=\{-69.33772,-87.02423,59.44602,4.30660\}$\\
$p_3=\{-91.88519,10.31462,20.61165,-46.63546\}$\\
$m_1 = m_2 =  m_3 = m_4 = 9.94045$
  \item[{\bf Point \ref{point23}}] $p_1=\{-72.75921,-40.60802,-46.57923,73.25257\}$\\
$p_2=\{-92.33823,-18.31403,-5.67422,-86.71053\}$\\
$p_3=\{-76.47950,44.67795,-51.72040,57.95624\}$\\
$m_1 = 44.76393, m_2 = 29.71647, m_3 = 48.95514, m_4 = 72.76312$
  \item[{\bf Point \ref{point24}}] $p_1=\{69.70234,62.68042,25.44429,-97.78603\}$\\
$p_2=\{-65.98494,-85.19920,98.05702,-70.89141\}$\\
$p_3=\{-26.75642,-30.42288,-26.84633,14.81944\}$\\
$p_4=\{-69.44800,56.74842,-32.23649,96.45829\}$\\
$m_1 = m_2 = m_3 = m_4 = m_5 = 87.00572$
  \item[{\bf Point \ref{point25}}] $p_1=\{-45.80756,95.63842,-55.04954,44.01174\}$\\
$p_2=\{36.09562,52.66752,-11.22354,-87.48918\}$\\
$p_3=\{-4.90798,41.11273,14.29379,2.15944\}$\\
$p_4=\{49.48233,40.26756,-23.16581,-96.89362\}$\\
$m_1 = m_2 = m_3 = m_4 = m_5 = 56.97318$
  \item[{\bf Point \ref{point26}}] $p_1=\{-18.90057,-97.14671,44.69176,-16.67528\}$\\
$p_2=\{-70.86315,-81.27489,-3.71628,18.79403\}$\\
$p_3=\{-89.53092,50.02356,33.39784,-51.66031\}$\\
$p_4=\{-96.59097,-34.80215,-83.24353,44.73888\}$\\
$m_1 = m_2 = m_3 = m_4 = m_5 = 43.87459$
  \item[{\bf Point \ref{point27}}] $p_1=\{-88.70322,37.98826,62.19352,-35.86433\}$\\
$p_2=\{-58.60617,-58.60074,-83.75298,61.78210\}$\\
$p_3=\{-83.73607,46.98912,67.44602,78.40612\}$\\
$p_4=\{-96.41508,71.69925,-14.47818,-61.82390\}$\\
$m_1 = m_2 = m_3 = m_4 = m_5 = 16.73899$
  \item[{\bf Point \ref{point28}}] $p_1=\{-3.43584,4.73492,17.31242,61.53467\}$\\
$p_2=\{12.12233,32.23256,87.57836,-58.25073\}$\\
$p_3=\{-38.67209,-54.27020,21.15570,79.15640\}$\\
$p_4=\{-90.90573,-79.70266,-88.26463,-66.00973\}$\\
$p_5=\{-34.40043,-88.73043,84.41781,-4.21221\}$\\
$m_1 = 54.36459, m_2 = 30.96600, m_3 = 51.03652,$\\
$m_4 = 16.03115, m_5 = 2.25657, m_6 = 59.45020$
  \item[{\bf Point \ref{point29}}] $p_1=\{-9.85384,15.70678,80.94234,-84.96387\}$\\
$p_2=\{90.11707,-74.59469,-70.73997,54.32748\}$\\
$p_3=\{-55.84212,-34.47531,-87.20597,-27.73882\}$\\
$p_4=\{16.72808,64.83574,-31.16733,63.94189\}$\\
$p_5=\{-42.62943,49.91058,-46.12974,59.76096\}$\\
$m_1 = 42.61768, m_2 = 22.13590, m_3 = 34.87263,$\\
$m_4 = 54.00634, m_5 = 79.54844, m_6 = 87.50131$
  \item[{\bf Point \ref{point30}}] $p_1=\{35.27512,36.08798,-89.66662,18.22907\}$\\
$p_2=\{-32.58939,14.45447,86.93898,-47.20827\}$\\
$p_3=\{-76.40210,-62.22587,-63.59955,41.03465\}$\\
$p_4=\{-2.30248,0.45058,-76.74256,-64.19292\}$\\
$p_5=\{-88.80252,18.06504,-6.53891,49.34535\}$\\
$m_1 = m_2 = m_3 = m_4 = m_5 = m_6 = 82.87370$
  \item[{\bf Point \ref{point31}}] $p_1=\{-99.20747,-68.16217,95.24772,68.87644\}$\\
$p_2=\{-95.09224,78.51258,-82.38270,20.36899\}$\\
$p_3=\{-56.04092,22.93681,-72.82681,96.81954\}$\\
$p_4=\{78.53840,-86.40143,-82.49674,-57.42855\}$\\
$p_5=\{13.70265,77.87278,99.79126,8.31677\}$\\
$m_1 = 63.23680, m_2 = 86.48449, m_3 = 44.51361,$\\
$m_4 = 79.73599, m_5 = 74.43246, m_6 = 70.11421$
\end{enumerate}
\newpage
\begin{enumerate}[resume*]
  \item[{\bf Point \ref{point32}}] $p_1=\{-70.26380,96.72681,21.66556,-37.40054\}$\\
$p_2=\{-13.45985,2.12040,3.20198,91.44246\}$\\
$p_3=\{-62.59164,-29.93690,-22.16595,-58.38466\}$\\
$p_4=\{-67.60797,-83.23480,18.49429,8.94427\}$\\
$p_5=\{-34.70936,-62.59326,-60.71318,2.77450\}$\\
$m_1 = 94.53242, m_2 = 64.45092, m_3 = 74.74299,$\\
$m_4 = 10.63129, m_5 = 31.77881, m_6 = 23.93819$
\end{enumerate}

\section{Scans}

\begin{enumerate}[resume*]
  \item[{\bf Figure \ref{fig:trithreshold}}] $p_1=\{27.95884,25.55639,-29.88288,-2.17433\}$\\
$p_2=\{27.45521,-7.81292,3.19651,6.05088\}$\\
$6.05088 \leq m_1 = m_2 = m_3 \leq 31.53414$
  \item[{\bf Figure \ref{fig:boxalleq}}] $p_1=\{67.40483,49.44993,-20.67085,48.63654\}$\\
$p_2=\{54.64295,-58.23071,9.55042,-16.59411\}$\\
$p_3=\{41.37620,11.75178,-40.77655,-8.25014\}$\\
$2.33822 \leq m_1 = m_2 = m_3 = m_4 \leq 70.14658$
  \item[{\bf Figure \ref{fig:box2adjm}}] $p_1=\{56.91015,-1.53764,58.47812,-64.49872\}$\\
$p_2=\{74.07761,77.46365,97.18347,-63.75152\}$\\
$p_3=\{32.85182,-13.87036,-68.91291,87.62289\}$\\
$3.29978 \leq m_1 = m_2 \leq 98.99340$\\
$m_3 = m_4 = 22.70133$
  \item[{\bf Figure \ref{fig:boxvart}}] $p_1=\{-44.68568,-76.98829,54.10542,-72.67533\}$\\
$p_2=\{30.54785,-3.09363,-50.03807,88.12743\}$\\
$p_{3,min}=\{71.46086,28.68398,47.82601,-93.53563\}$\\
$p_{3,max}=\{3265.93916,860.51940,1434.78028,-2806.06893\}$\\
$m_1 = 68.14502, m_2 = 85.10360, m_3 = 75.43127, m_4 = 50.79371$
  \item[{\bf Figure \ref{fig:pentalleq}}] $p_1=\{86.88039,29.81650,28.58637,88.04167\}$\\
$p_2=\{79.07341,18.41517,-37.32140,-80.98985\}$\\
$p_3=\{48.59046,-32.90460,-14.18821,85.89247\}$\\
$p_4=\{26.61931,-21.93050,21.22092,26.23208\}$\\
$3.16786 \leq m_1 = m_2 = m_3 = m_4 = m_5 \leq 95.03580$
\end{enumerate}
\newpage
\begin{enumerate}[resume*]
  \item[{\bf Figure \ref{fig:pentalldiff}}] $p_1=\{-97.64085,75.52538,15.99594,35.63304\}$\\
$p_2=\{-99.78509,-71.86820,-55.31804,-58.19254\}$\\
$p_3=\{-87.92245,-85.76444,19.12306,-59.11598\}$\\
$p_4=\{-87.70477,83.72569,-0.57040,35.79677\}$\\
$1.39548 \leq m_1 \leq 41.86441,$\\
$m_2 = 35.09895, m_3 = 74.42515, m_4 = 15.47168, m_5 = 41.24757$
  \item[{\bf Figure \ref{fig:pentachu}}] $p_{1,min}=\{-15.22437,-26.74156,6.65483,29.13661\}$\\
$p_{1,max}=\{-83.38733,-146.46961,36.45001,159.58778\}$\\
$p_2=\{-91.22611,-63.97875,55.07507,-52.90153\}$\\
$p_3=\{0.95105,75.90791,-10.13814,-88.40860\}$\\
$p_4=\{43.04908,77.11321,-50.69469,-7.60198\}$\\
$m_1 = 49.12560, m_2 = 57.87487, m_3 = 26.47098,$\\
$m_4 = 0.42094, m_5 = 62.31320$
  \item[{\bf Figure \ref{fig:trinum}}] $p_{1}=\{49.84064,5.20440,40.73789,-27.57262\}$\\
$p_{2}=\{61.25771,32.42640,-14.71631,-3.67014\}$\\
$1.92598 \leq m_1 = m_2 = m_3 \leq 57.77950$
  \item[{\bf Figure \ref{fig:boxnum}}] $p_{1}=\{-64.20450,59.44455,7.63891,57.71454\}$\\
$p_{2}=\{1.91777,-70.66999,4.99626,48.86490\}$\\
$p_{3,min}=\{-95.66288,-30.14530,14.13910,-68.75167\}$\\
$p_{3,max}=\{-2292.42604,-904.35922,424.17308,-2062.55008\}$\\
$m_1 = 67.86064, m_2 = 88.88038, m_3 = 74.26926, m_4 = 34.85441$
  \item[{\bf Figure \ref{fig:pentanum}}] $p_{1,min}=\{-51.76504,-81.75539,-46.42422,-40.15540\}$\\
$p_{1,max}=\{-283.52879,-447.79274,-254.27593,-219.94019\}$\\
$p_{2}=\{-63.76533,-2.53015,16.27485,69.16770\}$\\
$p_{3}=\{-78.50262,46.32052,13.19246,-54.00166\}$\\
$p_{4}=\{25.40582,81.48058,39.11105,93.24648\}$\\
$m_1 = 78.45208, m_2 = 42.71315, m_3 = 91.94256,$\\
$m_4 = 61.59730, m_5 = 16.75672$
\end{enumerate}

\addtocontents{toc}{\vspace{2em}} % Add a gap in the Contents, for aesthetics

\backmatter

%----------------------------------------------------------------------------------------
%	BIBLIOGRAPHY
%----------------------------------------------------------------------------------------

\label{Bibliography}

%\lhead{\emph{Bibliography}} % Change the page header to say "Bibliography"

\bibliographystyle{kp} % Use the "unsrtnat" BibTeX style for formatting the Bibliography
%unsrtnat
\bibliography{Bibliography} % The references (bibliography) information are stored in the file named "Bibliography.bib"

\end{document}